\documentclass[12pt]{report}
\usepackage{fsuthesis}
\usepackage{amsmath}
\usepackage{amssymb}
\usepackage{color}
\usepackage[nice]{nicefrac}
\usepackage[dvips]{graphicx} 
\usepackage{subfigure}
\DeclareSymbolFont{AMSb}{U}{msb}{m}{n}
\DeclareMathSymbol{\N}{\mathbin}{AMSb}{"4E}
\DeclareMathSymbol{\Z}{\mathbin}{AMSb}{"5A}
\DeclareMathSymbol{\R}{\mathbin}{AMSb}{"52}
\DeclareMathSymbol{\Q}{\mathbin}{AMSb}{"51}
\DeclareMathSymbol{\I}{\mathbin}{AMSb}{"49}
\newcommand{\WZbb}{W/Z\ b{\bar b}}
\newcommand{\Wbb}{W b{\bar b}}
\newcommand{\Zbb}{Z b{\bar b}}
\newcommand{\as}{\alpha_s}
\newcommand{\sss}{\scriptscriptstyle}
\newcommand{\Lg}{{\cal L}}
\newcommand{\Dcal}{{\cal D}}

\newcommand{\Slash}[1]{#1\negthickspace\!/}
\newcommand{\Zvert}[1]{\gamma^#1(g_{\sss V}^f+g_{\sss A}^f\gamma_5)}
\newcommand{\Zvertf}[2]{\gamma^#1(g_{\sss V}^#2+g_{\sss A}^#2\gamma_5)}
\newcommand{\Wvert}[1]{\gamma^#1(1-\gamma_5)}
\newcommand{\Ic}{{\cal I}}

\title{Next-to-Leading-Order Corrections to Weak Boson Production with
	a Massive Quark Jet Pair at Hadron Colliders}
\author{Fernando Febres Cordero}
\uauthor{FERNANDO FEBRES CORDERO}
\thesistype{Dissertation}
\department{Department of Physics}
\approvalneededtrue
\headofdept{Mark Riley}
\deptheadtitle{Chair}
\deanofschool{Joseph Travis}
\degree{Doctor of Philosophy}
\gradyear{2007}
\dept{Physics}
\majorprof{Laura Reina}
\majorprofdegree{Ph.D.}
\commmembera{Howard Baer}
\commmemberb{Harrison Prosper}
\commmemberc{Jorge Piekarewicz}
\outcommmember{Paolo Aluffi}

\semester{Fall}
\college{Arts and Sciences}
\defensedate{September 26, 2007}

\begin{document}
%%  front.tex is the file that contains your abstract and
%%  acknowledgements and dedication information -- the ``front
%%  matter'' 
\dedication{
\begin{flushright}
{\it a Kiko y Tefina...\hspace{3.0cm}}
\end{flushright}
}

\def\acknowledgementtext{\hskip\parindent
First I would like to thank all the members of the Florida State University Department of Physics
for giving me the opportunity to come to Tallahassee, where I have continued to learn and enjoy the
many dimensions of Science. 

I would like to give special thanks to the High Energy Physics Group, Faculty, Postdocs, Students
and Staff.  Throughout this long journey they have created a stimulating environment for learning
and discussing many subjects of the exciting field of particle physics. 

I would like to thank Doreen Wackeroth, with whom I have collaborated in the development of a good
part of the research presented in this dissertation.

I lack words to say how grateful I am to my thesis adviser, Laura Reina, so, at first order, I say
{\it thank you!} From the very beginning till the very end she has been extremely supportive, always
encouraging me to explore new areas with an acute and clear scientific mind.

I cannot finish without thanking my family: They have instilled in me the values of work,
persistence and righteousness, which have been keen for me to endure the sharp ups and downs found
along the way. {\it Much\'{\i}simas gracias, esta tesis va dedicada a ustedes...}
\\
\null\hfill --- Fernando. }

\def\abstracttext{
  \hskip\parindent
We present the calculation of Next-to-Leading-Order Quantum Chromo Dynamics corrections for the
production of a $W$ or $Z$ weak boson associated with a bottom anti-bottom quark pair at hadron
colliders ($p\bar p$, $pp \to\WZbb$), including the effects of a non-zero bottom-quark mass. We find
a considerable reduction of the renormalization and factorization scale dependence of our results
with respect to Leading-Order calculations.  In particular, we study the impact of the corrections
on the total cross section and invariant mass distributions of the bottom anti-bottom quark pair at
the Fermilab Tevatron $p\overline p$ collider. We perform a detailed comparison with a calculation
that considers massless bottom quarks and find significant deviations in regions of phase space with
small invariant mass of the bottom anti-bottom quark pair.

Our results will be relevant to ongoing and future searches at hadron colliders, as the $W/Z\
b\overline b$ production mode is the main background to important signals, such as light Standard
Model Higgs boson production or single top-quark production.
}

\dedicationtrue     % yes I've got one 
\frontmatterformat  % now use the layout specific to this material
%%%     2/27/03 
%%%  The Graduate Office only needs the DAI abstract page in 
%%%  hard copy  .... BUT ... needs two copies of it  There is 
%%%  another file called  hardcopy.tex which prints out the 
%%%  necessary paper pages that are required in addition to 
%%%  the electronic file  Comment out the next two lines when 
%%%  making your electronic copy.
%\makefsuabstract
%\endfsuabstractpage
\titlepage
\maketitle

%%  The signature page is only needed in hard copy - comment
%%  out the line when making your electronic copy.  Only the
%%  committee page is contained within the electronic file.
%\makesignaturepage
%\makecommitteepage

\dedicationpage

%Just use the next line if you have acknowledgements.  The command
%itself will do the work for you ;-)  If none, comment it out. 
%\acknowsection

\setcounter{secnumdepth}{4}
\setcounter{tocdepth}{3}
\tableofcontents

\listoftables
\listoffigures

\newpage
\doublespace
   \chapter*{ABSTRACT}
    \addtocontents{toc}{\protect\vspace{2ex}}
  \addcontentsline{toc}{chapter}{Abstract}  % lowercased 1/26/99
\medskip

\abstracttext  % text pulled from  front.tex 

\medskip

\newpage

\maintext    % Now switch to dissertation layout 

%%  I recommend including chapters one at a time, commenting out all
%%  unnecessary chapters until such time as you are ready to run the
%%  whole thing.  Cuts down om wasted paper, and time  

\chapter{INTRODUCTION}
\label{chap:intro}

The {\it Standard Model} (SM) of particle physics is today the best mathematical framework to
understand the dynamics of all elementary particles that have been observed in high-energy
collisions.  This quantum field theory has been amazingly successful in describing and predicting to
an unprecedented level of precision observations made in all high energy particle accelerator
experiments.

Extraordinary phenomenological efforts in the past few decades, both from the experimental side
producing impressive amounts of precision data, and from the theoretical side producing powerful
techniques to handle challenging calculations, show the SM to be a robust model of fundamental
interactions. So much so that, if we consider only collider data, the SM fits all observables, with
only a few showing statistical deviations of up to $2-3\ \sigma$~\footnote{See for example the
LEPEWWG website at http://lepewwg.web.cern.ch/LEPEWWG and the reviews on SM related topics of the
PDG~\cite{Yao:2006px}}. 

At the same time, however, we know that the SM is an incomplete theory, as it does not include
gravitational interactions. Moreover, from cosmological data we know that the SM falls short of
explaining the origin of Dark Matter and Dark Energy, and does not predict as large an asymmetry
between matter and antimatter as observed in the universe.  Another puzzle is the mechanism that
breaks the electroweak $SU(2)_W\times U(1)_Y$ symmetry. In the SM this breaking occurs as a doublet
of complex scalar fields acquires a non-zero vacuum expectation value (VEV), thus producing what is
called spontaneous symmetry breaking. This mechanism gives rise to effective mass terms for weak
force carriers, quarks, and leptons and leaves a physical scalar particle, the Higgs particle, which
has so far eluded observation\footnote{A more detailed description of this mechanism and a brief
introduction to the SM is given in Appendix~\ref{app:SMint}}.  Among others, these are the reasons
why nowadays there are many models that posit physics beyond the SM, which often embed different
mechanisms of electroweak symmetry breaking (EWSB).

The current and future hadron colliders, i.e. the Tevatron, a $p\bar p$ collider currently taking
data at $1.96$ TeV center of mass energy at Fermilab, near Chicago, and the Large Hadron Collider
(LHC), a $pp$ collider with $14$ TeV center of mass energy that will start in 2008 at CERN, Geneva,
have as a main goal the elucidation of the mechanism of EWSB as well as the exploration of the
energy spectrum beyond the weak scale, where physics beyond the SM (BSM) is expected.  The processes
studied in this thesis, i.e. the hadronic production of a weak force carrier with a
bottom-antibottom quark pair ($p\bar p, pp\to\WZbb$), play a crucial role in some of the current
studies of EWSB and BSM. They represent an important QCD background in the searches for a light
SM-like Higgs boson ($H$) and for single top-quark production.

\subsubsection*{The Necessity of Higher Order Corrections in perturbative QCD}

In order to improve our understanding of the behavior of fundamental particles at high energies,
theorists are faced with the necessity to calculate signal and, often, background processes with
high precision. This last task becomes essential when the signal to background ratio is small and
the background cannot be easily extracted from data.  Typically, this is the case for processes that
involve a large number of kinematic variables and that have broad kinematic distributions, as often
arises when final states consist of several jets and/or missing energy.

At hadron colliders, QCD effects are particularly important and must be taken into account to obtain
precise theoretical predictions. Since at high energies QCD is a perturbative quantum field theory
(pQFT), QCD effects at collider energies can be calculated order by order in the strong coupling
constant. The lowest order at which a process can be calculated, the Leading Order (LO), typically
has a large theoretical uncertainty associated with it. This is mainly due to the opening of new
production channels at higher orders of the perturbative series and to the large dependence of LO
calculations on renormalization and factorization scales, in certain renormalization prescriptions.
Adding the first order QCD corrections, that is, Next-to-Leading-Order (NLO) corrections, usually
improves the stability of theoretical predictions considerably and tests the behavior of the
perturbative expansion. Occasionally, when the NLO corrections are unusually large, the reliability
of the predictions can be improved by computing Next-to-Next-to-Leading Order (NNLO) QCD
corrections. In all known examples this is enough to reduce the theoretical uncertainty to an
acceptable level.

Today the standard for hard scattering cross section calculations is NLO, and since the early
nineties a wide set of processes have been studied at this level in perturbation theory (for an
up-to-date review of some of them and a look at state of the art techniques that have been developed
see Ref.~\cite{Bern:2007dw}).

However, while several programs exist that allow automated calculations of partonic differential
cross sections at LO (e.g. Madgraph~\cite{Murayama:1992gi,Stelzer:1994ta,Maltoni:2002qb},
CompHEP~\cite{Pukhov:1999gg}, AMEGIC++~\cite{Krauss:2001iv}), there are as yet no algorithms able to
deal with all processes at NLO in a completely automatic way. There is a bottleneck that occurs with
the calculation of virtual one-loop diagrams with many external partons or ``legs".  Basically,
following a traditional Feynman diagram approach, the complexity of the analytic expressions grows
exponentially with the number of legs and the number of massive internal/external particles. This is
mainly due to the number of Feynman diagrams, the increased number of kinematic variables and the
increased complexity of the tensor integrals appearing in each Feynman diagram.

It has been observed that certain amplitudes (for example the so called Maximally Helicity Violating
(MHV) amplitudes) show a surprising analytic simplicity, hidden by the cumbersome intermediate steps
of standard calculations.  A good part of the progress in the field is due to a better understanding
of such amplitudes (the literature on MHV amplitudes is now extensive, but for a brief review see
Refs.~\cite{Parke:1986gb,Berends:1987me,Dixon:1996wi,Witten:2003nn}).  Many new techniques have
appeared that exploit general properties of gauge field theories such as gauge invariance,
factorization, unitarity and the existence of representations in terms of Feynman
integrals~\cite{Dixon:1996wi,Bern:1996je,Bern:2007dw,Berger:2006uc}. An example of their success is
the recently completed one-loop calculation of the set of all helicity amplitudes with six external
gluons~\cite{Bern:1994zx,Bern:1994cg,Bidder:2004tx,Bidder:2005ri,Britto:2005ha,Britto:2006sj,
Berger:2006ci,Berger:2006vq,Xiao:2006vt}.  We have used generalized unitarity, specifically
quadruple cuts~\cite{Britto:2004nc}, as a means to make non-trivial cross checks of coefficients of
scalar box integrals of sets of Feynman diagrams. This represents the first direct application of
this techniques to a phenomenologically relevant computation including massive internal and external
particles.

In this dissertation we present the calculation of NLO QCD corrections to the production of a $W$ or
$Z$ weak gauge boson in association with a bottom-antibottom quark pair at hadron colliders ($p\bar
p,pp\to\WZbb$), including full bottom-quark mass effects. The main difficulty we encounter is the
calculation of virtual one-loop diagrams with up to five legs that include the full effects of
massive bottom-quarks.  The latter increases the complexity of the calculation due to the addition
of an extra kinematic invariant. We follow a traditional Feynman diagram approach to produce fully
analytical expressions for one-loop amplitudes, which will allow non-trivial cross checks with
on-shell recursion techniques, beyond the box coefficients checks presented in this work. 

\boldmath
\subsubsection*{$\WZbb$ \ Production at Hadron Colliders}
\unboldmath

The associated production of a $W/Z$ boson with a $b\bar b$ pair is by itself an interesting
signature for hadron colliders, since it can be precisely studied experimentally.  Sophisticated
techniques exist to detect weak bosons at hadron colliders, especially when they decay into leptons.
Even in the messy environment of a high energy hadron collider, Drell-Yan processes ($W,Z$ or
$\gamma$ production) are often used as tools for detector calibration and luminosity measurements.
Furthermore, $b$-quarks are very useful tools too, as the mesons in which they fragment have a
life-time long enough to allow tagging. Nowadays the efficiency of this tagging is close to $50\%$
for a large range of transverse momenta of the $b$-jets.  These sorts of studies, in the context of
$\WZbb$ production, allow non-trivial cross checks of experimental techniques, and provide further
constraints on the SM, as long as the theoretical uncertainty on such signals is equal or smaller
than the experimental precision.

\begin{figure}[tp]
  \begin{center}
    \subfigure[Associated $VH$ production]{\label{fig:Vbb_bg-a}\includegraphics[scale=0.3]{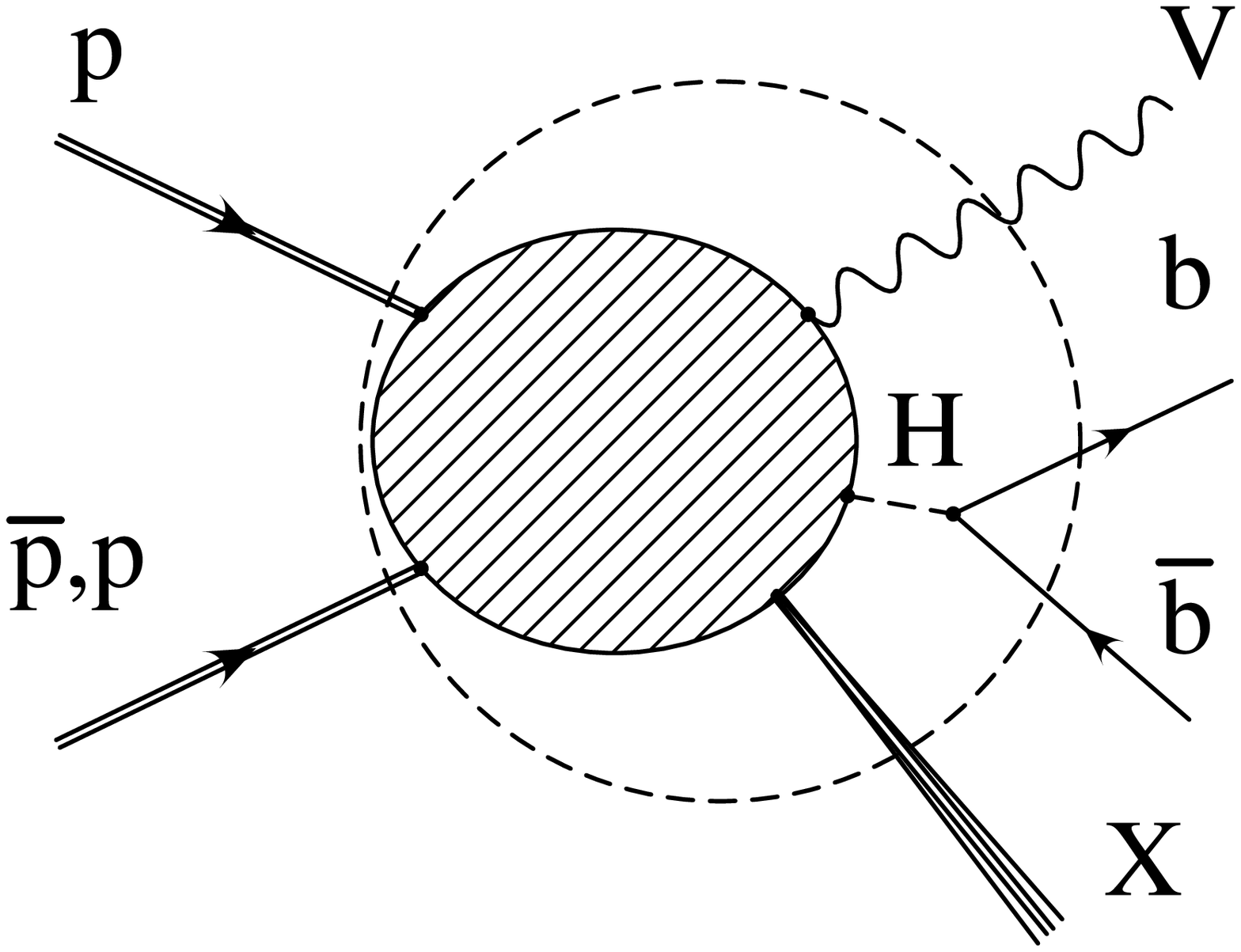}}\hspace{2.0cm}
    \subfigure[Single top-quark production]{\label{fig:Vbb_bg-b}\includegraphics[scale=0.3]{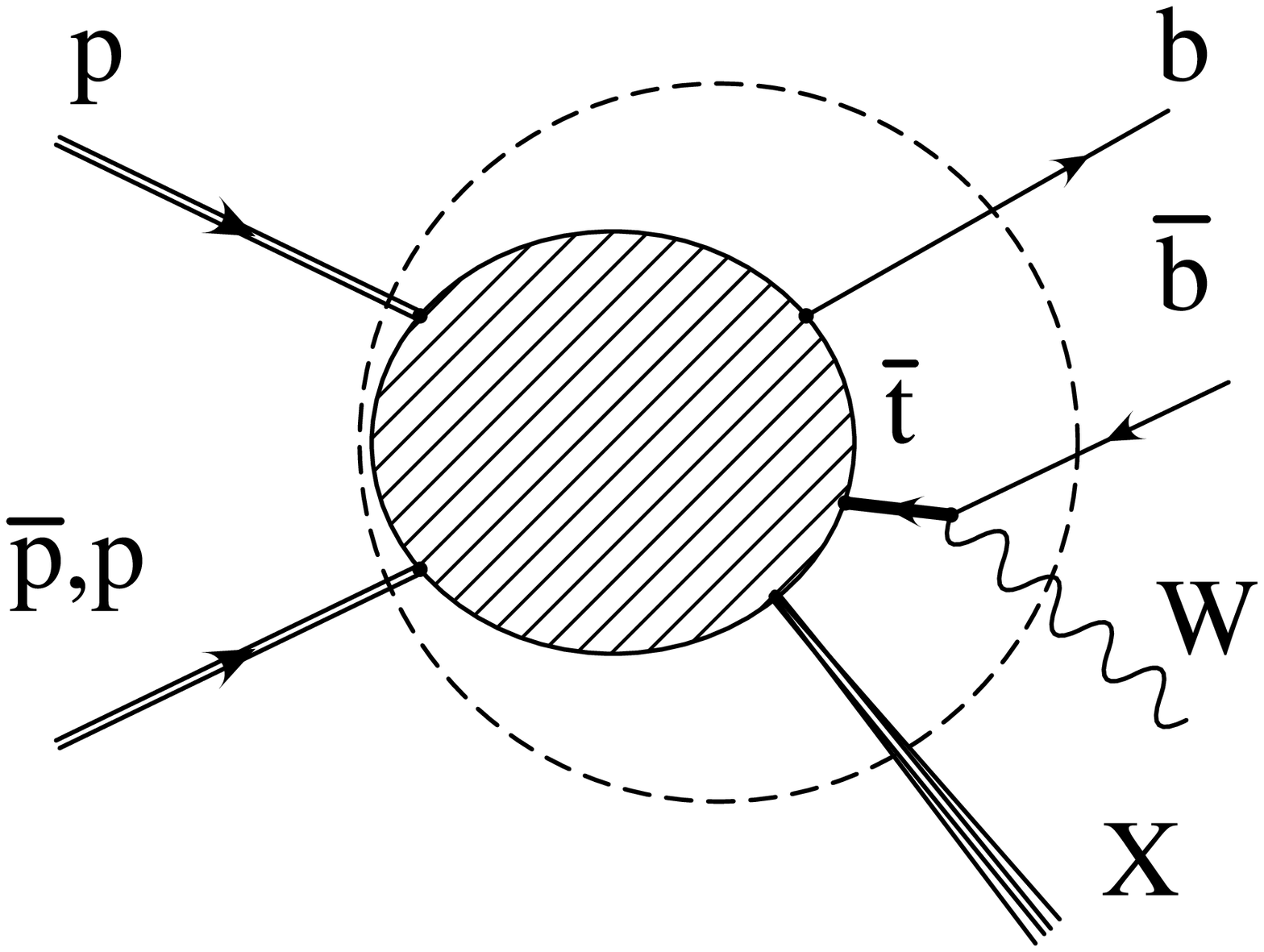}} \\
  \end{center}
  \caption{Example of processes to which $Vb\bar b$ ($V=W$ or $Z$) is a main background.}
  \label{fig:Vbb_bg}
\end{figure}
The interest in $\WZbb$ production at hadron colliders is increased by the characteristic features
of a light SM Higgs boson $H$. One of the strengths of the SM Higgs sector is that it is highly
predictive as only one of its parameters is not yet constrained by direct observation, the Higgs
boson mass $m_H$.  Electroweak precision measurements hint at the existence of a light Higgs boson,
with mass below $144$ GeV at 95\% confidence level~\footnote{For an update see the LEPEWWG website
at http://lepewwg.web.cern.ch/LEPEWWG}. For such a Higgs boson one of the main production channels
is the $W/Z\ H$ associated production, with the Higgs boson decaying most of the time into a $b\bar
b$ pair, as depicted in Figure~\ref{fig:Vbb_bg-a}, for which $\WZbb$ represents a major background. 

The production of a Higgs boson in association with an electroweak gauge boson, $p\bar p \to HV$
($V\!=\!Z,W$) with $H\to b\bar{b}$, is indeed the most sensitive production channel of a SM Higgs
boson at the Tevatron for a Higgs boson lighter than about
140~GeV~\cite{Abulencia:2005ep,CDFnote:2006Wbb,Abazov:2004jy,D0note:2006Wbb,Patwa:2006rd}.  The
Tevatron with an integrated luminosity of $6 \; {\rm fb}^{-1}$ will be able to exclude a Higgs boson
with $115 \; {\rm GeV}<M_H<180$ GeV at 95\% confidence level~\cite{sonnenschein}, and will provide
crucial guidance for the search strategy at the LHC.  

On the other hand, due to the huge hadronic activity, the experiments at the CERN LHC will probably
first look for a light SM-like Higgs boson in the $H\to\gamma\gamma$ decay channel. In spite of the
small branching ratio, $H\to\gamma\gamma$ produces a clear peak in the invariant mass distribution
of the pair of photons.  In order to fully identify a potential Higgs boson candidate, however, the
LHC experiments will have to measure $H\to b\bar b$ and this will have to be done when the Higgs
boson is produced via associated production, either $W/Z H$ or $t\bar tH$.  

Finally, $\Wbb$ is among the most relevant irreducible background processes for single-top quark
production~\cite{Acosta:2004bs,Abazov:2005zz,Gresele:2006se}, as illustrated in
Figure~\ref{fig:Vbb_bg-b}, both at the Tevatron and at the LHC. This is particularly relevant to the
Tevatron, where, via single-top production, the $Wbt$ vertex is being measured for the first
time~\cite{GarciaBellido:2007ui,Abazov:2006gd}.

The cross section for $p\bar{p}\rightarrow HV$ has been calculated including up to NNLO QCD
corrections~\cite{Han:1991ia,Mrenna:1997wp,Brein:2003wg} and $O(\alpha)$ electroweak
corrections~\cite{Ciccolini:2003jy}, while single-top production has been calculated at NLO in
QCD~\cite{Stelzer:1997ns,Stelzer:1998ni,Smith:1996ij,Harris:2002md,Sullivan:2004ie,
Cao:2004ky,Cao:2004ap,Cao:2005pq,Sullivan:2005ar}, and at one-loop of electroweak (SM and MSSM)
corrections~\cite{Beccaria:2006ir}. Thus, to fully exploit the Tevatron's and LHC's potentials to
detect the SM Higgs boson and to impose limits on its mass, as well as to test the third generation
quark coupling to the $W$ boson, it is crucial that the dominant background processes are also
precisely calculated. 

In the present experimental analyses\footnote{For updated results, see the CDF and $D0$ websites at
http://www-cdf.fnal.gov/physics/exotic/exotic.html and
http://www-d0.fnal.gov/Run2Physics/WWW/results/higgs.htm.}, the effects of NLO QCD corrections on
the total cross section and the dijet invariant mass distribution of the $W/Z\ b\bar{b}$ background
process have been taken into account by using the MCFM package~\cite{MCFM:2004}. In MCFM, the NLO
QCD predictions of both total and differential cross sections for the $p\bar{p}(p)\rightarrow W/Z\
b\bar{b}$ production processes have been calculated in the zero bottom-quark mass ($m_b=0$)
approximation~\cite{Ellis:1998fv,Campbell:2000bg,Campbell:2002tg}, using the analytical results
of~\cite{Bern:1997ka,Bern:1997sc}.  From a study of the Leading Order (LO) cross section, finite
bottom-quark mass effects are expected to affect both the total and differential $W/Z\ b\bar{b}$
cross sections mostly in the region of small $b\bar{b}$-pair invariant
masses~\cite{Campbell:2002tg}. Given the variety of experimental analyses involved both in the
search for $HW/Z$ associated production and single-top production, it is important to assess
precisely the impact of a finite bottom-quark mass over the entire kinematical reach of the process,
including complete NLO QCD corrections.

Using the MCFM package~\cite{MCFM:2004}, we compared our results with the corresponding results
obtained in the $m_b=0$ limit.  Numerical results are presented for the total cross section and the
invariant mass distribution of the $b\bar{b}$ jet pair ($m_{b\bar b}$), at the Tevatron $p\bar{p}$
collider, including kinematic cuts and a jet-finding algorithm. In particular, we apply the $k_T$
jet algorithm and require two tagged $b$-jets in the final state.

We have found that the NLO QCD corrections reduce considerably the theoretical uncertainty in the
total hadronic cross section for $\WZbb$ production. We have found that NLO corrections are
significant.  The shape of the $m_{b\bar b}$ invariant mass distribution is changed by the NLO QCD
corrections; that is, the effect does not amount to a simple NLO/LO rescaling factor ($K$-factor).
Finally, we have found that mass effects affect the total cross section by about 8\% to 10\% at NLO.
The influence is greatest in the small $m_{b\bar b}$ invariant mass region. 

\[
\cdots
\]

This dissertation is organized as follows. In Chapter~\ref{chap:calculation} we present all the
details of the full NLO calculation, including a thorough description of the techniques used to
obtain both virtual and real corrections, first for $Wb\bar b$ production and then for $Zb\bar b$
production.  We have included in the Appendices~\ref{app:SMint}-\ref{app:PSint} a few complementary
reviews and several collections of technical details, omitted for brevity and aesthetic reasons
from the body of this dissertation.  In Chapter~\ref{chap:results} we present  the results of our
calculation.  We show the LO and NLO dependence on renormalization and factorization scales, we
study the impact of the corrections on the total cross sections and invariant mass distributions of
the bottom quark-antiquark pair.  We show the $m_b$ effects by comparing to a calculation that
considers massless bottom-quarks.  In Chapter~\ref{chap:conclusion} we conclude with a summary of
the main results and discuss future studies of the $\WZbb$ production mode, as well as the natural
generalization of our calculation to other processes.

\chapter{NLO Calculation for $W/Z\ b\bar b$ Production at Hadron Colliders}
\label{chap:calculation}

In this Chapter we present the details of the NLO QCD calculation of $\Wbb$ and $\Zbb$ hadronic
production including full $b$-quark mass effects.

The NLO QCD calculation of the $\WZbb$ production cross section for massless $b$ quarks has been
available in the literature for quite some time~\cite{Ellis:1998fv,Campbell:2000bg,Campbell:2002tg}.
It was performed by using the analytic expression of the scattering amplitudes for a weak gauge
boson to four (massless) partons~\cite{Bern:1997ka,Bern:1997sc}, and simulating $b$-quark mass
effects by imposing the kinematic conditions:
\begin{equation}
(p_b+p_{\bar b})^2  >  4Q^2,\qquad
p^T_b  >  Q\; ,\qquad
p^T_{\bar b}  >  Q\; , 
\label{eq:masslesscond}
\end{equation}
where $Q$ is a scale of the order of the $b$-quark mass, $p_{b({\bar b})}$ is the momentum of the
$b({\bar b})$ quark and $p^T_{b({\bar b})}$ represents the $b({\bar b})$-quark transverse momentum.

We have improved on the massless calculation by considering a fully massive $b$ quark both at the
level of the scattering amplitude and in the integration over the final state phase space.  We keep
the weak bosons as on-shell particles, though the extension to include their leptonic decays does
not present in principle any special complications. The rest of this Chapter is organized as
follows. In Section~\ref{sec:FactTh} we present the main theoretical framework that allows the
calculation of hadronic cross sections. We introduce briefly the parton model and the QCD
factorization theorems.  In Section~\ref{sec:HadXsecW} we present the main characteristics of the
cross sections for $\WZbb$ hadronic production. Sections~\ref{sec:Wbbcalc} and \ref{sec:Zbbcalc}
present the core of the calculation for $W b\bar b$ and $Z b\bar b$ production at hadron colliders
respectively.  They are organized into subsections which discuss the LO and the NLO virtual and real
corrections of the calculation.  We show explicitly the cancellation of UV singularities, by
renormalization, and of IR singularities, by matching virtual and real corrections and by
consistently absorbing all long-distant physics into the renormalized parton distribution functions
(PDFs) (see Section~\ref{sec:FactTh}).

Due to the complexity of this calculation, all results have been cross checked using at least two
independent sets of codes.  The analytic calculation of the scattering amplitudes has been
implemented using, at different stages, FORM~\cite{Vermaseren:2000nd}, TRACER~\cite{Jamin:1991dp},
{\it Maple} and {\it Mathematica}. Final numerical results have been obtained with codes built in
$C$ and FORTRAN, and we have used the FF package~\cite{vanOldenborgh:1990yc} and
Madgraph~\cite{Murayama:1992gi,Stelzer:1994ta,Maltoni:2002qb} to cross check pieces of our code.
Some of the figures in this Chapter have been produced using AXODRAW~\cite{Vermaseren:1994je},
FeynDiagram~\cite{FeynDiag:2003} and Grace~\cite{Grace:2007}.

\section{Factorization Theorem for Hadronic Cross Sections}\label{sec:FactTh}
The {\it parton model} of hadron structure~\cite{Feynman:1969ej,Feynman:1973xc} paved the way to the
full formulation of QCD, as it created a framework to connect data from high energy hadronic
collisions with the {\it quark model} according to which quarks are the elementary constituents of
hadrons.  The parton model basically states that when hadrons interact via a high momentum transfer,
they appear as built of point-like, quasi-free constituents, the {\it partons} which we now know as
the quarks and gluons of QCD. To each parton $i$ in a given hadron $A$ is associated a parton
density function (PDF), ${\cal F}_i^A(x)$, which describes the probability of finding parton $i$ in
hadron $A$ with a fraction $x$ of the total momentum of the hadron (for more details on the
development of the model see for example Ref.~\cite{Feynman:1973xc,Peskin:1995ev,Weinberg:1996kr}).

Despite its success, the parton model remained without a firm theoretical basis until the 80's when
a series of QCD factorization theorems rigorously proved that most hadronic observables can be
calculated in QCD by disentangling the non-perturbative properties of hadrons and the asymptotically
free behavior of partons into well defined building blocks\footnote{Based on general physics
principles these theorems have been extended to a wide variety of processes, although full proofs
only exist for a few processes~\cite{Collins:1989gx}. Recently it has been
shown~\cite{Collins:2007nk,Collins:2007jp} that certain processes in colliders with unpolarized
beams can break factorization at Next-to-Next-to-Next-to-Leading Order (NNNLO), an order still far
from being tested.} (see for example Ref.~\cite{Collins:1989gx,Sterman:1995fz,Soper:1996fy}).  As a
result, a given observable ${\cal O}_A(Q)$ involving an initial hadronic state $A$ and a momentum
transfer $Q$ considerably larger than $\Lambda_{\rm QCD}$ can be decomposed as: 
\begin{equation}
{\cal O}_A(Q) = \sum_i f_A^i(Q,\mu_f)\otimes {\cal O}_i(Q,\mu_f)\; ,
\label{eq:Ofact}
\end{equation}
where the sum is over a full set of partons $i$, $\mu_f$ is the so called factorization scale and
the operation $\otimes$ is an integral convolution. The ${\cal O}_i(Q,\mu_f)$ are point-like
operators that describe the short-distance parton interactions which, as the momentum transfer is
large, can be calculated perturbatively. The functions $f_A^i(Q,\mu_f)$ contain all the
long-distance physics information related to hadron $A$.  They are universal (i.e. process
independent) non-perturbative objects whose $Q$-evolution can be determined perturbatively starting
from a data-driven non-perturbative core.

The factorization scale $\mu_f$ is introduced to define a boundary between the perturbative and
non-perturbative regimes. Although the LHS of Eq.~(\ref{eq:Ofact}) is in principle independent of
$\mu_f$, this is only true when considering all orders in the perturbative expansion. In practice,
the perturbative series is always truncated and that leaves a spurious $\mu_f$ dependence in the
calculated ${\cal O}_A(Q)$, which can be argued that asymptotically is of the order of the next
order in the perturbative expansion. That is why when a fixed-order calculation is performed in
perturbative QCD, the spurious dependence on the factorization scale (and similarly on the
renormalization scale) is used as an indicator of the theoretical uncertainty associated with the
calculation. For the reader interested in expanding on this subject, see for example the review
papers in Ref.~\cite{Sterman:1995fz,Soper:1996fy,Tung:2001cv} and references therein.
 
\boldmath
\section{Hadronic Cross Section for $\WZbb$ Production: General Structure at NLO}\label{sec:HadXsecW}
\unboldmath

Enforcing the factorization properties of QCD cross sections, we can write the NLO QCD total or
differential cross section for $p\bar p(pp)\to \WZbb$ as:
\begin{eqnarray}
\sigma(p{\bar p}(pp)\to W/Z\ b{\bar b}) & = &
\sum_{ij}\frac{1}{1+\delta_{ij}} \int dx_1dx_2 \nonumber\\
& &  \qquad \left[ {\cal F}_i^p(x_1,\mu) {\cal F}_j^{{\bar p}(p)}(x_2,\mu) 
{\hat\sigma}_{ij}(x_1,x_2,\mu) + (x_1\leftrightarrow x_2) \right],
\label{eq:HaddXsec}
\end{eqnarray}
where ${\cal F}_i^{p({\bar p})}$ are the PDFs for parton $i$ in a proton (antiproton). The sum runs
over all relevant subprocesses contributing to the hadronic differential cross section initiated by
partons $i$ and $j$.  The partonic cross section for the subprocess $ij\to W/Z\ b{\bar b}$ is
denoted by ${\hat\sigma}_{ij}$.  The hadronic process $p\bar p(pp)\to \Wbb$ receives contributions
from the initial state $q\bar q^\prime$ at LO (where $q$ and $\bar q^\prime$ represent quarks of
up-type and down-type respectively), and from $q\bar q^\prime$, $qg$ and $\bar qg$ at NLO, while
$p\bar p(pp)\to \Zbb$ receives contributions from $q\bar q$ and $gg$ at LO and from $q\bar q$, $gg$,
$qg$ and $\bar q g$ at NLO.  The scale $\mu$ corresponds to both the factorization scale ($\mu_f$)
and renormalization scale ($\mu_r$).  The factor in front of the integral is a symmetry factor that
accounts for the presence of identical particles in the initial state of a given subprocess
($\delta_{ij}$ is the Kronecker delta).  The partonic center-of-mass energy squared, $s$, is given
in terms of the hadronic center of mass energy square, $s_H$, by $s=x_1x_2s_H$.

We then write the NLO partonic cross section as follows:
\begin{eqnarray}
{\hat\sigma}_{ij}^{\rm NLO}(x_1,x_2,\mu) & = & \alpha_s^2(\mu)\left\{ 
			f_{ij}^{\rm LO}(x_1,x_2)+\frac{\alpha_s(\mu)}{4\pi} 
				f_{ij}^{\rm NLO}(x_1,x_2,\mu)\right\}\nonumber\\
		& \equiv & {\hat\sigma}_{ij}^{\rm LO}(x_1,x_2,\mu)
				+\delta {\hat\sigma}_{ij}^{\rm NLO}(x_1,x_2,\mu)\; ,
\label{eq:nloXsec}
\end{eqnarray}
where $\alpha_s(\mu)$ is the strong coupling constant evaluated at the scale $\mu$, ${\hat
\sigma}_{ij}^{\rm LO}(x_1,x_2,\mu)$ is the ${\cal O}(\alpha_s^2)$ LO partonic cross section and
$\delta {\hat\sigma}_{ij}^{\rm NLO}(x_1,x_2,\mu)$ contains the ${\cal O}(\alpha_s)$ corrections to
the partonic LO cross section. The $\delta {\hat\sigma}_{ij}^{\rm NLO}(x_1,x_2,\mu)$ corrections can
be decomposed in the following way:
\begin{eqnarray}
\delta {\hat\sigma}_{ij}^{\rm NLO} & = & \int d\left( PS_3\right) {\overline \sum}|
	{\cal A}_{\rm virt}(ij\to\WZbb)|^2 + \int d\left( PS_4\right) {\overline \sum}|
	{\cal A}_{\rm real}(ij\to\WZbb +k)|^2 \nonumber\\
&\equiv & {\hat\sigma}_{ij}^{\rm virt} +{\hat\sigma}_{ij}^{\rm real}\;,
\label{eq:virtrealXsec}
\end{eqnarray}
where the term integrated over the phase space measure $d\left(PS_3\right)$ corresponds to the
virtual one-loop corrections (three final-state particles), while the one integrated over the phase
space measure $d\left(PS_4\right)$ corresponds to the real tree level corrections with one
additional emitted parton (four final-state particles). The sum $\overline \sum$ indicates that the
corresponding amplitudes squared, $|{\cal A}_{\rm virt(real)}(ij\to\WZbb(+k))|^2$, have been
averaged over the initial-state degrees of freedom and summed over the final-state ones. The final
phase space integration have been performed using Monte Carlo techniques using the adaptive
multi-dimensional integration routine VEGAS~\cite{Lepage:1977sw}.

Intermediate  stages of the calculation of $\delta{\hat \sigma}^{\rm NLO}_{ij}$ contain both
ultraviolet (UV) and infrared (IR) divergences, whose calculation will be discussed in detail in
Sections~\ref{subsec:sigvirt}, \ref{subsec:sigreal}, \ref{sec:Wbbcalc} (for $\Wbb$ production) and
\ref{sec:Zbbcalc} (for $\Zbb$ production).

\subsection{Renormalization Scale Dependence of the NLO Cross Section}\label{subsec:murdepNLO}

We observe that the scale dependence of the total cross section at NLO is dictated by
renormalization group arguments. In order to assure the renormalization scale independence of the
total cross section at ${\cal O}(\as^3)$, $f_{ij}^{\rm NLO}(x_1,x_2,\mu)$ has to be of the form:
\begin{equation}
f_{ij}^{\rm NLO}(x_1,x_2,\mu) = f_{ij}^1(x_1,x_2)
		+{\tilde f}_{ij}^1(x_1,x_2)\ln\left(\frac{\mu^2}{s}\right),
\label{eq:muNLOdep}
\end{equation}
where, taking into account that other sources of renormalization scale dependence in
Eq.~(\ref{eq:nloXsec}) are $\as(\mu)$ and the PDFs ${\cal F}_i^{p,\bar p}(x,\mu)$, we can prove
that:
\begin{eqnarray}
{\tilde f}_{ij}^1(x_1,x_2) & = & 2\left\{4\pi b_0f_{ij}^{\rm LO}(x_1,x_2)
		-\sum_k\left[ \int_\rho^1dz_1P_{ik}(z_1)f_{kj}^{\rm LO}(x_1z_1,x_2)\right.\right.\nonumber\\
& & \left.\left.+ \int_\rho^1dz_2P_{jk}(z_2)f_{ik}^{\rm LO}(x_1,x_2z_2) \right]\right\},
\label{eq:fmuNLOdep}
\end{eqnarray}
where $\rho=(2m_b+M_V)^2/s$ ($V=W,\ Z$), $P_{ij}$ are the Altarelli-Parisi splitting functions as
presented in Eqs.~(\ref{eq:splitfuncqq}), (\ref{eq:splitfuncgq}), (\ref{eq:splitfuncgg}) and
(\ref{eq:splitfuncqg}), and $b_0$ is determined by the one-loop renormalization group evolution of
the strong coupling constant $\alpha_s$:
\begin{equation}
\frac{d\alpha_s(\mu)}{d\ln(\mu^2)}=-b_0\alpha_s^2+{\cal O}(\alpha_s^3),\qquad 
	b_0=\frac{1}{4\pi}\left(\frac{11}{3}N-\frac{2}{3}n_{lf} \right),
\label{eq:alphasRun}
\end{equation} 
where $N=3$ is the number of colors and $n_{lf}=5$ the number of light flavors. To write
Eq.~(\ref{eq:fmuNLOdep}) we have assumed that the $b$-quark mass does not run, given the mild
dependence of the cross section on it. The origin of the rest of terms in Eq.~(\ref{eq:fmuNLOdep})
will become clear in Sections~\ref{sec:Wbbcalc} and \ref{sec:Zbbcalc} where we describe in detail
the calculation of virtual, real and total cross section corrections for both $\Wbb$ and $\Zbb$
production respectively.

\boldmath
\subsection{Calculating the Virtual Cross Section $\hat\sigma^{\rm virt}$}\label{subsec:sigvirt}
\unboldmath
The ${\cal O}(\as)$ virtual corrections to the hadronic $\WZbb$ production tree level processes
consist of self-energy, vertex, box and pentagon diagrams. The contributions to $\hat\sigma^{\rm
virt}$ in Eq.~(\ref{eq:virtrealXsec}) can be written as:
\begin{equation}
{\overline \sum}|{\cal A}_{\rm virt}(ij\to \WZbb)|^2 
	= \sum_D{\overline \sum}\left( {\cal A}_0{\cal A}_D^\dag + {\cal A}_0^\dag{\cal A}_D\right)
	= \sum_D{\overline \sum}2{\cal R}e\left( {\cal A}_0{\cal A}_D^\dag \right)\;,
\label{eq:virtXsec}
\end{equation}
where ${\cal A}_0$ is the tree level amplitude and ${\cal A}_D$ denotes the amplitude for the
one-loop diagram $D$, with $D$ running over all self-energy, vertex, box and pentagon diagrams
corresponding to the $ij$-initiated subprocess.

The amplitude for each virtual diagram (${\cal A}_D$) is calculated as a linear combination of Dirac
structures with coefficients that depend on both tensor and scalar one-loop Feynman integrals with
up to five denominators. We solve the one-loop integrals in the coefficients either at the level of
the amplitude or at the level of the amplitude squared (see Eq.~(\ref{eq:virtXsec})). These two
independent approaches allow us to thoroughly cross check the calculation of each individual
diagram. Indeed, the tensor structures present in the one-loop integrals of the amplitude are
typically different from the ones present in the amplitude squared, as one can perform non-trivial
reductions of the latter by canceling dot products of the integration momentum in the numerator with
denominators in the Feynman integrals. In this way, the final analytical expression of a given
diagram ends up being represented in terms of different building blocks. A possible incorrect
relation between the building blocks would then naturally produce a discrepancy between the two
approaches.

Tensor and scalar one-loop integrals are treated as follows. Using the Passarino-Veltman (PV)
method~\cite{Passarino:1978jh,Denner:1993kt}, the tensor integrals are expressed as a linear
combination of tensor structures and coefficients, where the tensor structures depend on the
external momenta and the metric tensor, while the coefficients depend on scalar integrals,
kinematics invariants and the dimension of the integral (for a more detailed description of the
technique see Appendix~\ref{app:Intred}). Numerical stability issues may arise at this level as a
consequence of the proportionality of the tensor integral coefficients to powers of inverse Gram
Determinants (GDs), specially when considering a full set of independent momenta $\{p_{a_i}\}$
($i=1,\dots,4$) of the $ij\to\WZbb$ phase space, which is defined by ${\rm GD}=\det{(p_{a_i}\cdot
p_{a_k})}$.  The problem becomes more serious for higher rank tensor integrals, since the higher the
rank of the original tensor integral, the higher the inverse power of the GD that appears in the
coefficients of its tensor decomposition.

To illustrate the problem, we parametrize the GD appearing in pentagon tensor integrals in terms of
the $\WZbb$ phase space variables as
\begin{eqnarray}
\label{eq:gram_det}
{\rm GD}&=&-\frac{[s-(2 m_b+M_V)^2]}{64}
[M_V^4+(s-\bar s_{b\bar b})^2-2M_V^2(s+\bar s_{b\bar b})]\, s \, 
\bar s_{b\bar b}
\sin^2\theta_{b \bar b} \sin^2\phi_{b \bar b} \sin^2\theta\,\,\,, 
\nonumber \\
\end{eqnarray}
where $s\!=\!x_1x_2s_{\sss H}$ is the partonic center-of-mass energy squared, $M_V$ is the mass of
the weak vector boson ($V=W,Z$), and the $\WZbb$ phase space has been expressed in terms of a
time-like invariant $\bar s_{b\bar b}\!=\!(p_b+p_{\bar b})^2$, polar angles ($\theta, \theta_{b\bar
b}$) and azimuthal angles ($\phi, \phi_{b\bar b}$) in the center-of-mass frames of the incoming
partons and of the $b\bar b$ pair, respectively.  As can be seen in Eq.~(\ref{eq:gram_det}), the GD
vanishes when the set of momenta become degenerate or co-planar, for example at the boundaries of
phase space.  Near these regions of phase space it can become arbitrarily small, giving rise to
spurious divergences which cause serious numerical difficulties, since large cancellations then
appear between various parts of the calculation at the numerical level.  The probability that the
Monte Carlo integration hits a point close to these regions of phase space is not negligible and
these points cannot just be discarded.

The numerical instabilities we just discussed can be considered as ``spurious" or ``unphysical"
divergences, since it is well known that only two-particle invariants can give rise to a physical
singularity. Indeed, these spurious divergences cancel when large sets of diagrams are
combined~\cite{Bern:1997sc}, such as, for example, when one combines gauge invariant sets of color
amplitudes (i.e. amplitudes with a common color factor). Explicit analytic cancellations have been
found for example when using helicity amplitudes and the helicity product formalism (see for example
Refs.~\cite{Bern:1997ka,Bern:1997sc}), mainly because certain GD can be decomposed in terms of
helicity products. As we have expressed our calculation in terms of kinematical invariants, the full
cancellation only occurs between numerator and denominator at the numerical level, often between
fairly large expressions.  Nevertheless, when we consider gauge invariant sets of color amplitudes
(as the ones presented in Tables~\ref{tb:A1color}-\ref{tb:A5color}) and full analytical reductions
of all tensor integrals, we find cancellation of some powers of GD, which improves the behavior of
the numerical code so that we can integrate, using Monte Carlo techniques, over the entire phase
space, to obtain statistical errors from the numerical integration below $0.1\%$.

The fully reduced numerical amplitudes are often more demanding computationally, and because of that
we have built numerical codes that use them only when close to regions of phase space where certain
GD is small. With this the computer needs are reduced. All this was found particularly useful when
considering $E$-PV functions (see Appendix~\ref{app:Intred}), and probably it would break down if
one were to extend this technique to processes with even more legs, where probably using other
techniques would be necessary.

\begin{figure}[ht]
\begin{center}
\includegraphics[scale=1.0]{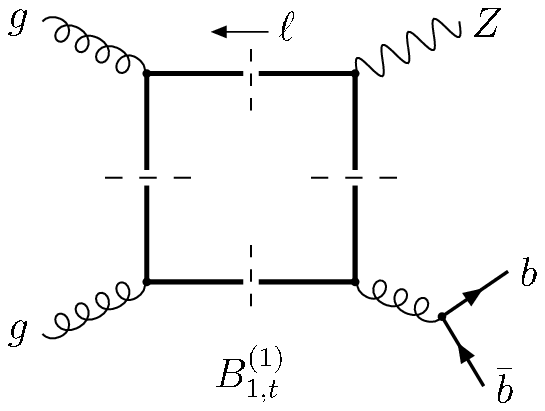}
\caption[Quadruple cut check of the calculation 
of a box diagram involving a top-quark loop.]{Quadruple cut~\cite{Britto:2004nc} check of the calculation 
of a box diagram involving a top-quark loop. It corresponds to
  two Feynman diagrams ($B_{1,t}^{(1)}$ in Fig.~\ref{fig:boxggZbb}) given by the two possible orientations of the
  fermion line.}
\label{fig:b1ttop}
\end{center}
\end{figure}

We also checked parts of our result by using unitarity techniques~\cite{Bern:1997sc}, specifically
the quadruple-cut technique~\cite{Britto:2004nc}. As shown by Britto, Cachazo and Feng (BCF), from
any set of Feynman diagrams (or more generally from any tensor integral~\cite{Ossola:2006us}) one
can extract the coefficient of a given scalar box integral by cutting the four corresponding
propagators (see Fig.~\ref{fig:b1ttop}), i.e. by replacing $i/(\ell^2-m^2+i\epsilon)\rightarrow
2\pi\delta^{(+)}(\ell^2-m^2)$ for each cutted propagator of momentum $\ell$ and mass $m$. This effectively
freezes the momentum integration, and replaces it by a set of algebraic equations which determine
the loop momentum entirely.  We solved this set of equations by using a BCF
ansatz~\cite{Britto:2004nc}, and then compared the result to the corresponding box coefficient
extracted from our analytic expression, and found agreement (for more details and specific solutions
for the topology in Fig.~\ref{fig:b1ttop} see Appendix~\ref{app:quadcuts}). This is a rather
non-trivial check for the set of $E$-PV and $D$-PV functions (see Appendix~\ref{app:Intred}) we have
employed at different stages, since they all contribute to the coefficients of the scalar
$D$-functions occurring in the one-loop $\WZbb$ amplitudes. For instance, it has been particularly
useful in the case of box diagrams like the one shown in Fig.~\ref{fig:b1ttop}, since this diagram
and related ones contain up to $D4$-PV functions that cannot be reduced even at the level of the
amplitude squared. Since they involve up to four powers of inverse GDs, they are particularly
subject to numerical instabilities and it is important to have their analytic expressions as compact
as possible.

After the tensor integral reduction is performed, the fundamental building blocks are one-loop
scalar integrals with up to five denominators. They may be finite or contain both ultraviolet (UV)
and infrared (IR) divergences. 

The UV singularities of the virtual cross section are regularized in $d\!=\!4-2 \epsilon_{\sss UV}$
dimensions and renormalized by introducing a suitable set of counterterms, while the residual
renormalization scale dependence is checked from first principles using renormalization group
arguments as in Eq.~(\ref{eq:muNLOdep}).  The IR singularities of the virtual cross section are
extracted in $d=4-2\epsilon_{\rm IR}$ dimensions and are canceled by analogous singularities in the
${\cal O}(\as^3)$ real cross section.

In our calculation we treat $\gamma_5$ according to the naive dimensional regularization approach,
i.e.  we enforce the fact that $\gamma_5$ anticommutes with all other $\gamma$ matrices in
$d=4-2\epsilon$ dimensions. This is known to give rise to inconsistencies when, at the same time,
the $d$-dimensional trace of four $\gamma$ matrices and one $\gamma_5$ is forced to be non-zero (as
in $d=4$, where
$Tr(\gamma^\mu\gamma^\nu\gamma^\rho\gamma^\sigma\gamma_5)=4i\epsilon^{\mu\nu\rho\sigma}$)
~\cite{Larin:1993tq}.  In our calculation, both UV and IR divergences are handled in such a way that
we never have to enforce simultaneously these two properties of the Dirac algebra in $d$ dimensions.
For instance, the UV divergences are extracted and canceled at the amplitude level, after which the
$d\rightarrow 4$ limit is taken and the renormalized amplitude is squared using $d=4$. Thus, all
fermion traces appearing at this point are computed in four dimensions and therefore have no
ambiguities.

The finite scalar integrals are evaluated using the method described in Ref.~\cite{Denner:1993kt}
and cross checked with the numerical package FF~\cite{vanOldenborgh:1990yc}. The singular scalar
integrals are calculated analytically by using dimensional regularization in $d\!=\!4-2\epsilon$
dimensions. The most difficult integrals arise from IR divergent pentagon diagrams with several
external and internal massive particles. We calculate them as linear combinations of box integrals
using the method of Ref.~\cite{Bern:1992em,Bern:1993kr} and of Ref.~\cite{Denner:1993kt}.  Details
of the box scalar integrals, including two that we calculated explicitly since they were not
previously in the literature, and of the pentagon reduction used in this calculation are given in
Appendix~\ref{app:Scint}, as well as the set of UV- and IR-divergent three and two point functions. 

We note that the tree level amplitude ${\cal A}_0$ in Eq.~(\ref{eq:virtXsec}) has generically to be
considered as a $d$-dimensional tree level amplitude. This matters when the ${\cal A}_D$ amplitudes
in Eq.~(\ref{eq:virtXsec}) are UV or IR divergent. Actually, as will be shown in the following, both
UV and IR divergences are always proportional to pieces of the tree level amplitudes and they can be
formally canceled without having to explicitly specify the dimensionality of the tree level
amplitude itself.  After UV and IR singularities have been canceled, everything is calculated in
$d=4$ dimensions.

\boldmath
\subsection{Calculating the Real Cross Section $\hat\sigma^{\rm real}$}\label{subsec:sigreal}
\unboldmath

The NLO real cross section $\hat\sigma^{\rm real}$ in Eq.~(\ref{eq:virtrealXsec}) corresponds to the
${\cal O}(\alpha_s)$ corrections to $ij\to \WZbb$ due to the emission of an additional real extra
parton, i.e. to the process $ij\to \WZbb+k$. It contains IR singularities which cancel the analogous
singularities present in the ${\cal O}(\alpha_s)$ virtual corrections and in the NLO PDFs.  These
singularities can be either \emph{soft}, when the emitted extra parton is a gluon and its energy
becomes very small, or \emph{collinear}, when the final state parton is emitted collinear to one of
particles in the initial state. There is no collinear singularity from the final $b$ and $\bar{b}$
quarks, because the $b$-quark mass regularize the collinear divergence. 

These IR singularities can be conveniently isolated by \emph{slicing} the phase space of the final
state particles into different regions defined by suitable cutoffs, a method which goes under the
general name of {\it Phase Space Slicing} (PSS).  The dependence on the arbitrary cutoffs introduced
in \emph{slicing} the final state phase space is not physical, and cancels at the level of the total
real hadronic cross section, i.e. in $\sigma^{\rm real}$, as well as at the level of the real cross
section for each separate subprocess.  This cancellation constitutes an important check of the
calculation and will be discussed in detail in Sections~\ref{subsec:totalWbb} and
\ref{subsec:totalZbb}.

We have calculated the cross section for the processes
\[
i(q_1)+j(q_2)\to b(p_b)+\bar{b}(p_{\bar b})+V(p_{\sss V})+h(k)\,\,\,,
\]
with $q_1+q_2=p_b+p_{\bar b}+p_{\sss V}+k$, using the \emph{two-cutoff} PSS method, which includes
cutoffs of the soft and collinear kind.  This implementation of the PSS method was originally
developed to study QCD corrections to dihadron production~\cite{Bergmann:1989zy} and has since then
been applied to a variety of processes. (A nice review about it can be found in
Ref.~\cite{Harris:2001sx}, to which we refer for more extensive references and details.)

In the following, we briefly review the structure of the real calculation using the two-cutoff PSS.
We mention that the soft and collinear kernels employed throughout our calculation have been used
also in the NLO calculation of $t\bar tH$ production at hadron colliders, where results have been
checked using a PSS method with one-cutoff and a dipole cancellation
method~\cite{Reina:2001bc,Dawson:2003zu,Beenakker:2001rj,Beenakker:2002nc}.  Although the processes
we are considering are different, the kinematics are equivalent, and the color structure and IR
behavior are the same, so necessarily their soft and collinear kernels are the same.

\subsubsection*{Phase Space Slicing method with two cutoffs}
%\label{subsec:two_cutoff}

The general implementation of the PSS method using two cutoffs proceeds in two steps. First, to
isolate the soft singularities of a final state extra gluon we introduce an arbitrarily small
\emph{soft} cutoff $\delta_s$ and we separate the overall integration over the phase space of that
gluon into two regions according to whether the energy of the final state gluon ($k^0\!=\!E_g$) is
\emph{soft}, i.e.  $E_g\le\delta_s\sqrt{s}/2$, or \emph{hard}, i.e.  $E_g>\delta_s\sqrt{s}/2$. The
partonic real cross section of Eq.~(\ref{eq:virtrealXsec}) can then be written as:
\begin{equation}
\label{eq:sigma_real_two_cutoff}
\hat{\sigma}^{\rm real} =
\hat{\sigma}^{soft}+\hat{\sigma}^{hard}\,\,\,,
\end{equation}
where $\hat{\sigma}^{soft}$ is obtained by integrating over the \emph{soft} region of the gluon
phase space, and contains all the IR soft divergences of $\hat\sigma^{\rm real}$.  To isolate the
remaining collinear divergences from $\hat{\sigma}^{hard}$, we further split the integration over
the phase space of any final state parton according to whether the parton is
($\hat{\sigma}^{hard/coll}$) or is not ($\hat{\sigma}^{hard/non-coll}$) emitted within an angle
$\theta$ from the initial state partons such that $(1-\cos\theta)<\delta_c$, for an arbitrary small
\emph{collinear} cutoff $\delta_c$:
\begin{equation}
\label{eq:sigma_gg_hard}
\hat{\sigma}^{hard}=\hat{\sigma}^{hard/coll}+
\hat{\sigma}^{hard/non-coll}\,\,\,.
\end{equation}
The hard non-collinear part of the real cross section, $\hat{\sigma}^{hard/non-coll}$, is finite and
can be computed numerically, using standard Monte Carlo techniques.

On the other hand, in the soft and collinear regions the integration over the phase space of the
emitted gluon or quark can be performed analytically, thus allowing us to isolate and extract the IR
divergences of $\hat{\sigma}^{\rm real}$.  More details on the calculation of $\hat{\sigma}^{soft}$
and $\hat\sigma^{hard}$ for each relevant subprocess will be given in Sections~\ref{subsec:realWbb}
and \ref{subsec:realZbb} for the hadronic production of $\Wbb$ and $\Zbb$ respectively.

The cross sections describing soft, collinear and IR-finite radiation depend on the two arbitrary
parameters $\delta_s$ and $\delta_c$.  However, in the total real hadronic cross section
$\sigma^{\rm real}$, after mass factorization, the dependence on these arbitrary cutoffs vanishes
for sufficiently small values of the cutoffs.

\begin{figure}[htb]
\begin{center}
\includegraphics[scale=0.65]{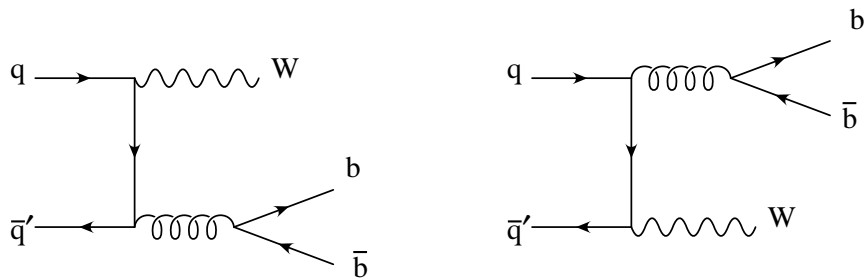}
\caption{Tree level Feynman diagrams for $q\bar{q}^\prime\rightarrow Wb\bar{b}$.}
\label{fig:Wbbtree_level}
\end{center}
\end{figure}

\section[NLO QCD Corrections to $W b\bar b$ Production at Hadron Colliders]{Calculation of NLO QCD Corrections
to $W b\bar b$ Production at Hadron Colliders}\label{sec:Wbbcalc}

In this Section we present in detail the calculation of the partonic total or differential cross
section ${\hat\sigma}^{\rm NLO}(ij\to Wb\bar b)$~\cite{FebresCordero:2006sj}, as defined in
Eqs.~(\ref{eq:nloXsec}) and (\ref{eq:virtrealXsec}).
 
There is only one subprocess contributing to $\Wbb$ at LO, $q{\bar q^\prime}\to \Wbb$, where $q$ and
$\bar q^\prime$ represent quarks or antiquarks of up-type and down-type, respectively.  We neglect
contributions from third generation initial quarks as they are suppressed by either the initial
state quark densities (PDFs) or by the corresponding Cabibbo-Kobayashi-Maskawa (CKM) matrix
elements. 

The contributing tree level Feynman diagrams are shown in Figure~\ref{fig:Wbbtree_level}. Given the
assignment of momenta
\[
q(q_1)\bar q^\prime(q_2)\to b(p_b)+\bar{b}(p_{\bar b})+W(p_{\sss W})\,\,\,,
\]
the LO amplitude can be written as
\begin{eqnarray}
{\cal A}_0(q\bar q^\prime\to\Wbb) & = & ig_s^2\frac{g_{\sss W}}{2\sqrt{2}}V_{q\bar q^\prime}\;\; \epsilon_\mu^*(p_{\sss
W})\;\; \frac{g_{\nu\rho}}{(p_b+p_{\bar b})^2}\;\; \bar{u}_b \gamma^\rho v_{\bar{b}}\;\;t_{ij}^at_{kl}^a \nonumber\\
	& & \Bigl[\bar v_{\bar q^\prime}\Wvert{\mu}\frac{-\Slash{q}_2+\Slash{p}_{\sss W}}{(-q_2+p_{\sss
	W})^2}\gamma^\nu u_q \nonumber\\
	& & +\bar v_{\bar q^\prime}\gamma^\nu\frac{\Slash{q}_1-\Slash{p}_{\sss W}}{(q_1-p_{\sss
	W})^2}\Wvert{\mu} u_q \Bigl]\; ,
\label{eq:LOWbb}
\end{eqnarray}
where $g_s$ and $g_{\sss W}$ are the strong and weak coupling constants, respectively,
$t^{a}=\lambda^{a}/2$ are given in terms of the Gell-Mann matrices $\lambda^{a}$ and $V_{q\bar
q^\prime}$ are the entries of the CKM mixing matrix.  (For more details see
Appendix~\ref{app:LOamp}.)

The partonic LO cross section is obtained by integrating $|{\cal A}_0|^2$ over the $\Wbb$ final
state phase space:
\begin{equation}
{\hat\sigma}^{\rm LO}  =  \int d\left( PS_3\right) {\overline \sum}|{\cal A}_0|^2\; ,
\label{eq:LOXsec}
\end{equation}
where the sums indicates average over initial and sum over final spins and colors of the fermion
lines, as well as sum over polarizations of the vector boson.  As we are considering an on-shell
gauge boson, we have summed over its polarizations according to the prescription used for massive
vector bosons, i.e.:
\begin{equation}
\sum \epsilon_\mu(k)\epsilon^*_\nu(k)=-g_{\mu\nu}+\frac{k_{\mu}k_{
\nu}}{M_{\sss V}^2}\; ,
\label{eq:polsum}
\end{equation}
where $M_{\sss V}$ is the mass of the $V$ weak boson ($V=W,Z$). For $\Wbb$ production the second
term in Eq.~(\ref{eq:polsum}) does not contribute, because in our calculation the $W$ boson only
couples to the initial massless fermion line.  The full expression has to be considered, however,
when calculating $\Zbb$ production.

At NLO one has to consider three processes: $q\bar q^\prime\to\Wbb$, which contributes both at
${\cal O}(\as^2)$ and at ${\cal O}(\as^3)$ through the one-loop ${\cal O}(\as)$ virtual corrections,
and the real ${\cal O}(\as)$ corrections, due to $q\bar q^\prime\to\Wbb+g$ and $q(\bar
q)g\to\Wbb+q^\prime(\bar q^\prime)$ which contribute at ${\cal O}(\as^3)$. In the following sections
we will discuss in detail the structure of both virtual and real ${\cal O}(\as)$ corrections.

\begin{figure}[htb]
\begin{center}
\includegraphics{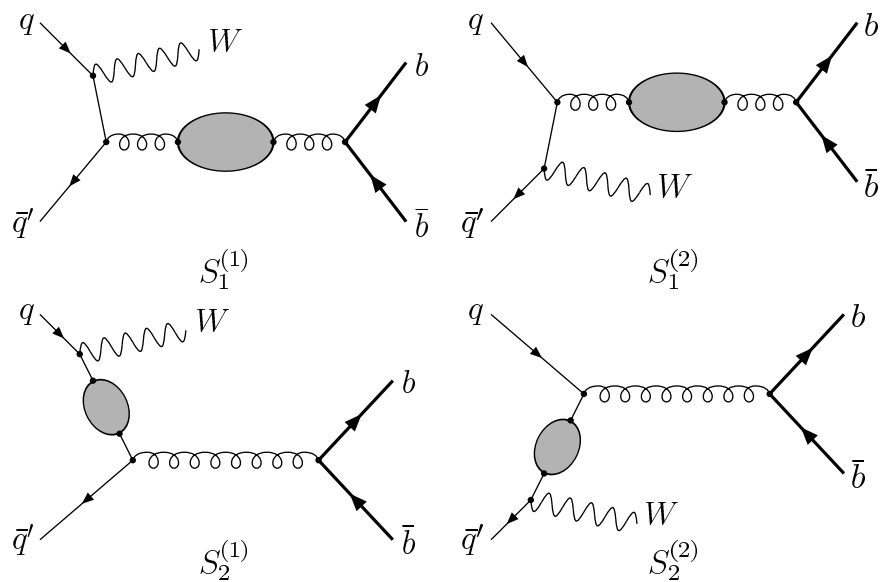}
\caption{Gluon ($S_1^{(1,2)}$) and quark ($S_2^{(1,2)}$) ${\cal O}(\as)$ self-energy corrections contributing to 
	the $q{\bar q^\prime}\to Wb\bar b$ subprocess at NLO. The shaded blobs denote standard one-loop QCD
	corrections to the gluon and quark propagators, respectively.}
\label{fig:selfWbb}
\end{center}
\end{figure}

\begin{figure}[htb]
\begin{center}
\includegraphics{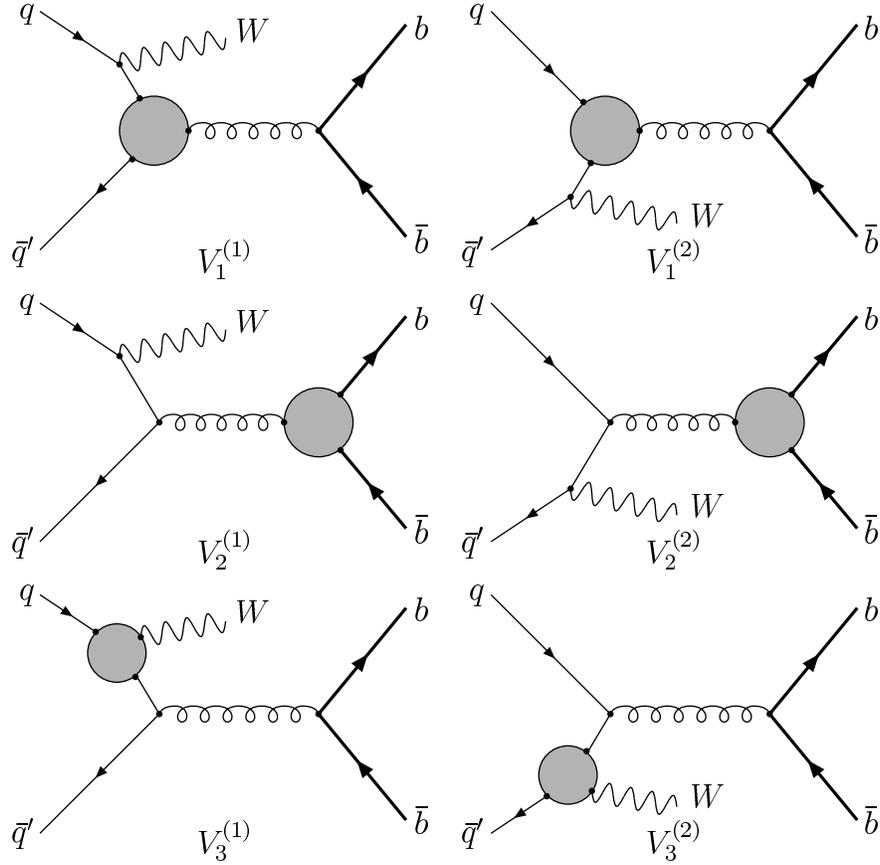}
\caption{${\cal O}(\as)$ vertex corrections contributing to the subprocess $q{\bar q^\prime}\to Wb\bar b$ at NLO. The shaded
	blobs denote standard one-loop QCD corrections to the $q\bar q g$, $b\bar b g$ and $q\bar q^\prime W$
	vertices, respectively.}
\label{fig:vertWbb}
\end{center}
\end{figure}
\subsection{Virtual Corrections to $q{\bar q^\prime}\to\Wbb$}\label{subsec:virtWbb}
\begin{figure}[htb]
\begin{center}
\includegraphics{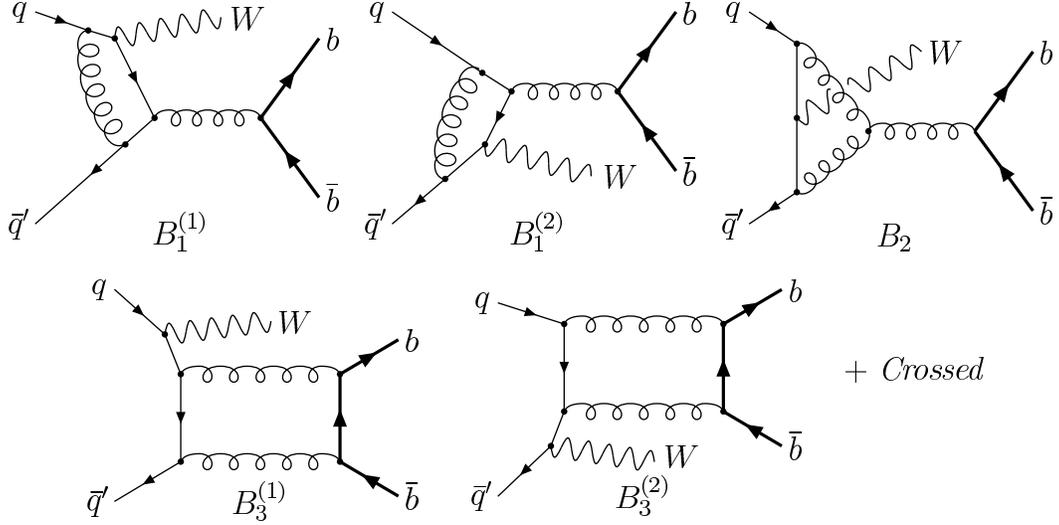}
\caption{${\cal O}(\as)$ box diagram corrections contributing to the $q{\bar q^\prime}\to Wb\bar b$ process at NLO. The
	term {\it Crossed} refers to the box diagrams, $B_{3c}^{(1,2)}$ obtained from the $B_3^{(1,2)}$ boxes by flipping the
	$b$-quark fermion line.}
\label{fig:boxWbb}
\end{center}
\end{figure}

\begin{figure}[htb]
\begin{center}
\includegraphics{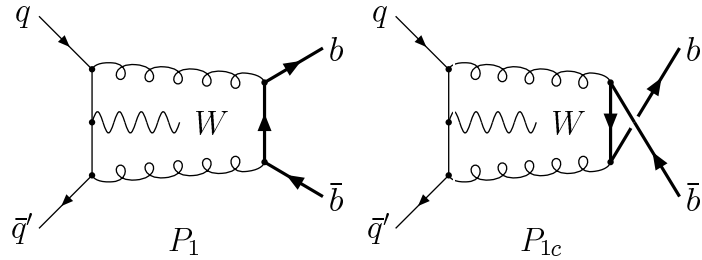}
\caption{${\cal O}(\as)$ pentagon diagram corrections contributing to the $q{\bar q^\prime}\to Wb\bar b$
	process at NLO.}
\label{fig:pentWbb}
\end{center}
\end{figure}

The ${\cal O}(\as)$ virtual corrections to the $q{\bar q^\prime}\to \Wbb$ tree level process consist
of the self-energy, vertex, box and pentagon one-loop diagrams illustrated in
Figures~\ref{fig:selfWbb}, \ref{fig:vertWbb}, \ref{fig:boxWbb} and \ref{fig:pentWbb}. The
contributions to the virtual amplitude squared of Eq.~(\ref{eq:virtrealXsec}) can then be written
as:
\begin{equation}
{\overline \sum}|{\cal A}_{\rm virt}(q{\bar q^\prime}\to \Wbb)|^2 
	= \sum_D{\overline \sum}\left( {\cal A}_0{\cal A}_D^\dag + {\cal A}_0^\dag{\cal A}_D\right)
	= \sum_D{\overline \sum}2{\cal R}e\left( {\cal A}_0{\cal A}_D^\dag \right),
\label{eq:virtWbb}
\end{equation}
where ${\cal A}_0$ is the tree level amplitude given in Eq.~(\ref{eq:LOWbb}) and corresponding to
the diagrams shown in Figure~\ref{fig:Wbbtree_level}, and ${\cal A}_D$ denotes the amplitude for the
one-loop diagram $D$, with $D$ running over all self-energy, vertex, box and pentagon diagrams
illustrated in Figures~\ref{fig:selfWbb}, \ref{fig:vertWbb}, \ref{fig:boxWbb} and \ref{fig:pentWbb}. 

The amplitude of each virtual diagram (${\cal A}_D$) is calculated as described in
Section~\ref{subsec:sigvirt}. 

Inserting all diagram contributions into Eq.~(\ref{eq:virtWbb}), we obtain the complete ${\cal
O}(\alpha_s^3)$ contribution to the virtual amplitude squared, and integrating over the final state
phase space we calculate $\hat\sigma_{qq}^{\rm virt}$ in Eq.~(\ref{eq:virtrealXsec}). 

Results for the renormalization of the one-loop corrections are shown in
Section~\ref{subsubsec:UVWbb}.  The structure of the IR singular part of the virtual cross section
is presented in Section~\ref{subsubsec:IRvirtWbb}, while the IR singularities of the real cross
section are discussed in Section~\ref{subsec:realWbb}. The explicit cancellation of IR singularities
in the total inclusive NLO cross section is outlined in Sections~\ref{subsec:realWbb} and
\ref{subsec:totalWbb}.

\subsubsection{Virtual corrections: UV singularities and counterterms}\label{subsubsec:UVWbb}

The UV singularities of the ${\cal O}(\alpha_s^3)$ total cross section originate from the
self-energy and vertex virtual corrections shown in Figures~\ref{fig:selfWbb} and \ref{fig:vertWbb}.
These singularities are renormalized by introducing counterterms for the wave function of the
external fields ($\delta Z_2^{(q)}$, $\delta Z_2^{(b)}$) and the strong coupling constant ($\delta
Z_{\alpha_s}$).  If we denote by $\Delta_{\rm UV}(S_i^{(1,2)})$ and $\Delta_{\rm UV}(V_i^{(1,2)})$
the UV-divergent contribution of each self-energy ($S_i^{(1,2)}$) or vertex diagram ($V_i^{(1,2)}$)
to the virtual amplitude squared (see Eq.~(\ref{eq:virtWbb})), we can write the UV-singular part of
the total virtual amplitude squared as:
\begin{eqnarray}
\label{eq:amp2_virt_uvWbb}
\overline{\sum}|{\cal A}_{\rm virt}^{\sss \rm UV}|^2&=& \overline{\sum} 
|{\cal A}_0|^2
\,\left\{\sum_{i=1}^2\Delta_{\rm UV}(S_i^{(1)}+S_i^{(2)})+
\,\sum_{i=1}^3\Delta_{\rm UV}(V_i^{(1)}+V_i^{(2)})\right.\\
&+&\left.2\,\left[\,\left(\delta Z_2^{(q)}\right)_{\rm UV}+\left(\delta Z_2^{(b)}\right)_{\rm UV}+
\delta Z_{\alpha_s}\right]\right\}\,\,\,.\nonumber
\end{eqnarray}
We denote by $|{\cal A}_0|^2$ the matrix element squared of the tree-level amplitude for $q\bar
q^\prime\rightarrow \Wbb$, computed in $d=4$ dimensions (see Eq.~(\ref{eq:LOWbb}) and
Section~\ref{subsec:sigvirt}).  

The UV-divergent contributions due to the individual diagrams are given by:
\begin{eqnarray}
\label{eq:virtual_uvWbb}
\Delta_{\rm UV}\left(S_1^{(1)}+S_1^{(2)}\right)&=&
\frac{\alpha_s}{2\pi}\biggl[ {\cal N}_s
\left(\frac{5}{3}N-\frac{2}{3}n_{lf}\right)-{\cal N}_b\frac{2}{3}\biggr]
\biggl({1\over \epsilon_{\sss \rm UV}}\biggr)\,\,\,,
\nonumber\\
\Delta_{\rm UV}\left(S_2^{(1)}+S_2^{(2)}\right)&=& 
\!\!\!-\frac{\alpha_s}{2\pi}{\cal N}_b\left(\frac{N}{2}-\frac{1}{2N}\right)
\biggl({1\over \epsilon_{\sss \rm UV}}\biggr)\,\,\,,\nonumber\\
\Delta_{\rm UV}\left(V_1^{(1)}+V_1^{(2)}\right)&=& \frac{\alpha_s}{2\pi}
{\cal N}_s\left(\frac{3N}{2}-{1\over 2 N}\right)
\biggl({1\over \epsilon_{\sss \rm UV}}\biggr)\,\,\,,\nonumber\\
\Delta_{\rm UV}\left(V_2^{(1)}+V_2^{(2)}\right)&=& \frac{\alpha_s}{2\pi}
{\cal N}_b\left(\frac{3N}{2}-{1\over 2N}\right)
\biggl({1\over \epsilon_{\sss \rm UV}}\biggr)\,\,\,,\nonumber\\
\Delta_{\rm UV}\left(V_3^{(1)}+V_3^{(2)}\right)&=& \frac{\alpha_s}{2\pi}
{\cal N}_b\left(\frac{N}{2}-\frac{1}{2N}\right)
\biggl({1\over \epsilon_{\sss \rm UV}}\biggr)\,\,\,,
\end{eqnarray}
where $N=3$ is the number of colors, $n_{lf}=5$ is the number of light flavors and ${\cal N}_s$ and
${\cal N}_b$ are standard normalization factors defined as:
\begin{equation}
\label{eq:nsnb}
{\cal N}_s=
\biggl({4 \pi \mu^2\over s}\biggr)^\epsilon
\Gamma(1+\epsilon)\,\,\,\,,\,\,\,\,
{\cal N}_b=
\biggl({4 \pi \mu^2\over m_b^2}\biggr)^\epsilon
\Gamma(1+\epsilon)\,\,\,\,.
\end{equation}

Moreover, we define the required counterterms according to the following convention.  For the
external fields, we fix the wave-function renormalization constants of the external fields
($Z_2^{(i)}=1+\delta Z_2^{(i)}$, $i\!=\!q,b$) using on-shell subtraction, i.e.:
\begin{eqnarray}
\label{eq:z2_ct}
\left(\delta Z_2^{(q)}\right)_{\rm UV}
&=&-\biggl({\alpha_s\over 4 \pi}\biggr){\cal N}_s
\biggl({N\over 2}-{1\over 2 N}\biggr)
 \biggl({1\over \epsilon_{\sss \rm UV}}\biggr)\,\,\,,\\
\left(\delta Z_2^{(b)}\right)_{\rm UV}
&=&-\biggl({\alpha_s\over 4 \pi}\biggr){\cal N}_b
\biggl({N\over 2}-{1\over 2 N}\biggr)
 \biggl({1\over \epsilon_{\sss \rm UV}}+4\biggr)\,\,\,.\nonumber
\end{eqnarray}
We notice that both $\delta Z_2^{(q)}$ and $\delta Z_2^{(b)}$, as well as some of the vertex
corrections ($V_1^{(1,2)}$ and $V_2^{(1,2)}$), have also IR singularities. In this section we limit
the discussion to the UV singularities only, while the IR structure of this counterterm will be
included in the IR-singularities shown in Section~\ref{subsubsec:IRvirtWbb}.

Finally, for the renormalization of $\alpha_s$ we use the $\overline{MS}$ scheme, modified to
decouple the top quark \cite{Collins:1978wz}.  The first $n_{lf}$ light flavors are subtracted using
the $\overline{MS}$ scheme, while the divergences associated with the top-quark loop are subtracted
at zero momentum:
\begin{equation}
\label{eq:alphas_ct}
\delta Z_{\alpha_s}=\biggl({\alpha_s\over 4 \pi}\biggr)({4 \pi})^\epsilon
\Gamma(1+\epsilon)
\biggl[ \biggl(\frac{2}{3}n_{lf} - {11\over 3} N\biggr)
{1\over \epsilon_{\sss \rm UV}}+
\frac{2}{3} \biggl( {1\over \epsilon_{\sss \rm UV}}+
\ln\biggl(\frac{\mu^2}{m_t^2}\biggr)
\biggr)\biggr]\,\,\,,
\end{equation}
such that, in this scheme, the renormalized strong coupling constant $\alpha_s$ evolves with
$n_{lf}=5$ light flavors, as justified by the energy scale of the processes under consideration.

It is easy to verify that the sum of all the UV-singular contributions as given in
Eq.~(\ref{eq:amp2_virt_uvWbb}) is finite. We also notice that the left over renormalization scale
dependence, due to the mismatch between the renormalization scale dependence of $\Delta_{\rm
UV}(S_1)$ and $\delta(Z_{\alpha_s})$, is given by:
\begin{equation}
\overline{\sum}|{\cal A}_0|^2 {\alpha_s(\mu)\over 2 \pi}
\left(-\frac{2}{3} n_{lf}+\frac{11}{3}N\right) 
\ln\left(\frac{\mu^2}{s}\right)\,\,\,,
\label{eq:mudep_res_uvWbb}
\end{equation}
and corresponds exactly to the first term of Eq.~(\ref{eq:fmuNLOdep}), as predicted by
renormalization group arguments. We note that the presence of $s$ in the argument of the logarithm
of Eq.~(\ref{eq:mudep_res_uvWbb}) has no particular relevance. Choosing a different argument would
amount to reabsorbing some $\mu$-independent logarithms in $f_1^{ij}$ of Eq.~(\ref{eq:muNLOdep}).

\subsubsection{IR singularities}\label{subsubsec:IRvirtWbb}

This section describes the structure of the IR singularities originating from the ${\cal
O}(\alpha_s)$ virtual corrections.  The virtual IR singularities come from the following set of
diagrams: vertex diagrams $V_1^{(1,2)}$ and $V_2^{(1,2)}$, box diagrams $B_1^{(1,2)}$, $B_2$,
$B_3^{(1,2)}$ and $B_{3c}^{(1,2)}$ and pentagon diagrams $P_1$ and $P_{1c}$, and from the wave
function renormalization of the external fields, $\delta Z_2^{(q)}$ and $\delta Z_2^{(b)}$. After
grouping all IR poles from this diagrams we obtain the total structure of the IR singularity of the
one-loop virtual corrections to $q\bar q^\prime\to\Wbb$.  Before writing such expressions, let us
introduce the following set of kinematical variables:
\begin{eqnarray}
\label{eq:kinematic_invariants}
s&=&(q_1+q_2)^2\,\,\,,\nonumber\\
\tau_1 &=& m_b^2-(q_1-p_b)^2=2\,q_1\cdot p_b\,\,\,, \nonumber\\
\tau_2 &=& m_b^2-(q_2-p_{\bar b})^2=2\,q_2\cdot p_{\bar b} \,\,\,,
\nonumber\\
\tau_3 &=& m_b^2-(q_2-p_b)^2=2\,q_2\cdot p_b \,\,\,,\nonumber\\
\tau_4 &=& m_b^2-(q_1-p_{\bar b})^2=2\,q_1\cdot p_{\bar b}\nonumber \,\,\,,\\
\bar{s}_{b\bar b} &=& (p_b+p_{\bar b})^2=2p_b\cdot p_{\bar b}+2m_b^2\,\,\,.
\end{eqnarray}

Summing all the IR-divergent contributions yields:
\begin{equation}
\overline{\sum}|{\cal A}_{\rm virt}^{\rm \sss IR}|^2 
=  
\left(\frac{\alpha_s}{2\pi}\right){\cal N}_b\,
\overline{\sum}|{\cal A}_0|^2
\left\{
\frac{X_{-2}^{\rm virt}}{\epsilon_{\rm \sss IR}^2}+
\frac{X_{-1}^{\rm virt}}{\epsilon_{\rm \sss IR}}
+\delta_{\rm virt}^{\rm \sss IR}\right\}\,\,\,,
\label{eq:IRpoles}
\end{equation}
with
\begin{eqnarray}
\label{eq:sigma_ir_poles}
X_{-2}^{\rm virt}&=& -\left(N-\frac{1}{N}\right)\,\,\, ,\\
X_{-1}^{\rm virt}&=& N\,\left[-\frac{5}{2}
+\ln\left(\frac{\tau_1}{m_b^2}\right)
+\ln\left(\frac{\tau_2}{m_b^2}\right)\right]\nonumber\\
&+&\frac{1}{N}\,\left[-\ln\left(\frac{s}{m_b^2}\right)+\frac{5}{2}
-\frac{\bar{s}_{b\bar b}-2m_b^2}{\bar{s}_{b\bar b}\beta_{b\bar b}}
\Lambda_{b\bar b}-2\ln\left(\frac{\tau_1\tau_2}
{\tau_4\tau_3}\right)\right]\,\,\,,\nonumber
\end{eqnarray}
where we have used the kinematical invariants presented in Eq.~(\ref{eq:kinematic_invariants}) while
$\beta_{b\bar b}$ and $\Lambda_{b\bar b}$ are defined by: 
\begin{eqnarray}
\label{eq:betadef}
\beta_{b\bar b} &=&\sqrt{1-{4 m_b^2\over \bar{s}_{b\bar b}}}\,\,\,, \nonumber\\
\Lambda_{b\bar b} &=& \ln\biggl({1+\beta_{b\bar b}
\over 1 -\beta_{b\bar b}}\biggr)\,\,\,,
\end{eqnarray}
and $\delta_{\rm virt}^{\rm IR}$ is a finite term that comes from having factored out a common
factor ${\cal N}_b$ and is given by:
\begin{equation}
\delta_{\rm virt}^{\rm \sss IR}=\left(N-\frac{1}{N}\right)
\left[\frac{3}{2}\ln\left(\frac{s}{m_b^2}\right)\right]+
\frac{1}{N}\left[\frac{1}{2}\ln^2\left(\frac{s}{m_b^2}\right)\right]\,\,\,.
\end{equation}

Before finishing let us write the IR-pole contributions to the counterterms $\delta Z_2^{(q)}$ and
$\delta Z_2^{(b)}$:
\begin{eqnarray}
\left(\delta Z_2^{(q)}\right)_{IR}&=&
\biggl(\frac{\alpha_s}{4\pi}\biggr){\cal N}_s\left(\frac{N}{2}-\frac{1}{2N}\right)
\left(\frac{1}{\epsilon_{\sss IR}}\right)\,\,\,,\\
\left(\delta Z_2^{(b)}\right)_{IR}&=&
\!\!\!-\biggl({\alpha_s\over 4 \pi}\biggr){\cal N}_b\left(\frac{N}{2}-\frac{1}{2N}\right)
\left(\frac{2}{\epsilon_{\rm \sss IR}}\right)\,\,\,.\nonumber
\end{eqnarray}

In Sec.~\ref{subsec:realWbb} we will show how the IR singularities of the real cross section exactly
cancel the IR poles of the virtual cross section (see
Eqs.~(\ref{eq:sigma_soft_totalWbb})-(\ref{eq:sigma_soft_polesWbb})), as predicted by the
Bloch-Nordsieck~\cite{PhysRev.52.54} and Kinoshita-Lee-Nauenberg~\cite{Kinoshita:1962ur,Lee:1964is}
theorems.

\begin{figure}[t]
\begin{center}
\includegraphics[scale=0.8]{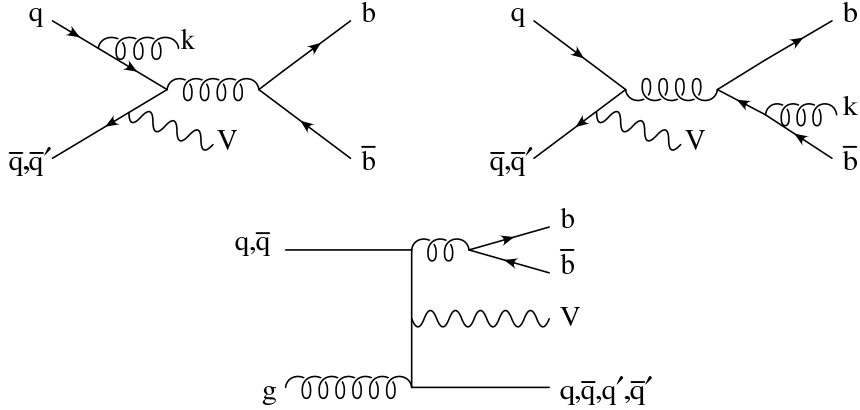}
\caption{${\cal O}(\as)$ real corrections: examples of initial and final real gluon emission and $q(\bar q)g$
	initiated subprocess.}
\label{fig:realdiagWbb}
\end{center}
\end{figure}

\subsection[Real Corrections to $\Wbb$ Production]{Real Corrections to $\Wbb$ Production}\label{subsec:realWbb}

The ${\cal O}(\alpha_s)$ corrections to $q\bar q^\prime\rightarrow \Wbb$ due to real gluon emission
(see Figure~\ref{fig:realdiagWbb}) give rise to IR singularities which cancel exactly the analogous
singularities present in the ${\cal O}(\alpha_s)$ virtual corrections (see
Sec.~\ref{subsubsec:IRvirtWbb}). We also have real contributions from the subprocess $q(\bar
q)g\to\Wbb +q^\prime(\bar q^\prime)$ that give rise to IR singularities of the collinear kind. We
present results for the latter at the end of this section.

We have calculated the cross section for the process 
\begin{equation} 
q (q_1)+{\overline q^\prime}(q_2) \rightarrow b (p_b)+{\overline b}(p_{\bar b}) + W(p_{\sss W})+g(k) 
\end{equation} 
using the
\emph{two-cutoff} PSS method, as presented in Section~\ref{subsec:sigreal}.  In the following
subsections we explain in detail how we have applied this method to the calculation of the real
contributions to $\Wbb$ hadronic production. We will present details of the calculation of the
pieces of $\sigma^{\rm real}$ introduced in Section~\ref{subsec:sigreal}, namely $\sigma^{soft}$ and
$\sigma^{hard}$.

\subsubsection*{Soft gluon emission}
%\label{subsubsec:two_cutoff_softWbb} 

The soft region of the $q\bar q^\prime\rightarrow \Wbb+g$ phase space is defined by requiring that
the energy of the gluon satisfies:
\begin{equation}
E_g <\delta_s {\sqrt{s}\over 2}\,\,\,,
\end{equation}
for an arbitrary small value of the \emph{soft} cutoff $\delta_s$.  In the limit when the energy of
the gluon becomes small, i.e. in the \emph{soft limit}, the matrix element squared for the real
gluon emission, $\overline{\sum} |{\cal A}_{\rm real}|^2$, assumes a very simple form, i.e.  it
factorizes into the LO matrix element squared times an eikonal factor $\Phi_{eik}$:
\begin{equation}
\label{eq:m2_soft_lim}
\overline{\sum}|{\cal A}_{\rm real}(q\bar q^\prime\rightarrow \Wbb+g)|^2 
\stackrel{soft}{\longrightarrow}
 (4 \pi \alpha_s) \overline{\sum}|{\cal A}_0|^2\,\Phi_{eik}\,\,\,,
\end{equation}
where the eikonal factor is given by:
\begin{eqnarray}
\label{eq:eik_factor}
\Phi_{eik}&=&
\frac{N}{2}\left[
-\frac{m_b^2}{(p_b\!\cdot\! k)^2}
-\frac{m_b^2}{(p_{\bar b}\!\cdot\! k)^2}
+\frac{\tau_1}{(q_1\!\cdot\! k)(p_b\!\cdot\! k)}
+\frac{\tau_2}{(q_2\!\cdot\! k)(p_{\bar b}\!\cdot\! k)}\right]
\\
&& +
\frac{1}{2N} \left[
\frac{m_b^2}{(p_b\!\cdot\! k)^2}
+\frac{m_b^2}{(p_{\bar b}\!\cdot\! k)^2}
-\frac{s}{(q_1\!\cdot\! k)(q_2\!\cdot\! k)}
-\frac{\bar{s}_{b\bar b}-2m_b^2}{(p_b\!\cdot\! k)(p_{\bar b}\!\cdot\! k)}
\right. \nonumber \\
&&\left.
+2\left( -\frac{\tau_1}{(q_1\!\cdot\! k)(p_b\!\cdot\! k)}
+\frac{\tau_4}{(q_1\!\cdot\! k)(p_{\bar b}\!\cdot\! k)}
+\frac{\tau_3}{(q_2\!\cdot\! k)(p_b\!\cdot\! k)}
-\frac{\tau_2}{(q_2\!\cdot\! k)(p_{\bar b}\!\cdot\! k)}
\right)\right]\,\,\,,\nonumber
\end{eqnarray}
where we have used the kinematical invariants defined in Eq.~(\ref{eq:kinematic_invariants}).
Moreover, in the soft region the $q\bar q^\prime\rightarrow \Wbb+g$ phase space also factorizes as:
\begin{eqnarray}
\label{eq:ps_soft_lim}
d(PS_4)(q\bar q^\prime\rightarrow \Wbb+g)
& \stackrel{soft}{\longrightarrow} & d(PS_3)(q\bar q^\prime\rightarrow \Wbb)
d(PS_g)_{soft}\\
&=& d(PS_3)(q\bar q^\prime\rightarrow \Wbb)
\frac{d^{(d-1)}k}{(2\pi)^{(d-1)}2E_g} \theta(\delta_s {\sqrt{s}\over 2}-E_g)
\nonumber\,\,\,,
\end{eqnarray}
where $d(PS_g)_{soft}$ denotes the integration over the phase space of the soft gluon. The parton
level soft cross section can then be written as:
\begin{equation}
\label{eq:sigma_softWbb}
\hat{\sigma}^{soft}=(4 \pi\alpha_s) \, \mu^{2 \epsilon} \,
\int d(PS_3) \overline{\sum}|{\cal A}_0|^2\int d(PS_g)_{soft} 
\Phi_{eik}\,\,\,.
\end{equation}
Since the contribution of the soft gluon is now completely factorized, we can perform the
integration over $d(PS_g)_{soft}$ in Eq.~(\ref{eq:sigma_softWbb}) analytically, and extract the soft
poles that will cancel $X_{-2}^{\rm virt}$ and $X_{-1}^{\rm virt}$ of Eq.~(\ref{eq:sigma_ir_poles}).
The integration over the gluon phase space in Eq.~(\ref{eq:sigma_softWbb}) can be performed using
standard techniques and we refer to Refs.~\cite{Harris:2001sx,Beenakker:1988bq} for more details.
For sake of completeness, in Appendix~\ref{app:PSint} we give explicit results for the soft
integrals used in our calculation.

Finally, the soft gluon contribution to 
$\hat{\sigma}^{\rm real}_{q\bar q}$ can be written as follows:
\begin{equation}
\label{eq:sigma_soft_totalWbb}
\hat{\sigma}^{soft}=
\frac{\alpha_s}{2\pi}{\cal N}_b
\int d(PS_3) \overline{\sum} |{\cal A}_0|^2 
\left\{
\frac{X_{-2}^{s}}{\epsilon^2}+\frac{X_{-1}^{s}}{\epsilon}+
N C_1^s+\frac{C_2^s}{N}\right\} \,\,\,,
\end{equation}
where $\epsilon$ corresponds to $\epsilon_{\rm IR}$ of Eq.~(\ref{eq:IRpoles}) and
\begin{eqnarray}
\label{eq:sigma_soft_polesWbb}
X_{-2}^{s} &=& -X_{-2}^{\rm virt}\,\,\,,\nonumber\\
X_{-1}^{s} &=& -X_{-1}^{\rm virt}-\left(N-\frac{1}{N}\right)\,\left[\frac{3}{2} 
+2 \ln\left(\delta_s\right)\right]\,\,\,,\nonumber\\
C_1^s&=&
\frac{3}{2}\ln\left(\frac{s}{\mu^2}\right) + 2\ln^2(\delta_s)
-2\ln(\delta_s)\left[1+\ln\left(
{m_b^2\mu^2\over \tau_1 \tau_2}\right)\right]
\nonumber \\
&+& {1\over 2} \ln^2\biggl({s\over m_b^2}\biggr)-{\pi^2\over 3}
-\ln\biggl({s\over m_b^2}\biggr) \biggl[{5\over 2} 
+\ln\biggl({sm_b^2\over \tau_1 \tau_2}\biggr)\biggr]
\nonumber \\
&+&\frac{1}{2}\frac{1}{\beta_{b}}\ln\left(\frac{1+\beta_{b}}{1-\beta_{b}}\right)
+\frac{1}{2}\frac{1}{\beta_{\bar{b}}}\ln\left(\frac{1+\beta_{\bar{b}}}{1-\beta_{\bar{b}}}\right)
\nonumber\\
&+&
{1\over 2} \big( F_{qb}
+F_{\bar q\bar b}\big)+\left[ \frac{3}{2} 
+2 \ln\left(\delta_s\right)\right]\,\ln\left(\frac{\mu^2}{m_b^2}\right)
\,\,\,,\nonumber \\
C_2^s&=&
-{3\over 2} \ln \biggl({ s\over \mu^2}\biggr)-2
\ln^2(\delta_s) -2
\ln( \delta_s)\biggl[\,-1 \nonumber\\
&+&
\frac{\bar{s}_{b\bar b}-2m_b^2}{\bar{s}_{b\bar b}\beta_{b\bar b}}
\Lambda_{b\bar b}
+\ln\biggl({s\over \mu^2}\biggr)+2
\ln\biggl( {\tau_1\tau_2\over \tau_4 \tau_3}\biggr)\biggr]
\nonumber \\
&-&
{1\over 2} \ln^2\biggl({s\over m_b^2}\biggr)
+{\pi^2\over 3} -\ln\biggl({s\over m_b^2}\biggr)\biggl[\,
-{5\over 2} 
\nonumber \\
&+&
\frac{\bar{s}_{b\bar b}-2m_b^2}{\bar{s}_{b\bar b}\beta_{b\bar b}}
\Lambda_{b\bar b}
+2\ln\biggl({\tau_1 \tau_2\over \tau_4
\tau_3}\biggr)
\biggr]-
\frac{1}{2}\frac{1}{\beta_{b}}\ln\left(\frac{1+\beta_{b}}{1-\beta_{b}}\right)
-\frac{1}{2}\frac{1}{\beta_{\bar{b}}}\ln\left(\frac{1+\beta_{\bar{b}}}{1-\beta_{\bar{b}}}\right)
\nonumber \\
&+&\frac{\bar{s}_{b\bar b}-2m_b^2}{\bar{s}_{b\bar b}\beta_{b\bar b}}
\left[
-\frac{1}{4}\ln^2\left(\frac{1+\beta_b}{1-\beta_b}\right)
+\frac{1}{4}\ln^2\left(\frac{1+\beta_{\bar{b}}}{1-\beta_{\bar{b}}}\right)
\right.\nonumber\\
&&
-\mathrm{Li}_2\left(1-\frac{\alpha_{b\bar{b}}}{v_{b\bar{b}}}p_b^0(1+\beta_b)\right)
-\mathrm{Li}_2\left(1-\frac{\alpha_{b\bar{b}}}{v_{b\bar{b}}}p_b^0(1-\beta_b)\right)
\nonumber\\
&&\left.
+\mathrm{Li}_2\left(1-\frac{1}{v_{b\bar{b}}}p_{\bar b}^0(1+\beta_{\bar{b}})\right)
+\mathrm{Li}_2\left(1-\frac{1}{v_{b\bar{b}}}p_{\bar b}^0(1-\beta_{\bar{b}})\right)
\right]\nonumber\\
&+&
-F_{qb}+F_{q\bar b}+F_{\bar qb}-F_{\bar q\bar b}
\nonumber \\
&-& \left[\frac{3}{2} 
+2 \ln\left(\delta_s\right)\right]\,\ln\left(\frac{\mu^2}{m_b^2}\right)
\,\,\, ,
\end{eqnarray}
while ${\cal N}_b$ is defined in Eq.~(\ref{eq:nsnb}), $\mbox{Li}_2$ denotes the dilogarithm as
described in Ref.~\cite{lewin} and $X_{-2}^{\rm virt}$ and $X_{-1}^{\rm virt}$ are given in
Eq.~(\ref{eq:sigma_ir_poles}). We have used the kinematical invariants defined in
Eq.~(\ref{eq:kinematic_invariants}), $\beta_{b\bar b}$ and $\Lambda_{b\bar b}$ are defined in
Eq.~(\ref{eq:betadef}), 
\begin{equation}
\alpha_{b\bar{b}}=\frac{1+\beta_{b\bar{b}}}{1-\beta_{b\bar{b}}}\,\,\,
\mbox{and}\,\,\,
v_{b\bar{b}}=\frac{m_b^2(\alpha_{b\bar{b}}^2-1)}{2(\alpha_{b\bar{b}}p_b^0-p_{\bar b}^0)}
\,\,\,,
\end{equation}
while, for any initial parton $i$ and final parton $f$, the function $F_{if}$ can be written as:
\begin{eqnarray}
\label{eq:chapf_if}
F_{if}&=&
\ln^2\biggl({1-\beta_f\over 1-\beta_f \cos \theta_{if}}\biggr)
-{1\over 2} \ln^2\biggl( {1+\beta_f\over 1-\beta_f}\biggr)\\
&&+2 \mbox{Li}_2\biggl(-{\beta_f (1-\cos \theta_{if})\over 
1-\beta_f}\biggr)
-2 \mbox{Li}_2\biggl(-{\beta_f
 (1+\cos \theta_{if})\over 1-\beta_f \cos\theta_{if}}\biggr)\,\,\,,
\nonumber
\end{eqnarray}
where $\theta_{if}$ is the angle between partons $i$ and $f$ in the center-of-mass frame of the
initial state partons, and
\begin{equation}
\beta_f=\sqrt{1-\frac{m_b^2}{(p_f^0)^2}}\,\,\,\,\,,\,\,\,\,\, 
1-\beta_f\cos{\theta_{if}}=\frac{s_{if}}{p_f^0\sqrt{s}}\,\,\,.
\label{eq:bf}
\end{equation}
All the quantities in Eq.~(\ref{eq:chapf_if}) can be expressed in terms of kinematic invariants, for
details see Appendix~\ref{app:PSint}.

As can be easily seen from Eqs.~(\ref{eq:sigma_ir_poles}) and (\ref{eq:sigma_soft_polesWbb}), the IR
poles of the virtual corrections are exactly canceled by the corresponding singularities in the soft
gluon contribution. The remaining IR poles in $\hat \sigma^{soft}$ will be canceled by the PDF
counterterms as described in detail in Sec.~\ref{subsec:totalWbb}.

\subsubsection*{Hard gluon emission} 
%\label{subsubsec:two_cutoff_hardWbb}

The hard region of the gluon phase space is defined by requiring that the energy of the emitted
gluon is above a given threshold.  As we discussed earlier this is expressed by the condition that
\begin{equation}
E_g >\delta_s {\sqrt{s}\over 2}\,\,\,,
\end{equation}
for an arbitrary small \emph{soft} cutoff $\delta_s$, which automatically assures that
$\hat{\sigma}^{hard}$ does not contain soft singularities.  However, a hard gluon can still yield
singularities when it is emitted at a small angle, i.e.  \emph{collinear}, to a massless incoming or
outgoing parton. In order to isolate these divergences and compute them analytically, we divide the
hard region of the $q\bar q^\prime\rightarrow \Wbb+g$ phase space into \emph{hard/collinear} and
\emph{hard/non-collinear} regions, by introducing a small \emph{collinear} cutoff $\delta_c$.  The
\emph{hard/non-collinear} region is defined by the conditions
\begin{equation}
\label{eq:deltac_cutsWbb}
%\mbox{both}\,\,\,\,\,\,\,\,
\frac{2 q_1\!\cdot\! k}{E_g \sqrt{s}}> \delta_c\,\,\,\,\,\,\,
\mbox{and}\,\,\,\,\,\,\,
\frac{2q_2\!\cdot\! k}{E_g\sqrt{s}}>\delta_c\;.
\end{equation}
The contribution from the \emph{hard/non-collinear} region, $\hat{\sigma}^{hard/non-coll}$, is
finite and we compute it numerically using standard Monte Carlo integration techniques.

In the $\emph{hard/collinear}$ region, one of the conditions in Eq.~(\ref{eq:deltac_cutsWbb}) is not
satisfied and the hard gluon is emitted collinear to one of the incoming partons.  In this region,
the initial-state parton $i$ ($i\!=\!q,\bar q$) is considered to split into a hard parton $i^\prime$
and a collinear gluon $g$, $i\rightarrow i^\prime g$, with $p_{i^\prime}\!=\!z p_i$ and
$k\!=\!(1-z)p_i$. The matrix element squared for $i j\rightarrow \Wbb+g$ factorizes into the LO
matrix element squared and the Altarelli-Parisi splitting function for $i\rightarrow i^\prime g$,
i.e.:
\begin{equation}
\label{eq:m2_coll_lim}
\overline{\sum}|{\cal A}_{\rm real}(ij \rightarrow \Wbb+g)|^2 
\stackrel{collinear}{\longrightarrow}
(4 \pi \alpha_s) \sum_{i^\prime}\overline{\sum}|{\cal A}_0
(i^\prime j\rightarrow \Wbb)|^2
\frac{2 P_{ii^\prime}(z)}{z \, s_{ig}} \,\,\,,
\end{equation}
with $s_{ig}=2 p_i\!\cdot\! k$, and $P_{ii^\prime}(z)$ is the unregulated Altarelli-Parisi splitting
function for $q\!\rightarrow\!q+g$ at lowest order, including terms of ${\cal O}(\epsilon)$ as given
by $P_{qq}(z)=P_{qq}^4+\epsilon P_{qq}^\prime$ with
\begin{eqnarray}
P_{qq}^4(z)& = & C_F \frac{1+z^2}{1-z}, \nonumber\\
P_{qq}^\prime(z)& = & -C_F (1-z), 
\label{eq:splitfuncqq}
\end{eqnarray}
where $C_F=1/2(N-1/N)$. Moreover, in the collinear limit, the $q\bar q^\prime\rightarrow \Wbb+g$
phase space also factorizes as:
\begin{eqnarray}
\label{eq:ps_coll_lim}
d(PS_4)(ij\rightarrow \Wbb+g)
&\stackrel{collinear}{\longrightarrow}& 
d(PS_3)(i^\prime j\rightarrow \Wbb)
\frac{z\,d^{(d-1)}k}{(2\pi)^{(d-1)}2E_g}
\theta\left(E_g-\delta_s {\sqrt{s}\over 2}\right)\times \\
&&\theta(\cos\theta_{ig}-(1-\delta_c))\nonumber\\
&&\!\!\!\!\!\!\!\!\!\!\!\!\!\!\!\!\!\!\!\!\!\! \stackrel{d=4-2 \epsilon}{=}
\frac{\Gamma(1-\epsilon)}{\Gamma(1-2\epsilon)}
\frac{\left(4\pi\right)^\epsilon}{16 \pi^2}\,z\,dz\,ds_{ig} 
\left[(1-z) s_{ig}\right]^{-\epsilon}
\theta\left({(1-z)\over z }s^\prime {\delta_c \over 2}-s_{ig}\right)
\nonumber\; ,
\end{eqnarray}
where the integration range for $s_{ig}$ in the collinear region is given in terms of the collinear
cutoff, and we have defined $s^\prime\!=\!2 p_{i^\prime}\cdot p_j$.  The integral over the collinear
gluon degrees of freedom can then be performed separately, and this allows us to extract explicitly
the collinear singularities of $\hat{\sigma}^{hard}$.  $\hat{\sigma}^{hard/coll}$ turns out to be of
the form \cite{Harris:2001sx,Baur:1998kt}:
\begin{eqnarray}
\label{eq:coll_pole}
\hat{\sigma}^{hard/coll}&=&
\left[\frac{\alpha_s}{2\pi}\frac{\Gamma(1-\epsilon)}{\Gamma(1-2\epsilon)}
\left(\frac{4\pi\mu^2}{m_b^2}\right)^\epsilon\right]
\left(-\frac{1}{\epsilon}\right)\delta_c^{-\epsilon}\times\\
&&\left\{
\int_{0}^{1-\delta_s} dz
\left[\frac{(1-z)^2}{2 z} \frac{s^\prime}{m_b^2}\right]^{-\epsilon} 
P_{ii^\prime}(z) \,
\hat{\sigma}^{\rm \sss LO}(i^\prime j\rightarrow \Wbb)
+ (i\leftrightarrow j)\right\}\ .\nonumber
\end{eqnarray}
The upper limit on the $z$ integration ensures the exclusion of the soft gluon region.  As usual,
these initial-state collinear divergences are absorbed into the parton distribution functions as
will be described in detail in the Sec.~\ref{subsec:totalWbb}.

\boldmath
\subsubsection*{The tree level processes $(q,\bar{q})g\to \Wbb+(q^\prime,\bar{q^\prime})$}
\unboldmath
%\label{subsubsec:two_cutoff_qg}

The extraction of the collinear singularities of $\hat \sigma^{qg}_{\rm real}$ is done in the same
way as described in the previous subsection for the $q\bar q^\prime$ initial state.  In the
collinear region, $\cos\theta_{iq}>1-\delta_c$, the initial state parton $i$ with momentum $q_i$ is
considered to split into a hard parton $i^\prime$ and a collinear quark $q$, $i\rightarrow i^\prime
q$, with $q_{i^\prime}\!=\!z q_i$ and $k\!=\!(1-z)q_i$. The matrix element squared for $ij\to
\Wbb+q$ factorizes into the unregulated Altarelli-Parisi splitting functions in $d$ dimensions:
$P_{ii'}= P_{ii'}^4+\epsilon P_{ii'}^\prime$ and the corresponding LO matrix elements squared.  The
$ij\to \Wbb+q$ phase space factorizes into the $i'j\to \Wbb$ phase space and the phase space of the
collinear quark. As a result, after integrating over the phase space of the collinear quark, the
collinear singularity of $\hat \sigma^{\rm real}_{qg}$ can be extracted as:
\begin{eqnarray}
\label{eq:qgcoll_poleWbb}
\hat{\sigma}^{coll}_{qg}&=&
\left[\frac{\alpha_s}{2\pi}\frac{1}{\Gamma(1-\epsilon)}
\left(\frac{4\pi\mu^2}{m_b^2}\right)^\epsilon\right]
\left(-\frac{1}{\epsilon}\right)\delta_c^{-\epsilon}
\int_{0}^{1} dz
\left[\frac{(1-z)^2}{2 z} \frac{s^\prime}{m_b^2}\right]^{-\epsilon} 
\times\nonumber\\
&&
\left[ P_{gq}(z) \,
\hat{\sigma}^{\rm \sss LO}_{q\bar{q}}(q(q_1)\bar{q}(q_{2^\prime})\to \Wbb)
\right]\,\,\,. 
\end{eqnarray}
The collinear radiation of an antiquark in $\bar{q}g\to \Wbb+\bar{q}$ is treated analogously. In the
case of $(q,\bar{q})g\to \Wbb+(q,\bar{q})$ we have the possible splitting $g\to q\bar{q}$.  The
$O(1)$ and $O(\epsilon)$ parts of the corresponding splitting function are: 
\begin{eqnarray}
P_{gq}^4(z)& = & \frac{1}{2}(z^2+(1-z)^2), \nonumber\\
P_{gq}^\prime(z)& = & -z(1-z). 
\label{eq:splitfuncgq}
\end{eqnarray}

Again, these initial state collinear divergences are absorbed into the parton distribution functions
as will be described in detail in Section~\ref{subsec:totalWbb}.

\subsection{Total Cross Section for $p{\bar p}(pp)\to\Wbb$}\label{subsec:totalWbb}

As described in Sec.~\ref{sec:FactTh}, the observable total cross section at NLO is obtained by
convoluting the NLO parton level cross section with the NLO parton distribution functions ${\cal
F}_q^{p,\bar p}(x,\mu)$, thereby absorbing the remaining initial-state singularities of
$\delta\hat\sigma^{\rm \sss NLO}_{q\bar q}$ into the quark distribution functions.  This can be
understood as follows.  First the parton cross section is convoluted with the {\em bare}
quark/antiquark distribution functions ${\cal F}_{q,\bar q}^{p,\bar p}(x)$ and subsequently ${\cal
F}_{q,\bar q}^{p,\bar p}(x)$ is replaced by the renormalized quark/antiquark distribution functions
${\cal F}_{q,\bar q}^{p,\bar p}(x,\mu)$ defined in some subtraction scheme. Using the
${\overline{MS}}$ scheme, the scale-dependent NLO quark distribution functions are given in terms of
${\cal F}_{q,\bar q}^{p,\bar p}(x)$ and the QCD NLO parton distribution function counterterms
\cite{Harris:2001sx} as follows:\\

\begin{itemize}
\item[(a)] For the case where an initial state gluon
  $g$ splits into a $q\bar{q}$ pair ($g\to q\bar  q$):
\begin{eqnarray}
\label{eq:pdfqg_muWbb}
{\cal F}_{q(\bar q)}^{p,\bar p}(x,\mu_f)&=& {\cal F}_{q(\bar q)}^{p,\bar p}(x) 
+\left[\frac{\alpha_s}{2\pi}
\left(\frac{4\pi\mu_r^2}{\mu_f^2}\right)^\epsilon
\frac{1}{\Gamma(1-\epsilon)}\right]
\int_{x}^{1} \frac{dz}{z}
\left(-\frac{1}{\epsilon}\right) P_{gq(\bar q)}^4(z) 
{\cal F}_{g}^{p,\bar p}\left(\frac{x}{z}\right)\,\,\,,\nonumber\\
\end{eqnarray}
where $P^4_{gq}$ is defined in Eq.~(\ref{eq:splitfuncgq}). This is relevant to process $q(\bar
q)g\to\Wbb+q^\prime(\bar q^\prime)$ when the gluon becomes collinear with the final massless parton.

\item[(b)] For the case of $q\to qg$ splitting:
\begin{eqnarray}
\label{eq:pdf_mu2}
{\cal F}_q^{p,\bar p}(x,\mu)&=&
{\cal F}_q^{p,\bar p}(x) \left[1- 
\frac{\alpha_s}{2\pi}
\frac{1}{\Gamma(1-\epsilon)}
\left(\frac{4 \pi\mu_r^2}{\mu_f^2} \right)^\epsilon
\left(\frac{1}{\epsilon}\right) C_F
\left(2\ln(\delta_s)+\frac{3}{2}\right)\right] \nonumber\\
&+&\left[\frac{\alpha_s}{2\pi}
\frac{1}{\Gamma(1-\epsilon)}
\left(\frac{4\pi\mu_r^2}{\mu_f^2}\right)^\epsilon\right]
\int_{x}^{1-\delta_s} \frac{dz}{z}
\left(-\frac{1}{\epsilon}\right) P_{qq}(z){\cal F}_q^{p,\bar p}
(\frac{x}{z})\,\,\,,
\end{eqnarray}
where the ${\cal O}(\alpha_s)$ terms in the previous equation are calculated from the ${\cal
O}(\alpha_s)$ corrections to the $q\rightarrow qg$ splitting, in the two-cutoff PSS formalism, and $P_{qq}(z)$
is the Altarelli-Parisi splitting function of Eq.~(\ref{eq:splitfuncqq}). This is relevant to
process $q\bar q^\prime\to\Wbb+g$ when the final gluon goes soft or when one of the initial partons
become collinear with the gluon.
\end{itemize}

When convoluting the parton cross section with the renormalized quark/antiquark distribution
functions of Eq.~(\ref{eq:pdf_mu2}), the IR singular counterterm, that is the first term of the RHS
of Eq.~(\ref{eq:pdf_mu2}), exactly cancels the remaining IR poles of $\hat{\sigma}^{\rm virt}_{q
\bar q}+\hat\sigma^{soft}$ and $\hat{\sigma}^{hard/coll}$.  Finally, the complete ${\cal
O}(\alpha_s^3)$ inclusive total cross section for $p\bar p(pp) \to \Wbb$ in the ${\overline{MS}}$
factorization scheme can be written as follows:
\begin{equation}
\sigma^{\rm \sss NLO}=\sigma^{\rm \sss NLO}_{q\bar q^\prime}+\sigma^{\rm \sss NLO}_{qg+\bar qg}\; ,
\end{equation}
with $\sigma^{\rm \sss NLO}_{q\bar q^\prime}$ and $\sigma^{\rm \sss NLO}_{qg}$ defined in the following,
Eqs.~(\ref{eq:sigmatot2}) and (\ref{eq:sigmatot2qgWbb}) respectively.
\begin{eqnarray}
\label{eq:sigmatot2}
\sigma^{\rm \sss NLO}_{q\bar q^\prime} &=&\sum_{q \bar q^\prime} \int dx_1 dx_2 {\cal F}_q^p(x_1,\mu)
{\cal F}_{\bar q^\prime}^{\bar p(p)}(x_2,\mu) \left[
\hat{\sigma}_{q\bar q^\prime}^{\rm \sss LO}(x_1,x_2,\mu)+
\hat{\sigma}^{\rm virt}_{q\bar q^\prime}(x_1,x_2,\mu)+
\hat \sigma^{\prime\ soft}(x_1,x_2,\mu)\right]\nonumber\\
&+&\frac{\alpha_s}{2\pi} C_F \sum_{q \bar q^\prime} \int dx_1 dx_2 \left\{
\int_{x_1}^{1-\delta_s}\frac{dz}{z}
\left[{\cal F}_q^p(\frac{x_1}{z},\mu) {\cal F}_{\bar q^\prime}^{\bar p(p)}(x_2,\mu)+
{\cal F}_q^{\bar p(p)}(x_2,\mu) {\cal F}_{\bar q^\prime}^p(\frac{x_1}{z},\mu)\right]
\right. \\
&&\times \left. \hat{\sigma}^{\rm \sss LO}_{q\bar q^\prime}(x_1, x_2,\mu)
\left[\frac{1+z^2}{1-z}
\ln\left(\frac{s}{\mu^2}\frac{(1-z)^2}{z}\frac{\delta_c}{2}\right)
+1-z\right]+(1\leftrightarrow 2) \right\}\nonumber\\
&+& \sum_{q \bar q^\prime} \int dx_1 dx_2 {\cal F}_q^p(x_1,\mu)
{\cal F}_{\bar q^\prime}^{\bar p(p)}(x_2,\mu) \, 
\hat{\sigma}^{hard/non-coll}(x_1,x_2,\mu) \,\,\, ,\nonumber
\end{eqnarray}
with 
\begin{equation}
\hat \sigma^{\prime\ soft}=\hat \sigma^{soft}+
\frac{\alpha_s}{2\pi}
\frac{1}{\Gamma(1-\epsilon)}
\left(4 \pi \right)^\epsilon
\left(\frac{1}{\epsilon}\right) C_F
\left[4\ln(\delta_s)+3\right]\,\,\, .
\end{equation}

\begin{eqnarray}
\label{eq:sigmatot2qgWbb}
\sigma_{qg+\bar qg}^{\rm \sss NLO} &=&
\frac{\alpha_s}{2\pi} \sum_{i=q,\bar{q}}\int dx_1dx_2 
\left\{ \int_{x_1}^{1}\frac{dz}{z} 
{\cal F}_g^p(\frac{x_1}{z},\mu) {\cal F}_{i}^{\bar p(p)}(x_2,\mu) 
\times \right. \nonumber\\
& &\left. \hat{\sigma}^{\rm \sss LO}_{q\bar{q}}(x_1, x_2,\mu) 
\left[
{P}^4_{gi}(z) \ln\left(\frac{s}{\mu^2}\frac{(1-z)^2}{z}
\frac{\delta_c}{2}\right)-
{P}^{\prime}_{gi}(z)\right] +(1\leftrightarrow 2)\right\}\nonumber\\
&+&\sum_{i=q,\bar{q}} \int dx_1 dx_2 
\left\{ {\cal F}_i^p(x_1,\mu) {\cal F}_{g}^{\bar p(p)}(x_2,\mu) \, 
\hat{\sigma}^{non-coll}_{ig}(x_1,x_2,\mu)+(1 \leftrightarrow 2)
\right\}\,\,\, .\nonumber\\
\end{eqnarray}
\begin{figure}[htb]
\begin{center}
\includegraphics*[scale=0.55,angle=-90]{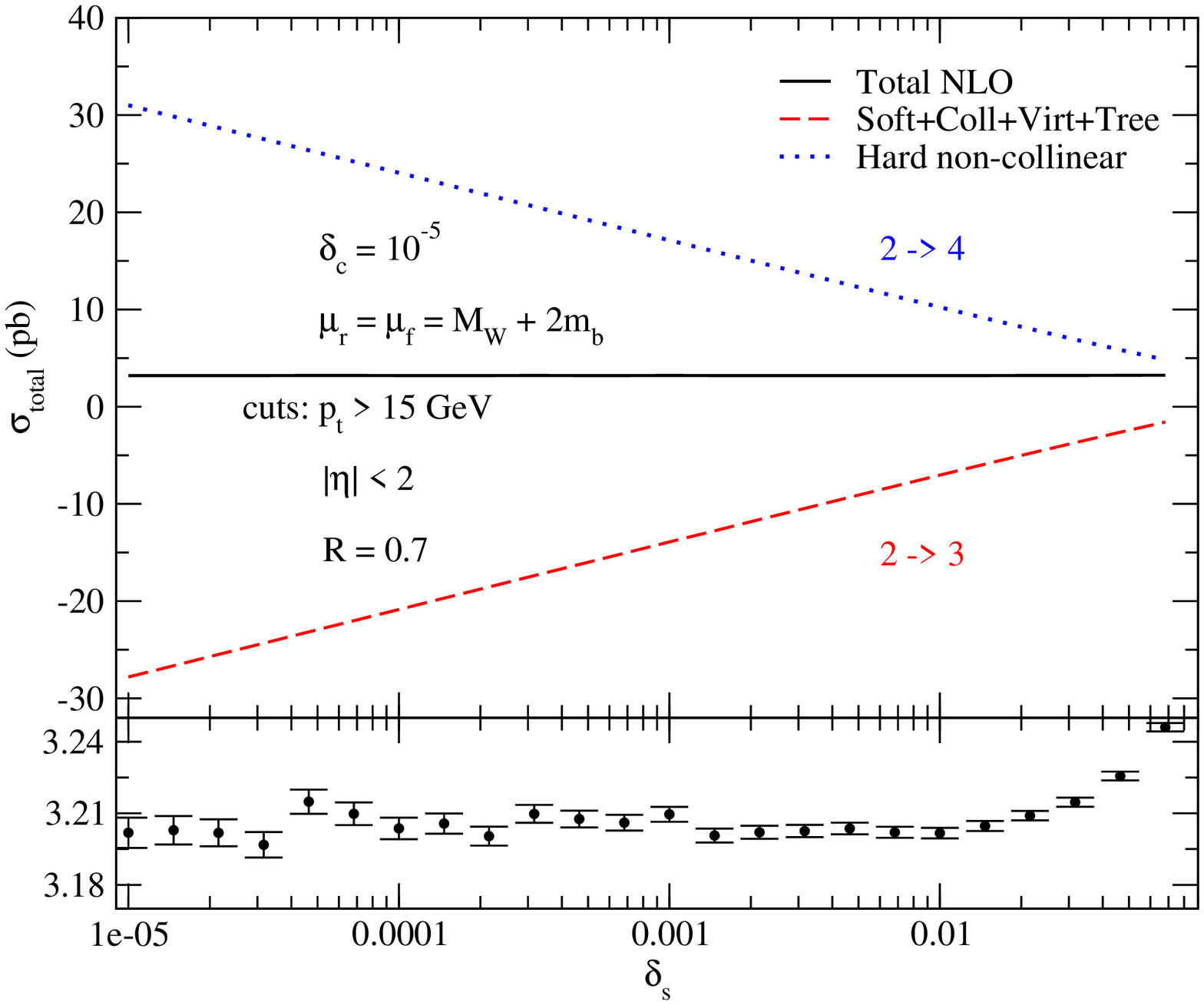}
\caption[Dependence of the total $\Wbb$ NLO QCD cross section on the
  $\delta_s$ PSS parameter.]{Dependence of $\sigma^{\rm \sss NLO}(p\bar{p}\to \Wbb)$ on the
  $\delta_s$ PSS parameter, when $\delta_c$ is fixed at
  $\delta_c=10^{-5}$. In the upper window we illustrate separately the
  cutoff dependence of the soft and hard-collinear part ($2\rightarrow
  3$, red dashed curve) and of the hard non-collinear part
  ($2\rightarrow 4$, blue dotted curve) of the real corrections to the
  total cross section.  The $2\rightarrow 3$ curve also includes those
  parts of the $2\rightarrow 3$ NLO cross section that do not depend
  on $\delta_c$ and $\delta_s$, i.e. the tree level and one-loop
  virtual contributions.  The sum of all the contributions corresponds
  to the black solid line. The lower window shows a blow-up of the
  black solid line in the upper plot, to illustrate the stability of
  the result. The error bars indicate the statistical uncertainty of
  the Monte Carlo integration.}
\label{fig:ds_dependenceWbb}
\end{center}
\end{figure}
\begin{figure}[htb]
\begin{center}
\includegraphics*[scale=0.55,angle=-90]{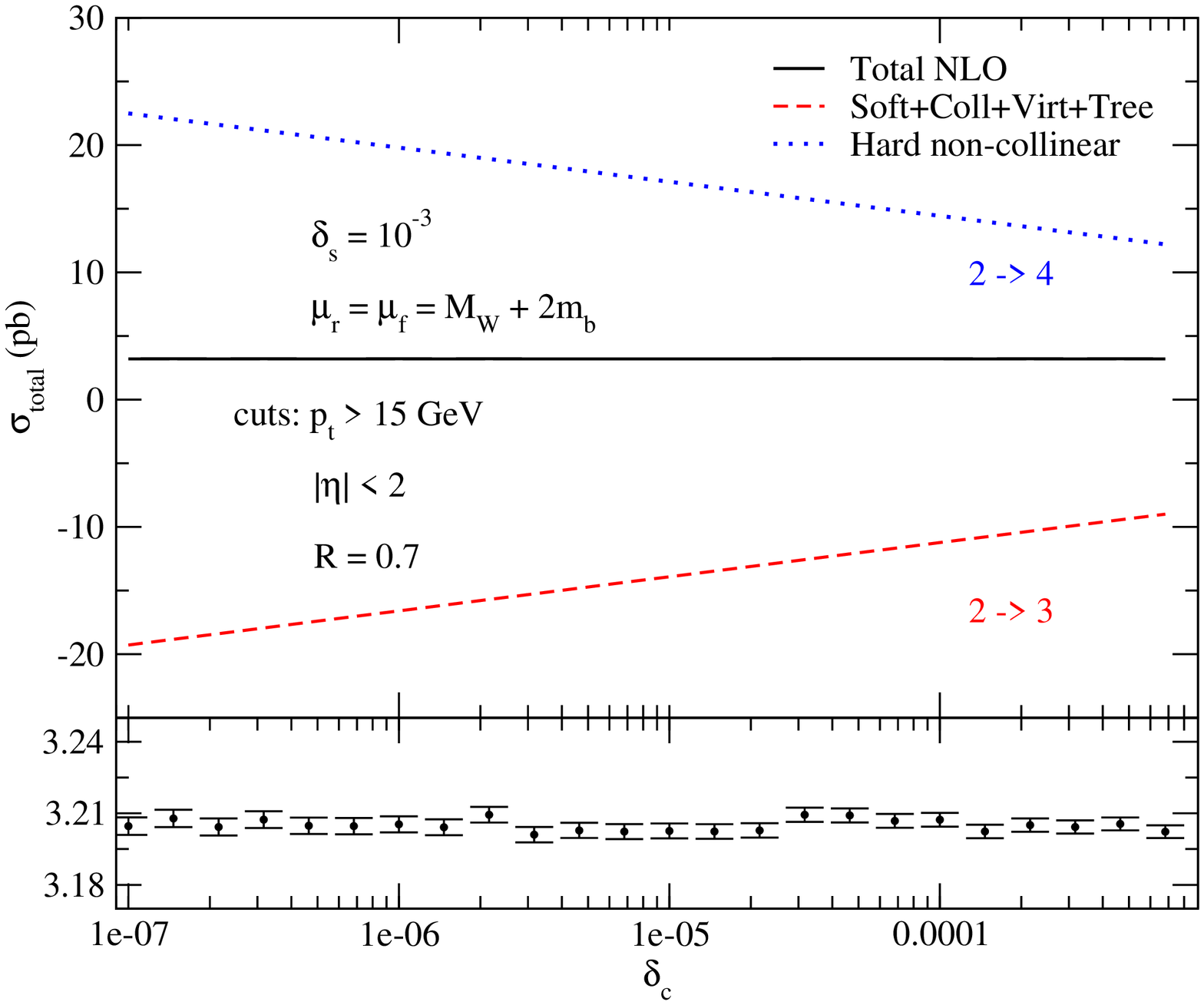}
\caption[Dependence of the total $\Wbb$ NLO QCD cross section on the
  $\delta_c$ PSS parameter.]{Dependence of $\sigma^{\rm \sss NLO}(p\bar{p}\to \Wbb)$ on the
  $\delta_c$ PSS parameter, when $\delta_s$ is fixed at
  $\delta_s=10^{-3}$. In the upper window we illustrate separately the
  cutoff dependence of the soft and hard-collinear part ($2\rightarrow
  3$, red dashed curve) and of the hard non-collinear part
  ($2\rightarrow 4$, blue dotted curve) of the real corrections to the
  total cross section. The $2\rightarrow 3$ curve also includes those
  parts of the $2\rightarrow 3$ NLO cross section that do not depend
  on $\delta_c$ and $\delta_s$, i.e. the tree level and one-loop
  virtual contributions.  The sum of all the contributions corresponds
  to the black solid line. The lower window shows a blow-up of the
  black solid line in the upper plot, to illustrate the stability of
  the result. The error bars indicate the statistical uncertainty of
  the Monte Carlo integration.}
\label{fig:dc_dependenceWbb}
\end{center}
\end{figure}

We note that $\sigma^{\rm \sss NLO}$ is finite, since, after mass factorization, both soft and
collinear singularities have been canceled between $\hat{\sigma}^{\rm virt}_{q\bar
q}+\hat{\sigma^\prime}^{soft}$ and $\hat{\sigma}^{hard/coll}$.  Note that the second term in
Eq.~(\ref{eq:sigmatot2}), which is proportional to $\ln\left(\frac{s}{\mu^2}\right)$, corresponds
exactly to the second and third terms of Eq.~(\ref{eq:fmuNLOdep}), as predicted by renormalization
group arguments.

To finish this Section, and before we discuss in detail in Chapter~\ref{chap:results} the numerical
results for the NLO total cross section for $p\bar p \to \Wbb$, we first demonstrate that
$\sigma^{\rm \sss NLO}$ does not depend on the arbitrary cutoffs of the PSS method, i.e.  on the
soft and hard/collinear cutoffs $\delta_s$ and $\delta_c$.  We note that the cancellation of the
cutoff dependence at the level of the total NLO cross section is a very delicate issue, since it
involves both analytical and numerical contributions. It is crucial to study the behavior of
$\sigma^{\rm \sss NLO}$ in a region where the cutoffs are small enough to justify the approximations
used in the analytical calculation of the IR-divergent part of $\hat\sigma^{\rm real}_{q\bar q}$,
but not so small to cause large numerical cancellations. The Monte Carlo phase space integration has
been performed using the adaptive multi-dimensional integration routine VEGAS \cite{Lepage:1977sw}.

Figures~\ref{fig:ds_dependenceWbb} and \ref{fig:dc_dependenceWbb} illustrate the dependence of the
total cross section on the two-cutoffs of the PSS method, using the setup outlined in
Section~\ref{sec:setup}. In Figure~\ref{fig:ds_dependenceWbb}, we
show the dependence of $\sigma^{\rm \sss NLO}$ on the soft cutoff, $\delta_s$, for a fixed value of
the hard/collinear cutoff, $\delta_c\!=\!10^{-5}$.  In Figure~\ref{fig:dc_dependenceWbb}, we show
the dependence of $\sigma^{\rm \sss NLO}$ on the hard/collinear cutoff, $\delta_c$, for a fixed
value of the soft cutoff, $\delta_s\!=\! 10^{-3}$.  In the upper window of
Figure~\ref{fig:ds_dependenceWbb}(\ref{fig:dc_dependenceWbb}) we illustrate the cancellation of the
$\delta_s$($\delta_c$) dependence between ${\sigma}_{soft}+{\sigma}_{hard/coll}$ and
${\sigma}_{hard/non-coll}$, while in the lower window we show, on a larger scale, ${\sigma}^{\rm
\sss NLO}$ with the statistical errors from the Monte Carlo integration.  As before, $\sigma^{\rm
\sss NLO}$ also includes the contribution from the LO and the virtual cross sections, which are both
cutoff-independent.  For $\delta_s$ in the range $10^{-5}-10^{-2}$ and $\delta_c$ in the range
$10^{-7}-10^{-3}$, a clear plateau is reached and the NLO total cross section is independent of the
arbitrary cutoffs of the two-cutoff PSS method.  All the results presented in
Chapter~\ref{chap:results} are obtained using the two-cutoff PSS method with $\delta_s=10^{-3}$ and
$\delta_c=10^{-5}$.

\begin{figure}[htb]
\begin{center}
\includegraphics[scale=0.75]{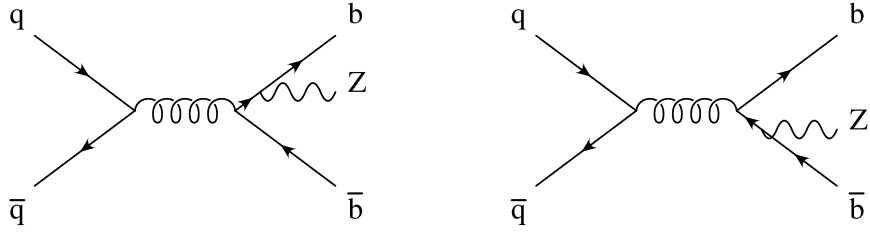}
\caption{Tree level Feynman diagrams for $q\bar{q}\to\Zbb$, with $Z$ emitted from final fermion line.}
\label{fig:finalqqZbbtree_level}
\end{center}
\end{figure}
\section[NLO QCD Corrections to $Z b\bar b$ Production at Hadron Colliders]{Calculation of NLO QCD Corrections
	to $Z b\bar b$ Production at Hadron Colliders}\label{sec:Zbbcalc}
In this Section we present in detail the calculation of the partonic total and differential cross
section ${\hat\sigma}^{\rm NLO}(ij\to\Zbb)$~\cite{FebresCordero:2008ci}, which can be decomposed as
in Eq.~(\ref{eq:nloXsec}).  We work throughout in the 4-flavor number scheme, where only 4 massless
quark flavors can be excited in the initial state, as we consider full $b$-quark mass contributions
to the partonic cross section.

There are two subprocesses contributing to ${\hat\sigma}^{\rm LO}$ in Eq.~(\ref{eq:nloXsec}), namely
$q{\bar q}\to\Zbb$ and $gg\to\Zbb$.

\begin{figure}[htb]
\begin{center}
\includegraphics[scale=0.9]{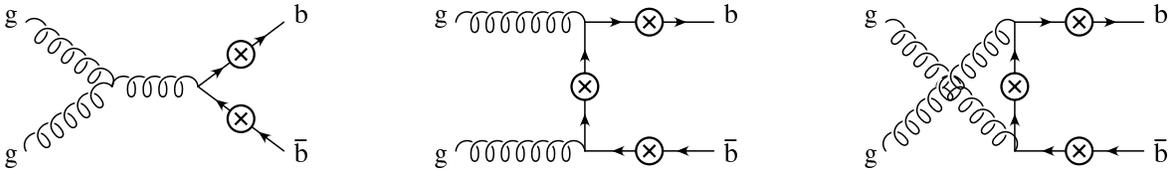}
\caption{Tree level Feynman diagrams for $gg\to\Zbb$. The circled crosses correspond to all possible insertions of the
	$Z$ boson, each one representing a different diagram.}
\label{fig:ggZbbtree_level}
\end{center}
\end{figure}

The tree level Feynman diagrams contributing to the LO cross section for subprocess $q\bar q\to\Zbb$
are shown in Figures~\ref{fig:Wbbtree_level} (with $V=Z$) and \ref{fig:finalqqZbbtree_level}, when
the $Z$ weak boson is emitted from initial and final fermion lines respectively. In
Figure~\ref{fig:ggZbbtree_level} we show the tree level Feynman diagrams contributing to the
$gg\to\Zbb$ subprocess. The corresponding amplitudes are presented in Section~\ref{sec:sigma_loZbb}
and more details are summarized in Appendix~\ref{app:LOamp}.

We notice that the tree level diagrams for $q\bar q^\prime\to\Wbb$ are a subset of the diagrams for
$q\bar q\to\Zbb$, namely the ones in which the $Z$ boson is emitted from the initial quarks legs.
The same holds at NLO: the NLO QCD corrections to $q\bar q^\prime\to\Wbb$ discussed in
Section~\ref{sec:Wbbcalc} are a subset of the NLO QCD corrections to $q\bar q\to\Zbb$ discussed in
this Section.  The results derived for $q\bar q^\prime\to\Wbb$ can be automatically translated to
the $q\bar q\to\Zbb$ case by keeping the contributions of the vector and axial vector parts of the
$Wq\bar q^\prime$ vertex separate and substituting the corresponding values of vector and axial
vector couplings of the $Zq\bar q$ vertex, as well as making the necessary changes to the PDFs of
the initial partons in the hadronic cross section. 

At NLO one has to consider five subprocesses: $q{\bar q}\to\Zbb$ and $gg\to\Zbb$ contributing at
tree level to ${\hat\sigma}^{\rm LO}$ in Eq.~(\ref{eq:nloXsec}) and at one-loop to
${\hat\sigma}^{\rm virt}$ in Eq.~(\ref{eq:virtrealXsec}), as well as $q{\bar q}\to\Zbb+g$, $q(\bar
q)g\to\Zbb+q(\bar q)$ and $gg\to\Zbb+g$ contributing at tree level to ${\hat\sigma}^{\rm real}$ in
Eq.~(\ref{eq:virtrealXsec}). 

The rest of this Section is organized as follows. We present in Section~\ref{sec:sigma_loZbb}
results for the LO amplitudes. In Sections~\ref{subsec:virtqqZbb} and \ref{subsec:virtggZbb} we
shall present results for the ${\cal O}(\as)$ virtual corrections to $\Zbb$ hadronic production.
For details on the ${\cal O}(\as^3)$ subprocess $q\bar q\to\Zbb+g$ we refer the reader to
Sections~\ref{subsec:realWbb} and \ref{subsec:totalWbb}, as results are analogous to the ${\cal
O}(\as^3)$ subprocess $q\bar q^\prime\to\Wbb+g$.  In Sections~\ref{subsec:realZbb} and
Section~\ref{subsec:totalZbb} we shall discuss the ${\cal O}(\as^3)$ subprocess $gg\to\Zbb+g$.
${\cal O}(\as^3)$ real subprocesses initiated by $q(\bar q)g$ for $\Zbb$ are slightly different in
structure with respect to the similar ones encountered in the $\Wbb$, as the LO structure of the
former is more complex than the latter.  For that reason we discuss their structure in detail in
Sections~\ref{subsec:realZbb} and ~\ref{subsec:totalZbb}.

\boldmath
\subsection{Tree Level Cross Section for $\Zbb$ Hadronic Production}
\unboldmath
\label{sec:sigma_loZbb}

The contributing tree level Feynman diagrams for the LO $q\bar q\to\Zbb$ process are shown in
Figures~\ref{fig:Wbbtree_level} (with $V=Z$) and \ref{fig:finalqqZbbtree_level} for subprocess
$q\bar q\to\Zbb$ with the $Z$ weak boson emitted from initial and final fermion lines, respectively.
Given the assignment of momenta:
\[
q(q_1)\bar q(q_2)\to b(p_b)+\bar{b}(p_{\bar b})+Z(p_{\sss Z})\,\,\,,
\]
the LO amplitude can be written as:
\begin{eqnarray}
{\cal A}_0(q\bar q\to\Zbb) & = & ig_s^2\frac{g_{\sss W}}{\cos{\theta_{\sss W}}}\;\; \epsilon_\mu^*(p_{\sss
Z})\;\; \frac{g_{\nu\rho}}{(p_b+p_{\bar b})^2}\;\; \bar{u}_b \gamma^\rho v_{\bar{b}}t^a_{ij}t^a_{kl} \nonumber\\
	& & \Bigl[\bar v_{\bar q}\Zvertf{\mu}{q}\frac{-\Slash{q}_2+\Slash{p}_{\sss Z}}{(-q_2+p_{\sss
	Z})^2}\gamma^\nu u_q \nonumber\\
	& & +\bar v_{\bar q}\gamma^\nu\frac{\Slash{q}_1-\Slash{p}_{\sss Z}}{(q_1-p_{\sss
	Z})^2}\Zvertf{\mu}{q} u_q \Bigl]\; \nonumber\\
&  &	+ ig_s^2\frac{g_{\sss W}}{\cos{\theta_{\sss W}}}\;\; \epsilon_\mu^*(p_{\sss
Z})\;\; \frac{g_{\nu\rho}}{(q_1+q_2)^2}\;\; \bar{v}_{\bar q} \gamma^\rho u_{q}t^a_{ij}t^a_{kl} \nonumber\\
	&& \left[ \bar{u}_b\Zvertf{\mu}{b}\frac{\Slash{p}_b+\Slash{p}_{\sss Z}+m_b}{\left[(p_b+p_{\sss
	Z})^2-m_b^2\right]} \gamma^\nu v_{\bar{b}} \right.\nonumber\\
	&& +\left. \bar{u}_b\gamma^\nu\frac{-\Slash{p}_{\bar b}-\Slash{p}_{\sss Z}+m_b}{\left[(-p_{\bar b}-p_{\sss
	Z})^2-m_b^2\right]} \Zvertf{\mu}{b} v_{\bar{b}} \right].
\label{eq:LOqqZbb}
\end{eqnarray}
where $g_s$ and $g_{\sss W}$ are the strong and weak coupling constant, respectively,
$t^{a}=\lambda^{a}/2$ are given in terms of the Gell-Mann matrices $\lambda^{a}$ and $\theta_{\sss
W}$ is the weak angle. The vector, $g_V^f$, and axial, $g_A^f$, couplings for the $Zff$ vertex are
given explicitly in Eq.~(\ref{eq:VAcoup}) (for more details see Appendices~\ref{app:SMint} and
\ref{app:LOamp}).

\[
\cdots
\]

The tree level amplitude for the process
\[
g^a(q_1)+g^b(q_2)\to b(p_b)+\bar{b}(p_{\bar b})+Z(p_Z) \; ,
\]
where $q_1+q_2=p_b+p_{\bar b} +p_Z$ and $a,b$ denote the color of the incoming gluons, is obtained
from the three classes of Feynman diagrams represented in Figure~\ref{fig:ggZbbtree_level},
identified as $s-$channel, $t-$channel, and $u-$channel diagrams, respectively. We find it
convenient to organize the color structure of both the tree level amplitude and the one-loop virtual
amplitude in terms of only two color factors, one symmetric and one antisymmetric in the color
indices of the initial gluons.  Following this prescription, the tree level amplitude for $gg\to
\Zbb$ can be written as:
\begin{equation}
  \label{eq:amp_treeggZbb}
  {\cal A}_0={\cal A}_0^{nab}[t^a,t^b]+{\cal A}_0^{ab}\{t^a,t^b\}\,\,\,,
\end{equation}
where $t^{a,b}=\lambda^{a,b}/2$ are given in terms of the Gell-Mann matrices
$\lambda^{a,b}~$\footnote{We note that the one-loop virtual amplitude can be expressed in terms of
the same antisymmetric color factor $[t^a,t^b]$ and a symmetric color factor made of $\{t^a,t^b\}$
and $\delta^{ab}$.}. ${\cal A}_0^{ab}$ and ${\cal A}_0^{nab}$ correspond to the terms in the
amplitude that are proportional, respectively, to the \emph{abelian} (or symmetric) and
\emph{non-abelian} (or antisymmetric) color factors and are given by:
\begin{equation}
  \label{eq:a0_ab_nabggZbb}
  {\cal A}_0^{ab}=\frac{1}{2}({\cal A}_{0,t}+{\cal A}_{0,u})\,\,\,,\,\,\,
  {\cal A}_0^{nab}={\cal A}_{0,s}+\frac{1}{2}({\cal A}_{0,t}-{\cal A}_{0,u})\,\,\,,
\end{equation}
where ${\cal A}_{0,s}$, ${\cal A}_{0,t}$, and ${\cal A}_{0,u}$ are the amplitudes corresponding to
the sum of the $s-$channel, $t-$channel, and $u-$channel tree level diagrams in
Figure~\ref{fig:ggZbbtree_level}. Explicit expressions for them are shown in
Eqs.~(\ref{eq:a0_stu})-(\ref{eq:A0_stu_munu_123}). 

Because of the \emph{orthogonality} between symmetric and antisymmetric color factors, the tree
level amplitude squared takes the very simple form:
\begin{equation}
\label{eq:a0_squareggZbb}
\overline{\sum}|{\cal A}_0|^2=
\overline{\sum}\left[\frac{N}{2}(N^2-1)\left(
|{\cal A}_0^{nab}|^2+|{\cal A}_0^{ab}|^2\right)-
\frac{1}{N}(N^2-1)|{\cal A}_0^{ab}|^2
\right]\,\,\,.
\end{equation}
The partonic LO cross section is obtained by integrating $|{\cal A}_0|^2$ over the $\Zbb$ final
state phase space, as shown in Eq.~(\ref{eq:LOXsec}). We perform the sum over polarizations of the
$Z$ gauge boson using the prescription shown in Eq.~(\ref{eq:polsum}), keeping both terms.

When averaging over the polarization states of the initial gluons (in the $gg\to\Zbb$ case), the
polarization sum of the gluon polarization vectors, $\epsilon_{\mu}(q_1,\lambda_1)$ and
$\epsilon_{\nu}(q_2,\lambda_2)$, has to be performed in such a way that only the physical
(transverse) polarization states of the gluons contribute to the matrix element squared. We adopt
the general prescription:
\begin{equation}
\sum_{\lambda_i=1,2}\epsilon_{\mu}(q_i,\lambda_i) 
\epsilon_{\nu}^*(q_i,\lambda_i)=
-g_{\mu\nu}+\frac{n_{i\mu}q_{i\nu}+q_{i\mu}n_{i\nu}}{n_i\cdot q_i}-
\frac{n_i^2q_{i\mu}q_{j\nu}}{(n_i\cdot q_i)^2}\,\,\,,
\end{equation}
where $i\!=\!1,2$ and the arbitrary vectors $n_i$ have to satisfy the relations:
\begin{equation}
n_i^\mu\sum_{\lambda_i=1,2} \epsilon_{\mu}(q_i,\lambda_i) 
\epsilon_{\nu}^*(q_i,\lambda_i)=0\,\,\,\,,\,\,\,\,
n_i^\nu\sum_{\lambda_i=1,2} \epsilon_{\mu}(q_i,\lambda_i) 
\epsilon_{\nu}^*(q_i,\lambda_i)=0\,\,\,,
\end{equation}
together with $n_i\cdot q_j\!\neq\!0$ and $n_1\!\neq\! n_2$. We choose
$n_1\!=\!q_2$ and $n_2\!=\!q_1$, such that:
\begin{equation}
\sum_{\lambda_i=1,2}\epsilon_{\mu}(q_i,\lambda_i) 
\epsilon_{\nu}^*(q_i,\lambda_i)=
-g_{\mu\nu} +\frac{q_{1\mu} q_{2\nu}+q_{2\mu} q_{1\nu}}{q_1\cdot q_2}\,\,\,.
\end{equation}
Finally, the entire calculation is performed using the Feynman gauge for both internal and external
gluons (for more details see Appendix~\ref{app:SMint}).

\begin{figure}[ht]
\begin{center}
\includegraphics[scale=0.65]{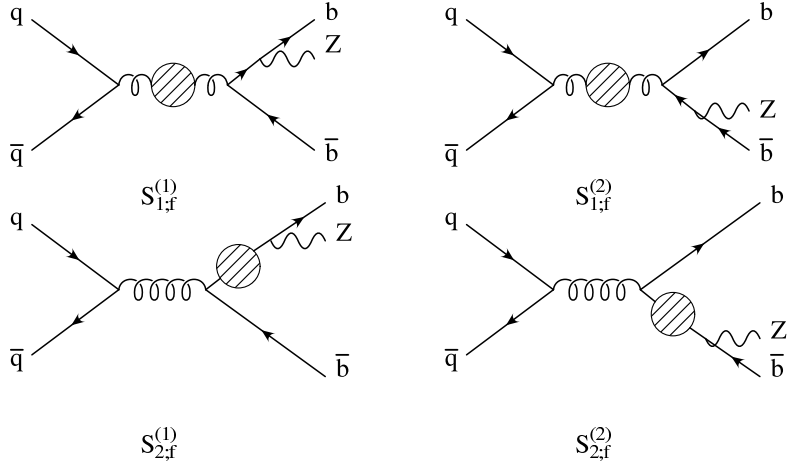}
\caption{Gluon ($S_{1\rm ; f}^{(1,2)}$) and $b$-quark ($S_{2\rm ;f}^{(1,2)}$) ${\cal O}(\as)$ self energy  corrections contributing to 
	the $q{\bar q}\to Zb\bar b$ subprocess at NLO, when the $Z$ boson is emitted from the final fermion line. The
	shaded blobs denote standard one-loop QCD corrections to the gluon and quark propagators respectively.}
\label{fig:selffinalqqZbb}
\end{center}
\end{figure}
\begin{figure}[htp]
\begin{center}
\includegraphics[scale=0.9]{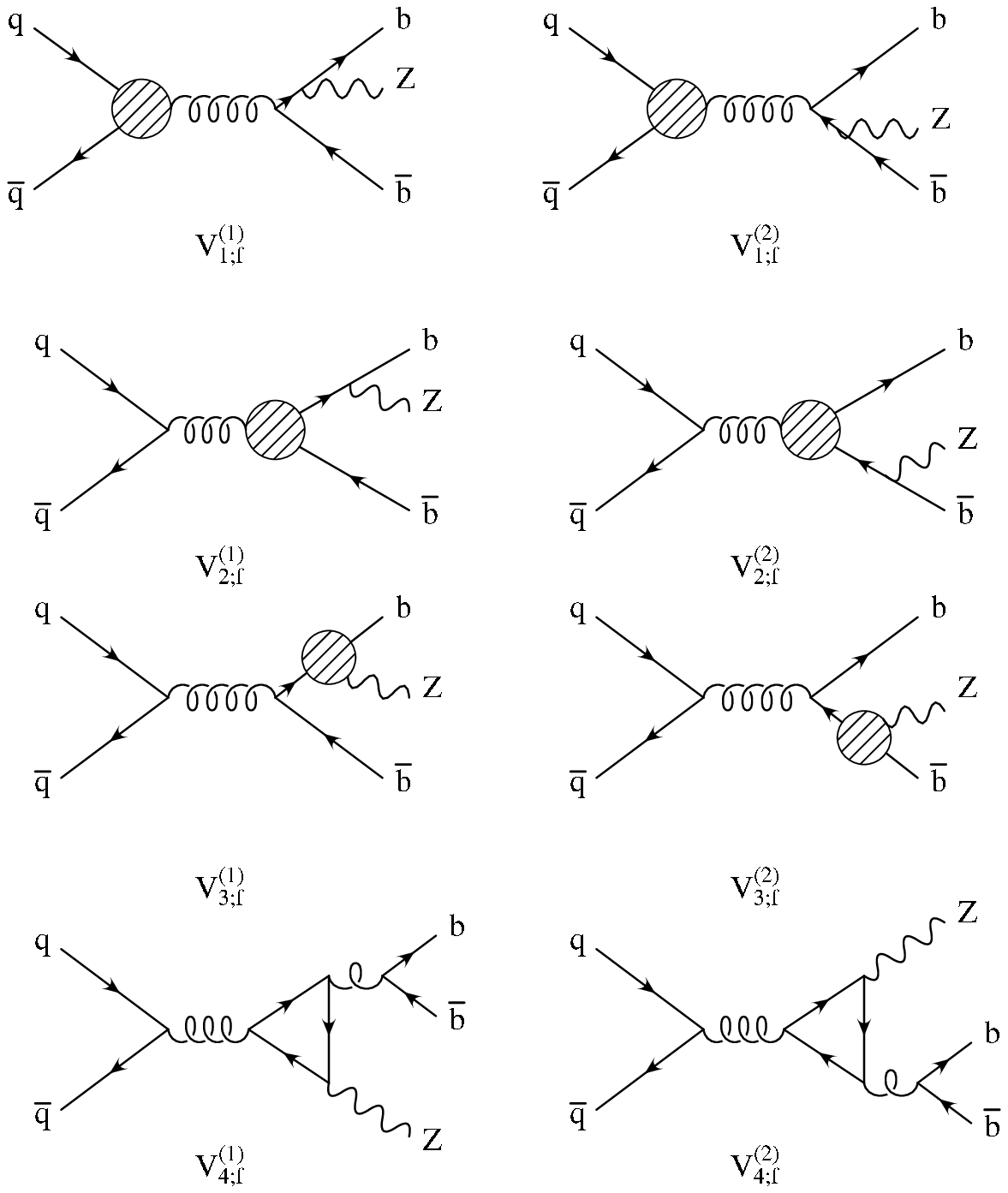}
\caption{${\cal O}(\as)$ vertex corrections contributing to the $q{\bar q}\to Zb\bar b$ subprocess at NLO when
	the $Z$ boson is emitted from the final fermion line
	or from a closed fermion line. The shaded blobs denote standard one-loop QCD
	corrections to the $q\bar qg$ ($V_{1\rm ;f}^{(1,2)}$), $b\bar bg$ ($V_{2\rm ;f}^{(1,2)}$) and
	$q{\bar q}Z$ ($V_{3\rm ;f}^{(1,2)}$) vertices respectively. $V_{4\rm ;f}^{(1,2)}$ are $b$- and $t$-fermion loop
	vertices which are UV and IR finite (contributions of quarks from first and second family vanish).}
\label{fig:vertfinalqqZbb}
\end{center}
\end{figure}
\boldmath
\subsection{Virtual Corrections to $q{\bar q}\to\Zbb$}\label{subsec:virtqqZbb}
\unboldmath

\begin{figure}[ht]
\begin{center}
\includegraphics[scale=0.63]{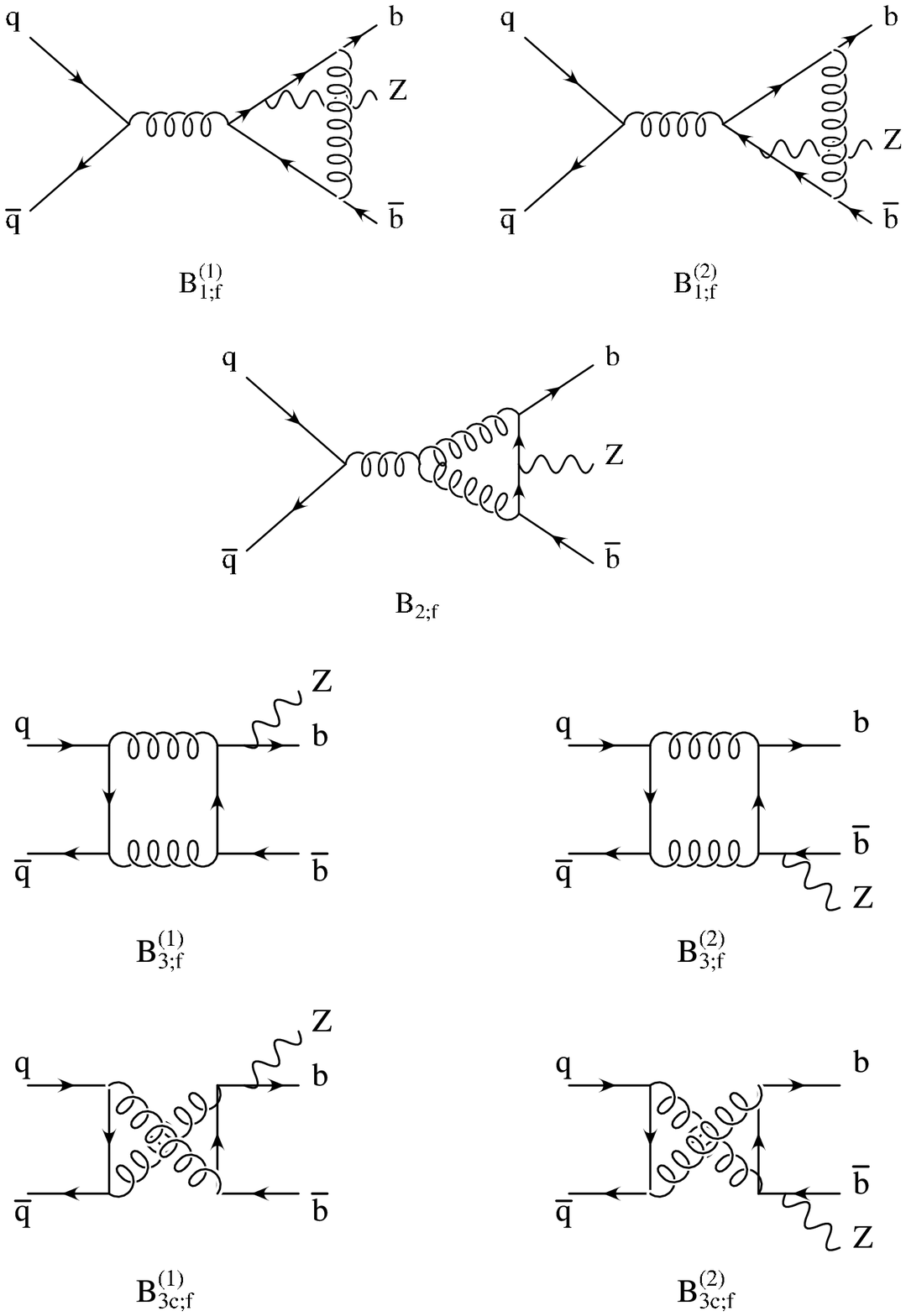}
\caption{${\cal O}(\as)$ box diagram corrections contributing to the $q{\bar q}\to\Zbb$ subprocess at NLO,
	when the $Z$ boson emitted from	the final fermion lines ($b$ or $\bar b$).}
\label{fig:boxfinalqqZbb}
\end{center}
\end{figure}

\begin{figure}[ht]
\begin{center}
\includegraphics[scale=0.7]{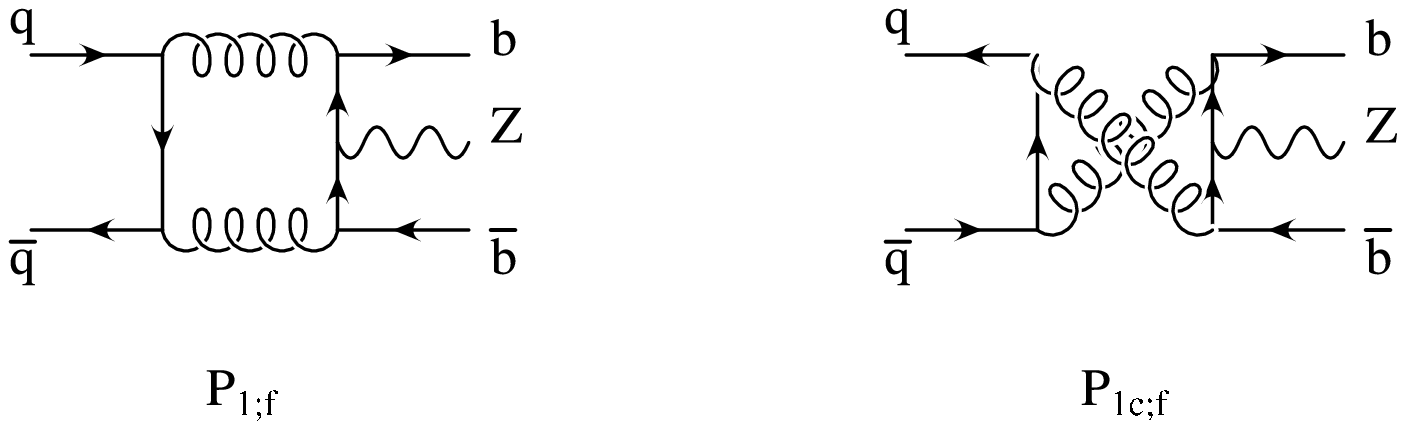}
\caption{${\cal O}(\as)$ pentagon diagram corrections contributing to the $q{\bar q}\to\Zbb$ subprocess at
	NLO, when the $Z$ boson emitted from the final fermion lines ($b$ or $\bar b$).}
\label{fig:pentfinalqqZbb}
\end{center}
\end{figure}

The ${\cal O}(\as)$ virtual corrections to the $q{\bar q}\to \Zbb$ tree level subprocess consist of
the self energy, vertex, box and pentagon diagrams illustrated in
Figures~\ref{fig:selfWbb}-\ref{fig:pentWbb}, when the $Z$ weak boson is emitted from the initial
fermion line ($q$ or $\bar q$), and in Figures~\ref{fig:selffinalqqZbb}-\ref{fig:pentfinalqqZbb},
when it is emitted from the final fermion line ($b$ or $\bar b$). The contributions to the virtual
amplitude squared of Eq.~(\ref{eq:virtrealXsec}) can then be written as:
\begin{equation}
{\overline \sum}|{\cal A}_{\rm virt}(q{\bar q}\to \Zbb)|^2 
	= \sum_{D_i}{\overline \sum}\left( {\cal A}_0{\cal A}_{D_i}^\dag + {\cal A}_0^\dag{\cal A}_{D_i}\right)
	= \sum_{D_i}{\overline \sum}2{\cal R}e\left( {\cal A}_0{\cal A}_{D_i}^\dag \right),
\label{eq:virtqqZbb}
\end{equation}
where ${\cal A}_0$ is the tree level amplitude, corresponding to the diagrams shown in
Figures~\ref{fig:Wbbtree_level} and \ref{fig:finalqqZbbtree_level}, and ${\cal A}_{D_i}$ denotes the
amplitude for a one-loop diagram, with $D_i$ running over all self-energy, vertex, box and pentagon
diagrams illustrated in Figures~\ref{fig:selfWbb}, \ref{fig:vertWbb}, \ref{fig:boxWbb},
\ref{fig:pentWbb}, \ref{fig:selffinalqqZbb}, \ref{fig:vertfinalqqZbb}, \ref{fig:boxfinalqqZbb} and
\ref{fig:pentfinalqqZbb}.

The calculation of each virtual diagram (${\cal A}_{D_i}$) is performed in the way explained in
Sections~\ref{subsec:sigvirt} and \ref{subsec:virtWbb}. As mentioned there, one of the most
challenging parts of the calculation is related to controlling the spurious divergences that appear
when reducing tensor Feynman integrals. This is specially true when considering pentagon diagrams,
like the ones in Figures~\ref{fig:pentWbb} and \ref{fig:pentfinalqqZbb}. There were only two
pentagon diagrams in the $\Wbb$ calculation, but that number is doubled just by considering the
one-loop diagrams for $q{\bar q}\to\Zbb$, and we will see in Section~\ref{subsec:virtggZbb} that
when considering one-loop diagrams for $gg\to\Zbb$ one has to add twelve more pentagon diagrams,
some of them containing up to $E4$-PV functions (see Appendix~\ref{app:Intred}). All this increases
considerably the stability problem.  In the case of pentagon diagrams it is convenient to reduce
consistently all $E$-PV functions by canceling systematically, at the level of the amplitude squared
in Eq.~(\ref{eq:virtqqZbb}), all possible vector products containing the loop momentum in the
numerator with some denominators. This is possible as, in the pentagon topology of our process, each
leg has an outgoing momentum which is on-shell, corresponding basically to one of the external
initial or final particles of the subprocess. One then ends with expressions for each pentagon
diagram containing purely scalar pentagon integrals, or tensor integrals with fewer than five
denominators, improving considerably the behavior of the numerical code. We checked analytically
these reductions with the non-reduced expressions by using the full reduction of all tensor
integrals to scalar integrals, and found agreement. Another gain we had by doing the aforementioned
reductions was a considerable speed up of all analytical and numerical computations.

In Section~\ref{subsubsec:UVqqZbb} we present the UV singularity structure of $\hat\sigma^{\rm
virt}(q\bar q\to\Zbb)$ as it has a slightly different structure to what is discussed in
Sections~\ref{subsubsec:UVWbb} due to the $Z$ boson emission from the final fermion lines.  The IR
divergent structure on the other side is analogous to that presented in
Section~\ref{subsubsec:IRvirtWbb}, to which we refer the reader for details. 

\subsubsection{Virtual corrections: UV singularities and counterterms}\label{subsubsec:UVqqZbb}

The UV singularities of the ${\cal O}(\alpha_s^3)$ total cross section for $q\bar q\to\Zbb$
originate from self-energy and vertex virtual corrections shown in Figures~~\ref{fig:selfWbb},
\ref{fig:vertWbb}, \ref{fig:selffinalqqZbb} and \ref{fig:vertfinalqqZbb}.  These singularities are
renormalized by introducing counterterms for the wave function of the external fields ($\delta
Z_2^{(q)}$, $\delta Z_2^{(b)}$), the bottom-quark mass ($\delta m_b$), and the strong coupling
constant ($\delta Z_{\alpha_s}$).  If we denote by $\Delta_{\rm UV}(S_i^{(1,2)})$ and $\Delta_{\rm
UV}(V_i^{(1,2)})$ the UV-divergent contribution of each self-energy ($S_i^{(1,2)}$) or vertex
diagram ($V_i^{(1,2)}$) to the virtual amplitude squared (see Eq.~(\ref{eq:virtqqZbb})), we can
write the UV-singular part of the total virtual amplitude squared as:
\begin{eqnarray}
\overline{\sum}|{\cal A}_{\rm virt}^{\sss \rm UV}|^2&=& \overline{\sum} 
|{\cal A}_0|^2
\,\left\{\sum_{i=1}^2\Delta_{\rm UV}\left(S_i^{(1)}+S_i^{(2)}
		+S_{i\rm ;f}^{(1)}+S_{i\rm ;f}^{(2)}\right)\right.\nonumber\\
&+&\sum_{i=1}^3\Delta_{\rm UV}\left(V_i^{(1)}+V_i^{(2)}
		+V_{i\rm ;f}^{(1)}+V_{i\rm ;f}^{(2)}\right)\nonumber\\
&+&\left.2\,\left[\,\left(\delta Z_2^{(q)}\right)_{\rm UV}+
\left(\delta Z_2^{(b)}\right)_{\rm UV}+
\delta Z_{\alpha_s}\right]\right\}\,\,\,.
\label{eq:amp2_virt_uvqqZbb}
\end{eqnarray}
We denote by $|{\cal A}_0|^2$ the matrix element squared of the tree-level amplitude for $q\bar
q\rightarrow \Zbb$, computed in $d=4$ dimensions, as presented in Eq.~(\ref{eq:LOqqZbb}).  

The UV-divergent contributions due to the individual diagrams are explicitly given by:
\begin{eqnarray}
\Delta_{\rm UV}\left(S_1^{(1)}+S_1^{(2)}
		+S_{1\rm ;f}^{(1)}+S_{1\rm ;f}^{(2)}\right)&=&
\frac{\alpha_s}{2\pi}\biggl[ {\cal N}_s
\left(\frac{5}{3}N-\frac{2}{3}n_{lf}\right)-{\cal N}_b\frac{2}{3}\biggr]
\biggl({1\over \epsilon_{\sss \rm UV}}\biggr)\,\,\,,
\nonumber\\
\Delta_{\rm UV}\left(S_2^{(1)}+S_2^{(2)}
		+S_{2\rm ;f}^{(1)}+S_{2\rm ;f}^{(2)}\right)&=& 
\!\!\!-\frac{\alpha_s}{2\pi}{\cal N}_b\left(\frac{N}{2}-\frac{1}{2N}\right)
\biggl({1\over \epsilon_{\sss \rm UV}}\biggr)\,\,\,,\nonumber\\
\Delta_{\rm UV}\left(V_1^{(1)}+V_1^{(2)}
		+V_{1\rm ;f}^{(1)}+V_{1\rm ;f}^{(2)}\right)&=& \frac{\alpha_s}{2\pi}
{\cal N}_s\left(\frac{3N}{2}-{1\over 2 N}\right)
\biggl({1\over \epsilon_{\sss \rm UV}}\biggr)\,\,\,,\nonumber\\
\Delta_{\rm UV}\left(V_2^{(1)}+V_2^{(2)}
		+V_{2\rm ;f}^{(1)}+V_{2\rm ;f}^{(2)}\right)&=& \frac{\alpha_s}{2\pi}
{\cal N}_b\left(\frac{3N}{2}-{1\over 2N}\right)
\biggl({1\over \epsilon_{\sss \rm UV}}\biggr)\,\,\,,\nonumber\\
\Delta_{\rm UV}\left(V_3^{(1)}+V_3^{(2)}
		+V_{3\rm ;f}^{(1)}+V_{3\rm ;f}^{(2)}\right)&=& \frac{\alpha_s}{2\pi}
{\cal N}_b\left(\frac{N}{2}-\frac{1}{2N}\right)
\biggl({1\over \epsilon_{\sss \rm UV}}\biggr)\,\,\,,
\label{eq:virtual_uvqqZbb}
\end{eqnarray}
where ${\cal N}_s$ and ${\cal N}_b$ are standard normalization factors defined in
Eq.~(\ref{eq:nsnb}).

We define the required counterterms according to the following convention.  For the external fields,
we fix the wave-function renormalization constants of the external fields ($Z_2^{(i)}=1+\delta
Z_2^{(i)}$, $i\!=\!q,b$) using on-shell subtraction, given expressions as in Eq.~(\ref{eq:z2_ct}).
We notice that both $\delta Z_2^{(q)}$ and $\delta Z_2^{(b)}$, as well as some of the vertex
corrections ($V_1^{(1,2)}, V_{1;f}^{(1,2)}$ and $V_2^{(1,2)}, V_{2;f}^{(1,2)}$), have also IR
singularities. In this section, we limit the discussion to the UV singularities only, while the IR
structure of these terms are analogous to what is shown in Section~\ref{subsubsec:IRvirtWbb}. 

We define the subtraction condition for the bottom-quark mass $m_b$ in such a way that $m_b$ is the
pole mass, in which case the bottom-mass counterterm is given by:
\begin{equation}
\label{eq:mb_ct}
\frac{\delta m_b}{m_b}=-\frac{\alpha_s}{4\pi}{\cal N}_b
\left(\frac{N}{2}-\frac{1}{2N}\right)
\left(\frac{3}{\epsilon_{\sss UV}}+4\right)\,\,\,,
\end{equation}
this counterterm have been already included into the term $\Delta_{\rm
UV}\left(S_2^{(1)}+S_2^{(2)}+S_{2\rm ;f}^{(1)}+S_{2\rm ;f}^{(2)}\right)$ in
Eq.~(\ref{eq:virtual_uvqqZbb}).

The structure of the remaining counterterms that appear in this case is the same as presented in
Section~\ref{subsubsec:UVWbb}, and we refer the reader to it.
\[
\]

\begin{figure}[ht]
\begin{center}
\includegraphics[scale=0.8]{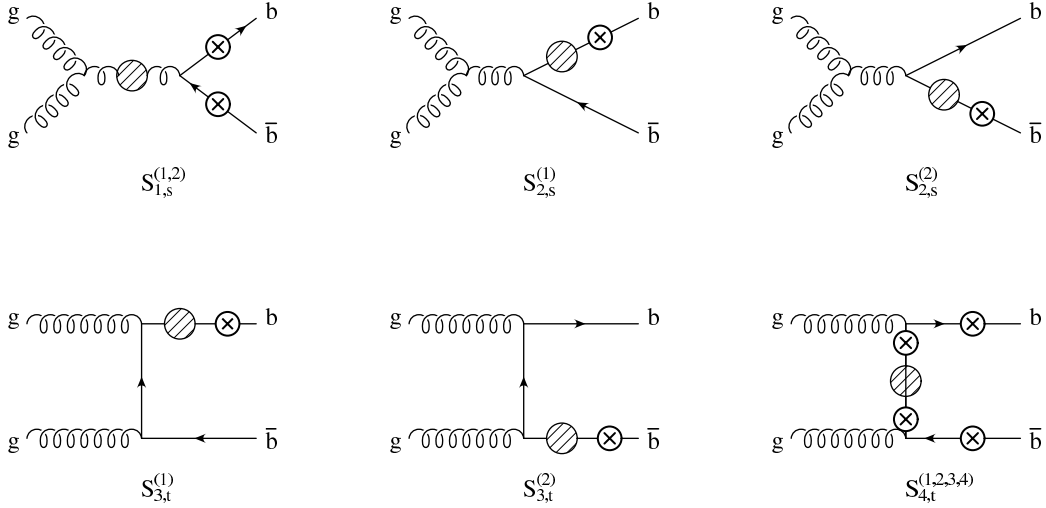}
\caption{Gluon ($S_{1,s}^{(1,2)}$) and quark ($S_{1,s}^{(1,2)}$, $S_{2,s}^{(1,2)}$, $S_{3,(t,u)}^{(1,2)}$,
	and $S_{4,(t,u)}^{(1,2,3,4)}$) ${\cal O}(\as)$ self energy  corrections contributing to the
	$gg\to Zb\bar b$ subprocess at NLO. The circled crosses correspond to all possible insertions of the
	$Z$ boson, each one representing a different diagram.}
\label{fig:selfggZbb}
\end{center}
\end{figure}

\begin{figure}[hp]
\begin{center}
\includegraphics[scale=0.8]{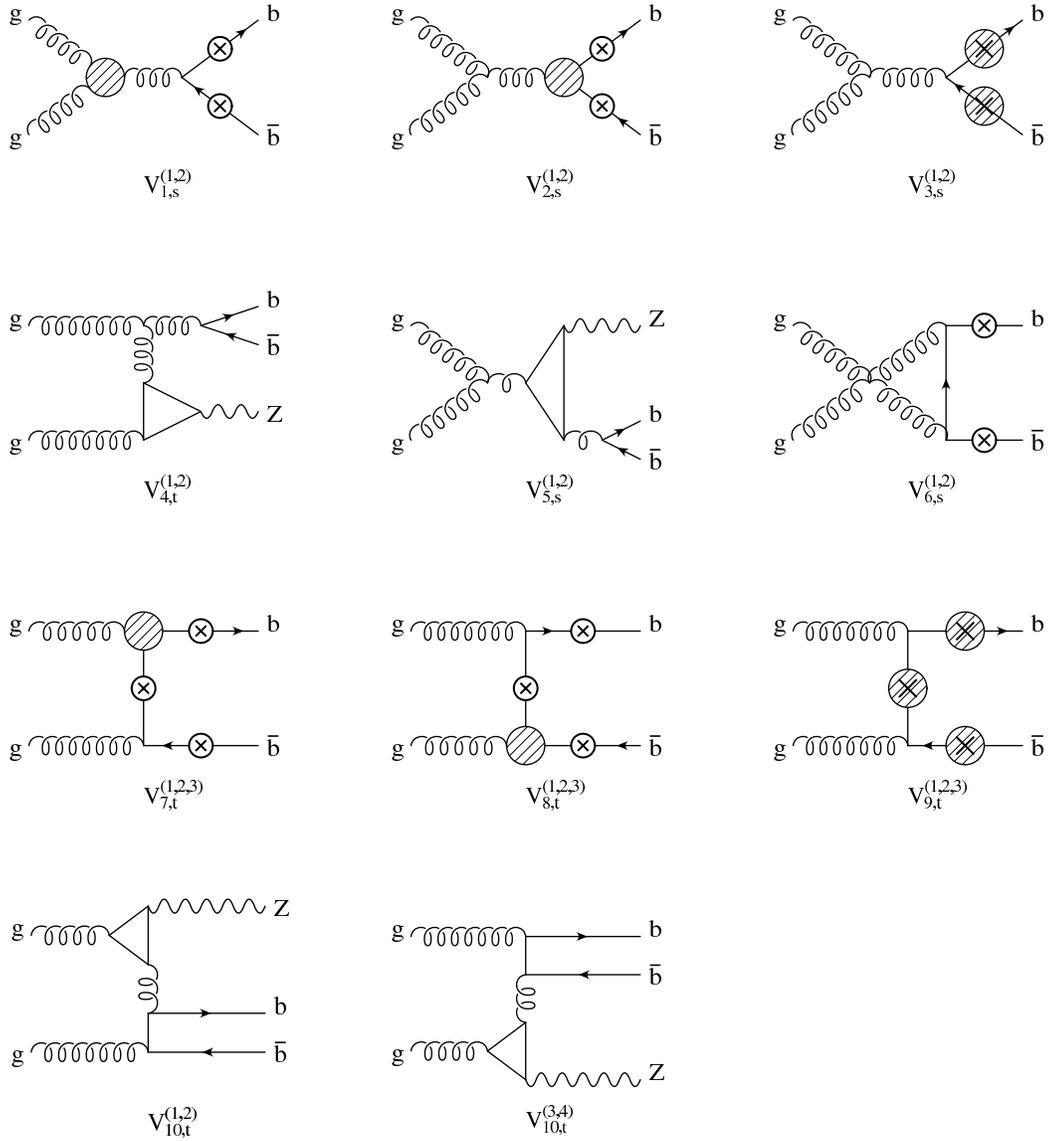}
\caption{${\cal O}(\as)$ vertex corrections contributing to the $gg\to\Zbb$ subprocess at NLO. The shaded
	blobs denote standard one-loop QCD corrections to the $ggg$ ($V_{1,s}^{(1,2)}$), $b\bar bg$
	($V_{2,s}^{(1,2)}$, $V_{7,(t,u)}^{(1,2,3)}$, and $V_{8,(t,u)}^{(1,2,3)}$) and $b{\bar b}Z$
	($V_{3,s}^{(1,2)}$ and $V_{9,(t,u)}^{(1,2,3)}$) vertices.  
	The circled crosses correspond to all possible insertions of the
	$Z$ boson, each one representing a different diagram.}
\label{fig:vertggZbb}
\end{center}
\end{figure}
\boldmath
\subsection{Virtual Corrections to $gg\to\Zbb$}\label{subsec:virtggZbb}
\unboldmath

\begin{figure}[htp]
\begin{center}
\includegraphics[scale=0.74]{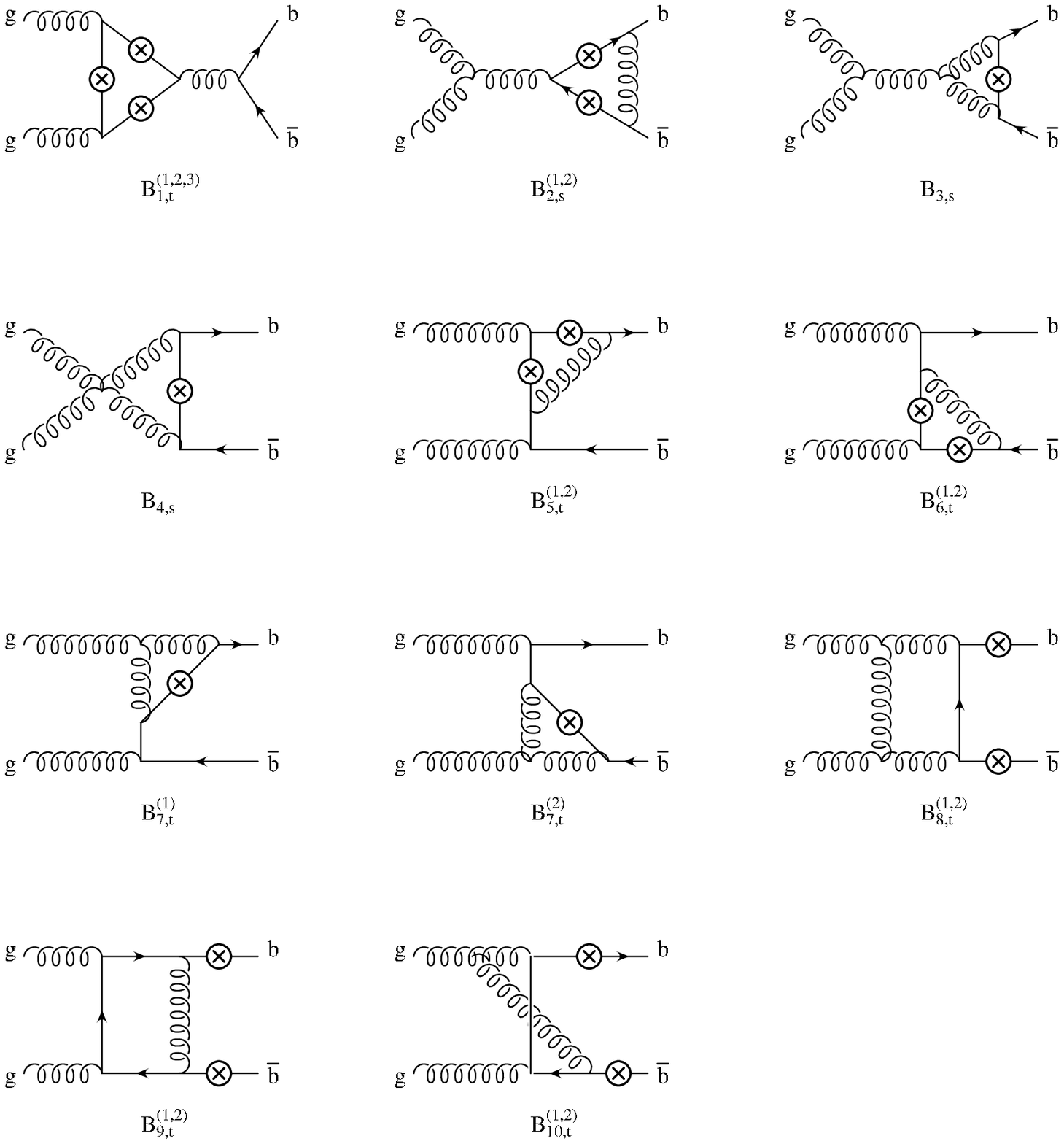}
\caption{${\cal O}(\as)$ box diagram corrections contributing to the $gg\to\Zbb$ subprocess at NLO.
	The circled crosses correspond to all possible insertions of the
	$Z$ boson, each one representing a different diagram.}
\label{fig:boxggZbb}
\end{center}
\end{figure}

\begin{figure}[ht]
\begin{center}
\includegraphics[scale=0.9]{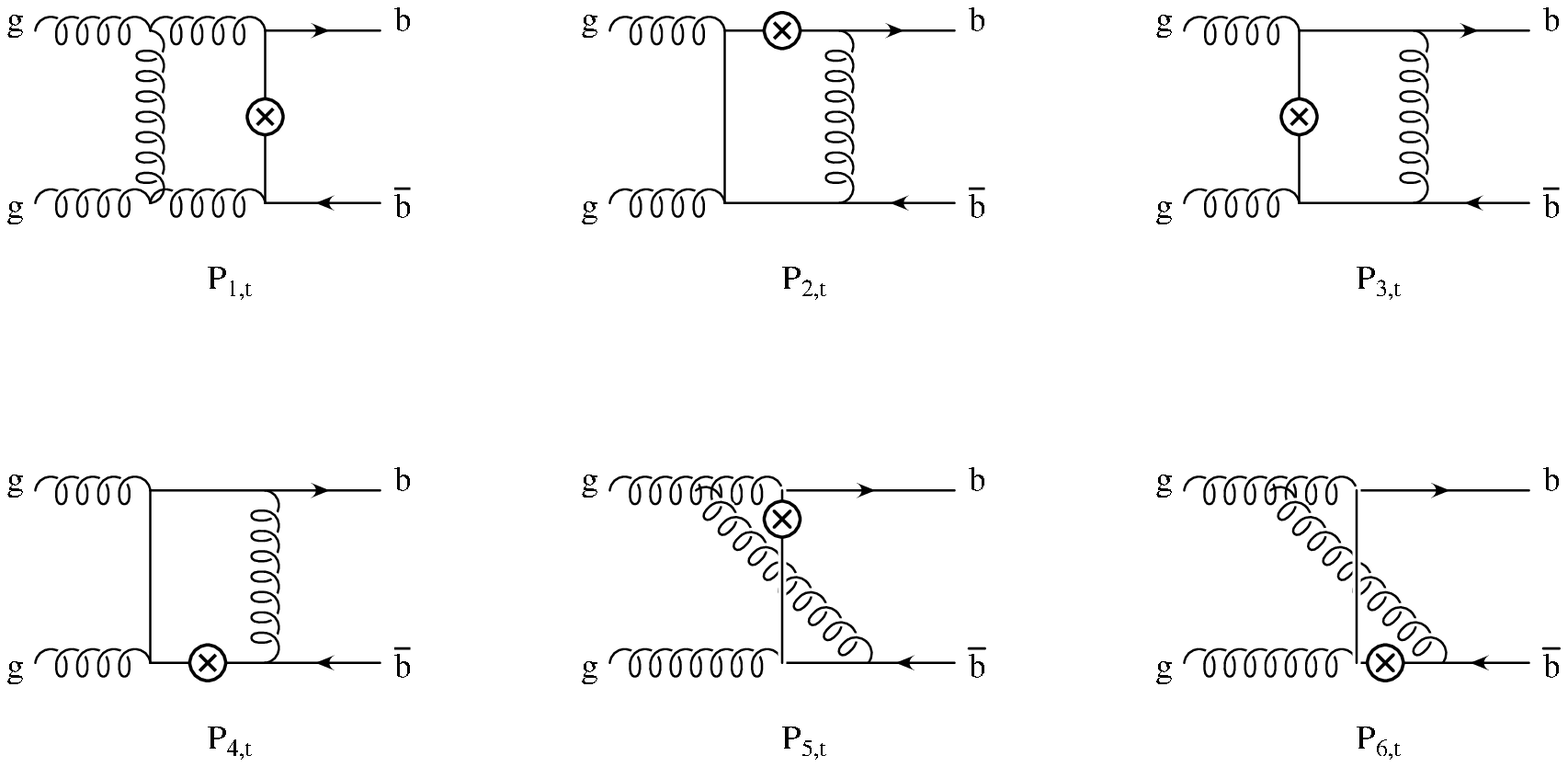}
\caption{${\cal O}(\as)$ pentagon diagram corrections contributing to the $gg\to\Zbb$ subprocess at NLO.
	The circled crosses correspond to all possible insertions of the
	$Z$ boson, each one representing a different diagram.}
\label{fig:pentggZbb}
\end{center}
\end{figure}

The ${\cal O}(\alpha_s)$ virtual corrections to the $gg\to \Zbb$ tree level subprocess consist of
the self-energy, vertex, box, and pentagon diagrams illustrated in
Figures~\ref{fig:selfggZbb}-\ref{fig:pentggZbb}. The ${\cal O}(\alpha_s^3)$ contribution to the
virtual amplitude squared of Eq.~(\ref{eq:virtrealXsec}) can then be written as:
\begin{equation}
\label{eq:amp2_virt_genggZbb}
\overline{\sum}|{\cal A}_{\rm virt}(gg\to \Zbb)|^2=
\sum_{D_{i,j}}\overline{\sum}\left({\cal A}_0 {\cal A}_{D_{i,j}}^*+
{\cal A}_0^* {\cal A}_{D_{i,j}}\right)=
\sum_{D_{i,j}}\overline{\sum}2\,{\cal R}e\left({\cal A}_0 
{\cal A}_{D_{i,j}}^*\right)\,\,\,,
\end{equation}
where ${\cal A}_0$ is the tree level amplitude given in Eq.~(\ref{eq:amp_treeggZbb}), while ${\cal
A}_{D_{i,j}}$ denotes the amplitude for a class of virtual diagrams that only differ by the
insertion of the final state $Z$ boson leg, i.e. $D_{i,j}=\sum_k D_{i,j}^{(k)}$ with
$D_i\!=\!S_i,V_i,B_i,P_i$, $j=s,t,u$, and $k$ running over all possible $Z$ boson insertions, as
illustrated in Figures~\ref{fig:selfggZbb}-\ref{fig:pentggZbb}.

The calculation of each virtual diagram (${\cal A}_{D_{i,j}}$) is performed in the way we have
explained in Sections~\ref{subsec:sigvirt}, \ref{subsec:virtWbb} and \ref{subsec:virtqqZbb}. We
refer to those sections for details.  We note that the greatest complexity for virtual diagrams in
our calculation is encountered in the $gg\to\Zbb$ subprocess. This is due to the higher number of
one-loop diagrams that need to be considered and to the presence of two external gluons. For this
reason we have chosen to organize the diagrams, at certain stages, into gauge invariant color
amplitudes, that is, into coefficients of the same color structure. This allows a better handling of
the spurious singularities and a natural way to make internal cross checks and cross checks with new
techniques (see Section~\ref{subsec:sigvirt} and Appendix~\ref{app:quadcuts}).

We introduce the leading and sub-leading color factors:
\begin{eqnarray}
\label{eq:color_factorsggZbb}
C_1&=&\frac{N^2}{4}(N^2-1)\,\,\,,\nonumber\\
C_2&=&-\frac{1}{4}(N^2-1)\,\,\,,\nonumber\\
C_3&=&\left(1+\frac{1}{N^2}\right)(N^2-1)\,\,\,,
\end{eqnarray}
and use them to group the different diagrams when interfered, as in
Eq.~(\ref{eq:amp2_virt_genggZbb}), with the LO amplitudes in Eq.~(\ref{eq:amp_treeggZbb}), ${\cal
A}_0^{ab}$ and ${\cal A}_0^{nab}$, independently. This leads to the seven sets of color amplitudes
$A^{ab/nab}_i$ ($i=1,\dots,5$), whose diagram content is shown in
Tables~\ref{tb:A1color}-\ref{tb:A5color}.

\begin{table}[h!t!p!]
\begin{center}
\caption{Diagram content of color amplitudes $A_1^{ab}$ and $A_1^{nab}$ whose color
factor, when interfered with ${\cal A}_0^{ab}$ or ${\cal A}_0^{nab}$ respectively, is $C_1$.}
\begin{tabular}{l|cccc}
\hline
Color Factor: $C_1$ & self energy & vertex & box & pentagon \\
\hline
\\[1pt]
$A_1^{ab}$ & $S_{3,(t,u)}^{(1,2)},S_{4,(t,u)}^{(1,2,3,4)}$ &
		$V_{6,s}^{(1,2)},V_{(7,8),(t,u)}^{(1,2,3)},$ & $B_{4,s},B_{(7,8),(t,u)}^{(1,2)}$ & $P_{1,(t,u)}$ \\
 &  & $V_{9,(t,u)}^{(1,2,3)}$ &  & \\
\\[1pt]
\hline
\\[1pt]
$A_1^{nab}$ &  $S_{1,s}^{(1,2)},S_{2,s}^{(1,2)},$ & $V_{1,s}^{(1,2)},V_{2,s}^{(1,2)},$ 
			& $B_{3,s}, B_{4,s},$ & $P_{1,(t,u)}$ \\
 &  $S_{3,(t,u)}^{(1,2)},S_{4,(t,u)}^{(1,2,3,4)}$ & $V_{3,s}^{(1,2)},V_{6,s}^{(1,2)},$ 
			& $B_{(7,8),(t,u)}^{(1,2)}$ & \\
 &  & $V_{(7,8),(t,u)}^{(1,2,3)},V_{9,(t,u)}^{(1,2,3)}$ &  & \\
\\[1pt]
\hline
\end{tabular}
\label{tb:A1color}
\end{center}
\end{table}
\begin{table}[h!t!p!]
\begin{center}
\caption{Diagram of content color amplitudes $A_2^{ab}$ and $A_2^{nab}$ whose color
factor, when interfered with ${\cal A}_0^{ab}$ or ${\cal A}_0^{nab}$ respectively, is $C_2$.}
\begin{tabular}{l|cccc}
\hline
Color Factor: $C_2$ & self energy & vertex & box & pentagon \\
\hline
\\[1pt]
$A_2^{ab}$ & $S_{3,(t,u)}^{(1,2)},S_{4,(t,u)}^{(1,2,3,4)}$ & $V_{(7,8),(t,u)}^{(1,2,3)},V_{9,(t,u)}^{(1,2,3)}$ 
		& $B_{(5,6),(t,u)}^{(1,2)}, B_{7,(t,u)}^{(1,2)},$ & $P_{(2,3,4),(t,u)},$ \\
 &  &  & $B_{(9,10),(t,u)}^{(1,2)}$ & $P_{(5,6),(t,u)}$ \\
\\[1pt]
\hline
\\[1pt]
$A_2^{nab}$ & $S_{2,s}^{(1,2)}$ & $V_{(2,3),s}^{(1,2)}$ 
		& $B_{2,s}^{(1,2)},B_{(5,6),(t,u)}^{(1,2)},$ & $P_{(2,3,4),(t,u)}$ \\
 & $S_{3,(t,u)}^{(1,2)},S_{4,(t,u)}^{(1,2,3,4)}$ & $V_{(7,8),(t,u)}^{(1,2,3)},V_{9,(t,u)}^{(1,2,3)}$ 
		& $B_{9,(t,u)}^{(1,2)}$ & \\
\\[1pt]
\hline
\end{tabular}
\label{tb:A2color}
\end{center}
\end{table}
\begin{table}[h!t!p!]
\begin{center}
\caption{Diagram content of color amplitude $A_3^{ab}$ whose color
factor, when interfered with ${\cal A}_0^{ab}$ or ${\cal A}_0^{nab}$ respectively, is $C_3$.}
\begin{tabular}{l|cccc}
\hline
Color Factor: $C_3$ & self energy & vertex & box & pentagon \\
\hline
\\[1pt]
$A_3^{ab}$ & $S_{3,(t,u)}^{(1,2)},S_{4,(t,u)}^{(1,2,3,4)}$ & $V_{(7,8),(t,u)}^{(1,2,3)},V_{9,(t,u)}^{(1,2,3)}$ 
		& $B_{(5,6),(t,u)}^{(1,2)}, B_{(9),(t,u)}^{(1,2)}$ & $P_{(2,3,4),(t,u)}$ \\
\\[1pt]
\hline
\hline
\end{tabular}
\label{tb:A3color}
\end{center}
\end{table}
\begin{table}[h!t!p!]
\begin{center}
\caption{Diagram content of color amplitude $A_4^{nab}$ whose color
factor, when interfered with ${\cal A}_0^{ab}$ or ${\cal A}_0^{nab}$ respectively, is $NC_2$.}
\begin{tabular}{l|cccc}
\hline
Color Factor: $NC_2$ & self energy & vertex & box & pentagon \\
\hline
\\[1pt]
$A_4^{nab}$ & $S_{1,s}^{(1,2)}$ & $V_{1,s}^{(1,2)},V_{4,(t,u)}^{(1,2)},$ 
		& $B_{1,t}^{(1,2,3)}$ & $-$ \\
 & & $V_{5,s}^{(1,2)},V_{10,(t,u)}^{(1,2,3,4)}$ & & \\
\\[1pt]
\hline
\hline
\end{tabular}
\label{tb:A4color}
\end{center}
\end{table}
\begin{table}[h!t!p!]
\begin{center}
\caption{Diagram content of color amplitude $A_5^{ab}$ whose color
factor, when interfered with ${\cal A}_0^{ab}$ or ${\cal A}_0^{nab}$ respectively, is $(N-2/N)C_2$.}
\begin{tabular}{l|cccc}
\hline
Color Factor: $(N-2/N)C_2$ & self energy & vertex & box & pentagon \\
\hline
\\[1pt]
$A_5^{ab}$ & $-$ & $V_{10,(t,u)}^{(1,2,3,4)}$ 
		& $B_{1,t}^{(1,2,3)}$ & $-$ \\
\\[1pt]
\hline
\hline
\end{tabular}
\label{tb:A5color}
\end{center}
\end{table}

The structure of the UV singularities for the virtual cross section of the $gg\to \Zbb$ subprocess
is presented in Section~\ref{subsubsec:virtual_uvggZbb}.  The structure of the IR singular part on
the other hand is presented in Section~\ref{subsubsec:virtual_irggZbb}, while the IR singularities
of the real cross section are discussed in Section~\ref{subsec:realZbb}. The explicit cancellation
of IR singularities in the total inclusive NLO cross section for $gg\to \Zbb$ is outlined in
Sections~\ref{subsec:realZbb} and \ref{subsec:totalZbb}.

\subsubsection{Virtual corrections: UV singularities and counterterms}
\label{subsubsec:virtual_uvggZbb}

Self-energy and vertex one-loop corrections to the tree level $gg\to \Zbb$ process (see
Figures~\ref{fig:selfggZbb} and \ref{fig:vertggZbb}) give rise to UV divergences. These
singularities are canceled by a set of counterterms fixed by well defined renormalization
conditions.  We need to introduce counterterms for the external field wave functions of bottom
quarks and gluons ($\delta Z_2^{(b)}$, $\delta Z_3$), for the bottom mass ($\delta m_b$), and for
the strong coupling constant ($\delta Z_{\alpha_s}$).  

By carefully grouping subsets of self-energy and vertex diagrams, we can factor out the UV
singularities of the ${\cal O}(\alpha_s^3)$ virtual amplitude and write them in terms of the tree
level partial amplitudes ${\cal A}_{0,s}$, ${\cal A}_{0,t}$, and ${\cal A}_{0,u}$ introduced in
Eq.~(\ref{eq:a0_ab_nabggZbb}). According to the notation introduced in
Figures~\ref{fig:selfggZbb}-\ref{fig:pentggZbb}, we denote by $D_{i,j}$ (with $D\!=\!S,V$,
$i\!=\!1,2,\ldots$, and $j\!=\!s,t,u$) a class of diagrams with a given self-energy or vertex
correction insertion, summed over all possible insertions of the external $Z$ boson, one for each
different diagram.  We now define $\Delta_{\rm UV}({\cal A}_{D_{i,j}})$ to be the UV pole part of
the corresponding amplitude.  Using this notation, we find:

\begin{eqnarray}
\label{eq:virtual_uvggZbb}
&&\Delta_{\rm UV}({\cal A}_{S_{1,s}})=\frac{\alpha_s}{4\pi}
\left[{\cal N}_s\,\left(\frac{5}{3}N-\frac{2}{3}n_{lf}\right)-
{\cal N}_b\,\frac{2}{3}\right]\left(\frac{1}{\epsilon_{\rm \sss UV}}\right)
[t^a,t^b]{\cal A}_{0,s}\,\,\,,
\nonumber\\
&&\Delta_{\rm UV}({\cal A}_{V_{1,s}})=\frac{\alpha_s}{4\pi}\left[
{\cal N}_s\,
\left(-\frac{2}{3}N+\frac{2}{3}n_{lf}\right)+{\cal N}_b\,\frac{2}{3}\right]
\left(\frac{1}{\epsilon_{\rm \sss UV}}\right)
[t^a,t^b]{\cal A}_{0,s}\,\,\,,
\nonumber\\
&&\Delta_{\rm UV}({\cal A}_{V_{2,s}}+{\cal A}_{V_{7,t}}+{\cal A}_{V_{7,u}})=
\frac{\alpha_s}{4\pi}{\cal N}_b\,
\left(\frac{3}{2}N-\frac{1}{2N}\right)
\left(\frac{1}{\epsilon_{\rm \sss UV}}\right){\cal A}_0\,\,\,,
\nonumber\\
&&\Delta_{\rm UV}({\cal A}_{V_{8,t}}+{\cal A}_{V_{8,u}})=
\frac{\alpha_s}{4\pi}{\cal N}_b\,
\left(\frac{3}{2}N-\frac{1}{2N}\right)
\left(\frac{1}{\epsilon_{\rm \sss UV}}\right)\times\nonumber\\
&&\quad\quad\quad\quad\quad
\left(\frac{1}{2}({\cal A}_{0,t}-{\cal A}_{0,u})[t^a,t^b]+
\frac{1}{2}({\cal A}_{0,t}+{\cal A}_{0,u})\{t^a,t^b\}\right)\,\,\,,
\nonumber\\
&&\Delta_{\rm UV}({\cal A}_{V_{3,s}}+{\cal A}_{V_{9,t}}+{\cal A}_{V_{9,u}})=
\frac{\alpha_s}{4\pi}{\cal N}_b\,
\left(\frac{N}{2}-\frac{1}{2N}\right)
\left(\frac{1}{\epsilon_{\rm \sss UV}}\right){\cal A}_0\,\,\,,
\nonumber\\
&&\Delta_{\rm UV}({\cal A}_{S_{2,s}}+
{\cal A}_{S_{3,t}}+{\cal A}_{S_{3,u}}+
{\cal A}_{S_{4,t}}+{\cal A}_{S_{4,u}})=
-\frac{\alpha_s}{4\pi}{\cal N}_b\,
\left(\frac{N}{2}-\frac{1}{2N}\right)
\left(\frac{1}{\epsilon_{\rm \sss UV}}\right)\times\nonumber\\
&&\quad\quad\quad\quad\quad
\left({\cal A}_0+\frac{1}{2}({\cal A}_{0,t}-{\cal A}_{0,u})[t^a,t^b]+
\frac{1}{2}({\cal A}_{0,t}+{\cal A}_{0,u})\{t^a,t^b\}\right)\,\,\,,
\nonumber\\
\end{eqnarray}
where $n_{lf}\!=\!5$ corresponds to the number of light quark flavors, $N\!=\!3$ is the number of
colors and ${\cal N}_s$ and ${\cal N}_b$ are defined in Eq.~(\ref{eq:nsnb}).

We notice that some of the UV divergent virtual corrections ($V_{1,s}^{(1,2)}$,
$V_{7,(t,u)}^{(1,2,3)}$, and $V_{8,(t,u)}^{(1,2,3)}$), as well as $\delta Z_2^{(b)}$ and $\delta Z_3$
in Eqs.~(\ref{eq:z2_ct}) and (\ref{eq:z3}) below, have also IR singularities. In this section, we
limit the discussion to the UV singularities only, while the IR structure of these terms will be
considered in Section~\ref{subsubsec:virtual_irggZbb}.  For this reason, we have explicitly denoted
by $\epsilon_{\rm \sss UV}$ the pole parameter.

The corresponding counterterms are defined as follows.  For the external fields, we fix the
wave-function renormalization constant of the external bottom quark fields using the on-shell
subtraction scheme, giving the expressions shown in Eq.~(\ref{eq:z2_ct}), while we renormalize the
wave-function of external gluons in the $\overline{MS}$ subtraction scheme:
\begin{equation}
\label{eq:z3}
\left(\delta Z_3\right)_{\rm UV}=\frac{\alpha_s}{4\pi}
(4\pi)^\epsilon\Gamma(1+\epsilon)
\left\{
\left(\frac{5}{3}N-\frac{2}{3}n_{lf}\right)\frac{1}{\epsilon_{\rm \sss UV}}-
\frac{2}{3}\left[\frac{1}{\epsilon_{\rm \sss UV}}+
\ln\left(\frac{\mu^2}{m_b^2}\right)\right]
\right\}\,\,\,,
\end{equation}
according to which we also need to consider the insertion of a finite self-energy correction on the
external gluon legs. This amounts to an extra contribution
\begin{equation}
\delta_{\rm \sss UV}=\frac{\alpha_s}{4\pi}
(4\pi)^\epsilon\Gamma(1+\epsilon)
\left(\frac{5}{3}N-\frac{2}{3}n_{lf}\right)
\ln\left(\frac{\mu^2}{m_b^2}\right)\,\,\,,
\label{eq:deltaUV}
\end{equation}
which is important in order to obtain the correct scale dependence of the NLO cross section.

The structure of the remaining counterterms that appear in this case is the same as presented in
Sections~\ref{subsubsec:UVWbb} and \ref{subsubsec:UVqqZbb}, and we refer the reader to them.

Using the results in Sections~\ref{subsubsec:UVWbb} and \ref{subsubsec:UVqqZbb} and in
Eqs.~(\ref{eq:virtual_uvggZbb})-(\ref{eq:deltaUV}), it is easy to verify that the UV pole part of
$\hat\sigma^{\rm virt}_{gg}$,
\begin{eqnarray}
\label{eq:sigma_virt_uv_poleggZbb}
(\hat\sigma^{\rm virt}_{gg})_{UV-pole}&=&
\int d(PS_3)\,\sum_{D_{i,j}} \overline{\sum}2\,{\cal R}e\left(
{\cal A}_0\,\Delta_{\rm \sss UV}({\cal A}_{D_{i,j}}^*)\right)+\nonumber\\
&&2 \hat{\sigma}^{\rm \sss LO}_{gg}\left[
\big(\delta Z_2^{(b)}\big)_{\rm \sss UV}+
\left(\delta Z_3\right)_{\rm \sss UV}+\delta_{\rm \sss UV}+
\frac{\delta m_b}{m_b}+\delta Z_{\alpha_s}\right]\;,
\end{eqnarray}
is free of UV singularities and has a residual renormalization scale dependence of the form:
\begin{equation}
\label{eq:mudep_res_uvggZbb}
\hat\sigma^{\rm \sss LO}_{gg} \frac{\alpha_s(\mu)}{2\pi}
\left(-\frac{2}{3}n_{lf}+\frac{11}{3} N\right)
\ln\left(\frac{\mu^2}{s}\right)\,\,\,,
\end{equation}
as expected by renormalization group arguments (see the first term of Eq.~(\ref{eq:fmuNLOdep})). We
note that the presence of $s$ in the argument of the logarithm of Eq.~(\ref{eq:mudep_res_uvggZbb})
has no particular relevance as described at the end of Section~\ref{subsubsec:UVWbb}. 
\subsubsection{Virtual corrections: IR singularities}
\label{subsubsec:virtual_irggZbb}

The structure of the IR singularities originating from the ${\cal O}(\alpha_s)$ virtual corrections
to the tree level amplitude for $gg\to \Zbb$ is more involved than that for the UV singularities.
However it simplifies considerably when given at the level of the amplitude squared, and this is
what we present in this section.

The IR divergent part of the ${\cal O}(\alpha_s^3)$ virtual amplitude squared of
Eq.~(\ref{eq:amp2_virt_genggZbb}) can be written in the following compact form:
\begin{equation}
\sum_{D_{i,j}} \overline{\sum}2{\cal R}e\left(
{\cal A}_0\,\Delta_{\rm \sss IR}({\cal A}_{D_{i,j}}^*)\right)=
\frac{\alpha_s}{2\pi}{\cal N}_b  \overline{\sum}
\left(C_1 {\cal M}_{V,\epsilon}^{(1)}+C_2 {\cal M}_{V,\epsilon}^{(2)}+
C_3 {\cal M}_{V,\epsilon}^{(3)}\right)\,\,\,,
\end{equation}
where ${\cal N}_b$ is defined in Eq.~(\ref{eq:nsnb}) and we denote by $\Delta_{\rm \sss IR}({\cal
A}_{D_{i,j}})$ the IR pole part of the amplitude of a given $D_{i,j}$ class of diagrams. The result
is organized in terms of leading and sub-leading color factors presented in
Eq.~(\ref{eq:color_factorsggZbb}), and the corresponding matrix elements squared
$M_{V,\epsilon}^{(1)}$, $M_{V,\epsilon}^{(2)}$, and $M_{V,\epsilon}^{(3)}$ are given by:
\begin{eqnarray}
\label{eq:m2_virt_ir_polesggZbb}
{\cal M}_{V,\epsilon}^{(1)}&=&
\left[-\frac{4}{\epsilon_{\rm \sss IR}^2}+\frac{2}{\epsilon_{\rm \sss IR}}
(-2+\Lambda_s)\right]
\left(|{\cal A}_0^{nab}|^2+|{\cal A}_0^{ab}|^2\right)\nonumber\\
&+&\frac{1}{\epsilon_{\rm \sss IR}}\left[
\left(\Lambda_{\tau_1}+\Lambda_{\tau_2}\right)
|{\cal A}_{0,s}+{\cal A}_{0,t}|^2+
\left(\Lambda_{\tau_3}+\Lambda_{\tau_4}\right)
|{\cal A}_{0,u}-{\cal A}_{0,s}|^2\right]\,\,\,,
\nonumber\\
{\cal M}_{V,\epsilon}^{(2)}&=&\left[-\frac{8}{\epsilon_{\rm \sss IR}^2}+
\frac{4}{\epsilon_{\rm \sss IR}}\left(-2+\Lambda_{\tau_1}+\Lambda_{\tau_2}+
\Lambda_{\tau_3}+\Lambda_{\tau_4}\right)\right]|{\cal A}_0^{ab}|^2\nonumber\\
&+&\frac{2}{\epsilon_{\rm \sss IR}}\frac{\bar s_{b\bar{b}}-2m_b^2}
{\bar s_{b\bar{b}}\beta_{b\bar{b}}} \Lambda_{b\bar{b}}
\left(|{\cal A}_0^{nab}|^2+|{\cal A}_0^{ab}|^2\right)\,\,\,,\nonumber\\
{\cal M}_{V,\epsilon}^{(3)}&=&\frac{1}{\epsilon_{\rm \sss IR}}
\frac{\bar{s}_{b\bar{b}}-2m_b^2}{\bar{s}_{b\bar{b}}\beta_{b\bar{b}}}
\Lambda_{b\bar{b}}
|{\cal A}_0^{ab}|^2 \,\,\,,
\end{eqnarray}
where the IR nature of the pole terms has been made explicit. ${\cal A}_0^{ab}$ and ${\cal
A}_0^{nab}$ are defined in Eq.~(\ref{eq:a0_ab_nabggZbb}), while ${\cal A}_{0,s}$, ${\cal A}_{0,t}$,
and ${\cal A}_{0,u}$ are given explicitly in Appendix~\ref{app:LOamp}.  We have introduced the
notation $\Lambda_s=\ln(s/m_b^2)$ and $\Lambda_{\tau_i}=\ln(\tau_i/m_b^2)$, where the invariants have
been defined in Eq.~(\ref{eq:kinematic_invariants}), and $\beta_{b\bar b}$ and $\Lambda_{b\bar b}$
are defined in Eq.~(\ref{eq:betadef}).

We write explicitly here the IR divergent structure of the $\left(\delta Z_3\right)$ counterterm, as
it is the only one that has not appeared previously. We get:
\begin{eqnarray}
\left(\delta Z_3\right)_{\rm\sss IR}&=&{\cal
N}_b\left(-\frac{5}{3}N+\frac{2}{3}n_{lf}\right)\frac{1}{\epsilon_{\rm\sss IR}}\;.
\end{eqnarray}

When we add the IR singularities coming from the counterterms we have introduced in
Section~\ref{subsubsec:virtual_uvggZbb}, we can write the complete pole part of the IR singular
${\cal O}(\alpha_s^3)$ virtual cross section as:
\begin{eqnarray}
\label{eq:sigma_virt_ir_polesggZbb}
(\hat\sigma^{\rm virt}_{gg})_{IR-pole}&=& \int d(PS_3) 
\sum_{D_{i,j}} \overline{\sum}2{\cal R}e\left(
{\cal A}_0\,\Delta_{\rm \sss IR}({\cal A}_{D_{i,j}}^*)\right)+
2\hat \sigma^{gg}_{\rm \sss LO} \left(
\big(\delta Z_2^{(b)}\big)_{\rm \sss IR}+
\left(\delta Z_3\right)_{\rm \sss IR}\right)\nonumber\\
&=& \int d(PS_3)\frac{\alpha_s}{2\pi}{\cal N}_b  \overline{\sum}
\left(C_1 {\cal M}_{V,\epsilon}^{(1)}+C_2 {\cal M}_{V,\epsilon}^{(2)}+
C_3 {\cal M}_{V,\epsilon}^{(3)}\right)\nonumber\\
&+&\frac{\alpha_s}{2\pi}{\cal N}_b
\left(\frac{2}{3}n_{lf}-\frac{8}{3}N+\frac{1}{N}\right)
\frac{1}{\epsilon_{\rm \sss IR}} \hat{\sigma}^{\rm \sss LO}_{gg}\,\,\,.
\end{eqnarray}
As will be demonstrated in Section~\ref{subsec:realZbb}, the IR singularities of $\hat \sigma^{\rm
virt}_{gg}$ are canceled by the corresponding IR singularities of $\hat \sigma^{\rm real}_{gg}$.

\begin{figure}[ht]
\begin{center}
\includegraphics[scale=0.9]{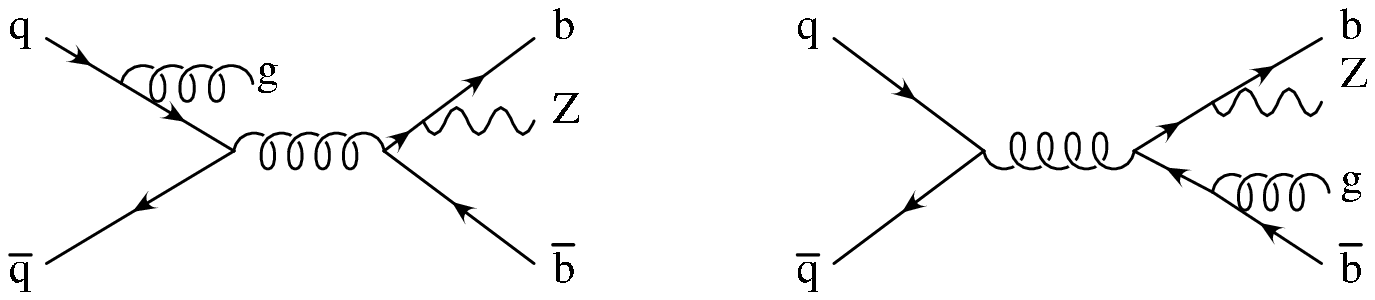}
\caption{Examples of ${\cal O}(\as)$ real corrections to $q\bar q\to\Zbb$ production, with $Z$ emitted from
	the final fermion line.}
\label{fig:realdiagfinalqqZbb}
\end{center}
\end{figure}

\subsection[Real Corrections to $\Zbb$ Production]{Real Corrections to $\Zbb$ Production}\label{subsec:realZbb}
\begin{figure}[ht]
\begin{center}
\includegraphics[scale=0.9]{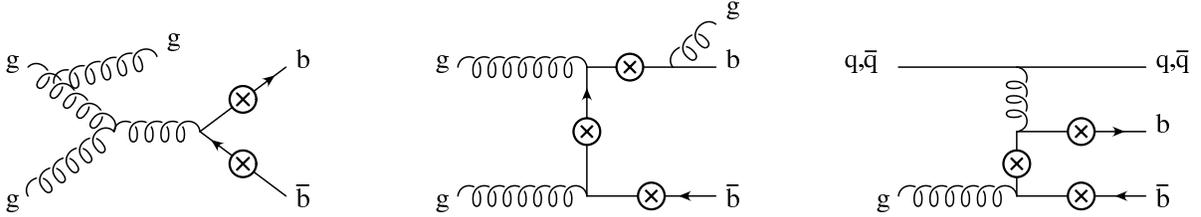}
\caption{Examples of ${\cal O}(\as)$ real corrections to $\Zbb$ production. The circled crosses denote all
	possible insertions of a $Z$ weak boson, each insertion corresponding to a different diagram.}
\label{fig:realdiagZbb}
\end{center}
\end{figure}
The real corrections to $\Zbb$ production is built up of three ${\cal O}(\as^3)$ subprocesses, namely $q\bar
q\to\Zbb+g$, $gg\to \Zbb+g$ and $(q,\bar{q})g\to \Zbb+(q,\bar{q})$.

The real cross section for the $q\bar q\to\Zbb+g$ subprocess is analogous to the $q\bar
q^\prime\to\Wbb+g$ subprocess, although it contains a set of diagrams with the $Z$ boson emitted from
the $b$-quark fermion line, as shown in Figure~\ref{fig:realdiagfinalqqZbb}. We refer the reader to
the discussion in Sections~\ref{subsec:realWbb} and \ref{subsec:totalWbb} for details.

The NLO real cross section $\hat\sigma^{\rm real}_{gg}$ in Eq.~(\ref{eq:virtrealXsec}) corresponds
to the ${\cal O}(\alpha_s)$ corrections to $gg\to \Zbb$ due to the emission of a real gluon, i.e. to
the process $gg\to \Zbb+g$, examples of which are illustrated in Figure~\ref{fig:realdiagZbb}. It
contains IR singularities that cancel the analogous singularities present in the ${\cal
O}(\alpha_s)$ virtual corrections (see Section~\ref{subsubsec:virtual_irggZbb}) and in the NLO
parton distribution functions.  These singularities can be either \emph{soft}, when the energy of
the emitted gluon becomes very small, or \emph{collinear}, when the final state gluon is emitted
collinear to one of the initial gluons. There is no collinear radiation from the final $b$ and
$\bar{b}$ quarks because they are massive. At the same order in $\alpha_s$, the $\hat\sigma^{\rm
real}_{qg}$ cross section corresponds to the tree level processes $(q,\bar{q})g\to
\Zbb+(q,\bar{q})$, an example of which is also illustrated in Figure~\ref{fig:realdiagZbb}. This
part of the NLO cross section develops IR singularities entirely due to the collinear emission of a
final state quark or antiquark from one of the initial state massless partons.  The IR singularities
can be conveniently isolated by \emph{slicing}  the $gg\to \Zbb+g$ and $(q,\bar{q})g\to
\Zbb+(q,\bar{q})$ phase spaces into different regions defined by suitable cutoffs (as was done in
Section~\ref{subsec:realWbb}).  The dependence on the arbitrary cutoffs introduced in \emph{slicing}
the phase space of the final state particles is not physical, and cancels at the level of the total
real hadronic cross section, i.e. in $\sigma^{\rm real}$, as well as at the level of the real cross
section for each separate channel, i.e., in $\sigma^{\rm real}_{q\bar{q}}$, $\sigma^{\rm
real}_{gg}$, and $\sigma^{\rm real}_{qg}$.  This cancellation constitutes an important check of the
calculation and will be discussed in detail in Section~\ref{subsec:totalZbb}.

We have calculated the cross section for the processes
\[
g(q_1)+g(q_2)\to b(p_b)+\bar{b}(p_{\bar b})+Z(p_Z)+g(k)\,\,\,,
\]
and
\[
(q,\bar{q})(q_1)+g(q_2)\to
b(p_b)+\bar{b}(p_{\bar b})+Z(p_Z)+(q,\bar{q})(k)\,\,\,, 
\]
with $q_1+q_2=p_b+p_{\bar b}+p_Z+k$, using the PSS method with \emph{two cutoffs}, as in
Section~\ref{subsec:realWbb}.

In the next section we discuss the application of the two-cutoff PSS method to the $gg\to\Zbb+g$ and $q(\bar
q)g\to\Zbb+q(\bar q)$ subprocesses. 

\subsubsection*{Real gluon emission, $gg\to \Zbb+g$: soft region}
%\label{subsubsec:two_cutoff_soft}

The soft region of the phase space for the gluon emission process
\begin{equation}
\label{eq:gg_Zbbg}
g^a(q_1)+g^b(q_2)\to b(p_b)+\bar{b}(p_{\bar b})+Z(p_Z)+g^c(k)
\end{equation}
is defined by demanding that the energy of the emitted gluon ($k^0\!=\!E_g$) satisfies the condition
\begin{equation}
\label{eq:eg_cutZbb}
E_g\le\delta_s\frac{\sqrt{s}}{2}
\end{equation}
for an arbitrary small value of the \emph{soft} cutoff $\delta_s$.  In the \emph{soft limit}
($E_g\to 0$), the amplitude for this process can be written as:
\begin{eqnarray}
&&{\cal A}_{soft}(gg\to \Zbb+g)=\nonumber\\
&&t^ct^at^b
\left(\frac{p_b\!\cdot\!\epsilon^*}{p_b\!\cdot\!k}-
\frac{q_1\!\cdot\!\epsilon^*}{q_1\!\cdot\!k}\right)
\left({\cal A}_{0,t}+{\cal A}_{0,s}\right)  
+
t^ct^bt^a
\left(\frac{p_b\!\cdot\!\epsilon^*}{p_b\!\cdot\!k}-
\frac{q_2\!\cdot\!\epsilon^*}{q_2\!\cdot\!k}\right)
\left({\cal A}_{0,u}-{\cal A}_{0,s}\right)\nonumber\\
&-&
t^at^bt^c
\left(\frac{p_{\bar b}\!\cdot\!\epsilon^*}{p_{\bar b}\!\cdot\!k}-
\frac{q_2\!\cdot\!\epsilon^*}{q_2\!\cdot\!k}\right)
\left({\cal A}_{0,t}+{\cal A}_{0,s}\right) 
-
t^bt^at^c
\left(\frac{p_{\bar b}\!\cdot\!\epsilon^*}{p_{\bar b}\!\cdot\!k}-
\frac{q_1\!\cdot\!\epsilon^*}{q_1\!\cdot\!k}\right)
\left({\cal A}_{0,u}-{\cal A}_{0,s}\right)\nonumber\\
&+&
t^at^ct^b
\left(\frac{q_1\!\cdot\!\epsilon^*}{q_1\!\cdot\!k}-
\frac{q_2\!\cdot\!\epsilon^*}{q_2\!\cdot\!k}\right)
\left({\cal A}_{0,t}+{\cal A}_{0,s}\right)
+
t^bt^ct^a
\left(\frac{q_2\!\cdot\!\epsilon^*}{q_2\!\cdot\!k}-
\frac{q_1\!\cdot\!\epsilon^*}{q_1\!\cdot\!k}\right)
\left({\cal A}_{0,u}-{\cal A}_{0,s}\right)\;, 
\nonumber\\
\end{eqnarray}
where $a,b$, and $c$ are the color indices of the external gluons, while $\epsilon^{\mu}(k,\lambda)$
(for $\lambda\!=\!1,2$) is the polarization vector of the emitted soft gluon. Moreover, in the soft
region the $gg\to \Zbb+g$ phase space factorizes as:
\begin{eqnarray}
\label{eq:ps_soft_limZbb}
d(PS_4)(gg\to \Zbb+g)
& \stackrel{soft}{\longrightarrow} & d(PS_3)(gg\to \Zbb)
d(PS_g)_{soft}\nonumber\\
&=& d(PS_3)(gg\to \Zbb)
\frac{d^{(d-1)}k}{(2\pi)^{(d-1)}2E_g} 
\theta\left(\delta_s\frac{\sqrt{s}}{2}-E_g\right)
\nonumber\,\,\,,\\
\end{eqnarray}
where $d(PS_4)$ and $d(PS_3)$ have been defined in Section~\ref{sec:FactTh}, while $d(PS_g)_{soft}$
denotes the the phase space measure of the soft gluon.  Since the contribution of the soft gluon is
now completely factorized, we can perform the integration over $d(PS_g)_{soft}$ analytically, using
dimensional regularization in $d\!=\!4-2\epsilon$ to extract the soft poles that will cancel the
corresponding singularities in Eq.~(\ref{eq:sigma_virt_ir_polesggZbb}).  For completeness, the
integrals that we have used to perform the integration over the phase space of the soft gluon are
collected in Appendix~\ref{app:PSint}.

After squaring the soft amplitude ${\cal A}_{soft}$, summing over the polarization of the radiated
soft gluon, and integrating over the soft gluon momentum, the pole part of the parton level soft
cross section reads
\begin{eqnarray}
\label{eq:soft_ir_polesZbb}
(\hat{\sigma}^{soft}_{gg})_{pole}&=&
\int d(PS_3)
\left(\int d(PS_g)_{soft}\overline{\sum}
|{\cal A}_{soft}(gg\to \Zbb+g)|^2\right)_{pole}
\nonumber\\ 
&=&\int d(PS_3) \frac{\alpha_s}{2\pi}{\cal N}_b \overline{\sum} 
\left(C_1 {\cal M}_{S,\epsilon}^{(1)}+C_2 {\cal M}_{S,\epsilon}^{(2)}+
C_3 {\cal M}_{S,\epsilon}^{(3)}\right)\,\,\,,
\end{eqnarray}
where $C_1$, $C_2$, and $C_3$ are defined in Eq.~(\ref{eq:color_factorsggZbb}), while ${\cal
M}_{S,\epsilon}^{(1)}$, ${\cal M}_{S,\epsilon}^{(2)}$, and ${\cal M}_{S,\epsilon}^{(3)}$ represent
the IR pole parts of the corresponding matrix elements squared, and can be written as:
\begin{eqnarray}
\label{eq:soft_a2_polesZbb}
{\cal M}_{S,\epsilon}^{(1)}&=&-{\cal M}_{V,\epsilon}^{(1)}-
\frac{2}{\epsilon}(1+4 \ln(\delta_s))
\left(|{\cal A}_0^{nab}|^2+|{\cal A}_0^{ab}|^2\right)\,\,\,,\nonumber\\
{\cal M}_{S,\epsilon}^{(2)}&=&-{\cal M}_{V,\epsilon}^{(2)}-
\frac{16}{\epsilon} \ln(\delta_s) |{\cal A}_0^{ab}|^2
+\frac{2}{\epsilon} \left(|{\cal A}_0^{nab}|^2+|{\cal A}_0^{ab}|^2\right) 
\,\,\,,\nonumber\\
{\cal M}_{S,\epsilon}^{(3)}&=&-{\cal M}_{V,\epsilon}^{(3)}+
\frac{1}{\epsilon}|{\cal A}_0^{ab}|^2~.
\end{eqnarray}
Note that in this section we do not explicitly denote the IR poles as poles in $\epsilon_{\rm \sss
IR}$, since all singularities present in $\sigma_{gg,qg}^{\rm real}$ are of IR origin.

After adding the IR divergent part of the parton level virtual cross section of
Eq.~(\ref{eq:sigma_virt_ir_polesggZbb}) we obtain:
\begin{eqnarray}\label{eq:irsvggZbb}
\hat{\sigma}^{s+v}_{gg}\equiv(\hat{\sigma}^{soft}_{gg})_{pole}+
(\hat{\sigma}^{\rm virt}_{gg})_{IR-pole}&=&
\frac{\alpha_s}{2\pi}{\cal N}_b 
\left[-4N\ln(\delta_s)-\frac{1}{3}(11N-2n_{lf})\right] 
\frac{1}{\epsilon} \hat{\sigma}^{\rm \sss LO}_{gg}\,\,\,,\nonumber\\
\end{eqnarray}
where we can see that the IR poles of the parton level virtual cross section are exactly canceled by
the corresponding singularities in the parton level soft gluon emission cross section. The residual
divergences will be canceled by the soft+virtual part of the PDF counterterm when convoluting with
the gluon PDFs, as will be demonstrated in Section~\ref{subsec:totalZbb}.  The finite contribution to
the parton level soft cross section is finally given by
\begin{eqnarray}
\label{eq:soft_ir_finiteggZbb}
(\hat \sigma^{soft}_{gg})_{finite}&=&
\int d(PS_3) \left(\int d(PS_g)_{soft}\overline{\sum}
|{\cal A}_{soft}(gg\to \Zbb+g)|^2\right)_{finite}
\nonumber\\ 
&=&\int d(PS_3) \frac{\alpha_s}{2\pi}{\cal N}_b\overline{\sum}
\left(C_1{\cal M}_S^{(1)}+C_2{\cal M}_S^{(2)}+
C_3{\cal M}_S^{(3)}\right)\,\,\,,
\end{eqnarray}
where the finite parts of the ${\cal M}_S^{(1)}$, ${\cal M}_S^{(2)}$, and ${\cal M}_S^{(3)}$ matrix
element squared are explicitly given by:

\begin{eqnarray}
\label{eq:soft_a2_finiteggZbb}
{\cal M}_S^{(1)}&=&\left[
-\frac{4}{3}\pi^2
+4\Lambda_s\ln(\delta_s)+8\ln^2(\delta_s)
-2\Lambda_s-4\ln(\delta_s)\right.\nonumber\\
&&\left.
+\frac{1}{\beta_b}\ln\left(\frac{1+\beta_b}{1-\beta_b}\right)
+\frac{1}{\beta_{\bar{b}}}\ln\left(\frac{1+\beta_{\bar{b}}}{1-\beta_{\bar{b}}}\right)
\right]\left(|{\cal A}_0^{nab}|^2+|{\cal A}_0^{ab}|^2\right)
\nonumber\\
&+&\left[
\left(\Lambda_s+2\ln(\delta_s)\right)
\left(\Lambda_{\tau_1}+\Lambda_{\tau_2}\right)
+\frac{1}{2}F(q_1,p_b)+\frac{1}{2}F(q_2,p_{\bar b})\right]
|{\cal A}_0^{nab}+{\cal A}_0^{ab}|^2
\nonumber\\
&+&\left[
\left(\Lambda_s+2\ln(\delta_s)\right)
\left(\Lambda_{\tau_3}+\Lambda_{\tau_4}\right)
+\frac{1}{2}F(q_2,p_b)+\frac{1}{2}F(q_1,p_{\bar b})\right]
|{\cal A}_0^{nab}-{\cal A}_0^{ab}|^2\,\,\,,
\nonumber\\
{\cal M}_S^{(2)}&=&\left\{
\frac{\bar s_{b\bar b}-2m_b^2}{\bar s_{b\bar{b}}}\left[
\left(2\Lambda_s+4\ln(\delta_s)\right)\frac{1}{\beta_{b\bar{b}}} 
\Lambda_{b\bar{b}}
+\frac{1}{2}\ln^2\left(\frac{1+\beta_b}{1-\beta_b}\right)
-\frac{1}{2}\ln^2\left(\frac{1+\beta_{\bar{b}}}{1-\beta_{\bar{b}}}\right)
\right.\right.\nonumber\\
&&
+2\mathrm{Li}_2\left(1-\frac{\alpha_{b\bar{b}}}{v_{b\bar{b}}}p_b^0(1+\beta_b)\right)
+2\mathrm{Li}_2\left(1-\frac{\alpha_{b\bar{b}}}{v_{b\bar{b}}}p_b^0(1-\beta_b)\right)
\nonumber\\
&&\left.
-2\mathrm{Li}_2\left(1-\frac{1}{v_{b\bar{b}}}p_{\bar b}^0(1+\beta_{\bar{b}})\right)
-2\mathrm{Li}_2\left(1-\frac{1}{v_{b\bar{b}}}p_{\bar b}^0(1-\beta_{\bar{b}})\right)
\right]\nonumber\\
&&\left.\phantom{\frac{1}{2}}-2\Lambda_s-4\ln(\delta_s)+
\frac{2}{\beta_{b\bar{b}}}\Lambda_{b\bar{b}}\right\}
\left(|{\cal A}_0^{nab}|^2+|{\cal A}_0^{ab}|^2\right)\nonumber\\
&+&2\left[-\frac{4}{3}\pi^2-2\Lambda_s^2+8\ln^2(\delta_s)
+2\left(\Lambda_s+2\ln(\delta_s)\right)
\left(\Lambda_{\tau_1}+\Lambda_{\tau_2}+\Lambda_{\tau_3}+\Lambda_{\tau_4}\right)
\right.\nonumber\\
&&\quad\,
+F(q_1,p_b)+F(q_2,p_{\bar b})+F(q_2,p_b)+F(q_1,p_{\bar b})\nonumber\\
&&\left.\phantom{\frac{1}{2}}-4\Lambda_s-8\ln(\delta_s)+
\frac{2}{\beta_b}\ln\left(\frac{1+\beta_b}{1-\beta_b}\right)
+\frac{2}{\beta_{\bar{b}}}\ln\left(\frac{1+\beta_{\bar{b}}}{1-\beta_{\bar{b}}}\right)
\right]|{\cal A}_0^{ab}|^2\,\,\,,
\nonumber\\
{\cal M}_S^{(3)}&=&\frac{1}{2}\left\{
\frac{\bar s_{b\bar b}-2m_b^2}{\bar s_{b\bar b}}\left[
\left(2\Lambda_s+4\ln(\delta_s)\right)\frac{1}{\beta_{b\bar{b}}} 
\Lambda_{b\bar{b}}
+\frac{1}{2}\ln^2\left(\frac{1+\beta_b}{1-\beta_b}\right)
-\frac{1}{2}\ln^2\left(\frac{1+\beta_{\bar{b}}}{1-\beta_{\bar{b}}}\right)
\right.\right.\nonumber\\
&&
+2\mathrm{Li}_2\left(1-\frac{\alpha_{b\bar{b}}}{v_{b\bar{b}}}p_b^0(1+\beta_b)\right)
+2\mathrm{Li}_2\left(1-\frac{\alpha_{b\bar{b}}}{v_{b\bar{b}}}p_b^0(1-\beta_b)\right)
\nonumber\\
&&\left.
-2\mathrm{Li}_2\left(1-\frac{1}{v_{b\bar{b}}}p_{\bar b}^0(1+\beta_{\bar{b}})\right)
-2\mathrm{Li}_2\left(1-\frac{1}{v_{b\bar{b}}}p_{\bar b}^0(1-\beta_{\bar{b}})\right)
\right]\nonumber\\
&&\left.\phantom{\frac{1}{2}}
-2\Lambda_s-4\ln(\delta_s)+
\frac{1}{\beta_b}\ln\left(\frac{1+\beta_b}{1-\beta_b}\right)
+\frac{1}{\beta_{\bar{b}}}\ln\left(\frac{1+\beta_{\bar{b}}}{1-\beta_{\bar{b}}}\right)
\right\}|{\cal A}_0^{ab}|^2\,\,\,,
\end{eqnarray}
where 
\begin{equation}
\beta_i=\sqrt{1-\frac{m_b^2}{(p_i^0)^2}}\,\,\,,
\end{equation}
\begin{equation}
\alpha_{b\bar{b}}=\frac{1+\beta_{b\bar{b}}}{1-\beta_{b\bar{b}}}\ , \qquad
v_{b\bar{b}}=\frac{m_b^2(\alpha_{b\bar{b}}^2-1)}{2(\alpha_{b\bar{b}}p_b^0-p_{\bar b}^0)}
\,\,\,,
\label{eq:alphaandvbb}
\end{equation}
$\Lambda_\delta=\ln{(\delta/m_b^2)}$, with $\delta$ a given kinematic invariant, and
$\bar{s}_{b\bar{b}}$, $\beta_{b\bar{b}}$, and $\Lambda_{b\bar{b}}$ are defined in
Eq.~(\ref{eq:betadef}), while the function $F(p_i,p_f)$ can be found in Eq.~(\ref{eq:f_if}) (see
Appendix~\ref{app:PSint} for more details). We have used the set of kinematic invariants presented
in Eq.~(\ref{eq:kinematic_invariants}).

\boldmath
\subsubsection*{Real gluon emission, $gg\to \Zbb+g$: hard region}
%\label{subsubsec:two_cutoff_hard}
\unboldmath

The matrix element squared for $ij\to \Zbb+g$ factorizes into the LO matrix element squared and the
unregulated Altarelli-Parisi splitting function $P_{ii'}= P_{ii'}^4+\epsilon P_{ii'}^\prime$ for
$i\to i^\prime g$, i.e.:
\begin{equation}
\label{eq:m2_coll_limggZbb}
\overline{\sum}|{\cal A}_{\rm real}(ij\to \Zbb+g)|^2 
\stackrel{collinear}{\longrightarrow}
(4\pi\alpha_s)\overline{\sum}|{\cal A}_0(i^\prime j\to \Zbb)|^2
\frac{2P_{ii^\prime}(z)}{z \, s_{ig}} \,\,\,,
\end{equation}
where $s_{ig}\!=\!2q_i\!\cdot\!k$ and
$P_{ii^\prime}^4$ and $P_{ii^\prime}^\prime$ are the Altarelli-Parisi splitting functions, which in
the $gg$ case are given by:
\begin{eqnarray}
P_{gg}^4(z)& = & 2N\left(\frac{z}{1-z}+\frac{1-z}{z}+z(1-z) \right)\ , \nonumber\\
P_{gg}^\prime(z)& = & 0.
\label{eq:splitfuncgg}
\end{eqnarray}

Moreover, in the collinear limit, the $ij\rightarrow \Zbb+g$ phase space also factorizes as:
\begin{eqnarray}
\label{eq:ps_coll_limggZbb}
d(PS_4)(ij\to \Zbb+g)
&\stackrel{collinear}{\longrightarrow}& 
d(PS_3)(i^\prime j\to \Zbb)
\frac{z\,d^{(d-1)}k}{(2\pi)^{(d-1)}2E_g}
\theta\left(E_g-\delta_s\frac{\sqrt{s}}{2}\right)\times\nonumber \\
&&\theta\left(\cos\theta_{ig}-(1-\delta_c)\right)\nonumber\\
&&\!\!\!\!\!\!\!\!\!\!\!\!\!\!\!\!\!\!\!\!\!\! \stackrel{d=4-2 \epsilon}{=}
d(PS_3)(i^\prime j\to \Zbb)
\frac{1}{\Gamma(1-\epsilon)}
\frac{\left(4\pi\right)^\epsilon}{16 \pi^2}\,z\,dz\,ds_{ig} 
\left[(1-z) s_{ig}\right]^{-\epsilon}\times\nonumber\\
&&
\theta\left(\frac{(1-z)}{z}s^\prime\frac{\delta_c}{2}-s_{ig}\right)
\nonumber\,\,\,,\\
\end{eqnarray}
where $d(PS_4)$ and $d(PS_3)$ have been defined in Section~\ref{sec:FactTh}, while the integration
range for $s_{ig}$ in the collinear region is given in terms of the collinear cutoff, and we have
defined $s^\prime\!=\!2q_{i^\prime}\!\cdot\!q_j$.  The integral over the collinear gluon degrees of
freedom can then be performed analytically, and this allows us to extract explicitly the collinear
singularities of
$\hat{\sigma}^{hard}_{gg}$ \cite{Harris:2001sx,Baur:1998kt}, which can be written as:
\begin{eqnarray}
\label{eq:coll_poleggZbb}
\hat{\sigma}^{hard/coll}_{gg}&=&
\left[\frac{\alpha_s}{2\pi}\frac{1}{\Gamma(1-\epsilon)}
\left(\frac{4\pi\mu^2}{m_b^2}\right)^\epsilon\right]
\left(-\frac{1}{\epsilon}\right)\delta_c^{-\epsilon}\times\nonumber\\
&&\left\{
\int_{0}^{1-\delta_s} dz
\left[\frac{(1-z)^2}{2 z} \frac{s^\prime}{m_b^2}\right]^{-\epsilon} 
P_{ii^\prime}(z) \,
\hat{\sigma}^{\rm \sss LO}_{gg}(i^\prime j\rightarrow \Zbb)
+ (i\leftrightarrow j)\right\}\,\,\,,
\end{eqnarray}
where $i,i^\prime$, and $j$ are all gluons.  As usual, these initial state collinear divergences are
absorbed into the gluon distribution functions as will be described in detail in
Section~\ref{subsec:totalZbb}.

\subsubsection*{The tree level processes $(q,\bar{q})g\to \Zbb+(q,\bar{q})$}
%\label{subsubsec:two_cutoff_qg}

The extraction of the collinear singularities of $\hat \sigma_{qg}^{\rm real}$ is done in the same
way as described in the previous subsection for the $gg$ initial state.  In the collinear region,
$\cos\theta_{iq}>1-\delta_c$, the initial state parton $i$ with momentum $q_i$ is considered to
split into a hard parton $i^\prime$ and a collinear quark $q$, $i\rightarrow i^\prime q$, with
$q_{i^\prime}\!=\!z q_i$ and $k\!=\!(1-z)q_i$. The matrix element squared for $ij\to \Zbb+q$
factorizes into the unregulated Altarelli-Parisi splitting functions in $d$ dimensions, $P_{ii'}=
P_{ii'}^4+\epsilon P_{ii'}^\prime$, shown in Eq.(\ref{eq:splitfuncgq}) for the $g\to q\bar{q}$
splitting and in Eq.~(\ref{eq:splitfuncqg}) for the $(q,\bar{q})\rightarrow g(q,\bar{q})$ splitting,
and the corresponding LO matrix elements squared.  The $ij\to \Zbb+q$ phase space factorizes into
the $i'j\to \Zbb$ phase space and the phase space of the collinear quark. As a result, after
integrating over the phase space of the collinear quark, the collinear singularity of $\hat
\sigma^{qg}_{\rm real}$ can be extracted as:
\begin{eqnarray}
\label{eq:qgcoll_poleZbb}
\hat{\sigma}^{coll}_{qg}&=&
\left[\frac{\alpha_s}{2\pi}\frac{1}{\Gamma(1-\epsilon)}
\left(\frac{4\pi\mu^2}{m_b^2}\right)^\epsilon\right]
\left(-\frac{1}{\epsilon}\right)\delta_c^{-\epsilon}
\int_{0}^{1} dz
\left[\frac{(1-z)^2}{2 z} \frac{s^\prime}{m_b^2}\right]^{-\epsilon} 
\times\nonumber\\
&&
\left[ P_{qg}(z) \,
\hat{\sigma}^{\rm \sss LO}_{gg}(g(q_{1^\prime})g(q_2)\to \Zbb)
+ P_{gq}(z) \,
\hat{\sigma}^{\rm \sss LO}_{q\bar{q}}(q(q_1)\bar{q}(q_{2^\prime})\to \Zbb)
\right]\,\,\,. 
\end{eqnarray}
The collinear radiation of an antiquark in $\bar{q}g\to \Zbb+\bar{q}$ is treated analogously.  The
$O(1)$ and $O(\epsilon)$ parts of the $P_{qg}$ splitting function are given by:
\begin{eqnarray}
P_{qg}^4(z)& = & C_F\left(\frac{1+(1-z)^2}{z} \right)\;, \nonumber\\
P_{qg}^\prime(z)& = & -C_Fz,
\label{eq:splitfuncqg}
\end{eqnarray}
and by Eq.~(\ref{eq:splitfuncgq}) for  $P_{gq}$.

Again, these initial state collinear divergences are absorbed into the parton distribution functions
as will be described in detail in Section~\ref{subsec:totalZbb}.

\subsection{Total Cross Section of $p{\bar p}(pp)\to \Zbb$}\label{subsec:totalZbb}

The total inclusive hadronic cross section for $p\bar{p}(pp)\to \Zbb$ is the sum of the contributions
from the $q\bar{q}$, the $gg$ and the $(q,\bar{q})g$ initial states:
\begin{equation}
\label{eq:sigma_nlo_ij}
\sigma^{\rm \sss NLO}(p\bar{p}(pp)\to \Zbb)=
\sigma^{\rm \sss NLO}_{q\bar q}(p\bar{p}(pp)\to \Zbb)+
\sigma^{\rm \sss NLO}_{gg}(p\bar{p}(pp)\to \Zbb)+
\sigma^{\rm \sss NLO}_{qg}(p\bar{p}(pp)\to \Zbb)\,\,\,.
\end{equation}
As described in Section~\ref{sec:FactTh}, $\sigma^{\rm \sss NLO}_{ij}(p\bar{p}(pp)\to \Zbb)$ is obtained
by convoluting the parton level NLO cross section $\hat\sigma^{\rm \sss NLO}_{ij}(p\bar{p}(pp)\to \Zbb)$
with the NLO PDFs ${\cal F}_i^{p,\bar p}(x,\mu)$ ($i=q,g$), thereby absorbing the remaining initial
state singularities of $\delta\hat\sigma^{\rm \sss NLO}_{ij}$ into the renormalized PDFs. In the
following we demonstrate in detail how this cancellation works in the case of the $gg$ and
$(q,\bar{q})g$ initiated processes. The case of the $q\bar{q}$ initiated process is discussed in
Section~\ref{subsec:totalWbb}, where we presented in detail the contribution of the $q\bar{q}$
initial state to $\sigma^{\rm\sss NLO}(p\bar{p}(pp)\to \Wbb)$. $\sigma^{\rm \sss
NLO}_{q\bar{q}}(p\bar{p}(pp)\to \Zbb)$ can be obtained from there with obvious modifications, and will
not be repeated here.

Similarly to the discussion in Section~\ref{subsec:totalWbb}, first the parton level cross section
is convoluted with the {\em bare} parton distribution functions ${\cal F}_i^{p,\bar p}(x)$ and
subsequently the ${\cal F}_i^{p,\bar p}(x)$ are replaced by the renormalized parton distribution
functions, ${\cal F}_i^{p,\bar p}(x,\mu_f)$, defined in some subtraction scheme at a given
factorization scale $\mu_f$. Using the ${\overline{MS}}$ scheme, the scale-dependent NLO parton
distribution functions for the two-cutoff PSS are given in terms of the bare ${\cal F}_i^{p,\bar
p}(x)$ and the QCD NLO parton distribution function counterterms \cite{Harris:2001sx} as follows:
\begin{itemize}
\item[(a)] For the case where an initial state gluon, quark or antiquark
  ($k=g,(q,\bar{q})$) splits, respectively, into a $q\bar{q}$ or
  $(q,\bar{q})g$ pair ($k'=(q,\bar{q}),g$):
\begin{eqnarray}
\label{eq:pdfqg_mu}
{\cal F}_{k'}^{p,\bar p}(x,\mu_f)&=& {\cal F}_{k'}^{p,\bar p}(x) 
+\left[\frac{\alpha_s}{2\pi}
\left(\frac{4\pi\mu_r^2}{\mu_f^2}\right)^\epsilon
\frac{1}{\Gamma(1-\epsilon)}\right]
\int_{x}^{1} \frac{dz}{z}
\left(-\frac{1}{\epsilon}\right) P_{kk'}^4(z) 
{\cal F}_{k}^{p,\bar p}\left(\frac{x}{z}\right)\,\,\,,\nonumber\\
\end{eqnarray}
where $P^4_{ij}$ are defined in Eqs.~(\ref{eq:splitfuncgq}) and (\ref{eq:splitfuncqg}).
\item[(b)] For the case of $g\to gg$ splitting:
\begin{eqnarray}
\label{eq:pdfgg_mu2}
{\cal F}_g^{p,\bar p}(x,\mu_f)&=&
{\cal F}_g^{p,\bar p}(x) \left[1- 
\frac{\alpha_s}{2\pi}
\left(\frac{4\pi\mu_r^2}{\mu_f^2}\right)^\epsilon
\frac{1}{\Gamma(1-\epsilon)}
\left(\frac{1}{\epsilon}\right) N
\left(2 \ln(\delta_s)+\frac{11}{6}-\frac{1}{3}\frac{n_{lf}}{N}\right)
\right]\nonumber\\
&&+\left[\frac{\alpha_s}{2\pi}
\left(\frac{4\pi\mu_r^2}{\mu_f^2}\right)^\epsilon
\frac{1}{\Gamma(1-\epsilon)}\right]
\int_{x}^{1-\delta_s} \frac{dz}{z}
\left(-\frac{1}{\epsilon}\right) P_{gg}(z)
{\cal F}_g^{p,\bar p}\left(\frac{x}{z}\right)\,\,\,,\nonumber\\
\end{eqnarray}
where $P_{gg}$ is Altarelli-Parisi splitting function given in Eq.~(\ref{eq:splitfuncgg}).
\end{itemize}
The ${\cal O}(\alpha_s)$ terms in the previous equations are calculated from the ${\cal
O}(\alpha_s)$ corrections to the $k\rightarrow k' j$ splittings, in the two-cutoffs PSS formalism.
Moreover, note that in Eqs.~(\ref{eq:pdfqg_mu}) and (\ref{eq:pdfgg_mu2}) we have carefully separated
the dependence on the factorization ($\mu_f$) and renormalization scale $(\mu_r$). It is understood
that $\alpha_s\!=\!\alpha_s(\mu_r)$. The definition of the subtracted PDFs is indeed the only place
where both scales play a role, and the only place where we have a dependence on $\mu_f$. In the rest
of this section, we have always set $\mu_r\!=\!\mu_f\!=\!\mu$ and we will also give the master
formulas for the total NLO cross section, Eq.~(\ref{eq:sigmatot_gg2}), using
$\mu_r\!=\!\mu_f\!=\!\mu$.  In Chapter~\ref{chap:results} we will study the explicit dependence of
the total hadronic cross section on both scales $\mu_r$ and $\mu_f$.

When convoluting the parton $gg$ cross section with the renormalized gluon distribution function of
Eq.~(\ref{eq:pdfgg_mu2}), the IR singular counterterm of Eq.~(\ref{eq:pdfgg_mu2}) exactly cancels
the remaining IR poles of $\hat{\sigma}^{\rm virt}_{gg}+\hat\sigma^{soft}_{gg}$ and
$\hat{\sigma}^{hard/coll}_{gg}$, shown in Eq.~(\ref{eq:irsvggZbb}).  Finally, the complete ${\cal
O}(\alpha_s^3)$ inclusive total cross section for $p\bar{p}(pp)\to \Zbb$ in the ${\overline{MS}}$
factorization scheme when only the $gg$ initial state is included, i.e.  $\sigma^{\rm \sss
NLO}_{gg}(p\bar p(pp)\to \Zbb)$ of Eq.~(\ref{eq:sigma_nlo_ij}), can be written as follows:\\
\begin{eqnarray}
\label{eq:sigmatot_gg2}
\sigma^{\rm \sss NLO}_{gg} &=&\frac{1}{2}\int dx_1 dx_2 
\left\{ {\cal F}_g^p(x_1,\mu)
{\cal F}_{g}^{\bar p(p)}(x_2,\mu) \left[
\hat{\sigma}^{\rm \sss LO}_{gg}(x_1,x_2,\mu)+
(\hat{\sigma}^{\rm virt}_{gg})_{finite}(x_1,x_2,\mu) \right. \right.\nonumber \\ 
&+&\left. \left. (\hat \sigma^{soft}_{gg})_{finite}(x_1,x_2,\mu) +
\hat \sigma^{s+v+ct}_{gg}(x_1,x_2,\mu) + (1\leftrightarrow 2) \right] \right\} 
\nonumber\\
&+&\frac{1}{2}\int dx_1 dx_2 \left\{
\int_{x_1}^{1-\delta_s}\frac{dz}{z}
\left[{\cal F}_g^p(\frac{x_1}{z},\mu) {\cal F}_{g}^{\bar p(p)}(x_2,\mu)+
{\cal F}_g^{p}(x_2,\mu) {\cal F}_{g}^{\bar p(p)}(\frac{x_1}{z},\mu)\right]
\right. \nonumber\\
&&\times \left. \hat{\sigma}^{\rm \sss LO}_{gg}(x_1, x_2,\mu)
\frac{\alpha_s}{2\pi} \ln\left(\frac{s}{\mu^2}\frac{(1-z)^2}{z}
\frac{\delta_c}{2}\right) P_{gg}(z)
+(1\leftrightarrow 2) \right\}\nonumber\\
&+& \frac{1}{2} \int dx_1 dx_2 \left\{ {\cal F}_g^p(x_1,\mu)
{\cal F}_{g}^{\bar p(p)}(x_2,\mu) \, 
\hat{\sigma}^{hard/non-coll}_{gg}(x_1,x_2,\mu)+(1 \leftrightarrow 2) 
\right\}\,\,\,,\nonumber\\
\end{eqnarray}
where $\hat{\sigma}^{s+v+ct}_{gg}$ is obtained from the sum of $(\hat{\sigma}^{\rm
virt}_{gg})_{UV-pole}$ of Eq.~(\ref{eq:sigma_virt_uv_poleggZbb}), $\hat\sigma^{s+v}_{gg}$ of
Eq.~(\ref{eq:irsvggZbb}), and the PDF counterterm in Eq.~(\ref{eq:pdfgg_mu2}) as follows
\begin{equation}
\hat{\sigma}^{s+v+ct}_{gg}=
\frac{\alpha_s}{2\pi} \left[4 N \ln(\delta_s) \ln\left(\frac{s}{\mu^2}\right) 
+\left(\frac{11}{3} N-\frac{2 n_{lf}}{3}+4N \ln(\delta_s)\right) 
\ln\left(\frac{m_b^2}{s}\right)\right]\hat{\sigma}^{\rm \sss LO}_{gg}\; .
\end{equation}

We note that $\sigma^{\rm \sss NLO}_{gg}$ is finite, since, after mass factorization, both soft and
collinear singularities have been canceled between $\hat{\sigma}^{\rm
virt}_{gg}+\hat\sigma^{soft}_{gg}$ and $\hat{\sigma}^{hard/coll}_{gg}$.  The last terms respectively
describe the finite real gluon emission of Eq.~(\ref{eq:sigma_gg_hard}).  Note that when collecting
all the terms in Eq.~(\ref{eq:sigmatot_gg2}) that are proportional to $\ln(\mu^2/s)$, one obtains
exactly the last two terms in Eq.~(\ref{eq:fmuNLOdep}), as predicted by renormalization group
arguments.

For the $(q,\bar{q})g$ initiated processes we find
\begin{eqnarray}
\label{eq:sigmatot_qg2}
\sigma^{qg}_{\rm \sss NLO} &=&
\frac{\alpha_s}{2\pi} \sum_{i=q,\bar{q}}\int dx_1dx_2 
\left\{ 
\int_{x_1}^{1}\frac{dz}{z} 
{\cal F}_i^p(\frac{x_1}{z},\mu) {\cal F}_{g}^{\bar p(p)}(x_2,\mu) 
\times \right. \nonumber \\
&&\left. \hat{\sigma}^{\rm \sss LO}_{gg}(x_1, x_2,\mu) 
\left[P^4_{ig}(z) 
\ln\left(\frac{s}{\mu^2}\frac{(1-z)^2}{z}\frac{\delta_c}{2}\right)-
P^{\prime}_{ig}(z) \right]\right.\nonumber\\
&+& \left. \int_{x_1}^{1}\frac{dz}{z} 
{\cal F}_g^p(\frac{x_1}{z},\mu) {\cal F}_{i}^{\bar p(p)}(x_2,\mu) 
\times \right. \nonumber\\
&&\left. \hat{\sigma}^{\rm \sss LO}_{q\bar{q}}(x_1, x_2,\mu) 
\left[
{P}^4_{gi}(z) \ln\left(\frac{s}{\mu^2}\frac{(1-z)^2}{z}
\frac{\delta_c}{2}\right)-
{P}^{\prime}_{gi}(z)\right] +(1\leftrightarrow 2)\right\}\nonumber\\
&+&\sum_{i=q,\bar{q}} \int dx_1 dx_2 
\left\{ {\cal F}_i^p(x_1,\mu) {\cal F}_{g}^{\bar p(p)}(x_2,\mu) \, 
\hat{\sigma}^{non-coll}_{ig}(x_1,x_2,\mu)+(1 \leftrightarrow 2)
\right\}\,\,\,,\nonumber\\
\end{eqnarray}

We would like to conclude this section by showing explicitly that the total NLO cross section,
$\sigma^{\rm \sss NLO}$, does not depend on the arbitrary cutoffs introduced by the PSS method, i.e.
on $\delta_s$ and $\delta_c$. The cancellation of the PSS cutoff dependence is realized in
$\sigma^{\rm real}$ by matching contributions that are calculated either analytically, in the
IR-unsafe region below the cutoffs, or numerically, in the IR-safe region above the cutoffs.  While
the analytical calculation in the IR-unsafe region reproduces the form of the cross section in the
soft or collinear limits and is therefore only accurate for small values of the cutoffs, the
numerical integration in the IR-safe region becomes unstable for very small values of the cutoffs.
Therefore, obtaining a convincing cutoff independence involves a delicate balance between the
previous antagonistic requirements and ultimately dictates the choice of values that are neither too
large nor too small for the cutoffs. The Monte Carlo phase space integration has been performed
using the adaptive multi-dimensional integration routine VEGAS \cite{Lepage:1977sw}.
\begin{figure}[ht]
\begin{center}
\includegraphics*[scale=0.55]{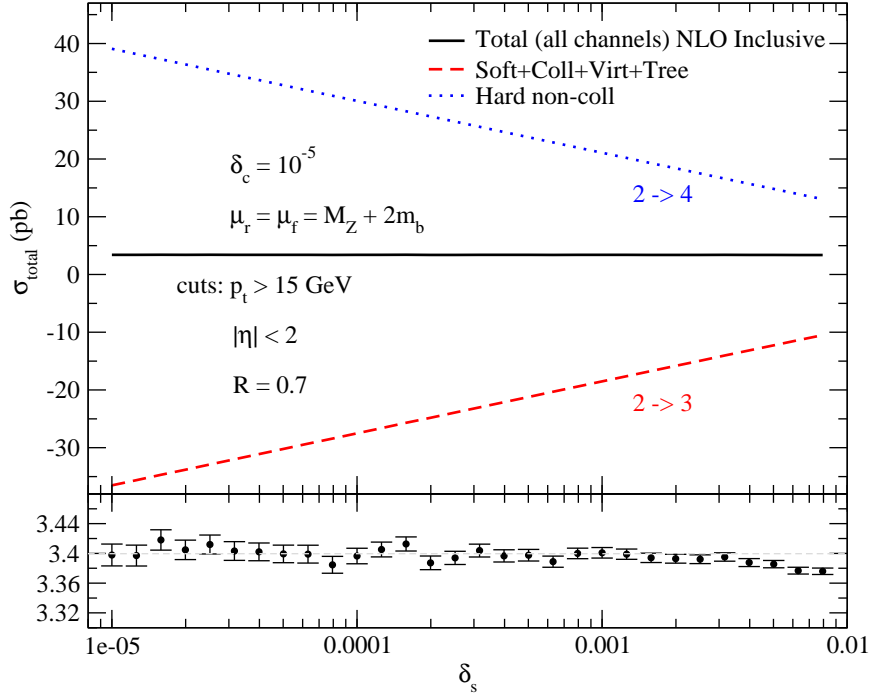}
\caption[Dependence of the total $\Zbb$ NLO QCD cross section on the
  $\delta_s$ PSS parameter.]{Dependence of $\sigma^{\rm \sss NLO}(p\bar{p}\to \Zbb)$ on the
  soft cutoff $\delta_s$ of the two-cutoff PSS method for
  $\mu\!=\!M_{\sss Z}+2m_b$, and $\delta_c\!=\!10^{-5}$. The upper plot
  shows the cancellation of the $\delta_s$-dependence between
  $\sigma^{soft}+\sigma^{hard/coll}$ and
  $\sigma^{hard/non-coll}$. The lower plot shows, on an enlarged
  scale, the dependence of the full $\sigma^{\rm \sss NLO}=\sigma^{\rm\sss
    NLO}_{gg}+\sigma^{\rm \sss NLO}_{q\bar{q}}+\sigma^{\rm \sss NLO}_{qg}$ on
  $\delta_s$ with the corresponding statistical errors.}
\label{fig:ds_dependenceZbb}
\end{center}
\end{figure}
\begin{figure}[ht]
\begin{center}
\includegraphics*[scale=0.55]{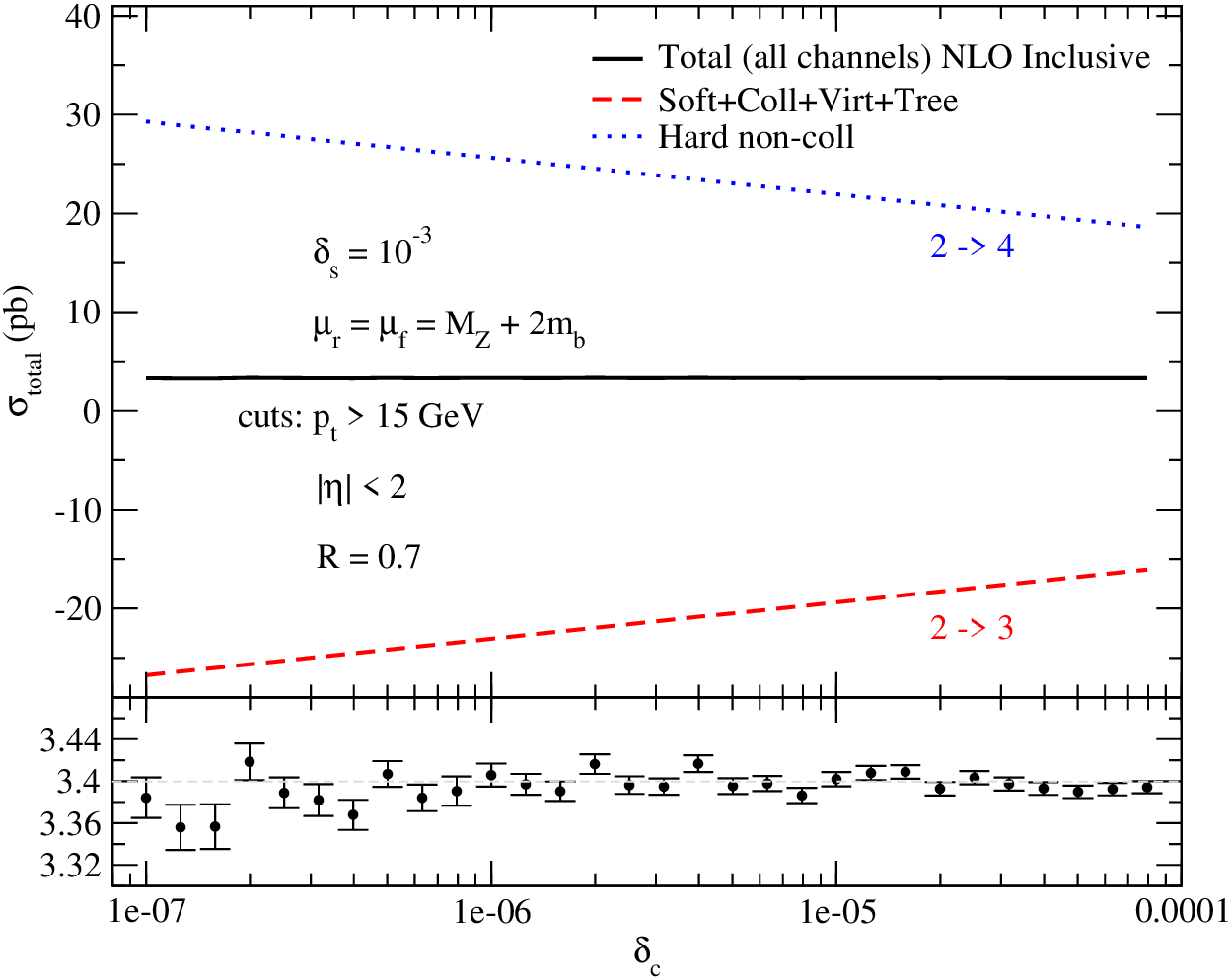}
\caption[Dependence of the total $\Zbb$ NLO QCD cross section on the
  $\delta_c$ PSS parameter.]{Dependence of $\sigma^{\rm \sss NLO}(p\bar{p}\to \Zbb)$ on the
  collinear cutoff $\delta_c$ of the two-cutoff PSS method, 
  for $\mu\!=\!M_{\sss Z}+2m_b$, and $\delta_s\!=\!10^{-3}$. The upper plot
  shows the cancellation of the $\delta_s$-dependence between
  $\sigma^{soft}+\sigma^{hard/coll}$, and
  $\sigma^{hard/non-coll}$. The lower plot shows, on an enlarged
  scale, the dependence of the full $\sigma^{\rm \sss NLO}=\sigma^{\rm\sss
    NLO}_{gg}+\sigma^{\rm \sss NLO}_{q\bar{q}}+\sigma^{\rm \sss NLO}_{qg}$ on
  $\delta_c$ with the corresponding statistical errors.}
\label{fig:dc_dependenceZbb}
\end{center}
\end{figure}

Figures~\ref{fig:ds_dependenceZbb} and \ref{fig:dc_dependenceZbb} illustrate the dependence of the
total cross section $\sigma^{\rm \sss NLO}(p\bar{p}\to \Zbb)$ on the two-cutoffs of the PSS method, using the setup outlined in
Section~\ref{sec:setup}. 
In
Figure~\ref{fig:ds_dependenceZbb}, $\delta_s$ is varied between $10^{-5}$ and $10^{-2}$ with
$\delta_c\!=\!10^{-5}$, while in Figure~\ref{fig:dc_dependenceZbb}, $\delta_c$ is varied between
$10^{-7}$ and $10^{-4}$ with $\delta_s\!=\!10^{-3}$. In both plots, we show in the upper window the
overall cutoff dependence cancellation between $\sigma^{soft}+\sigma^{hard/coll}$ and
$\sigma^{hard/non-coll}$ in $\sigma^{\rm real}$ including all channels, $gg$, $q\bar q$
and $qg$. We include too contributions from the LO and the virtual cross sections which are cutoff
independent.  In the lower window of the same plots we complement this information by reproducing
the full $\sigma^{\rm \sss NLO}$, including all channels, on a larger scale that magnifies the
details of the cutoff dependence cancellation. The statistical errors from the Monte Carlo phase
space integration are also shown.  Both Figures~\ref{fig:ds_dependenceZbb} and
\ref{fig:dc_dependenceZbb} show a clear plateau over a wide range of $\delta_s$ and $\delta_c$ and
the NLO cross section is proven to be cutoff independent. The results presented in
Chapter~\ref{chap:results} have been obtained by using $\delta_s\!=\!10^{-3}$ and
$\delta_c\!=\!10^{-5}$.

\chapter{Numerical Results}
\label{chap:results}
In this Chapter, we present numerical results for the total cross sections and distributions for
$\WZbb$ production including NLO QCD corrections and complete bottom-quark mass
effects~\cite{FebresCordero:2006sj,FebresCordero:2008ci}.  We specialize our discussion to the case
of the Tevatron collider, because this is at the moment the most interesting phenomenological
environment (see introduction in Chapter~\ref{chap:intro}).  We also investigate the stability of
the NLO QCD results by studying the dependence of the total cross section on the renormalization
($\mu_r$) and factorization ($\mu_f$) scales.  Finally, we carefully compare our results with
results obtained from a NLO calculation that considers massless bottom quarks by using the MCFM
code~\cite{MCFM:2004}.

In Section~\ref{sec:setup} we specify the setup used to produce the plots, while in
Sections~\ref{sec:resWbb} and \ref{sec:resZbb} we present and discuss results for $\Wbb$ and $\Zbb$
production respectively.

\section{The Setup}\label{sec:setup}

We present NLO QCD results for $\WZbb$ production at the Tevatron using a non-zero bottom-quark mass
fixed at $m_b$=4.62 GeV. The $W$ and $Z$ bosons are considered on-shell and their masses are taken
to be $M_W=80.410$~GeV and $M_Z=91.1876$~GeV.  The mass of the top quark, entering in virtual
corrections, is set to $m_t=170.9$~GeV. The LO results use the one-loop evolution of $\alpha_s$ and
the CTEQ6L1 set of PDF~\cite{Pumplin:2002vw}, with $\alpha_s^{\rm LO}(M_Z)=0.130$ , while the NLO
results use the two-loop evolution of $\alpha_s$ and the CTEQ6M set of PDF, with $\alpha_s^{\rm
NLO}(M_Z)=0.118$.  The $W$ boson coupling to quarks is proportional to the Cabibbo-Kobayashi-Maskawa
(CKM) matrix elements.  We take $V_{ud}=V_{cs}=0.975$ and $V_{us}=V_{cd}=0.222$, while we neglect
the contribution of the third generation of quarks, since it is suppressed either by the initial
state quark densities or by the corresponding CKM matrix elements.

Partons cannot be detected as they are always confined in hadrons. For this reason, any
phenomenological collider study, including final hadronic states, must implement a jet algorithm to
recombine partons into jets, in a way consistent with factorization (see Section~\ref{sec:FactTh})
and with experimental techniques.  The jet algorithm basically assigns a ``separation" between
partons, and, based on it, defines criteria to decide whether to group a set of partons in the final
state into a ``proto-jet". Finally, kinematic cuts are applied depending on the experimental setup,
to decide if a proto-jet is in an observable region, in which case the proto-jet is promoted to a
jet. We will consider in our study $b$-type and \emph{light}-type jets, where the former contains
either a $b$ or a $\bar b$ quark and the latter can only contain massless quarks or gluons.

We implement the $k_T$ jet algorithm~\cite{Catani:1992zp,Catani:1993hr,Ellis:1993tq,Kilgore:1996sq}
with a pseudo-cone size $R=0.7$ and we recombine the parton momenta within a jet using the so called
covariant $E$-scheme~\cite{Catani:1993hr}. We checked that our implementation of the $k_T$ jet
algorithm coincides with the one in MCFM.  We require all events to have a $b\bar{b}$ jet pair in
the final state, with each jet having a transverse momentum larger than $15$~GeV
($p_T^{b,\bar{b}}>15$~GeV) and a pseudorapidity that satisfies $|\eta^{b,\bar{b}}|<2$. We impose the
same $p_T$ and $|\eta|$ cuts also on the extra jet that may arise due to hard non-collinear real
emission of a parton, i.e. in the processes $\WZbb+g$ or $\WZbb+q(\bar{q})$. This hard non-collinear
extra parton is treated either \emph{inclusively} or \emph{exclusively}.  In the \emph{inclusive}
case we include both two- and three-jet events, while in the \emph{exclusive} case we require
exactly two jets in the event. Two-jet events consist of a bottom-quark jet pair that may also
include a final-state light parton (gluon or quark) due to the applied jet algorithm. Results in the
massless bottom-quark approximation have been obtained using the MCFM code~\cite{MCFM:2004}.

\begin{figure}[t]
\begin{center}
\includegraphics*[scale=0.4]{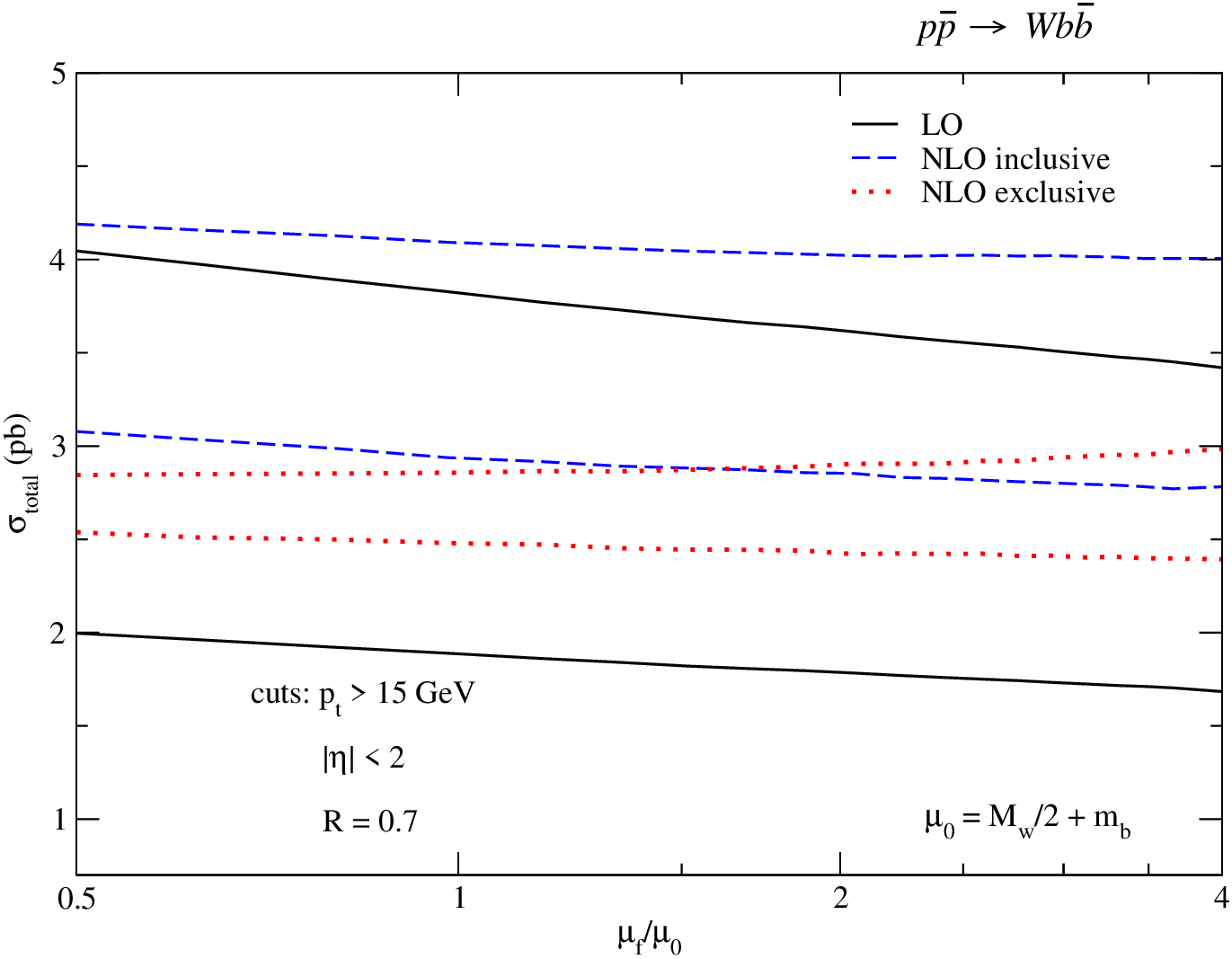} 
\caption{Dependence of the LO (black solid band), NLO
\emph{inclusive} (blue dashed band), and NLO
\emph{exclusive} (red dotted band) $\Wbb$ total cross sections on the
renormalization/factorization scales, including full bottom-quark
mass effects. The bands are obtained by independently varying both $\mu_r$ and
$\mu_f$ between $\mu_0/2$ and $4\mu_0$ (with $\mu_0=m_b+M_W/2$).}
\label{fig:mu_dependence_band_Wbb}
\end{center}
\end{figure}

\begin{table}[t]
\begin{center}
  \caption{LO and NLO total $\Wbb$ cross sections at the Tevatron for massive and
    massless bottom quarks, using $\mu_r=\mu_f=M_W+2m_b$. The numbers
    in square brackets are the ratios of the NLO and
    LO cross sections, the so called $K$-factors. Statistical errors of the
MC integration amount to about 0.1\%.}
\begin{tabular}{l|cc}
\hline
Cross Section & $m_b\ne 0$ (pb) [ratio] & $m_b=0$ (pb) [ratio] \\
\hline
$\sigma_{\rm LO}$ & $2.20 [-] $ & $2.38 [-] $ \\
$\sigma_{\rm NLO}$ \emph{inclusive} & $3.20 [1.45] $& $3.45 [1.45]$ \\
$\sigma_{\rm NLO}$ \emph{exclusive} & $2.64 [1.20] $& $2.84 [1.19]$ \\
\hline
\end{tabular}
\label{tb:WbbXsecs}
\end{center}
\end{table}
\section{$\Wbb$ Production at the Tevatron}\label{sec:resWbb}

Let us first consider the influence of the NLO QCD corrections on the total cross section. In
Table~\ref{tb:WbbXsecs} we present the results obtained for both LO and NLO total cross sections, at
a reference scale $\mu_r=\mu_f=M_W+2m_b$, both in our fully massive calculation and in the massless
approximation.

It can be seen that, given the setup explained in Section~\ref{sec:setup}, the NLO QCD corrections
increase considerably the total cross section, with NLO/LO ratios ($K$-factors) of about 1.45 and
1.2 for the inclusive and exclusive case respectively (for both the massive and massless
calculations). We can see also that, in general, the massless approximation overestimates the total
cross section.  In the following we will study in detail where these corrections are more important,
and we will show that, in the case of distributions, a global rescaling (or $K$-factor) does not
properly simulate the NLO corrections.

\begin{figure}[hp]
 \begin{center}
  \subfigure[\emph{Inclusive} case]{\label{fig:mu_dependence_inc_Wbb}\includegraphics*[scale=0.7]{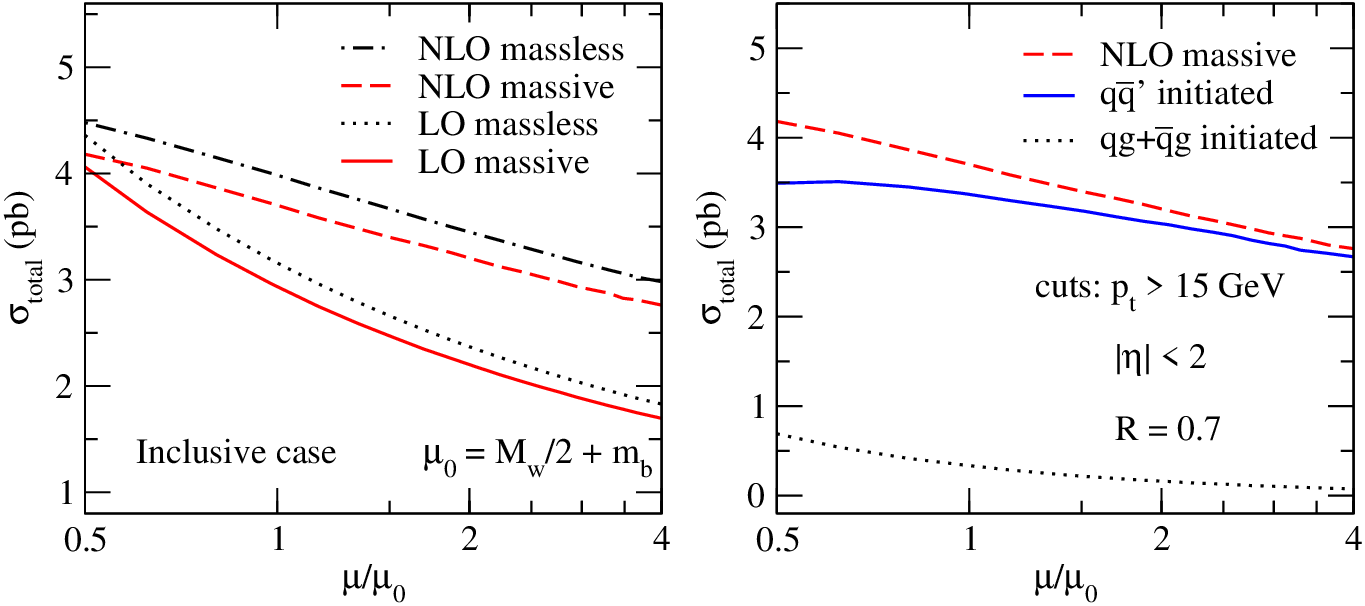}}\\
  \vspace{0.8cm}
  \subfigure[\emph{Exclusive} case]{\label{fig:mu_dependence_exc_Wbb}\includegraphics*[scale=0.7]{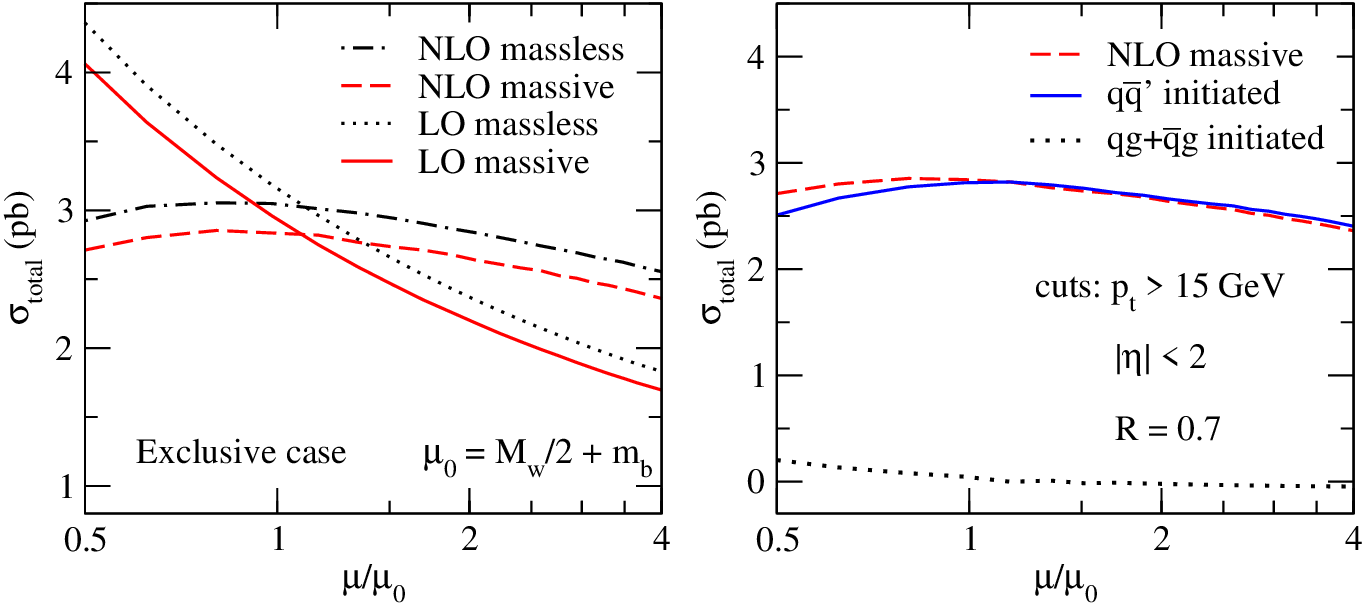}}
 \end{center}
\caption{Dependence of the LO and NLO \emph{inclusive} and \emph{exclusive} $p\bar{p}\rightarrow\Wbb$ total cross section
on the renormalization/factorization scale, when
$\mu_r=\mu_f=\mu$. The left hand side plot compares both LO and NLO
total cross sections for the case in which the bottom quark is treated
as massless (MCFM) or massive (our calculation).  The right hand side
plot shows separately, for the massive case only, the scale dependence
of the $q\bar{q}^\prime$ and $qg+\bar{q}g$ contributions, as well as
their sum.}
\label{fig:mu_dependence_Wbb}
\end{figure}

\begin{figure}[ht]
\begin{center}
\includegraphics*[scale=0.5]{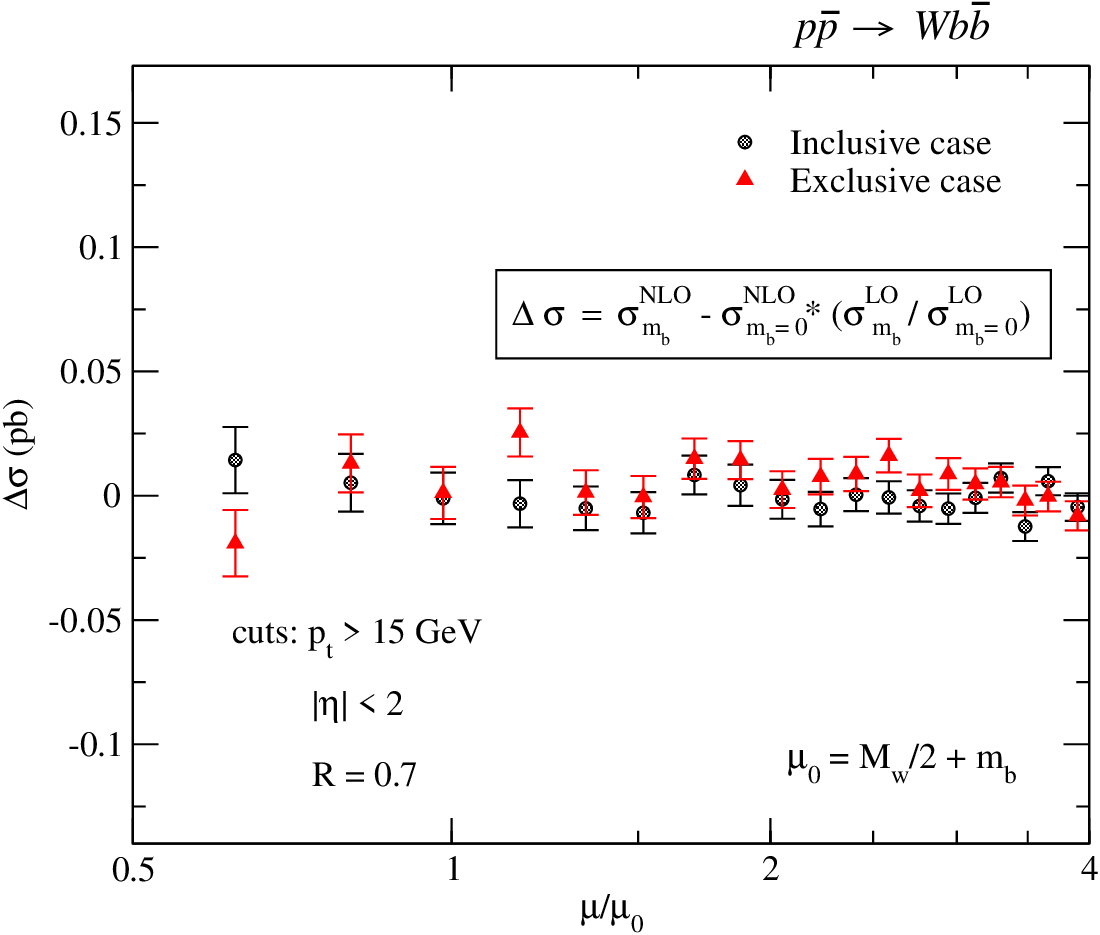} 
\caption{Dependence 
on the renormalization/factorization scale of the rescaled difference 
between our NLO calculation (with $m_b\ne 0$) of the total $\Wbb$ cross section 
and the corresponding result computed using MCFM (with
$m_b=0$) for the \emph{inclusive} and \emph{exclusive} cases
(with $\mu_r\!=\!\mu_f\!=\!\mu_0$) respectively. The error bars indicate the statistical uncertainty of
  the Monte Carlo integration.}
\label{fig:sigma_ratio_NLO_Wbb}
\end{center}
\end{figure}

In Figures~\ref{fig:mu_dependence_band_Wbb} and \ref{fig:mu_dependence_Wbb}, we illustrate the
renormalization and factorization scale dependence of the LO and NLO total cross sections, both in
the \emph{inclusive} and \emph{exclusive} case.  Figure~\ref{fig:mu_dependence_band_Wbb} shows the
overall scale dependence of both LO, NLO \emph{inclusive} and NLO \emph{exclusive} total cross
sections, when both $\mu_r$ and $\mu_f$ are varied independently between $\mu_0/2$ and $4\mu_0$
(with $\mu_0=m_b+M_W/2$), including full bottom-quark mass effects. We notice that the NLO cross
sections have a reduced scale dependence over the range of scales shown, and the \emph{exclusive}
NLO cross section is more stable than the \emph{inclusive} one especially at low scales.  This is
consistent with the fact that the \emph{inclusive} NLO cross section integrates over the entire
phase space of the $qg(\bar{q}g)\rightarrow b\bar{b}W + q(\bar{q})$ channels that are evaluated with
NLO $\alpha_s$ and NLO PDFs, but are actually tree-level processes and retain therefore a strong
scale dependence. In the \emph{exclusive} case only the $2\rightarrow 3$ collinear kinematic of
these processes is retained, since 3-jets events are discarded, and this makes the overall
renormalization and factorization scale dependence milder.  To better illustrate this point, we show
in the right hand side plots of Figures~\ref{fig:mu_dependence_inc_Wbb} and
\ref{fig:mu_dependence_exc_Wbb} the $\mu$-dependence of the total cross section and of the partial
cross sections corresponding to the $q\bar{q}^\prime$ and the $qg+\bar{q}g$ initiated channels
separately, for $\mu_r=\mu_f$, both for the \emph{inclusive} and for the \emph{exclusive} case.  It
is clear that the low scale behavior of the \emph{inclusive} cross section is mainly driven by the
$qg+\bar{q}g$ contribution.  In the left hand side plots of Figures~\ref{fig:mu_dependence_inc_Wbb}
and \ref{fig:mu_dependence_exc_Wbb}, we also compare the scale dependence of our results to the
scale dependence of the corresponding results obtained with $m_b=0$ (using MCFM), both at LO and at
NLO. Using a non-zero value of $m_b$ is not expected to have any impact on the scale dependence of
the result\footnote{Note that we always use $m_b=4.62$~GeV in the determination of the scales in
terms of $\mu_0=m_b+M_W/2$ even in the results obtained with $m_b=0$.} and, indeed, the scale
dependence of the LO and NLO pair of curves is very similar, with a shift due to the bottom-quark
mass effects. 

\begin{figure}[htp]
 \begin{center}
  \subfigure[\emph{Inclusive} case]{\label{fig:mbb_dist_LO_vs_NLO_inc_Wbb}
	\hspace{-0.5cm}\includegraphics*[scale=0.7,angle=-90]{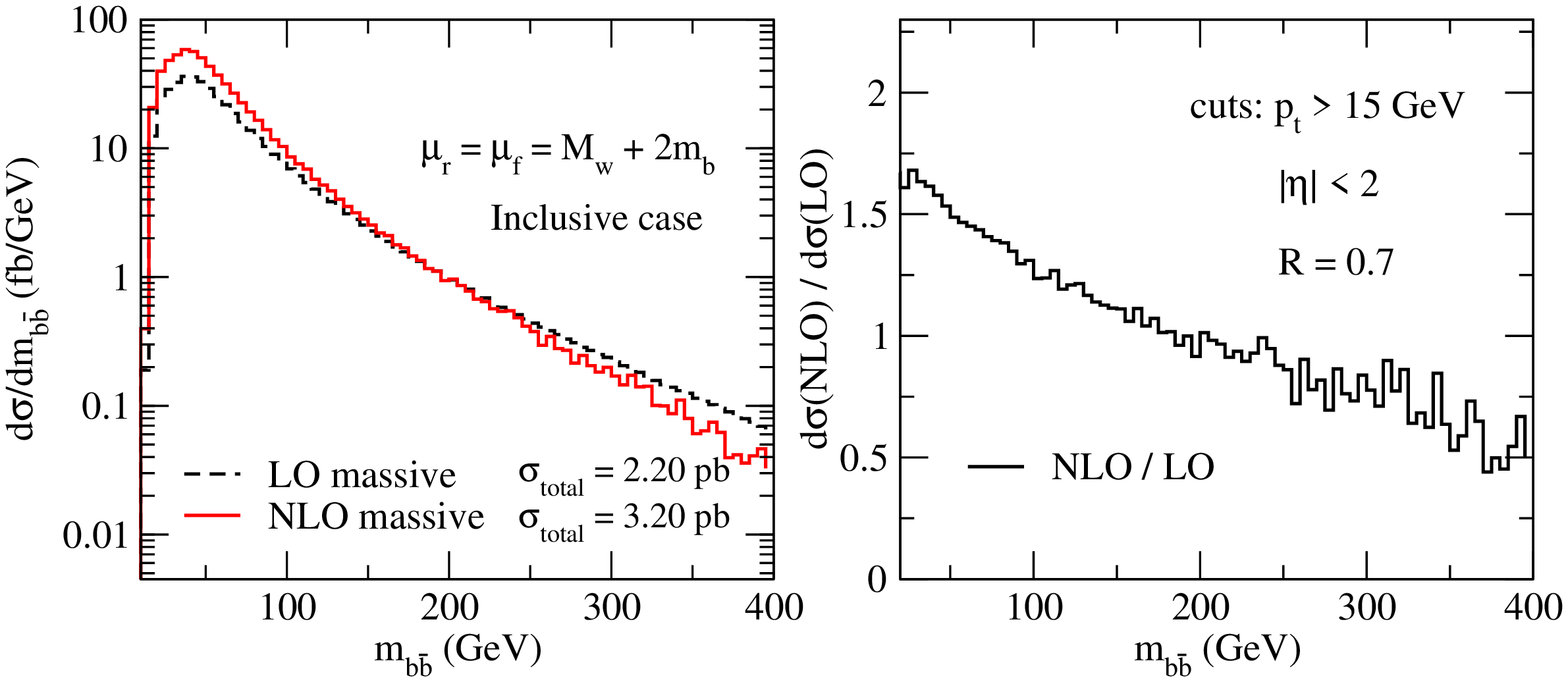}}\\
  \vspace{-0.4cm}
  \subfigure[\emph{Exclusive} case]{\label{fig:mbb_dist_LO_vs_NLO_exc_Wbb}
	\hspace{-0.5cm}\includegraphics*[scale=0.7,angle=-90]{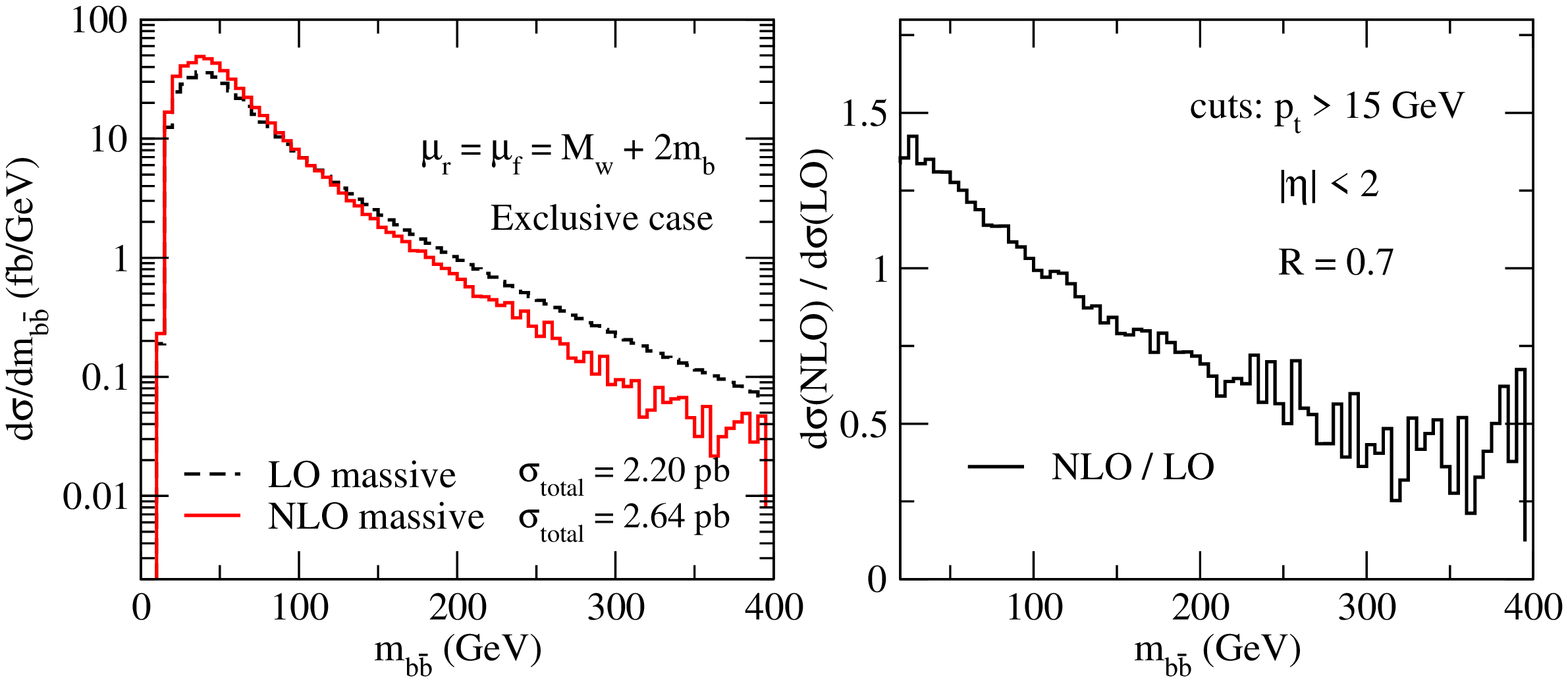}}
 \end{center}
   \caption{The distribution $d\sigma(p\bar{p}\rightarrow\Wbb)/dm_{b\bar{b}}$
      in LO and NLO QCD. The right hand side plot shows the ratio of
      the LO and NLO distributions.}
    \label{fig:mbb_dist_LO_vs_NLO_Wbb}
\end{figure}

While the LO cross section still has a 40\% uncertainty due to scale dependence, this uncertainty is
reduced at NLO to about 20\% for the \emph{inclusive} and to about 10\% for the \emph{exclusive}
cross sections. The uncertainties have been estimated as the positive/negative deviation with
respect to the mid-point of the bands plotted in Figure~\ref{fig:mu_dependence_band_Wbb}, where each
band range is defined by the minimum and maximum value in the band.  We notice incidentally that the
difference due to finite bottom-quark mass effects is less significant than the theoretical
uncertainty due to the residual scale dependence in the \emph{inclusive} case, but is comparable in
size in the \emph{exclusive} case.  Indeed, the finite bottom-quark mass effects amount to about 8\%
in both \emph{inclusive} and \emph{exclusive} cases.

\begin{figure}[hp]
 \begin{center}
  \subfigure[\emph{Inclusive} case]{\label{fig:mbb_dist_NLO_inc_Wbb}
	\hspace{-0.5cm}\includegraphics*[scale=0.7,angle=-90]{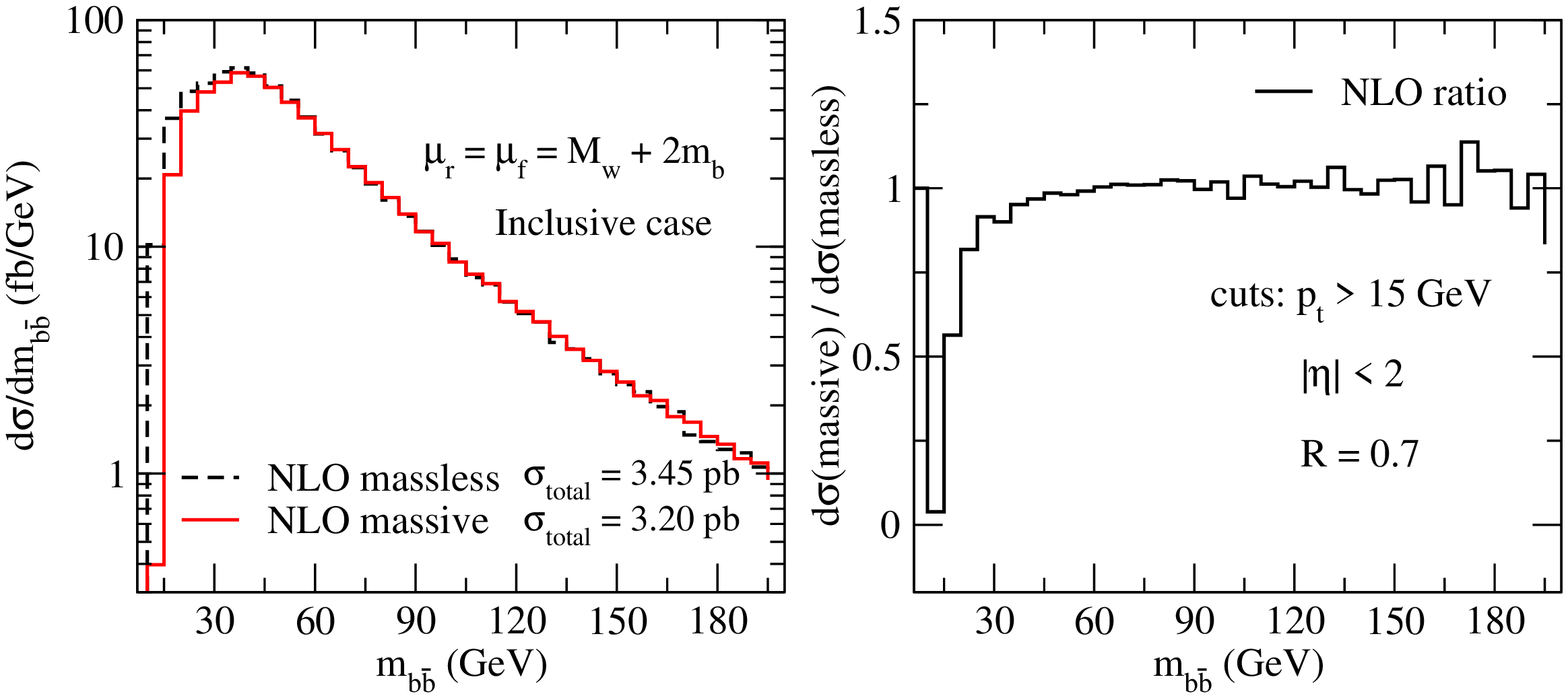}}\\
  \vspace{-0.4cm}
  \subfigure[\emph{Exclusive} case]{\label{fig:mbb_dist_NLO_exc_Wbb}
	\hspace{-0.5cm}\includegraphics*[scale=0.7,angle=-90]{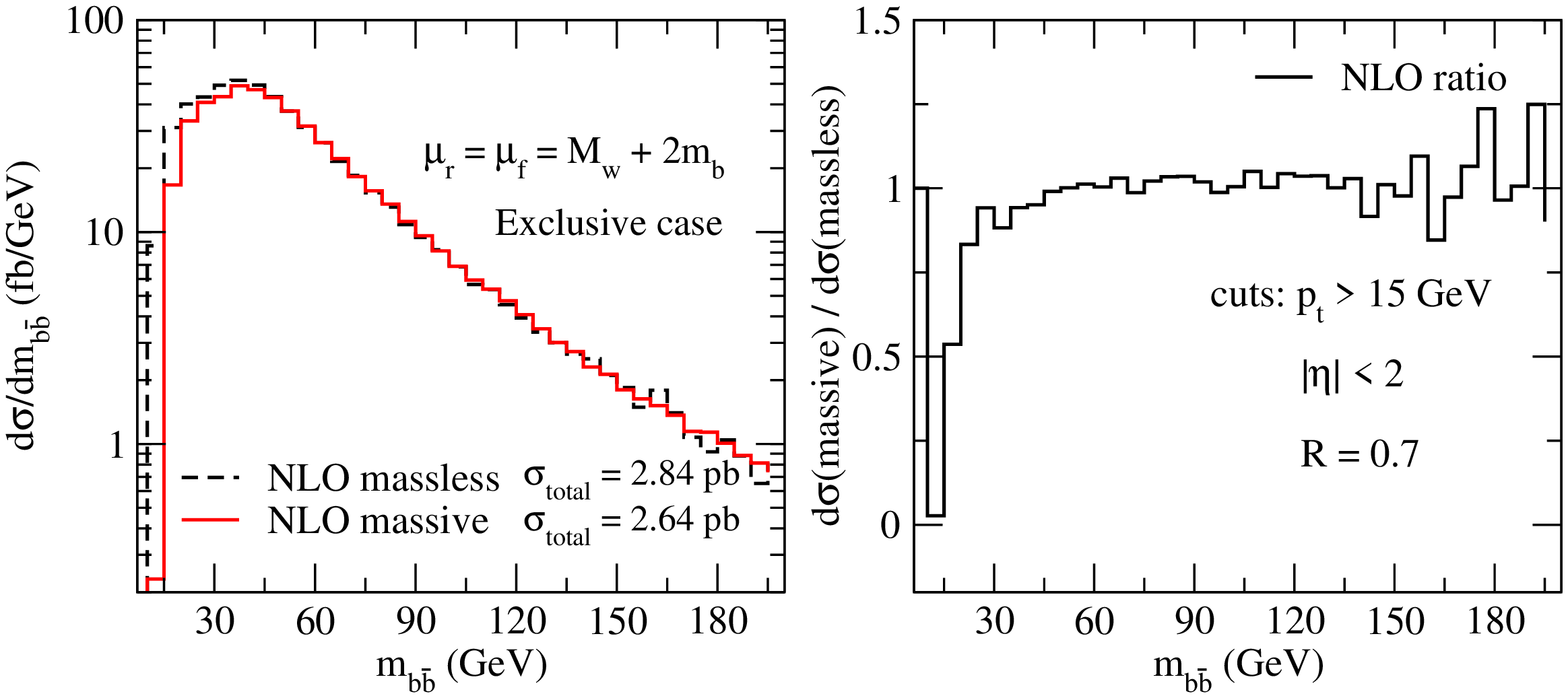}}
 \end{center}
\caption{The \emph{inclusive} and \emph{exclusive} distributions $d\sigma(p\bar{p}\rightarrow\Wbb)/dm_{b\bar{b}}$
derived from our calculation (with $m_b\ne 0$) and from MCFM (with
$m_b=0$).  The right hand side plot shows the ratio of the two
distributions, $d\sigma(m_b\neq 0)/d\sigma(m_b=0)$.}
\label{fig:mbb_dist_NLO_Wbb}
\end{figure}

In Figure~\ref{fig:sigma_ratio_NLO_Wbb} we show the rescaled difference between the NLO total cross
sections obtained from our calculation (with $m_b\ne 0$) and with MCFM (with $m_b=0$) defined as
follows: \begin{equation} \Delta\sigma=\sigma^{\rm NLO}(m_b\ne 0)-\sigma^{\rm NLO}(m_b=0) \;
\frac{\sigma^{\rm LO}(m_b\ne 0)}{\sigma^{\rm LO}(m_b=0)} \; .  \label{eq:mXsecrescaling}
\end{equation} As can be seen, within the statistical errors of the Monte Carlo integration, the
finite bottom-quark mass effects on the total cross sections at NLO are well described by the
corresponding effects at LO.

\begin{figure}[hp]
\begin{center}
\includegraphics*[scale=0.65,angle=-90]{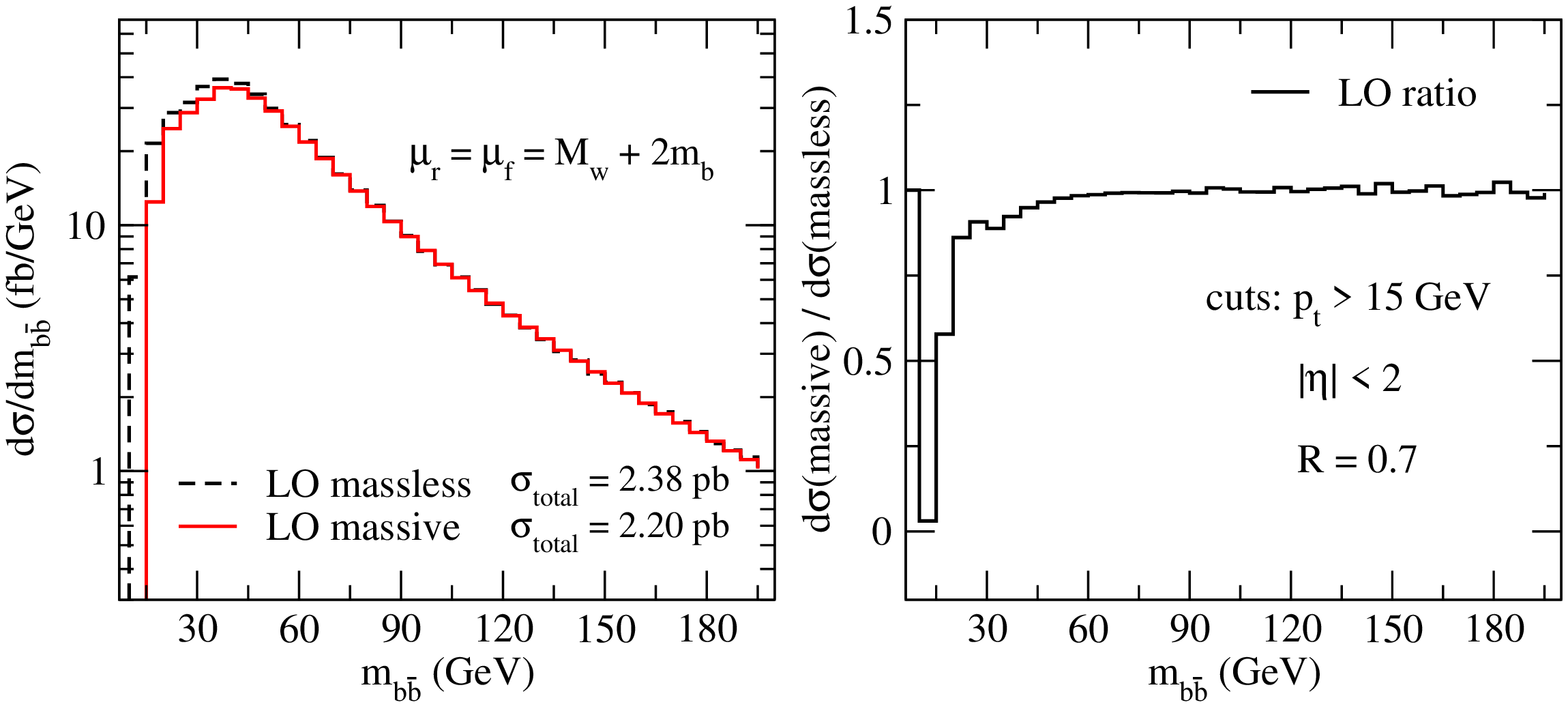} 
\caption{The LO distribution $d\sigma(p\bar{p}\rightarrow\Wbb)/dm_{b\bar{b}}$
derived from our calculation (with $m_b\ne 0$) and from MCFM (with
$m_b=0$). The right hand side plot shows the ratio of the two
distributions, $d\sigma(m_b\neq 0)/d\sigma(m_b=0)$.}
\label{fig:mbb_dist_LO_Wbb}
\end{center}
\end{figure}

\begin{figure}[hp]
\begin{center}
\includegraphics*[scale=0.37]{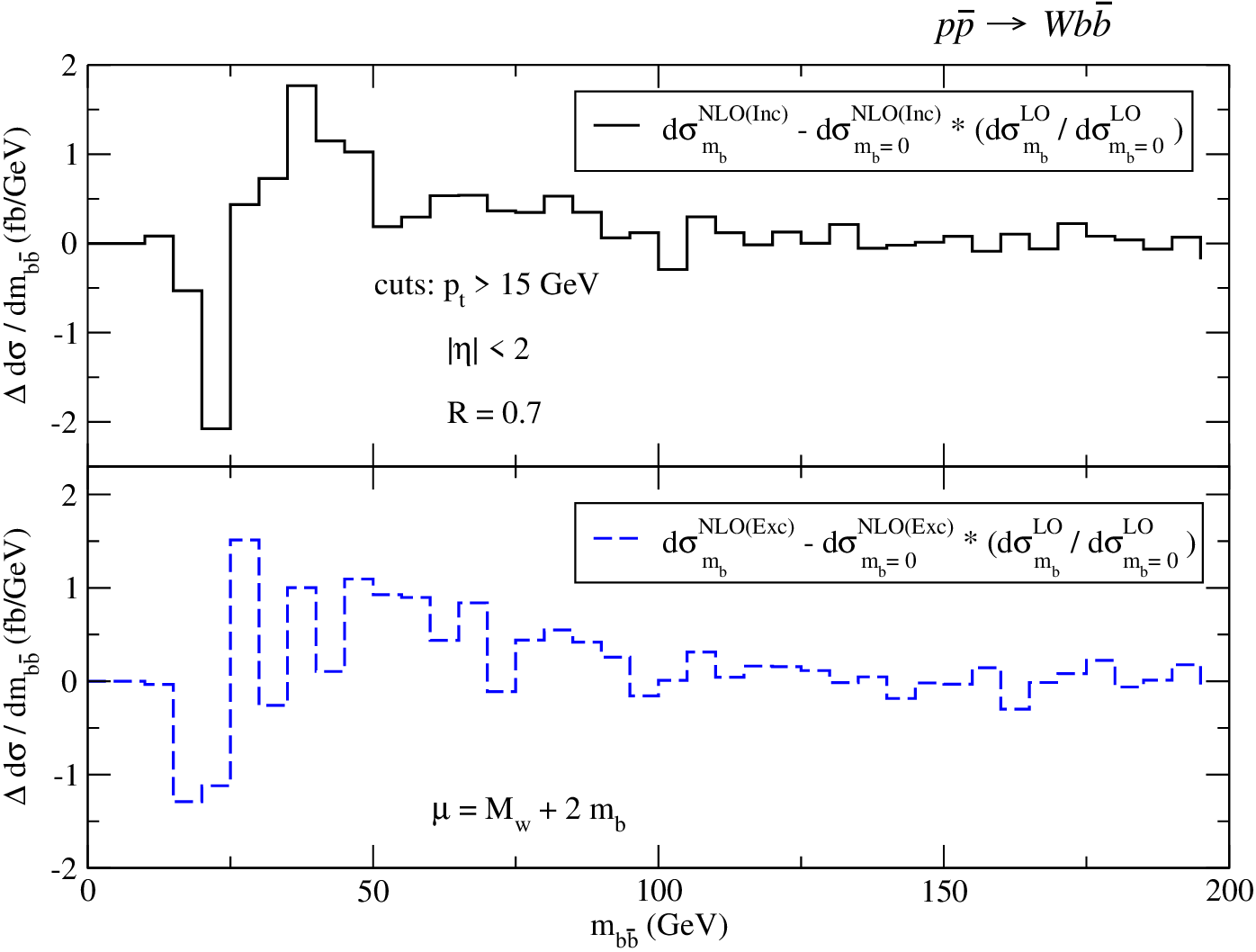} 
\caption{The $m_{b\bar b}$ distribution of the rescaled difference 
between our NLO calculation (with $m_b\ne 0$) and MCFM (with
$m_b=0$) for the \emph{inclusive} (upper plot) and \emph{exclusive} (lower plot)
$p\bar{p}\rightarrow\Wbb$ production.}
\label{fig:mbb_ratio_NLO_Wbb}
\end{center}
\end{figure}

Finally, in Figures~\ref{fig:mbb_dist_LO_vs_NLO_Wbb}-\ref{fig:mbb_dist_LO_Wbb} we study the
distribution $d\sigma/dm_{b\bar{b}}$, where $m_{b\bar{b}}$ is the invariant mass of the $b\bar{b}$
jet pair. The impact of NLO QCD corrections on this distribution is illustrated in
Figures~\ref{fig:mbb_dist_LO_vs_NLO_inc_Wbb} and \ref{fig:mbb_dist_LO_vs_NLO_exc_Wbb} for the
\emph{inclusive} and \emph{exclusive} case, respectively. We see that the NLO QCD corrections
affects the cross section quite substantially in particular for low values of $m_{b\bar{b}}$. In
each figure the right hand side plot gives the ratio of the NLO and LO distributions, providing a
sort of $K$-factor bin by bin. Figures~\ref{fig:mbb_dist_NLO_inc_Wbb} and
\ref{fig:mbb_dist_NLO_exc_Wbb} compare the NLO $d\sigma/dm_{b\bar{b}}$ distributions obtained from
the massive and massless bottom-quark calculations. The results with $m_b=0$ have been obtained
using MCFM.  As expected, most of the difference between the massless and massive bottom-quark cross
sections is coming from the region of low invariant mass $m_{b\bar{b}}$, both for the inclusive and
exclusive case, where the cross sections for $m_b\ne 0$ are consistently below the ones with
$m_b=0$.  For completeness, we also show in Figure~\ref{fig:mbb_dist_LO_Wbb} the comparison between
massive ($m_b\neq 0$) and massless ($m_b=0$) calculations at LO in QCD. The LO $m_{b\bar{b}}$
distribution for massive bottom-quarks has been obtained both from our calculation and from MCFM,
which implements the $m_b\neq 0$ option at tree level, and both results have been found in perfect
agreement. As can be seen by comparing
Figures~\ref{fig:mbb_dist_NLO_inc_Wbb}-\ref{fig:mbb_dist_NLO_exc_Wbb} and
Figure~\ref{fig:mbb_dist_LO_Wbb}, the impact of a non-zero bottom-quark mass is almost not affected
by including NLO QCD corrections.  To illustrate this in more detail, we show in
Figure~\ref{fig:mbb_ratio_NLO_Wbb} the rescaled difference between the $m_{b\bar b}$ distributions
obtained with our NLO calculation (with $m_b\ne 0$) and with MCFM (with $m_b=0$) defined as follows:
\begin{equation}
\Delta \frac{d\sigma}{d m_{b \bar b}}=\frac{d\sigma^{\rm NLO}}{d m_{b \bar b}}(m_b\ne 0)
-\frac{d\sigma^{\rm NLO}}{d m_{b \bar b}} (m_b=0)\; \frac{d\sigma^{\rm LO}(m_b\ne 0)}{d\sigma^{\rm LO}(m_b=0)} \; .
\label{eq:mDistrescaling}
\end{equation}
We notice that finite bottom-quark mass effects are particularly relevant for values of the
$m_{b\bar b}$ invariant mass below about 60 GeV and that they appear to be of the same order at LO
and NLO.

\begin{figure}[h]
\begin{center}
\includegraphics*[scale=0.4]{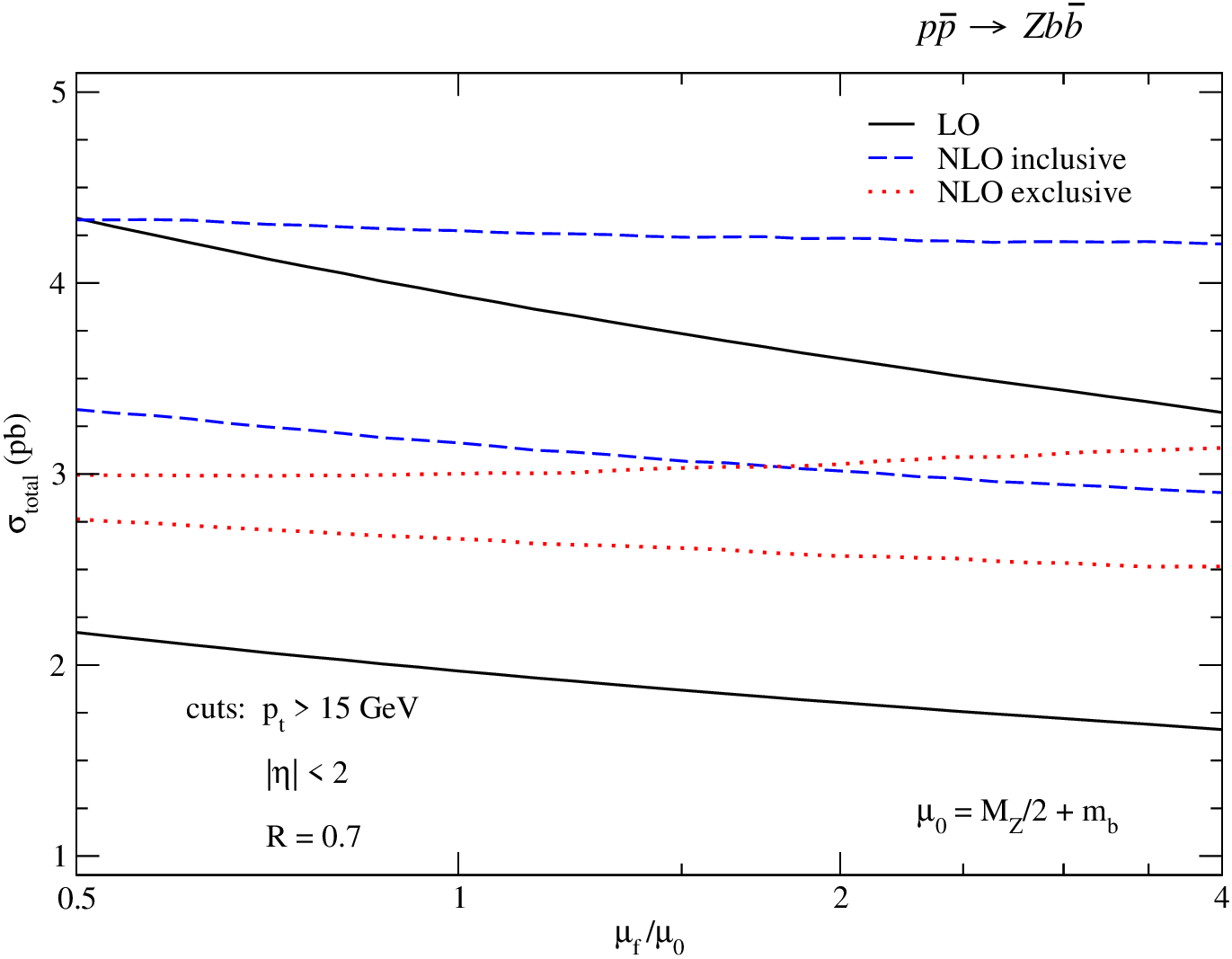} 
\caption{Dependence of the LO (black solid band), NLO
\emph{inclusive} (blue dashed band), and NLO
\emph{exclusive} (red dotted band) $\Zbb$ total cross sections on the
renormalization/factorization scales, including full bottom-quark
mass effects. The bands are obtained by independently varying both $\mu_r$ and
$\mu_f$ between $\mu_0/2$ and $4\mu_0$ (with $\mu_0=m_b+M_Z/2$).}
\label{fig:mu_dependence_band_Zbb}
\end{center}
\end{figure}
\section{$\Zbb$ Production at the Tevatron}\label{sec:resZbb}

To start, let us have a look at the influence of the NLO QCD corrections on the total cross section.
In Table~\ref{tb:ZbbXsecs} we present the values obtained with the scale $\mu_r=\mu_f=M_Z+2m_b$,
considering LO and NLO total cross sections, both in our fully massive calculation and in the
massless approximation.
\begin{table}[htp]
\begin{center}
  \caption{LO and NLO total $\Zbb$ cross sections at the Tevatron for massive and
    massless bottom quarks, using $\mu_r=\mu_f=M_Z+2m_b$. The numbers
    in square brackets are the ratios of the NLO and
    LO cross sections, the so called $K$-factors. Statistical errors of the
MC integration amount to about 0.1\%.}
\begin{tabular}{l|cc}
\hline
Cross Section & $m_b\ne 0$ (pb) [ratio] & $m_b=0$ (pb) [ratio] \\
\hline
$\sigma_{\rm LO}$ & $2.21 [-] $ & $2.37 [-] $ \\
$\sigma_{\rm NLO}$ \emph{inclusive} & $3.40 [1.54] $& $3.64 [1.54]$ \\
$\sigma_{\rm NLO}$ \emph{exclusive} & $2.80 [1.27] $& $3.01 [1.27]$ \\
\hline
\end{tabular}
\label{tb:ZbbXsecs}
\end{center}
\end{table}

It can be seen that, given the setup explained in Section~\ref{sec:setup}, the NLO QCD corrections
increase considerably the total cross section, with NLO vs. LO ratios ($K$-factors) of about 1.5 and
1.27 for the inclusive and exclusive case respectively (for both the massive and massless
calculations). We can also see that, in general, the massless approximation overestimates the total
cross section.  In the following we will study in detail where these corrections are more important,
and especially we show that, in the case of distributions, a global rescaling (or $K$-factor) does
not properly simulate the NLO corrections.

\begin{figure}[hp]
 \begin{center}
  \subfigure[\emph{Inclusive} case]{\label{fig:mu_dependence_inc_Zbb}\includegraphics*[scale=0.7]{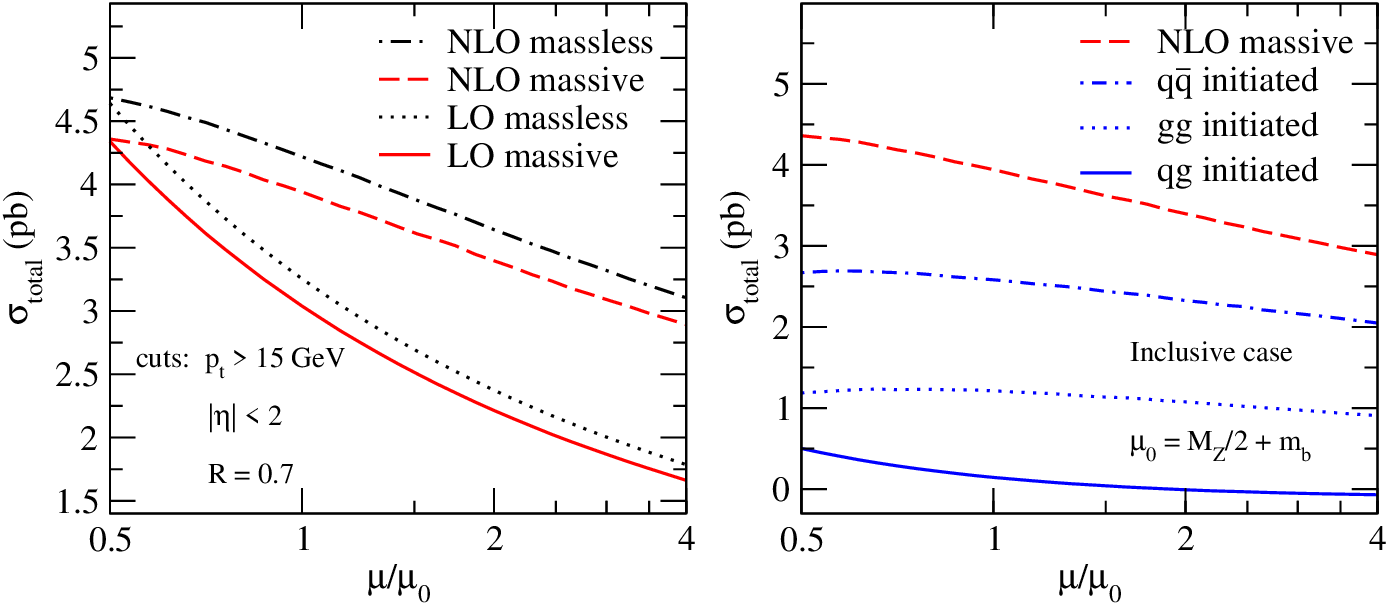}}\\
  \vspace{0.8cm}
  \subfigure[\emph{Exclusive} case]{\label{fig:mu_dependence_exc_Zbb}\includegraphics*[scale=0.7]{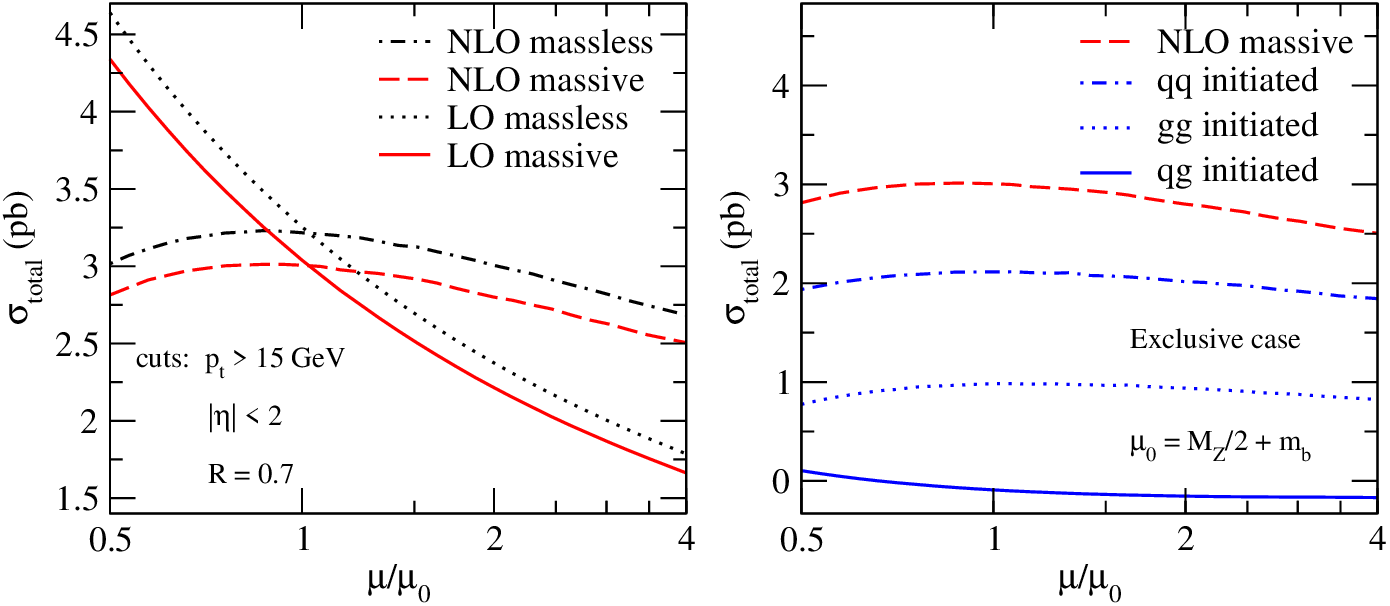}}
 \end{center}
 \caption{Dependence of the LO and NLO \emph{inclusive} and
   \emph{exclusive} $p\bar{p}\rightarrow\Zbb$ total cross section on the
   renormalization/factorization scale, when $\mu_r=\mu_f=\mu$. The
   LHS plots compare both LO and NLO total cross sections for the case
   in which the bottom quark is treated as massless (MCFM) or massive
   (our calculation).  The RHS plots show separately, for the massive
   case only, the scale dependence of the $q\bar{q},gg$ and
   $qg+\bar{q}g$ contributions, as well as their sum.}
\label{fig:mu_dependence_Zbb}
\end{figure}

In Figures~\ref{fig:mu_dependence_band_Zbb} and \ref{fig:mu_dependence_Zbb} we illustrate the
renormalization and factorization scale dependence of the LO and NLO total cross sections, both in
the \emph{inclusive} and \emph{exclusive} case.  Figure~\ref{fig:mu_dependence_band_Zbb} shows the
overall scale dependence of both LO, NLO \emph{inclusive} and NLO \emph{exclusive} total cross
sections, when both $\mu_r$ and $\mu_f$ are varied independently between $\mu_0/2$ and $4\mu_0$
(with $\mu_0=m_b+M_Z/2$), including full bottom-quark mass effects. We notice that the NLO cross
sections have a reduced scale dependence over the range of scales shown, and the \emph{exclusive}
NLO cross section is more stable than the \emph{inclusive}.  Similarly to what we have discussed in
the $\Wbb$ case, this effect is mainly driven by the tree level subprocess $q(\bar q)g\to\Zbb+q(\bar
q)$ contributing to the real corrections. In the $\Zbb$ case, we also have a new initial state,
namely $gg$. Its scale dependence behavior is similar to the $q\bar q$ initiated subprocess.  To
illustrate the independent contributions, we show in the right hand side plots of
Figures~\ref{fig:mu_dependence_inc_Zbb} and \ref{fig:mu_dependence_exc_Zbb} the $\mu$-dependence of
the total cross section and of the partial cross sections corresponding to the $q\bar{q}$,
$qg+\bar{q}g$ and $gg$ initiated channels separately, for $\mu_r=\mu_f$, both for the
\emph{inclusive} and for the \emph{exclusive} case.  It is clear that the low scale behavior of the
\emph{inclusive} cross section is considerably affected by the $qg+\bar{q}g$ contribution, which
show a monotonic dependence on $\mu$ (i.e. with no plateau) characteristic of tree level processes.
In the left hand side plots of Figures~\ref{fig:mu_dependence_inc_Zbb} and
\ref{fig:mu_dependence_exc_Zbb} we also compare the scale dependence of our results to the scale
dependence of the corresponding results obtained with $m_b=0$ (using MCFM), both at LO and at NLO.
Using a non-zero value of $m_b$ is expected to have a mild impact on the scale dependence of the
results, as the only modification to the renormalization scale dependence comes from the bottom
quark mass renormalization, as shown in the subsections of Section~\ref{sec:Zbbcalc} in
Chapter~\ref{chap:calculation}.  Indeed, the scale dependence of the LO and NLO curves is very
similar.

\begin{figure}[ht]
\begin{center}
\includegraphics*[scale=0.5]{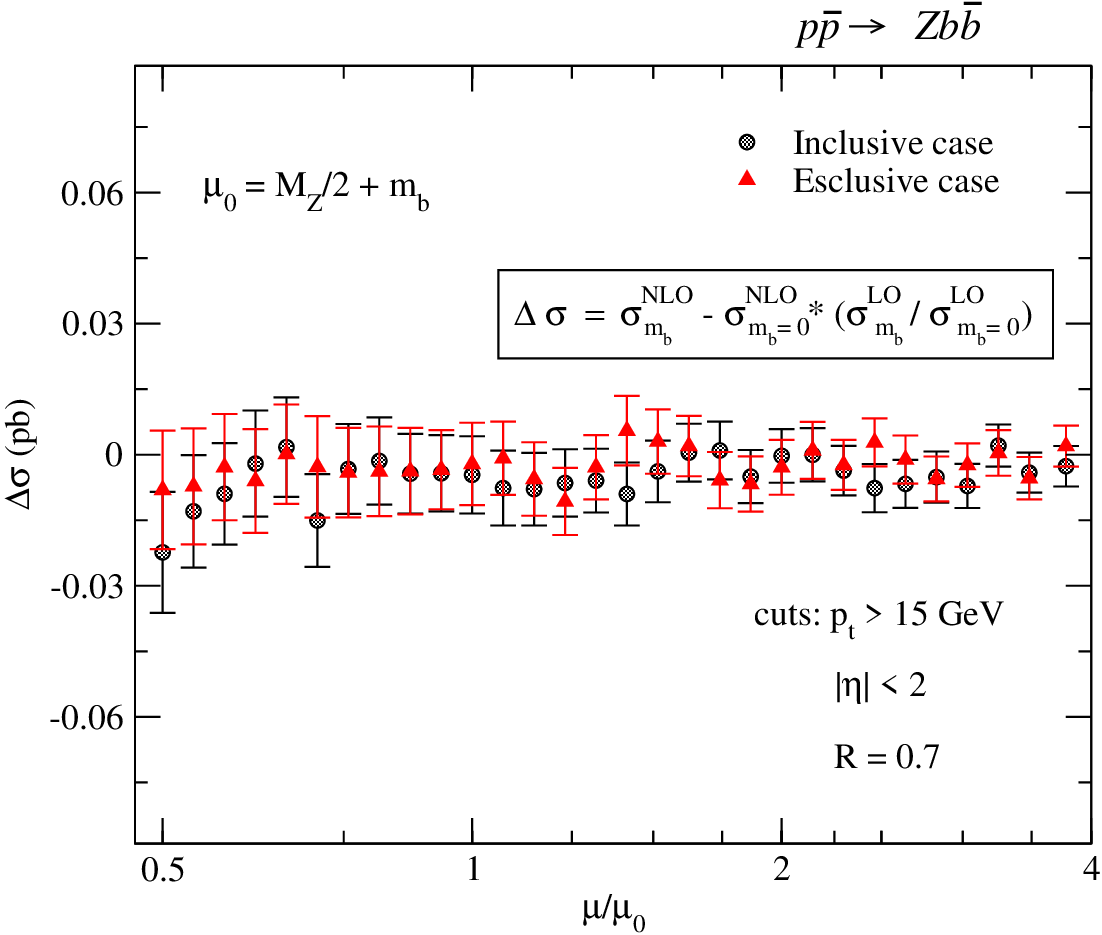} 
\caption{Dependence 
on the renormalization/factorization scale of the rescaled difference 
between our NLO calculation (with $m_b\ne 0$) of the total $\Zbb$ cross section 
and the corresponding result computed using MCFM (with
$m_b=0$) for the \emph{inclusive} and \emph{exclusive} cases
(with $\mu_r\!=\!\mu_f\!=\!\mu_0$) respectively. The error bars indicate the statistical uncertainty of
  the Monte Carlo integration.}
\label{fig:sigma_ratio_NLO_Zbb}
\end{center}
\end{figure}

While the LO cross section still has a 45\% uncertainty due to scale dependence, this uncertainty is
reduced at NLO to about 20\% for the \emph{inclusive} and to about 11\% for the \emph{exclusive}
cross sections. As before, the uncertainties have been estimated as the positive/negative deviation
with respect to the mid-point of the bands plotted in Figure~\ref{fig:mu_dependence_band_Zbb}, where
each band range is defined by the minimum and maximum value in the band.  We notice incidentally
that the difference due to finite bottom-quark mass effects is less significant than the theoretical
uncertainty due to the residual scale dependence in the \emph{inclusive} case, but is comparable in
size in the \emph{exclusive} case. Indeed, the finite bottom-quark mass effects amount to a
reduction of the total cross sections by about 7\% compared to the massless case at both LO and NLO
QCD.

\begin{figure}[htp]
 \begin{center}
  \subfigure[\emph{Inclusive} case]{\label{fig:mbb_dist_LO_vs_NLO_inc_Zbb}
	\hspace{-0.5cm}\includegraphics*[scale=0.7]{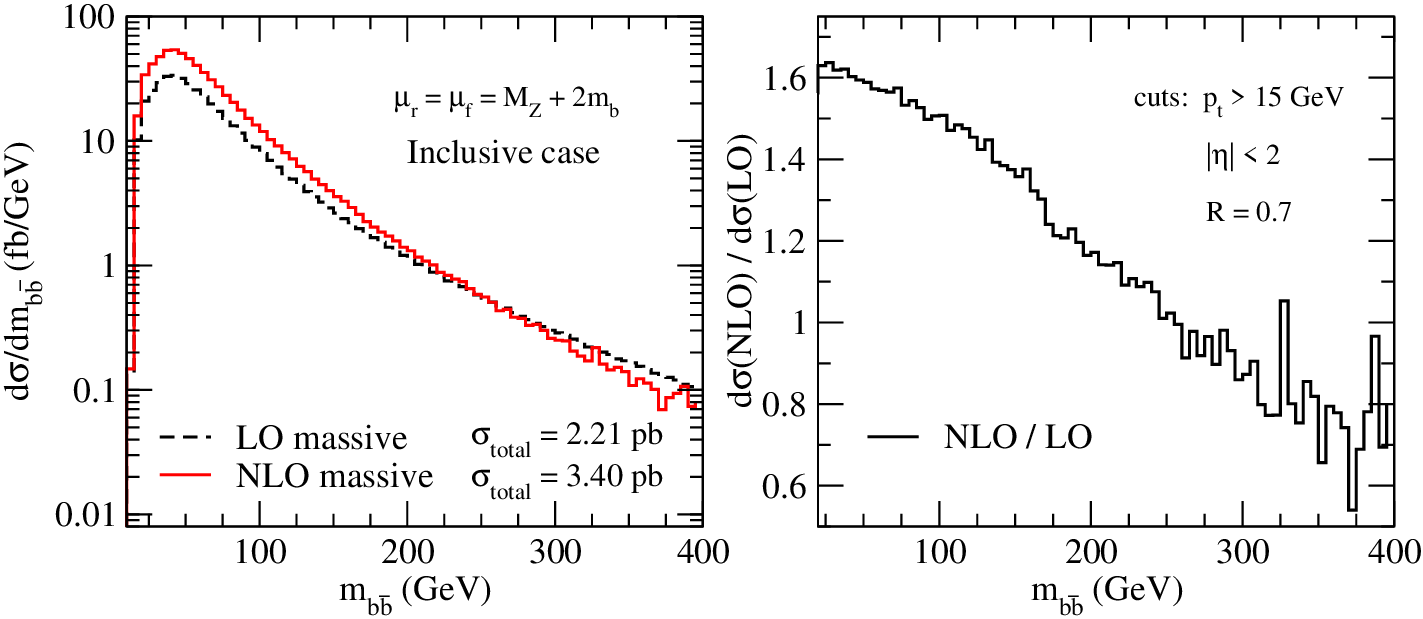}}\\
  \vspace{0.8cm}
  \subfigure[\emph{Exclusive} case]{\label{fig:mbb_dist_LO_vs_NLO_exc_Zbb}
	\hspace{-0.5cm}\includegraphics*[scale=0.7]{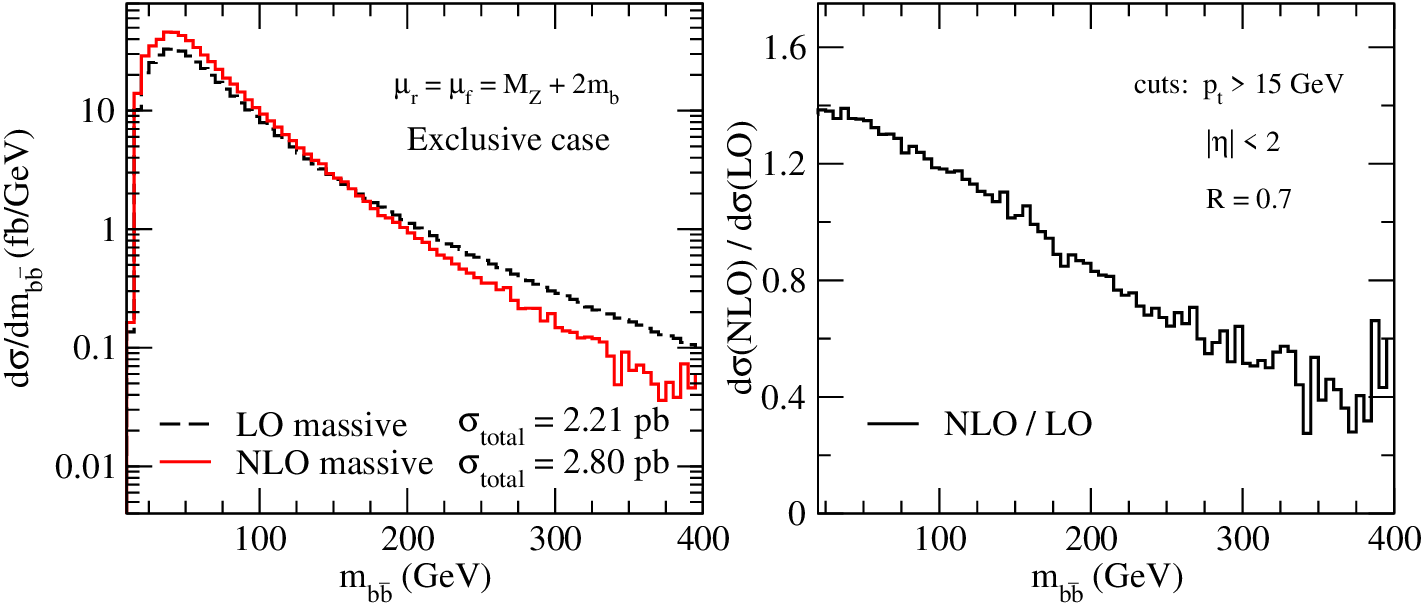}}
 \end{center}
   \caption{The distribution $d\sigma(p\bar{p}\rightarrow\Zbb)/dm_{b\bar{b}}$
      in LO and NLO QCD. The right hand side plot shows the ratio of
      the LO and NLO distributions.}
    \label{fig:mbb_dist_LO_vs_NLO_Zbb}
\end{figure}

In Figure~\ref{fig:sigma_ratio_NLO_Zbb}, we show the rescaled difference between the total cross
sections obtained from our calculation (with $m_b\ne 0$) and with MCFM (with $m_b=0$) defined as in
Eq.~(\ref{eq:mXsecrescaling}). As can be seen, within the statistical errors of the MC integration,
the finite bottom-quark mass effects on the total cross sections at NLO are well described by the
corresponding effects at LO, similarly to what is observed in the $\Wbb$ case.

\begin{figure}[hp]
 \begin{center}
  \subfigure[\emph{Inclusive} case]{\label{fig:mbb_dist_NLO_inc_Zbb}
	\hspace{-0.5cm}\includegraphics*[scale=0.7]{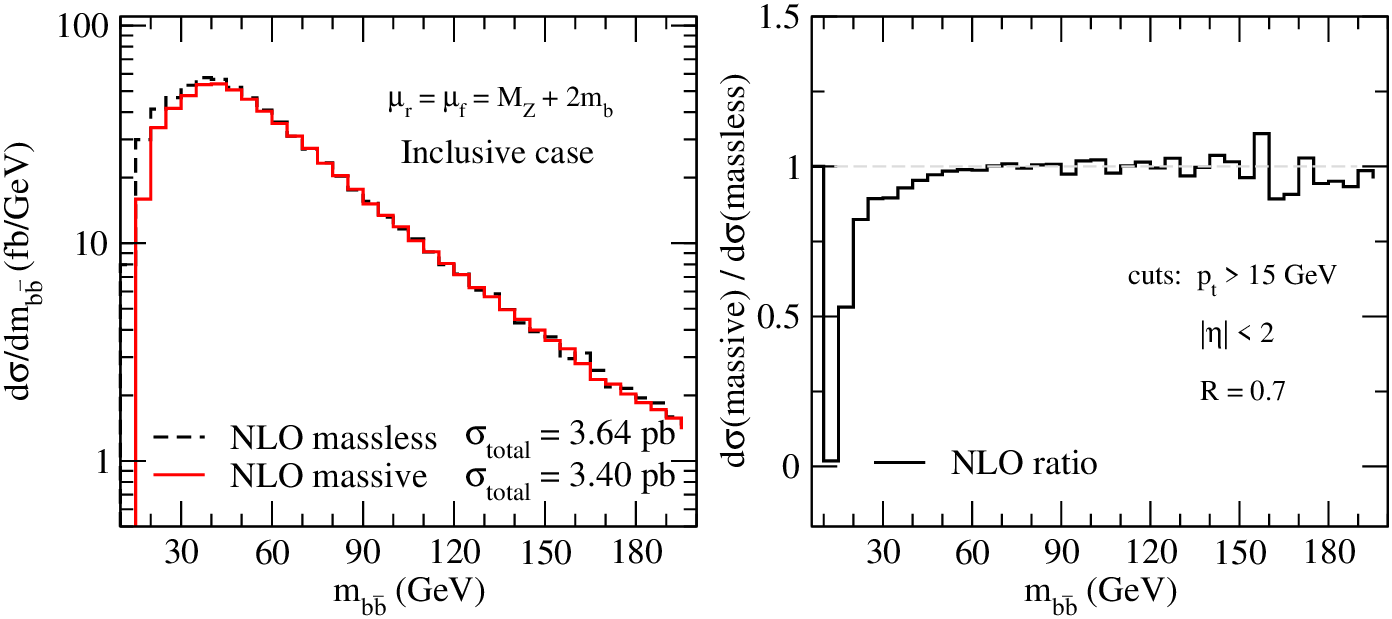}}\\
  \vspace{0.8cm}
  \subfigure[\emph{Exclusive} case]{\label{fig:mbb_dist_NLO_exc_Zbb}
	\hspace{-0.5cm}\includegraphics*[scale=0.7]{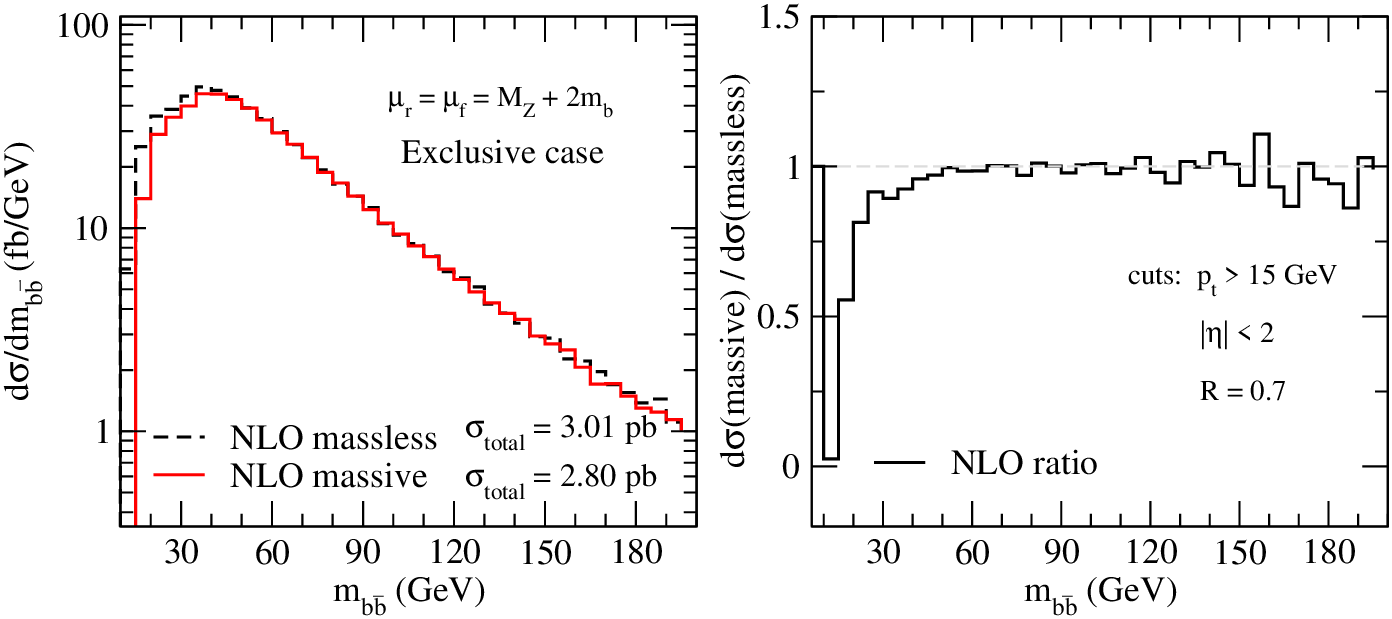}}
 \end{center}
\caption{The \emph{inclusive} and \emph{exclusive} distributions $d\sigma(p\bar{p}\rightarrow\Zbb)/dm_{b\bar{b}}$
derived from our calculation (with $m_b\ne 0$) and from MCFM (with
$m_b=0$).  The right hand side plot shows the ratio of the two
distributions, $d\sigma(m_b\neq 0)/d\sigma(m_b=0)$.}
\label{fig:mbb_dist_NLO_Zbb}
\end{figure}

\begin{figure}[t]
\begin{center}
\includegraphics*[scale=0.65]{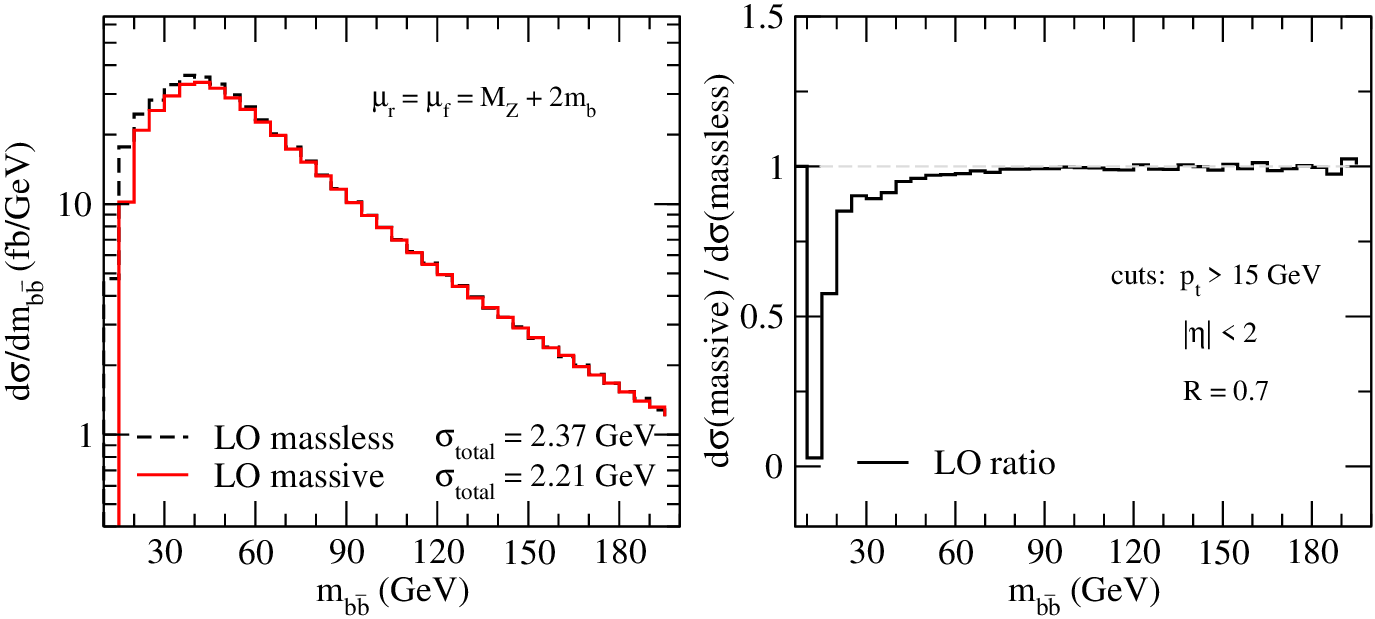} 
\caption{The LO distribution $d\sigma(p\bar{p}\rightarrow\Zbb)/dm_{b\bar{b}}$
derived from our calculation (with $m_b\ne 0$) and from MCFM (with
$m_b=0$). The right hand side plot shows the ratio of the two
distributions, $d\sigma(m_b\neq 0)/d\sigma(m_b=0)$.}
\label{fig:mbb_dist_LO_Zbb}
\end{center}
\end{figure}

Finally, in Figures~\ref{fig:mbb_dist_LO_vs_NLO_Zbb} to \ref{fig:mbb_dist_LO_Zbb} we study the
distribution $d\sigma/dm_{b\bar{b}}$, where $m_{b\bar{b}}$ is the invariant mass of the $b\bar{b}$
jet pair. The impact of NLO QCD corrections on this distribution is illustrated in
Figures~\ref{fig:mbb_dist_LO_vs_NLO_inc_Zbb} and \ref{fig:mbb_dist_LO_vs_NLO_exc_Zbb} for the
\emph{inclusive} and \emph{exclusive} case, respectively. We see that the NLO QCD corrections
affects the cross section quite substantially, in particular, for low values of the $m_{b\bar{b}}$
invariant mass. In each figure the right hand side plot gives the ratio of the NLO and LO
distributions.  We stress the fact that the LO and NLO distributions are not just rescaled, which is
clear from the RHS plots of Figures~\ref{fig:mbb_dist_LO_vs_NLO_Zbb}.

\begin{figure}[hp]
\begin{center}
\includegraphics*[scale=0.37]{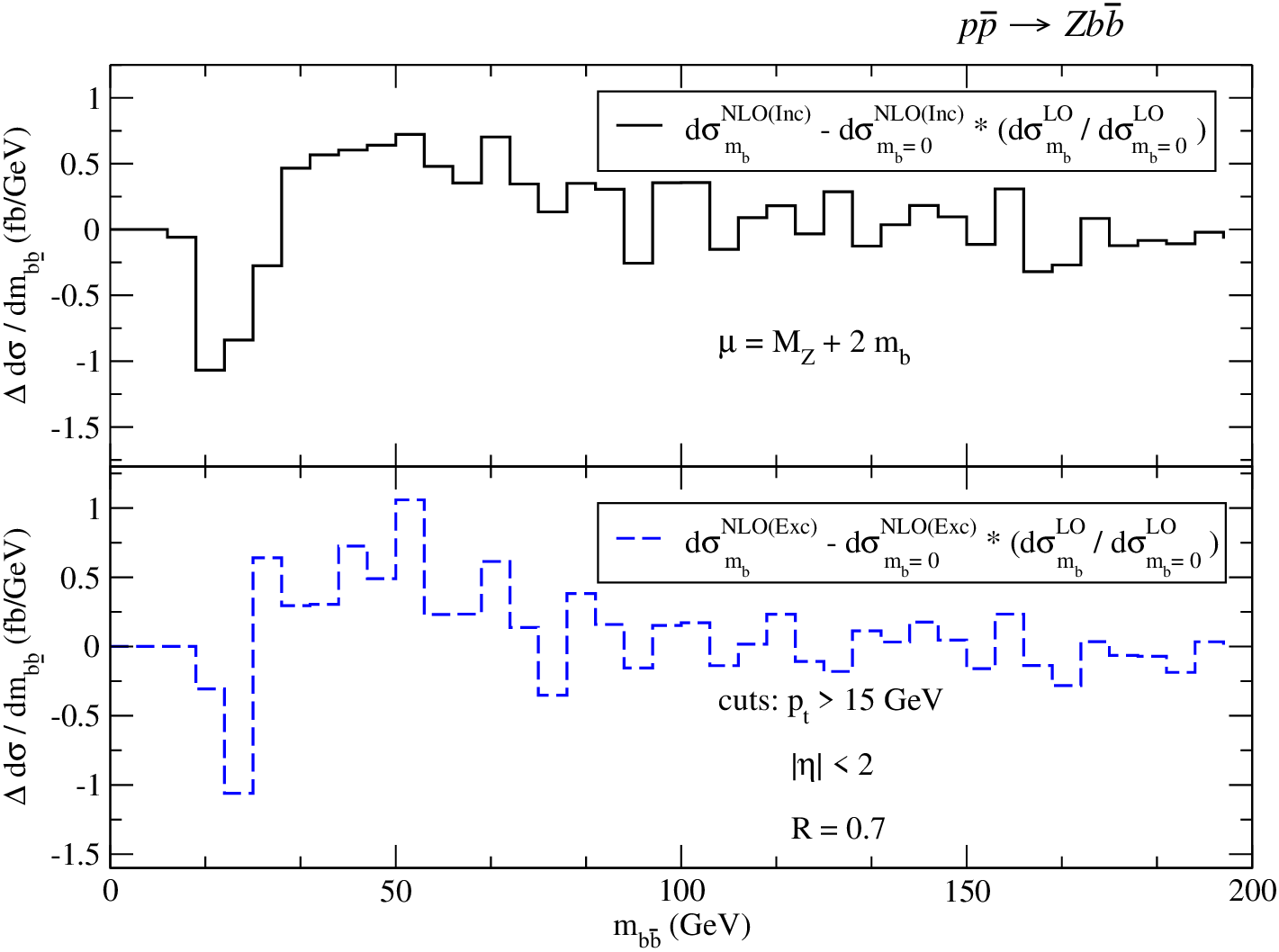} 
\caption{The $m_{b\bar b}$ distribution of the rescaled difference 
between our NLO calculation (with $m_b\ne 0$) and MCFM (with
$m_b=0$) for the \emph{inclusive} (upper plot) and \emph{exclusive} (lower plot)
$p\bar{p}\rightarrow\Zbb$ production.}
\label{fig:mbb_ratio_NLO_Zbb}
\end{center}
\end{figure}

Figures~\ref{fig:mbb_dist_NLO_inc_Zbb} and \ref{fig:mbb_dist_NLO_exc_Zbb} compare the NLO
$d\sigma/dm_{b\bar{b}}$ distributions obtained from the massive and massless bottom-quark
calculations. The results with $m_b=0$ have been obtained using MCFM.  As expected, most of the
difference between the massless and massive bottom-quark cross sections is coming from the region of
low $m_{b\bar{b}}$ invariant mass, both for the inclusive and exclusive case, where the cross
sections for $m_b\ne 0$ are consistently below the ones with $m_b=0$.  This is better emphasized in
the right hand side plots, where we show the ratio of the two distributions, $d\sigma(m_b\neq
0)/d\sigma(m_b=0)$.  For completeness, we also show in Figure~\ref{fig:mbb_dist_LO_Zbb} the
comparison between massive ($m_b\neq 0$) and massless ($m_b=0$) calculations at LO in QCD. The LO
$m_{b\bar{b}}$ distribution for massive bottom-quarks has been obtained both from our calculation
and from MCFM, which implements the $m_b\neq 0$ option at tree level, and both results agree
perfectly.  In general, mass effects are similar at LO and NLO. 
To illustrate
this in more detail we show in Figure~\ref{fig:mbb_ratio_NLO_Zbb} the rescaled difference between
the $m_{b\bar b}$ distributions obtained with our NLO calculation (with $m_b\ne 0$) and with MCFM
(with $m_b=0$) defined as in Eq.~(\ref{eq:mDistrescaling}).  
We notice that, in the $\Zbb$ case, finite bottom-quark mass effects
are relevant up to values of the $m_{b\bar b}$ invariant mass
around 50 GeV.

For ongoing searches of a light SM Higgs boson, regions with small $m_{b\bar b}$ invariant mass are
of relevance, as in many cases only one $b$-jet is tagged (semi-inclusive studies) in order to
increase the experimental statistics. In such studies $m_{b\bar b}$ invariant mass distributions are
produced by using the leading two jets in the event, one of them being the only tagged $b$-jet. In
such case, a possible signal can come from a real emission in which the $b\bar b$ quark pair is
recombined into a single b-jet (that is the partonic $m_{b\bar b}$ invariant mass is small) and the
extra light parton is seen as another jet.

\chapter{CONCLUSION}
\label{chap:conclusion}

We have presented a full review of our calculation of NLO QCD corrections to $\WZbb$ production at
hadron colliders including full bottom-quark mass
effects~\cite{FebresCordero:2006sj,FebresCordero:2008ci}. We have shown results for total cross
sections and $b\bar b$ invariant mass ($m_{b\bar b}$) distributions at the Tevatron Fermilab
collider. We have found that, for such collider, the NLO QCD corrections reduce considerably the
dependence on factorization and renormalization scales, in particular when considering exclusive
cross sections where exactly two $b$-quark jets are tagged in the final state. This then reduces the
theoretical uncertainty of the cross section, from about 40\% at LO to 20\% and 10\% at NLO for the
inclusive and exclusive cases, respectively.  Even more importantly, we have found that NLO
corrections change considerably the shape of LO distributions; that is, the NLO distributions are
not simply a rescaling of those at LO.

We have systematically compared our results to a calculation that considers massless bottom quarks
and have found that this approximation overestimates the total cross section by about 8\% for $\Wbb$
production and 10\% for $\Zbb$ production. The mass effects are particularly relevant in regions
with small $m_{b\bar b}$ invariant mass.  On the other hand, the massless calculation shows very
similar dependence on the factorization and renormalization scales, as including mass effects is
expected to affect very mildly such dependence. 

\section{Outlook}

As we have stressed, our results are of relevance to the search for a SM-like Higgs particle in the
$VH$ ($V=W,Z$) associated production channel and to the measurement of single-top production, both
processes of great interest to the high energy physics community.  The low $m_{b\bar b}$ invariant
mass region, where bottom-quark mass effects are most relevant, is important when a light SM-like
Higgs particle ($M_H\sim 100-140$ GeV) is searched semi-inclusively and by tagging only one
$b$-quark, in order to increase the experimental statistics. The low $m_{b\bar b}$ invariant mass
region is of course relevant for single-top production since in this case the whole $m_{b\bar b}$
spectrum is relevant, as the kinematics of the process is broader and semi-inclusive searches are
essential.

We are currently studying the impact of our calculation on searches for single-top production, where
we also consider final states with fewer than two $b$-quarks. We study modifications in the total
cross sections and implications for $b$-tagging efficiency\footnote{Since the completion of this
Dissertation, we have shown in Ref.~\cite{Campbell:2008hh} explicit results for NLO QCD corrections
to $Wb$ production both at the Tevatron and at the LHC.}.

The next natural step is to implement our calculation for the LHC and study its phenomenological
impact.  Since at the LHC gluon initiated processes are enhanced, we expect some fundamental
differences to appear. In particular, $q(\bar q)g$ initiated subprocesses will play a bigger role
and, given their tree level nature, will increase the dependence on renormalization and
factorization scales and, to some extent, increase the theoretical uncertainty of the cross
sections.

\vspace{0.5cm}

Finally, we want to emphasize that the calculation performed can be naturally extended to other
important processes, such as $\gamma t\bar t$ production, which might be studied at the LHC and
could give a direct measurement of the electric charge of the $t$-quark.  In the same direction, we
can study $\gamma b\bar b$ production at NLO in QCD. This process has considerable phenomenological
implications, as it can put direct constraints on the $b$-quark PDF, which so far has only been
derived from the gluon PDF evolution.

We can also study the associated production of a pseudo-scalar with heavy quarks, which is of
relevance to searches for physics beyond the SM, particularly models with an extended Higgs sector,
like supersymmetric models.

On the theoretical side, we expect to extend the checks performed on box coefficients by using
generalized unitarity, to triangle, bubble and tadpole coefficients. Even more, we can extract
analytically expressions for the contributing rational pieces. This will be of importance in the
development of efficient new techniques for their extraction, as they will represent a playground
to, for example, recursion relation techniques when massive external and internal particles are
present in multi-leg processes.

\newpage
\appendix
%\phantomsection
 \addtocontents{toc}{\protect\vspace{2ex}}  
\addcontentsline{toc}{chapter}{APPENDICES}%

%%%%%%%%%%%%%%%%%%%%%%%%%%%%%%%%%%%%%%%%%%%%%%%%%%%%%%%%%%%%%%%%%%%%%%%
\chapter{Standard Model of Particle Physics}\label{app:SMint}

%%%%%%%%%%%%%%%%%%%%%%%%%%%%%%%%%%%%%%%%%%%%%%%%%%%%%%%%%%%%%%%%%%%%%%%

The Standard Model of particle physics is a quantum field theory based on the gauge groups $SU_3$
for color, $SU_2$ for weak isospin, and $U_1$ for hypercharge, as dictated by the local gauge
symmetry invariance observed in the behavior of fundamental particles.  The color quantum number is
associated with the dynamics of the strong interactions, which by itself is the subject of Quantum
Chromodynamics (QCD), while the weak isospin and hypercharge quantum numbers are fundamental to the
dynamics of electroweak interactions.
 
The SM Lagrangian can be written as
\begin{equation}
\Lg_{SM}=\Lg_{YM}+\Lg_{f}+\Lg_{H}+\Lg_{\rm Yuk}\ ,
\label{smLg}
\end{equation}
where $\Lg_{YM}$ is the Yang-Mills Lagrangian, $\Lg_f$ the fermion Lagrangian, $\Lg_H$ the Higgs
Lagrangian and $\Lg_{\rm Yuk}$ contains the Yukawa interactions of the theory.  $\Lg_{YM}$ describes
the dynamics of the gauge fields (kinetic terms + self-interactions) and includes the following
terms
\begin{eqnarray}
\Lg_{YM} & = & \Lg_{QCD}+\Lg_{I_w}+\Lg_Y\nonumber\\
 	 & = & -\frac{1}{4}\sum_{a=1}^8 G_{\mu \nu}^a G^{a\mu\nu}
	-\frac{1}{4}\sum_{i=1}^3 F_{\mu \nu}^i F^{i\mu\nu}
	-\frac{1}{4}B_{\mu\nu}B^{\mu\nu}.
\end{eqnarray}
The color field strength tensor is given by
\begin{equation}
G_{\mu\nu}^a=\partial_\mu A_\nu^a-\partial_\nu A_\mu^a +g_1f^{abc} A_\mu^b A_\nu^c\ , \qquad 
	a,b,c=1,\ldots,8 \ ,
\end{equation}
with $A_\mu^b$ the eight color gauge fields (so called gluons), $g_1$ the dimensionless strong
coupling constant and $f^{abc}$ the structure constants of $SU_3$.  Analogously, the weak isospin,
$F_{\mu\nu}^i$, and hypercharge, $B_{\mu\nu}$, field strength tensors are given by
\begin{equation}
F_{\mu\nu}^i=\partial_\mu W_\nu^i-\partial_\nu W_\mu^i+g_2\epsilon^{ijk}W_\mu^j W_\nu^k\ , 
	\qquad i,j,k=1,2,3 \ ,
\end{equation}
\begin{equation}
B_{\mu\nu}=\partial_\mu B_\nu-\partial_\nu B_\mu\ ,
\end{equation}
where $W_\mu^i$ and $B_\mu$ are the 4 electroweak gauge bosons (a linear combination of which will
become the weak $W^\pm_\mu$ and $Z^0_\mu$ weak gauge bosons plus the photon $A_\mu$, as shown in
Eq.~(\ref{eq:weakbosons})), $g_2$ is the dimensionless weak isospin coupling constant and
$\epsilon^{ijk}$ are the structure constants of $SU_2$. 

Throughout the body of this dissertation we have denoted the strong coupling $g_1$ as $g_s$, and the
weak isospin coupling $g_2$ as $g_{\sss W}$. We also use the conventional definition
$\as=g_s^2/(4\pi)$.

The second part of the SM Lagrangian in Eq.~(\ref{smLg}), $\Lg_{f}$, describes the fermion fields and
their interactions with the gauge bosons. The fermion fields are classified as quarks, which are
triplets under the color gauge group, and leptons, which have no color. Taking into account the fact
that the $W$ boson couples only to left-handed helicity states of quarks and leptons, this part of
the Lagrangian is built such that right-handed and left-handed components of the fermion fields
couple independently to the gauge bosons. Using the notation $(SU_2,SU_3)_Y$, to denote weak
isospin, color, and hypercharge quantum number assignments of the fermion fields, we can write that
a quark weak doublet, $\mathbf{Q}_L={\mathbf{u}\choose \mathbf{d}}_L$, is a
$(\mathbf{2},\mathbf{3})_{y_1}$ and a quark weak singlet, ${\mathbf{u}}_R$, is a
$(\mathbf{1},\mathbf{3})_{y_2}$ (and similar for ${\mathbf{d}}_R$). On the other side,  a weak
doublet of leptons, $L_L={\nu\choose e}_L$, is a $(\mathbf{2},\mathbf{1})_{y_4}$ and a lepton weak
singlet, $e_R$, is a $(\mathbf{1},\mathbf{1})_{y_5}$. The fermion Lagrangian $\Lg_{f}$ then can be
written as
\begin{eqnarray}
\Lg_f & = &\mathbf{\overline Q}_L\sigma^\mu \Dcal_\mu \mathbf{Q}_L+\mathbf{\overline u}_R 
\sigma^\mu
\Dcal_\mu \mathbf{u}_R +\mathbf{\overline d}_R\sigma^\mu
\Dcal_\mu \mathbf{d}_R\nonumber\\
& & +{\overline L}_L\sigma^\mu \Dcal_\mu L_L+\overline e_R 
\sigma^\mu
\Dcal_\mu e_R +\cdots\ , 
\label{termLff}
\end{eqnarray} 
where the dots stand for similar terms for the remaining quarks and leptons.  In Eq.~(\ref{termLff})
$\sigma^\mu$ are the Pauli matrices ($\sigma^0=\mathbf{1}$), and $\Dcal_\mu$ are the covariant
derivatives corresponding to each field,
\begin{eqnarray} 
\Dcal_\mu \mathbf{Q}_L & = & (\partial_\mu+g_1\frac{i}{2}A_\mu^a\lambda^a+
			g_2\frac{i}{2}W_\mu^i\tau^i+g_3\frac{i}{2}y_1B_\mu)\mathbf{Q}_L\ ,\nonumber\\
\Dcal_\mu \mathbf{u}_R & = & (\partial_\mu+g_1\frac{i}{2}A_\mu^{a}\lambda^{a}
			+g_3\frac{i}{2}y_2B_\mu)\mathbf{u}_R\ ,\nonumber\\
\Dcal_\mu \mathbf{d}_R & = & (\partial_\mu+g_1\frac{i}{2}A_\mu^{a}\lambda^{a}
			+g_3\frac{i}{2}y_3B_\mu)\mathbf{d}_R\ ,\nonumber\\
\Dcal_\mu L_L & = & (\partial_\mu+g_2\frac{i}{2}W_\mu^i\tau^{i}
			+g_3\frac{i}{2}y_4B_\mu)L_L\ ,\nonumber\\
\Dcal_\mu e_R & = & (\partial_\mu
			+g_3\frac{i}{2}y_5B_\mu)e_R\ ,
\end{eqnarray}
where $g_3$ is the dimensionless hypercharge coupling constant, and $\tau^i$ and $\lambda^a$ are the
Pauli and Gell-Mann matrices for $SU_2$ and $SU_3$ respectively. 

Notice that a mass term for the fermion fields and for the vector boson fields (as needed for the
weak vector bosons $W^\pm_\mu$ and $Z^0_\mu$) is not allowed by gauge invariance.  The last two
terms in the Standard Model Lagrangian shown in Eq.~(\ref{smLg}) are introduced to remedy this
problem. Indeed, the simplest way to preserve the gauge symmetry of the SM while generating massive
electroweak gauge bosons is the so called Higgs mechanism, which we explain in the following. A
separate step needs to be taken to introduce massive fermions, and we will discuss this below.

The Higgs mechanism, in its simplest version
\cite{EnglertBrout:BS,Higgs:BS,Guralnik:1964eu,Kibble:1967sv}, starts by adding to the model another
field, called the Higgs field $H$, which transforms as a weak isospin doublet, a color singlet, and
it has hypercharge $y_h$:
\begin{equation}
H={H_1 \choose H_2}\sim (\mathbf{2},\mathbf{1})_{y_h}\ .
\end{equation}
Its dynamics is dictated by the $\Lg_H$ term in Eq.~(\ref{smLg}), which can be written as
\begin{equation}
\Lg_H=(\Dcal_\mu H)^\dag (\Dcal^\mu H)-V(H)\ ,
\label{higgsLg}
\end{equation}
where $\Dcal_\mu H=(\partial_\mu +g_2\frac{i}{2}W_\mu^i\tau^i+g_3\frac{i}{2}y_hB_\mu)H$ is the
covariant derivative of $H$ and $V(H)$ is the most general renormalizable potential invariant under
$SU_2\times U_1$,
\begin{equation}
V(H)=\mu^2H^\dag H+\lambda (H^\dag H)^2\ ,
\label{higgsPot}
\end{equation}
with $\mu^2$ and $\lambda$ real parameters. $\lambda$ is a dimensionless parameter.

If $\mu^2 < 0$ the field configurations that minimize the potential $V(H)$ has to satisfy:
\begin{equation}
H_{vac}^\dag H_{vac}=\frac{-\mu^2}{2\lambda}\equiv \frac{v^2}{2}\ .
\end{equation}
So, once $\mu^2<0$ the Higgs field develops a vacuum expectation value (VEV), which is degenerate
over the sphere defined in the last equation. Picking one configuration breaks this degeneracy,
causing the vacuum of the theory not to be $SU_2\times U_1$ symmetric anymore. To illustrate the
consequences, let us choose:
\begin{equation}
\langle H\rangle =\frac{v}{\sqrt{2}}{0\choose 1}\ .
\label{vevHiggs}
\end{equation}
Indeed one can verify that this choice breaks the original gauge symmetry:
\begin{equation}
SU_2\times U_1\rightarrow U_1^{EM}\ ,
\end{equation}
where $U_1^{EM}$ is the electromagnetic $U_1$ symmetry.

When $\Lg_H$ is expanded in the vicinity of the chosen minimum, by shifting the Higgs field as
follows
\begin{equation}
H=\frac{1}{\sqrt{2}}{0\choose v+h}\ ,
\label{Hshift}
\end{equation}
$\Lg_H$ becomes the Lagrangian of a real scalar field with mass $m_h=2v^2\lambda$, the physical
Higgs boson. Moreover, a mass term for the gauge bosons is generated by the first term in
Eq.~(\ref{higgsLg}), coming from
\begin{equation}
\frac{1}{2}(0,v)\left | \frac{1}{2}g_2W_\mu^i\tau^i+\frac{1}{2}g_3B_\mu\right |^2{0\choose v}\ .
\end{equation}
The corresponding mass eigenstates, i.e. the physical gauge fields, are obtained by diagonalizing
the mass matrix of the vector fields $W_\mu^a$ and $B_\mu$. The EW gauge bosons $W^\pm_\mu$ and
$Z^0_\mu$, as well as the photon $A_\mu$, are expressed as:
\begin{eqnarray}
W_\mu^\pm & = & \frac{1}{\sqrt{2}}(W_\mu^1\mp iW_\mu^2)\ ,\nonumber\\
Z_\mu & = & \frac{-g_3B_\mu+g_2W_\mu^3}{\sqrt{g_2^2+g_3^2}}\ ,\nonumber\\
A_\mu & = & \frac{g_2B_\mu+g_3W_\mu^3}{\sqrt{g_2^2+g_3^2}}\ ,
\label{eq:weakbosons}
\end{eqnarray}
with the associated masses:
\begin{eqnarray}
M_W^2 & = & \frac{1}{4}g_2^2v^2\ ,\nonumber\\
M_Z^2 & = & \frac{1}{4}(g_2^2+g_3^2)v^2\ ,\nonumber\\
M_A^2 & = & 0\ .
\end{eqnarray}
These simple relations are found to agree with experiment ($M_W=80.4\ {\rm GeV}$, $M_Z=91.2\ 
{\rm GeV}$) with $v\approx 174\sqrt{2}\ \rm GeV$. This is, of course, approximate as they are 
results based on the classical, or leading order level of the theory.

Finally let us focus on the last part of the SM Lagrangian presented in 
Eq.~(\ref{smLg}). This term, $\Lg_{\rm Yuk}$, couples massive fermion fields
to the Higgs field via Yukawa type interaction. For example, the gauge invariant
Yukawa coupling of the Higgs boson to the down quark, $\mathbf{d}$, is
\[
-\lambda_d \overline{\mathbf{Q}}_L H\mathbf{d}_R+h.c. \ ,
\]
where $\overline{\mathbf{Q}}_L=(\overline{\mathbf{u}},\overline{\mathbf{d}})_L$ 
and $\lambda_d$ is the Yukawa coupling
for the down quark. After the shift of Eq.~(\ref{Hshift}) this term gives the effective coupling
\begin{equation}
-\lambda_d\frac{1}{\sqrt{2}}\overline{\mathbf{Q}_L}{0\choose v+h}\mathbf{d}_R+h.c.\ ,
\end{equation}
which gives a mass term to the down quark with
\begin{equation}
m_d=\frac{\lambda_d v}{\sqrt{2}}\ ,
\end{equation} 
and defines the coupling between the down quark and the physical Higgs particle to be
$-\lambda_d/\sqrt{2}$.

Similar terms are added for each massive fermion field.  Then $\Lg_{\rm Yuk}$ will contain 9
arbitrary parameters, the Yukawa couplings, standing for 6 quark masses and 3 lepton masses.  This
completes the classical SM Lagrangian, which is the main theory used throughout this dissertation.
Several extensions of the SM follow the same prescription and ultimately break the EW symmetry
spontaneously by introducing several Higgs fields. On the other hand, alternative mechanisms to
break the EW symmetry have also been proposed like ``Technicolor", where scalars are seen as
strongly bound states of fermions, ``Little Higgs" models, where more complex global symmetries of
the fundamental fields are considered to justify the existence of a relatively light Higgs, or more
speculative theories that explain the breaking of the EW gauge symmetry in terms of extra
dimensions.  However, most alternatives to the Higgs mechanism encounter phenomenological
difficulties and till now the most successful theory in describing experimental data is still the SM
with a weakly coupled Higgs boson.

\subsection*{Feynman Rules}

From the Lagrangian one can readily extract the Feynman rules of the theory. Now we present the set
of rules we have employed throughout our calculation.

Let us start by writing the propagators for quarks and gluons, this last one in the Feynman gauge:

\begin{tabular}{rl}
\hspace{2.0cm}
\begin{minipage}{0.2\linewidth}{
\vspace{0.5cm}
\begin{center}
\includegraphics*{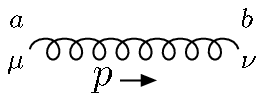}
\end{center}}
\end{minipage}
&
\hspace{-3.0cm}
\begin{minipage}{0.5\linewidth}{
\begin{center}
{\[\qquad = \;\; \frac{-ig_{\mu\nu}}{p^2+i\epsilon}\delta^{ab} \; , \]}
\end{center}}
\end{minipage}
\end{tabular}
\vspace{-1.0cm}
\begin{equation}\label{eq:FRg}\end{equation}

\begin{tabular}{rl}
\hspace{2.0cm}
\begin{minipage}{0.2\linewidth}{
\vspace{0.7cm}
\begin{center}
\includegraphics*{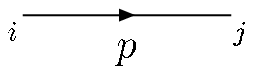}
\end{center}}
\end{minipage}
&
\hspace{-3.0cm}
\begin{minipage}{0.5\linewidth}{
\begin{center}
{\[\qquad = \;\; \frac{i(p\negthickspace\!/+m)}{p^2-m^2+i\epsilon}\delta_{ij}\; .  \]}
\end{center}}
\end{minipage}
\end{tabular}
\vspace{-1.0cm}
\begin{equation}\label{eq:FRq}\end{equation}

For the coupling between quarks and gluons and gluon self-interactions we have the vertices:

\begin{tabular}{rl}
\hspace{1.0cm}
\begin{minipage}{0.2\linewidth}{
\vspace{0.7cm}
\begin{center}
\includegraphics*{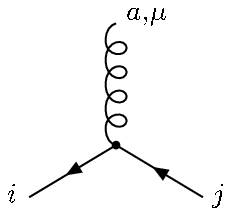}
\end{center}}
\end{minipage}
&
\hspace{-3.0cm}
\begin{minipage}{0.5\linewidth}{
\begin{center}
{\[\qquad = \;\;  ig_1\gamma^\mu t^a_{ij} \; , \]}
\end{center}}
\end{minipage}
\end{tabular}
\vspace{-1.8cm}
\begin{equation}\label{eq:FRqqg}\end{equation}

\begin{tabular}{rl}
\hspace{1.0cm}
\begin{minipage}{0.2\linewidth}{
\vspace{0.7cm}
\begin{center}
\includegraphics*{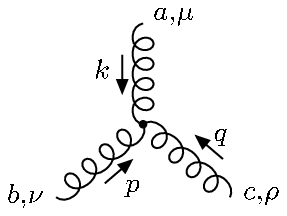}
\end{center}}
\end{minipage}
&
\hspace{-0.4cm}
\begin{minipage}{0.5\linewidth}{
\begin{center}
{\[\qquad  =   g_1f^{abc}\left[ g^{\mu\nu}(k-p)^\rho+
			g^{\nu\rho}(p-q)^\mu+g^{\mu\rho}(q-k)^\nu \right]\; , \]}
\end{center}}
\end{minipage}
\end{tabular}
\vspace{-0.8cm}
\begin{equation}\label{eq:FRggg}\end{equation}

\vspace{-0.5cm}
\begin{tabular}{rl}
\hspace{1.0cm}
\begin{minipage}{0.2\linewidth}{
\vspace{0.7cm}
\begin{center}
\includegraphics*{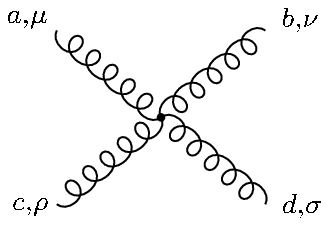}
\end{center}}
\end{minipage}
&
\hspace{-0.4cm}
\begin{minipage}{0.5\linewidth}{
\begin{center}
{\begin{eqnarray} & = & -ig_1^2\left[
	f^{abc}f^{cde}(g^{\mu\rho}g^{\nu\sigma}-g^{\mu\sigma}g^{\nu\rho}) \right.\nonumber\\
 & & \qquad+ f^{ace}f^{bde}(g^{\mu\nu}g^{\rho\sigma}-g^{\mu\sigma}g^{\nu\rho}) \nonumber\\
 & & \qquad\left. + f^{ade}f^{bce}(g^{\mu\nu}g^{\rho\sigma}-g^{\mu\rho}g^{\nu\sigma})\nonumber \right]\; ,
			 \end{eqnarray}}
\end{center}}
\end{minipage}
\end{tabular}
\vspace{-1.8cm}
\begin{equation}\label{eq:FRgggg}\end{equation}

\vspace{1.0cm}
where $2t^a_{ij}$ are the Gell-Mann matrices and $f^{abc}$ the structure constants of $SU(3)$.

The weak boson couplings to fermions are given by the vertices (stripped of color indices):

\begin{tabular}{rl}
\hspace{1.0cm}
\begin{minipage}{0.2\linewidth}{
\vspace{0.7cm}
\begin{center}
\includegraphics*{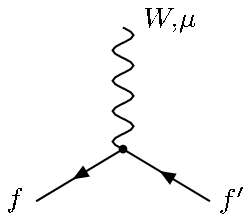}
\end{center}}
\end{minipage}
&
\hspace{-2.0cm}
\begin{minipage}{0.5\linewidth}{
\begin{center}
{\[\qquad = \;\;  \frac{-ig_2}{2\sqrt{2}} \gamma^\mu(1-\gamma_5) V_{ff^\prime}\; , \]}
\end{center}}
\end{minipage}
\end{tabular}
\vspace{-1.8cm}
\begin{equation}\label{eq:FRqqW}\end{equation}

\newpage
\begin{tabular}{rl}
\hspace{1.0cm}
\begin{minipage}{0.2\linewidth}{
\vspace{0.7cm}
\begin{center}
\includegraphics*{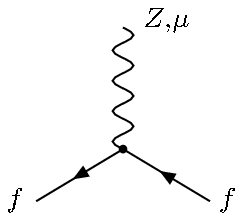}
\end{center}}
\end{minipage}
&
\hspace{-2.0cm}
\begin{minipage}{0.5\linewidth}{
\begin{center}
{\[\qquad = \;\; \frac{-ig_2}{\cos{\theta_{\sss W}}}\Zvert{\mu}\; ,  \]}
\end{center}}
\end{minipage}
\end{tabular}
\vspace{-1.8cm}
\begin{equation}\label{eq:FRqqZ}\end{equation}

\vspace{1.5cm}
\noindent where $V_{ff^\prime}$ are the entries of the CKM mixing matrix, and the vector, $g_V^f$, and axial, $g_A^f$,
couplings for the $Zff$ vertex are given by:
\begin{eqnarray}
g_V^f & = & \frac{1}{2}T_3^f-\sin{\theta_{\sss W}}^2 Q_f\; ,\\
g_A^f & = & -\frac{1}{2}T_3^f\; ,\nonumber
\label{eq:VAcoup}
\end{eqnarray}
where $\sin{\theta_{\sss W}}=g_3/\sqrt{g_2^2+g_3^2}$, $T_3^f$ if the third component of the weak isospin matrix for
the fermion $f$, and $Q_f$ is the charge of the fermion.

%%%%%%%%%%%%%%%%%%%%%%%%%%%%%%%%%%%%%%%%%%%%%%%%%%%%%%%%%%%%%%%%%%%%%%%
\chapter{Tree Level Amplitudes for $\WZbb$ Production}\label{app:LOamp}

%%%%%%%%%%%%%%%%%%%%%%%%%%%%%%%%%%%%%%%%%%%%%%%%%%%%%%%%%%%%%%%%%%%%%%%

In this Appendix we present explicit expressions for the tree level amplitudes that appear in the full
calculation of $\WZbb$ production. This is a straightforward application of the Feynman rules shown in the
last section of Appendix~\ref{app:SMint}. As we mentioned there,
we denote the strong coupling $g_1$ as $g_s$, and the weak isospin
coupling $g_2$ as $g_{\sss W}$.

\boldmath
\subsubsection*{Tree level amplitude for $q\bar q^\prime\to \Wbb$}
\unboldmath

The contributing tree level Feynman diagrams are shown in Figure~\ref{fig:Wbbtree_level}.

Given the momenta
assignment: 
\[
q(q_1)\bar q^\prime(q_2)\to b(p_b)+\bar{b}(p_{\bar b})+W(p_{\sss W})\,\,\,,
\]
the LO amplitude can be written as:
\begin{eqnarray}
{\cal A}_0(q\bar q^\prime\to\Wbb) & = & ig_s^2\frac{g_{\sss W}}{2\sqrt{2}}V_{q\bar q^\prime}\;\; \epsilon_\mu^*(p_{\sss
W})\;\; \frac{g_{\nu\rho}}{(p_b+p_{\bar b})^2}\;\; \bar{u}_b \gamma^\rho v_{\bar{b}}\;\;t_{ij}^at_{kl}^a \nonumber\\
	& & \Bigl[\bar v_{\bar q^\prime}\Wvert{\mu}\frac{-\Slash{q}_2+\Slash{p}_{\sss W}}{(-q_2+p_{\sss
	W})^2}\gamma^\nu u_q \nonumber\\
	& & +\bar v_{\bar q^\prime}\gamma^\nu\frac{\Slash{q}_1-\Slash{p}_{\sss W}}{(q_1-p_{\sss
	W})^2}\Wvert{\mu} u_q \Bigl]\; ,
\label{eq:apLOWbb}
\end{eqnarray}
where $V_{q\bar q^\prime}$ is the CKM mixing matrix, $\epsilon_\mu(p_{\sss W})$ is the polarization vector of
the $W$ boson and we denote by $v$ and $u$ the spinors for the external fermionic fields (for more details look
at the Feynman rules in Appendix~\ref{app:SMint}).
%where a bunch of things...

\boldmath
\subsubsection*{Tree level amplitude for $q\bar q\to \Zbb$}
\unboldmath

The contributing tree level Feynman diagrams are shown in Figures~\ref{fig:Wbbtree_level} (with
$V=Z$) and \ref{fig:finalqqZbbtree_level} for subprocess $q\bar q\to\Zbb$ with the $Z$ weak boson
emitted from initial and final fermion lines respectively.

Given the momenta
assignment: 
\[
q(q_1)\bar q(q_2)\to b(p_b)+\bar{b}(p_{\bar b})+Z(p_{\sss Z})\,\,\,,
\]
the LO amplitude can be written as:
\begin{eqnarray}
{\cal A}_0(q\bar q\to\Zbb) & = & ig_s^2\frac{g_{\sss W}}{\cos{\theta_{\sss W}}}\;\; \epsilon_\mu^*(p_{\sss
Z})\;\; \frac{g_{\nu\rho}}{(p_b+p_{\bar b})^2}\;\; \bar{u}_b \gamma^\rho v_{\bar{b}}t^a_{ij}t^a_{kl} \nonumber\\
	& & \Bigl[\bar v_{\bar q}\Zvertf{\mu}{q}\frac{-\Slash{q}_2+\Slash{p}_{\sss Z}}{(-q_2+p_{\sss
	Z})^2}\gamma^\nu u_q \nonumber\\
	& & +\bar v_{\bar q}\gamma^\nu\frac{\Slash{q}_1-\Slash{p}_{\sss Z}}{(q_1-p_{\sss
	Z})^2}\Zvertf{\mu}{q} u_q \Bigl]\; \nonumber\\
&  &	+ ig_s^2\frac{g_{\sss W}}{\cos{\theta_{\sss W}}}\;\; \epsilon_\mu^*(p_{\sss
Z})\;\; \frac{g_{\nu\rho}}{(q_1+q_2)^2}\;\; \bar{v}_{\bar q} \gamma^\rho u_{q}t^a_{ij}t^a_{kl} \nonumber\\
	&& \left[ \bar{u}_b\Zvertf{\mu}{b}\frac{\Slash{p}_b+\Slash{p}_{\sss Z}+m_b}{\left[(p_b+p_{\sss
	Z})^2-m_b^2\right]} \gamma^\nu v_{\bar{b}} \right.\nonumber\\
	&& +\left. \bar{u}_b\gamma^\nu\frac{-\Slash{p}_{\bar b}-\Slash{p}_{\sss Z}+m_b}{\left[(-p_{\bar b}-p_{\sss
	Z})^2-m_b^2\right]} \Zvertf{\mu}{b} v_{\bar{b}} \right].
\label{eq:apLOqqZbb}
\end{eqnarray}
where $g_{\sss V}^{f}$ and $g_{\sss A}^{f}$ are the vector and axial coupling constants for fermion $f$,
$\epsilon_\mu(p_{\sss Z})$ is the polarization vector of
the $Z$ boson and we denote by $v$ and $u$ the spinors for the external fermionic fields (for more details look
at the Feynman rules in Appendix~\ref{app:SMint}).
%where a bunch of things...

\boldmath
\subsubsection*{Tree level amplitude for $gg\to \Zbb$}
%\label{sec:app_tree_levelZbb}
\unboldmath

The amplitudes ${\cal A}_{0,s}$, ${\cal A}_{0,t}$, and ${\cal A}_{0,u}$
introduced in Section~\ref{sec:sigma_loZbb} can be written as:
\begin{eqnarray}
\label{eq:a0_stu}
{\cal A}_{0,s}&=& ig_s^2\,\frac{g_{\sss W}}{\cos{\theta_{\sss W}}}\,
	\epsilon_\mu(q_1)\,\epsilon_\nu(q_2)\,\epsilon_\rho^*(p_{\sss Z})\, 
\bar{u}_b {\cal A}_{0,s}^{\mu\nu\rho}v_{\bar{b}}\,\,\,,\nonumber\\
{\cal A}_{0,t}&=& ig_s^2\,\frac{g_{\sss W}}{\cos{\theta_{\sss W}}}\,
	\epsilon_\mu(q_1)\,\epsilon_\nu(q_2)\,\epsilon_\rho^*(p_{\sss Z})\,
\bar{u}_b {\cal A}_{0,t}^{\mu\nu\rho} v_{\bar{b}}\nonumber\,\,\,,\\
{\cal A}_{0,u}&=& ig_s^2\,\frac{g_{\sss W}}{\cos{\theta_{\sss W}}}\,
	\epsilon_\mu(q_1)\,\epsilon_\nu(q_2)\,\epsilon_\rho^*(p_{\sss Z})\,
\bar{u}_b {\cal A}_{0,u}^{\mu\nu\rho} v_{\bar{b}}\,\,\,, 
\end{eqnarray}
where 
$\epsilon_\mu(p_{\sss Z})$ is the polarization vector of
the $Z$ boson, $v$ and $u$ are the spinors for the external fermionic fields, and
${\cal A}_{0,s}^{\mu\nu\rho}$, ${\cal A}_{0,t}^{\mu\nu\rho}$, and ${\cal A}_{0,u}^{\mu\nu\rho}$ represent the total
$s-$channel, $t-$channel, and
$u-$channel amplitudes, corresponding to the diagrams in
Figure~\ref{fig:ggZbbtree_level}. More explicitly:
\begin{eqnarray}
\label{eq:A0_stu_munu}
{\cal A}_{0,s}^{\mu\nu\rho}&=&{\cal A}_{0,s}^{(1)\mu\nu\rho}+
{\cal A}_{0,s}^{(2)\mu\nu\rho}\,\,\,,\nonumber\\
{\cal A}_{0,t}^{\mu\nu\rho}&=&{\cal A}_{0,t}^{(1)\mu\nu\rho}+
{\cal A}_{0,t}^{(2)\mu\nu\rho}+{\cal A}_{0,t}^{(3)\mu\nu\rho}\,\,\,,\nonumber\\
{\cal A}_{0,u}^{\mu\nu\rho}&=&{\cal A}_{0,u}^{(1)\mu\nu\rho}+
{\cal A}_{0,u}^{(2)\mu\nu\rho}+{\cal A}_{0,u}^{(3)\mu\nu\rho}\,\,\,,
\end{eqnarray}
where
\begin{eqnarray}
\label{eq:A0_stu_munu_123}
{\cal A}_{0,s}^{(1),\mu\nu\rho}&=& 
\frac{1}{s}\Zvertf{\rho}{b}\frac{\Slash{p}_b+\Slash{p}_{\sss Z}+m_b}{[(p_b+p_{\sss Z})^2-m_b^2]} 
\gamma_{\alpha} V^{\mu\nu\alpha}\,\,\,,\nonumber\\
{\cal A}_{0,s}^{(2),\mu\nu\rho}&=& 
\frac{1}{s}\gamma_{\alpha}\frac{-\Slash{p}_{\bar b}-\Slash{p}_{\sss Z}+m_b} 
{[(p_{\bar b}+p_{\sss Z})^2-m_b^2]}\Zvertf{\rho}{b} V^{\mu\nu\alpha}\,\,\,,\nonumber\\
{\cal A}_{0,t}^{(1),\mu\nu\rho}&=& 
\Zvertf{\rho}{b}\frac{\Slash{p}_b+\Slash{p}_{\sss Z}+m_b}{[(p_b+p_{\sss Z})^2-m_b^2]} \gamma^\mu 
\frac{\Slash{q}_2-\Slash{p}_{\bar b}+m_b}{[(q_2-p_{\bar b})^2-m_b^2]} 
\gamma^\nu\,\,\,,\nonumber\\
{\cal A}_{0,t}^{(2),\mu\nu\rho}&=& 
\gamma^\mu \frac{\Slash{p}_b-\Slash{q}_1+m_b}{[(p_b-q_1)^2-m_b^2]}\Zvertf{\rho}{b} 
\frac{\Slash{q}_2-\Slash{p}_{\bar b}+m_b}{[(q_2-p_{\bar b})^2-m_b^2]} 
\gamma^\nu\,\,\,, \nonumber\\
{\cal A}_{0,t}^{(3),\mu\nu\rho}&=& 
\gamma^\mu\frac{\Slash{p}_b-\Slash{q}_1+m_b}{[(p_b-q_1)^2-m_b^2]} 
\gamma^\nu\frac{-\Slash{p}_{\bar b}-\Slash{p}_{\sss Z}+m_b}
{[(p_{\bar b}+p_{\sss Z})^2-m_b^2]}\Zvertf{\rho}{b}\,\,\,,\nonumber\\
{\cal A}_{0,u}^{(1),\mu\nu\rho}&=&{\cal A}_{0,t}^{(1),\mu\nu\rho} 
(\mu\leftrightarrow\nu,q_1\leftrightarrow q_2)\,\,\,,\nonumber\\ 
{\cal A}_{0,u}^{(2),\mu\nu\rho}&=&{\cal A}_{0,t}^{(2),\mu\nu\rho} 
(\mu\leftrightarrow\nu,q_1\leftrightarrow q_2)\,\,\,,\nonumber\\ 
{\cal A}_{0,u}^{(3),\mu\nu\rho}&=&{\cal A}_{0,t}^{(3),\mu\nu\rho} 
(\mu\leftrightarrow\nu,q_1\leftrightarrow q_2)\,\,\,, 
\end{eqnarray}
with
\[
V^{\mu\nu\alpha}=(q_1-q_2)^\alpha g^{\mu\nu}+(q_1+2 q_2)^\mu
g^{\nu\alpha}-(2 q_1+q_2)^\nu g^{\mu\alpha}\,\,\,,
\]
are the individual amplitudes for the $s-$channel, $t-$channel, and
$u-$channel diagrams in Figure~\ref{fig:ggZbbtree_level}.

%%%%%%%%%%%%%%%%%%%%%%%%%%%%%%%%%%%%%%%%%%%%%%%%%%%%%%%%%%%%%%%%%%%%%%%
\chapter{Scalar Integrals for $\WZbb$ Production}\label{app:Scint}

%%%%%%%%%%%%%%%%%%%%%%%%%%%%%%%%%%%%%%%%%%%%%%%%%%%%%%%%%%%%%%%%%%%%%%%

\begin{figure}[b]
\begin{center}
\includegraphics[scale=0.9]{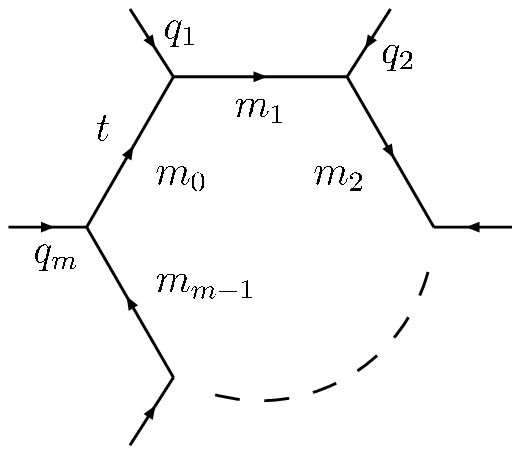}
\caption{Topology of one-loop Feynman integrals. We denote by $t$ the loop momentum, by $\{q_i\}$ the set of
incoming momenta and by $\{m_i\}$ the set of masses in the internal propagators.}
\label{fig:ScIntTop}
\end{center}
\end{figure}

In this Appendix we present a collection of all the one-loop IR-divergent scalar integrals that
appear in the $\WZbb$ NLO QCD calculation. For completeness we will also include the UV-structure of
the UV-divergent scalar integrals, namely tadpoles and self-energies. We will use the following
notation for the scalar $m$-point integral:
\begin{eqnarray}
\lefteqn{I0(q_1,\dots,q_{m-1};m_0,\dots,m_{m-1})=}\nonumber\\
&& \mu^{4-d}\int\frac{d^dt}{\left(2\pi\right)^d}
	\frac{1}{[t^2-m_0^2][(t+q_1)^2-m_1^2]\cdots
		[(t+q_1+\cdots+q_{m-1})^2-m_{m-1}^2]}\; ,
\label{eq:genscint}
\end{eqnarray}
which corresponds to the topology illustrated in Figure~\ref{fig:ScIntTop}.
We have denoted by $t$ the loop momentum,  by $\{q_i\}$ ($i=1,\dots,m$, $q_m=-\sum_{j=1}^{m-1}q_j$)
the set of incoming momenta to the diagram and by $\{m_i\}$ ($i=0,\dots,m-1$) the set of masses
corresponding to the propagators in the loop.  The integration over the loop momentum is performed
over $d=4-2\epsilon$ ($\epsilon=\epsilon_{\sss \rm IR}$ unless stated differently) dimensions.  We
will denote integrals $I0$ with $m=1,2,3,4$ and $5$ denominators by $A0$, $B0$, $C0$, $D0$ and $E0$
respectively, as conventional.

Most of the expressions presented in this Appendix exist in the literature (see for
example~\cite{Beenakker:1988jr,Bern:1992em,Bern:1993kr,Oleari:1997az,Reina:2001bc}), except for the
box integrals labelled in the following as II.2 and II.3, which as far as we know have not been
presented explicitly\footnote{Since the completion of this Dissertation a full set of IR-divergent
scalar integrals have been presented in Ref.~\cite{Ellis:2007qk}, and we found agreement with the
corresponding box expressions presented there.\label{ftnote:boxint}}.  We have calculated these
integrals using two independent techniques and found agreement. For IR-finite integrals we have used
expressions from~\cite{Beenakker:1988jr} and cross checked them with the
FF-package~\cite{vanOldenborgh:1990yc}.

The expressions presented below are correct up to ${\cal O}(\epsilon)$, since this is what is needed
for a full NLO calculation. We include only the real pieces~\footnote{By real pieces we mean the
real pieces of the integrals after factorizing $i/(16\pi^2)$.} of the integrals (according to the
kinematics we use) as the imaginary parts do not contribute to the NLO partonic cross section. We
will use the kinematics of the process:
\[
i(q_1)+j(q_2)\to V(p_{\sss V})+b(p_b)+\bar b(p_{\bar b})\; ,
\]
with $p_{\sss V}=q_1+q_2-p_b-p_{\bar b}$
and the on-shell conditions $q_1^2=q_2^2=0$ and $p_b^2=p_{\bar b}^2=m_b^2$. 
Although in this dissertation we treat
the $V$ vector boson as on-shell (i.e. $p_{\sss V}^2=M_{\sss V}^2$), 
the integrals in this Appendix are correct 
also for for the case of an
off-shell $V$ boson, as
we always eliminate $p_{\sss V}$ 
by using conservation of momentum. 
Therefore they can be used also
if one had to include leptonic decays of the $V$ boson. 
Moreover, all the following results are expressed in terms of
the invariants $s_{ij}=(q_i+q_j)^2$, where $q_i$ and $q_j$ are two of the external momenta.

Scalar integrals can be related
by rotation or reflection of the integration momentum:
\begin{eqnarray}
I0(q_1,\dots,q_{m-1};m_0,\dots,m_{m-1}) & =&
	I0(-\sum_{i=1}^{m-1}q_i,q_1,\dots,q_{m-2};m_{m-1},m_0,\dots,m_{m-2})\nonumber\\
&=&	I0(q_{m-1},\dots,q_1,m_{m-1},\dots,m_0)\; .
\label{eq:rotrefrel}
\end{eqnarray}
In the following we therefore give the minimal set of scalar integrals appearing in our calculation.

Finally we will always factor out of the integrals the factor:
\begin{equation}
\label{eq:nb}
{\cal N}_b = \left(\frac{4\pi\mu^2}{m_b^2}\right)^\epsilon
\Gamma(1+\epsilon)\,\,\,.\,
\end{equation}

\boldmath
\subsection*{$A0$ integral}
\unboldmath
$A0$ and $B0$ integrals are
the only scalar UV-divergent integrals. 
The 1-point integral is IR-finite, and it is given by: 
\begin{equation}
A0(m_b)=\frac{i}{16\pi^2}{\cal N}_b\;\; m_b^2\left(\frac{1}{\epsilon_{\sss \rm UV}}+1\right)\; .
\end{equation}
We notice that $A0(m\to0)=0$, as it can be inferred from the fact that such dimensionful integral
cannot be built from any kinematic invariant. We mention that the only other 1-point integral
appearing in our calculation is $A0(m_t)$, which can be obtained from the previous expression
replacing $m_b$ and ${\cal N}_b$ by $m_t$ and ${\cal N}_t$ respectively.

\boldmath
\subsection*{$B0$ integrals}
\unboldmath
All $B0$ integrals are UV-divergent and their corresponding UV-pole part is:
\begin{equation}
B0(p;m_0,m_1)\Bigl|_{UV-pole} =\frac{i}{16\pi^2}{\cal N}_b\;\; \frac{1}{\epsilon_{\sss\rm UV}}\; .
\end{equation}

The only IR-divergent $B0$ integral is the one with zero internal masses and with a light-like
($q_1^2=0$) external momentum $B0(q_1;0,0)$. Indeed, there is no invariant available to build this
integral, so it should vanish. One can understand this vanishing as a cancellation between the UV
and IR behavior of this integral of the form:
\unboldmath
\begin{equation}
B0(q_1;0,0) = \frac{i}{16\pi^2}{\cal N}_b\;\; \left(\frac{1}{\epsilon_{\sss \rm UV}} -\frac{1}{\epsilon_{\sss
\rm IR}}\right)\; ,
\end{equation}
where we have made explicitly the UV or IR nature of the two single poles. Notice that the overall
constant in front of the integral is a matter of convention.

\boldmath
\subsection*{IR-divergent $C0$ integrals}
\unboldmath
Let us organize the integrals by the number of internal masses.

\subsubsection*{\bf I. All internal masses equal to zero}
\begin{enumerate}
\boldmath
\item $C0(q_1,q_2;0,0,0)$, two on-shell massless legs:
\unboldmath
\begin{equation}
C0(q_1,q_2;0,0,0)=\frac{i}{16\pi^2}{\cal N}_b\;\; \frac{1}{s_{12}}\left(\frac{1}{\epsilon^2}+\frac{1}{\epsilon}
	\ln{\frac{m_b^2}{s_{12}}}+\frac{1}{2}\ln^2{\frac{m_b^2}{s_{12}}}-\frac{\pi^2}{6}\right)\;.
\end{equation}
\boldmath
\item $C0(q_1,-p_b-p_{\bar b};0,0,0)$, one on-shell massless leg:
\unboldmath
\begin{eqnarray}
\lefteqn{C0(q_1,-p_b-p_{\bar b};0,0,0)=\frac{i}{16\pi^2}{\cal N}_b\;\;}\nonumber\\
&& \hspace{-1.8cm}\frac{1}{2\ q_1\cdot (-p_b-p_{\bar b})}
	\left(\frac{1}{\epsilon}\left[\ln{\frac{m_b^2}{|s_{12}|}}-\ln{\frac{m_b^2}{(-p_b-p_{\bar b})^2}}\right]
	+\frac{1}{2}\left[\ln^2{\frac{m_b^2}{|s_{12}|}}-\pi^2-\ln^2{\frac{m_b^2}{(-p_b-p_{\bar b})^2}}\right]\right)\;
,\nonumber\\
\end{eqnarray}
notice that for this case, in our kinematics, $s_{12}<0$ so we have included absolute values in the logarithm
arguments to avoid negative arguments.
\end{enumerate}
\subsubsection*{\bf II. One non-zero internal mass}
\begin{enumerate}
\boldmath
\item $C0(q_1,-p_b;0,0,m_b)$, One on-shell massless leg and one on-shell massive leg:
\unboldmath
\begin{eqnarray}
\hspace{-1.0cm}\lefteqn{C0(q_1,-p_b;0,0,m_b)=\frac{i}{16\pi^2}{\cal
N}_b\;\;\frac{1}{s_{12}-m_b^2}\Biggl\{\frac{1}{2\epsilon^2}+\frac{1}{\epsilon}\ln\left|\frac{m_b^2}{s_{12}-m_b^2}\right| } \nonumber\\
&&
+\frac{1}{2}\ln^2\left(1-\frac{s_{12}}{m_b^2}\right)+{\rm Li}_2\left(\frac{1}{1-\frac{s_{12}}{m_b^2}}\right)
+\ln\left({1-\frac{m_b^2}{s_{12}}}\right)\ln\left(1-\frac{s_{12}}{m_b^2}\right)-\frac{\pi^2}{6}\Biggl\}\;
.\nonumber\\
\end{eqnarray}
\boldmath
\item $C0(q_1,q_2-p_b;0,0,m_b)$, One on-shell massless leg:
\unboldmath
\begin{eqnarray}
\lefteqn{C0(q_1,q_2-p_b;0,0,m_b)=\frac{i}{16\pi^2}{\cal N}_b\;\; \frac{1}{s_{12}-(q_2-p_b)^2}}\nonumber\\
&& \Biggl\{\frac{1}{\epsilon}\left(\ln\frac{m_b^2}{s_{12}-m_b^2}-\ln\frac{m_b^2}{|(q_2-p_b)^2-m_b^2|}\right)
 +\ln^2\frac{m_b^2}{s_{12}-m_b^2}-\ln^2\frac{m_b^2}{|(q_2-p_b)^2-m_b^2|}\nonumber\\
&&-\ln\left(\frac{m_b^2}{|(q_2-p_b)^2-m_b^2|}\right)\ln\left(1-\frac{(q_2-p_b)^2-m_b^2}{m_b^2}\right)
+{\rm Li}_2\left(1+\frac{s_{12}-m_b^2}{m_b^2}\right)\nonumber\\
&&\hspace{2.0cm}+{\rm Li}_2\left(\frac{m_b^2-(q_2-p_b)^2}{m_b^2}\right)-\frac{7}{6}\pi^2\Biggl\}\; .
\end{eqnarray}
\end{enumerate}
\subsubsection*{\bf III. Two non-zero internal masses}
\begin{enumerate}
\boldmath
\item $C0(-p_b,p_b+p_{\bar b};0,m_b,m_b)$, Two on-shell massive legs:
\unboldmath
\begin{eqnarray}
\lefteqn{C0(-p_b,p_b+p_{\bar b};0,m_b,m_b)=\frac{i}{16\pi^2}{\cal N}_b\;\; \frac{1}{\beta (p_b+p_{\bar b})^2}}\nonumber\\
&& \Biggl\{\frac{1}{\epsilon}\ln\left(\frac{1-\beta}{1+\beta}\right)\ -\ln\left(\frac{(p_b+p_{\bar
b})^2}{m_b^2}\right)\ln\left(\frac{1-\beta}{1+\beta}\right)-\frac{1}{2}I_2-\pi^2\Biggl\}\;,
\end{eqnarray}
where we have defined:
\[
\beta=\sqrt{1-\frac{4\ m_b^2}{(p_b+p_{\bar b})^2}}\;,
\]
and
\begin{eqnarray}
I_2 &= &-\ln^2\left(\frac{1}{2}(1+\beta)\right)+\ln^2\left(\frac{1}{2}(1-\beta)\right)-\pi^2
+2\ln\beta\ln\left(\frac{1-\beta}{1+\beta}\right)\nonumber\\
&& -2{\rm Li}_2\left(-\frac{1-\beta}{2\beta}\right)+2 {\rm Li}_2\left(\frac{1+\beta}{2\beta}\right)\;.\nonumber
\end{eqnarray}
\end{enumerate}

\boldmath
\subsection*{IR-divergent $D0$ integrals}
\unboldmath

\subsubsection*{\bf I. All internal masses equal to zero}
We write the massless box expressions following the results of~\cite{Bern:1993kr}.
\begin{enumerate}
\boldmath
\item $D0(q_1,q_2,-p_b-p_{\bar b};0,0,0,0)$, two adjacent on-shell massless legs:
\unboldmath
\begin{eqnarray}
\lefteqn{D0(q_1,q_2,-p_b-p_{\bar b};0,0,0,0)=\frac{i}{16\pi^2}{\cal N}_b\;\; \frac{1}{s_{12}s_{23}}\Biggl\{}\nonumber\\
&& \frac{2}{\epsilon^2}\left[\left(\frac{m_b^2}{-s_{12}}\right)^\epsilon +
\left(\frac{m_b^2}{-s_{23}}\right)^\epsilon - \left(\frac{m_b^2}{-(p_b+p_{\bar b})^2}\right)^\epsilon 
- \left(\frac{m_b^2}{-(-q_1-q_2+p_b+p_{\bar b})^2}\right)^\epsilon\ \right]\nonumber\\
&&+\frac{1}{\epsilon^2}\left[\left(-\frac{m_b^2 s_{12}}{(p_b+p_{\bar b})^2(-q_1-q_2+p_b+p_{\bar
b})^2}\right)^\epsilon\ \right]\nonumber\\
&& -2{\rm Li}_2\left(1-\frac{(-q_1-q_2+p_b+p_{\bar b})^2}{s_{23}}\right) -2{\rm
Li}_2\left(1-\frac{(-q_1-q_2+p_b+p_{\bar
b})^2}{s_{12}}\right)-\ln^2\left(\frac{-s_{12}}{-s_{23}}\right)\nonumber\\
&&\hspace{3.0cm}-\frac{\pi^2}{6}\Biggl\}\; .
\end{eqnarray}
\boldmath
\item $D0(q_1,-p_b-p_{\bar b},q_2;0,0,0,0)$, two opposite on-shell massless legs:
\unboldmath
\begin{eqnarray}
\lefteqn{D0(q_1,-p_b-p_{\bar b},q_2;0,0,0,0)=\frac{i}{16\pi^2}{\cal N}_b\;\;
\frac{1}{s_{12}s_{23}-M_2^2M_4^2}\Biggl\{}\nonumber\\
&& \frac{2}{\epsilon^2}\left[\left(\frac{m_b^2}{-s_{12}}\right)^\epsilon +
\left(\frac{m_b^2}{-s_{23}}\right)^\epsilon - \left(\frac{m_b^2}{-M_2^2}\right)^\epsilon 
- \left(\frac{m_b^2}{-M_4^2}\right)^\epsilon\ \right]\nonumber\\
&& -2\ {\rm Li}_2\left(1-\frac{M_2^2}{s_{12}}\right) 
-2\ {\rm Li}_2\left(1-\frac{M_2^2}{s_{23}}\right)
-2\ {\rm Li}_2\left(1-\frac{M_4^2}{s_{12}}\right) 
-2\ {\rm Li}_2\left(1-\frac{M_4^2}{s_{23}}\right)\nonumber\\
&&+2\ {\rm Li}_2\left(1-\frac{M_2^2 M_4^2}{s_{12}s_{23}}\right)
-\ln^2\left(\frac{-s_{12}}{-s_{23}}\right)\Biggl\}\; ,\nonumber\\
\end{eqnarray}
with $M_2^2=(p_b+p_{\bar b})^2$ and $M_4^2=(q_1+q_2-p_b-p_{\bar b})^2$.
\end{enumerate}

\subsubsection*{\bf II. One non-zero internal mass}
\begin{enumerate}
\boldmath
\item $D0(q_1,q_2,-p_{\bar b};0,0,0,m_b)$, two adjacent on-shell massless legs and one one-shell massive leg:
\unboldmath
\begin{eqnarray}
\lefteqn{D0(q_1,q_2,-p_{\bar b};0,0,0,m_b)=\frac{i}{16\pi^2}{\cal N}_b\;\; \frac{1}{s_{12}(s_{23}-m_b^2)}\Biggl\{
}\nonumber\\
&&\frac{3}{2\epsilon^2}+\frac{1}{\epsilon}\left[2\ln\frac{m_b^2}{m_b^2-s_{23}}+\ln\frac{m_b^2}{s_{12}}
-\ln\frac{m_b^2}{(q_1+q_2-p_{\bar b})^2-m_b^2}\right]\nonumber\\
&& +2\ln\left(\frac{m_b^2}{m_b^2-s_{23}}\right)
\ln\left(\frac{m_b^2}{\sigma}\right)
-\ln^2\left(\frac{m_b^2}{(q_1+q_2-p_{\bar b})^2-m_b^2}\right)\nonumber\\ 
&&-2{\rm Li}_2\left(1+\frac{(q_1+q_2-p_{\bar
b})^2-m_b^2}{m_b^2-s_{23}}\right)+\frac{\pi^2}{3}\Biggl\}\;.
\end{eqnarray}
\boldmath
\item $D0(q_1,-q_1+p_b+p_{\bar b},-p_{\bar b};0,0,0,m_b)$, one on-shell massless leg and two one-shell
massive legs:
\unboldmath
\begin{eqnarray}
\lefteqn{D0(q_1,-q_1+p_b+p_{\bar b},-p_{\bar b};0,0,0,m_b)=\frac{i}{16\pi^2}{\cal N}_b\;\;
\frac{1}{s_{12}(s_{23}-m_b^2)}\Biggl\{}\nonumber\\
&& \frac{1}{2\epsilon^2}+\frac{1}{\epsilon}\Biggl[\ln\left(\frac{m_b^2}{-s_{12}}\right)
+\ln\left(\frac{m_b^2}{m_b^2-s_{23}}\right)
-\ln\left(\frac{m_b^2}{-M_2^2}\right)
\Biggl]\nonumber\\
&&
\ln^2\left(\frac{m_b^2}{{-s_{12}}}\right)- \ln^2\left(\frac{{-M_2^2}}{{m_b^2-s_{23}}}\right)
			+\frac{\pi^2}{2}\nonumber\\
&&		+\ln\left(\frac{-M_2^2+s_{12}}{m_b^2-s_{23}}\right)
			\ln\left(\frac{\left(m_b^2\right)^2}{s_{12}^2}\right)	
		-2\ln\left(1-\frac{{-s_{12}}}{{-M_2^2}}\right)
			\ln\left(\frac{m_b^2}{{m_b^2-s_{23}}}\right)\nonumber\\
&&		+2\Biggl[\ln\left(\frac{-M_2^2}{m_b^2-s_{23}}\right) 
		\ln\left(1-\frac{-M_2^2(-s_{12}+m_b^2-s_{23}+M_2^2)}{(m_b^2-s_{23}) (-s_{12})}
			\right)\nonumber\\
&&	- \ln\left(\frac{-M_2^2}{m_b^2-s_{23}}\right) \ln\left(\frac{-M_2^2-m_b^2+s_{23}}{-M_2^2}\right)
	+ \ln\left(\frac{-M_2^2+s_{12}}{m_b^2-s_{23}}\right)
				\ln\left(1-\frac{m_b^2-s_{23}}{-M_2^2+s_{12}}\right)\nonumber\\
&&		 
		- {\rm Li}_2\left(1-\frac{m_b^2-s_{23}}{-M_2^2+s_{12}}\right)
		- {\rm Li}_2\left(1-\frac{-M_2^2+s_{12}}{{m_b^2-s_{23}}}\right)
		-{\rm Li}_2\left(1-\frac{-M_2^2-m_b^2+s_{23}}{-s_{12}}\right)\nonumber\\
&&		+{\rm Li}_2\left(\frac{-M_2^2 (-s_{12}+m_b^2-s_{23}+M_2^2)}{(m_b^2-s_{23}) (-s_{12})}\right)
		-{\rm Li}_2\left(\frac{m_b^2-s_{23}}{-M_2^2+s_{12}}\right)
		+ {\rm Li}_2\left(\frac{m_b^2-s_{23}}{-M_2^2}\right)
		\Biggl]\Biggl\}\;,\nonumber\\
\end{eqnarray}
where we have defined the mass square of the second leg as $M_2^2=(-q_1+p_b+p_{\bar b})^2$.
\boldmath
\item $D0(q_1,-q_1-q_2+p_b+p_{\bar b},q_2-p_{\bar b};0,0,0,m_b)$\unboldmath\footnote{ During the completion of
this manuscript a compilation of one-loop scalar integrals has appeared on http://qcdloop.fnal.gov/,
written and maintained by R.K.~Ellis. The box integrals presented in sections II.2 and II.3 of this
Appendix are presented in there with one expression, corresponding to $B_9$ in their notation.  We
have compared analytically the pole structure and numerically the finite real part of our
expressions II.2 and II.3 with the expression of $B_9$, and we found perfect agreement. (See
footnote~\ref{ftnote:boxint} in this Appendix).}, one on-shell massless leg and one one-shell
massive leg:
\begin{eqnarray}
\lefteqn{D0(q_1,-q_1-q_2+p_b+p_{\bar b},q_2-p_{\bar b};0,0,0,m_b)=\frac{i}{16\pi^2}{\cal N}_b\;\;
\frac{1}{s_{12}(s_{23}-m_b^2)}\Biggl\{}\nonumber\\
&& \frac{1}{2\epsilon^2}+\frac{1}{\epsilon}\Biggl[\ln\left(\frac{m_b^2}{-s_{12}}\right)
+\ln\left(\frac{m_b^2}{m_b^2-s_{23}}\right)
-\ln\left(\frac{m_b^2}{M_2^2}\right)
\Biggl]\nonumber\\
&& +\ln^2\left(\frac{-s_{12}}{m_b^2}\right)+\ln^2\left(\frac{m_b^2-s_{23}}{m_b^2}\right)
	-\ln^2\left(\frac{M_2^2}{m_b^2}\right)+\frac{4}{3}\pi^2\nonumber\\
&&\hspace{-1.5cm}+2\ln\left(\frac{M_2^2}{m_b^2}\right)\ln\left(\frac{(-s_{12}+M_2^2)(m_b^2-s_{23}+M_2^2)}{s_{12}(s_{23}-m_b^2)}\right)
	-2\ln\left(\frac{m_b^2-s_{23}}{m_b^2}\right)\ln\left(-\frac{m_b^2-s_{23}+M_2^2}{s_{12}}\right)\nonumber\\
&&\hspace{-1.5cm}+2\ln\left(-\frac{s_{12}}{m_b^2}\right)\ln\left(\frac{-s_{12}+M_2^2}{m_b^2-s_{23}}\right)
	+\ln\left(\frac{m_b^2-M_3^2}{m_b^2}\right)\ln\left(-\frac{M_2^2M_3^2}{-m_b^2M_2^2-(s_{23}-m_b^2)(M_3^2-m_b^2)}\right)\nonumber\\
&&+2{\rm Li}_2\left(1-\frac{(-s_{12}+M_2^2)(m_b^2-s_{23}+M_2^2)}{s_{12}(s_{23}-m_b^2)}\right)
	-2{\rm Li}_2\left(1-\frac{m_b^2-s_{23}+M_2^2}{s_{12}}\right)\nonumber\\
&&-2{\rm Li}_2\left(1+\frac{-s_{12}+M_2^2}{m_b^2-s_{23}}\right)-{\rm
Li}_2\left(1-\frac{s_{23}-m_b^2}{M_2^2}\right)\nonumber\\
&&\hspace{-1.5cm}-{\rm Li}_2\left(1-\frac{(m_b^2-s_{23})M_3^2}{-m_b^2M_2^2-(s_{23}-m_b^2)(M_3^2-m_b^2)}\right)
	+{\rm Li}_2\left(1-\frac{M_2^2M_3^2}{m_b^2M_2^2+(s_{23}-m_b^2)(M_3^2-m_b^2)}\right)\nonumber\\
&&-{\rm Li}_2\left(1-\frac{(-M_2^2+s_{12})M_3^2}{m_b^2(-M_2^2+s_{12})-(s_{23}-m_b^2)(M_3^2-m_b^2)}\right)
	+{\rm Li}_2\left(1-\frac{m_b^2-M_3^2}{m_b^2}\right)\nonumber\\
&&+{\rm Li}_2\left(1+\frac{(-s_{12}+M_2^2)M_3^2}{m_b^2(-M_2^2+s_{12})-(s_{23}-m_b^2)(M_3^2-m_b^2)}\right)\Biggl\}\;,
\end{eqnarray}
where we have defined $M_2^2=(-q_1-q_2+p_b+p_{\bar b})^2$ and $M_3^2=(q_2-p_{\bar b})^2$.
\vspace{3.5cm}
\end{enumerate}

\subsubsection*{\bf III. Two non-zero internal masses}
\begin{enumerate}
\boldmath
\item $D0(q_1,-p_{\bar b},q_2;0,0,m_b,m_b)$, two opposite on-shell massless legs and one on-shell massive leg:
\unboldmath
\begin{eqnarray}
\lefteqn{D0(q_1,-p_{\bar b},q_2;0,0,m_b,m_b)=\frac{i}{16\pi^2}{\cal N}_b\;\;
\frac{1}{(s_{12}-m_b^2)(s_{23}-m_b^2)}\Biggl\{}\nonumber\\
&& \frac{1}{2\epsilon^2}+\frac{1}{\epsilon}\left[\ln\left(\frac{m_b^2}{m_b^2-s_{23}}\right)
+\ln\left(\frac{m_b^2}{m_b^2-s_{12}}\right) -\ln\left(\frac{m_b^2}{(q_1+q_2-p_{\bar
b})^2-m_b^2}\right)\right]\nonumber\\
&& + \ln^2\left(\frac{m_b^2-s_{23}}{m_b^2}\right)
+\ln^2\left(\frac{m_b^2-s_{12}}{m_b^2}\right)
-\ln^2\left(\frac{(q_1+q_2-p_{\bar b})^2-m_b^2}{m_b^2}\right)+\frac{3}{2}\pi^2 \nonumber\\
&&+2\ln\left(\frac{-s_{23}+(q_1+q_2-p_{\bar b})^2}{m_b^2-s_{12}}\right)
  \ln\left(\frac{m_b^2-s_{12}}{m_b^2-s_{23}-s_{12}+(q_1+q_2-p_{\bar b})^2}\right)\nonumber\\
&&+2\ln\left(\frac{-s_{12}+(q_1+q_2-p_{\bar b})^2}{m_b^2-s_{23}}\right)
  \ln\left(\frac{m_b^2-s_{23}}{m_b^2-s_{23}-s_{12}+(q_1+q_2-p_{\bar b})^2}\right)\nonumber\\
&&-2\,\mbox{Li}_2\left(\frac{m_b^2-s_{23}-s_{12}+(q_1+q_2-p_{\bar
	b})^2}{m_b^2-s_{12}}\right)\nonumber\\
&&-2\,\mbox{Li}_2\left(\frac{m_b^2-s_{23}-s_{12}+(q_1+q_2-p_{\bar
	b})^2}{m_b^2-s_{23}}\right)\nonumber\\
&&-2\,\mbox{Li}_2\left(\frac{(-s_{23}+(q_1+q_2-p_{\bar b})^2)(-s_{12}+(q_1+q_2-p_{\bar b})^2)}
                        {(m_b^2-s_{23})(m_b^2-s_{12})}\right) \Biggl\}\;.\nonumber\\
\end{eqnarray}
\boldmath
\item $D0(q_1,-p_{\bar b},-q_1+p_b+p_{\bar b};0,0,m_b,m_b)$, one on-shell massless legs and two opposite
on-shell massive legs:
\unboldmath
\begin{eqnarray}
\lefteqn{D0(q_1,-p_{\bar b},-q_1+p_b+p_{\bar b};0,0,m_b,m_b)=\frac{i}{16\pi^2}{\cal N}_b\;\;
\frac{1}{(s_{12}-m_b^2)(s_{23}-m_b^2)}\Biggl\{}\nonumber\\
&& \frac{1}{\epsilon^2}+\frac{1}{\epsilon}\left[\ln\left(\frac{m_b^2}{m_b^2-s_{12}}\right)+
\ln\left(\frac{m_b^2}{m_b^2-s_{23}}\right)\right]+\Biggl[\ln^2\left(\frac{m_b^2}{m_b^2-s_{12}}\right)+
\ln^2\left(\frac{m_b^2}{m_b^2-s_{23}}\right)\nonumber\\
&& +\ln^2\left(\frac{m_b^2-s_{12}}{m_b^2-s_{23}}\right)-\frac{2\pi^2}{3}+2{\rm Li}_2\left(\frac{1}{z_+}\right)
+2{\rm Li}_2\left(\frac{1}{z_-}\right)\Biggl]\Biggl\}\;,
\end{eqnarray}
with
\[
z_{\pm}=\frac{1}{2}\left(1\pm\sqrt{1-\frac{4\ m_b^2}{(-q_1+p_b+p_{\bar b})^2}}\right)\;.
\]
\boldmath
\item $D0(q_2,-p_{\bar b},-q_1-q_2+p_b+p_{\bar b};0,0,m_b,m_b)$, one on-shell massless leg and one on-shell massive
leg:
\unboldmath
\begin{eqnarray}
\lefteqn{D0(q_2,-p_{\bar b},-q_1-q_2+p_b+p_{\bar b};0,0,m_b,m_b)=\frac{i}{16\pi^2}{\cal N}_b\;\;
\frac{1}{(s_{12}-m_b^2)(s_{23}-m_b^2)} \Biggl\{}\nonumber\\
&& \frac{1}{2\epsilon^2}+\frac{1}{\epsilon}\left[\ln\left(\frac{m_b^2}{m_b^2-s_{12}}\right)
+\ln\left(\frac{m_b^2}{s_{23}-m_b^2}\right)-\ln\left(\frac{m_b^2}{m_b^2-M_4^2}\right)\right]\nonumber\\
&&+ Re\Biggl[-\frac{5}{6}\pi^2+
 \ln^2\left(\frac{s_{23}-m_b^2}{m_b^2}\right)+
 \ln^2\left(\frac{m_b^2-s_{12}}{m_b^2}\right)-
 \ln^2\left(\frac{m_b^2-M_4^2}{m_b^2}\right)\nonumber\\
&&+ 2\,\ln\left(\frac{s_{23}-M_4^2}{m_b^2-s_{12}}\right)
    \ln\left(\frac{m_b^2-M_4^2}{s_{23}-m_b^2}\right)+
 2\,\ln\left(\frac{-s_{12}+M_4^2}{s_{23}-m_b^2}\right)
    \ln\left(\frac{m_b^2-M_4^2}{m_b^2-s_{12}}\right)\nonumber\\
&&- 2\,\mbox{Li}_2\left(\frac{-s_{12}+M_4^2-s_{23}+m_b^2}{m_b^2-s_{12}}\right)-
 2\,\mbox{Li}_2\left(\frac{s_{23}-M_4^2-m_b^2+s_{12}}{s_{23}-m_b^2}\right)\nonumber\\
&&+2\,\mbox{Li}_2\left(\frac{(m_b^2-M_4^2)(s_{23}-M_4^2-m_b^2+s_{12})}{(s_{23}-m_b^2)(m_b^2-s_{12})}
\right)-
 I_0\Biggl]\Biggl\}\;,\nonumber\\
\end{eqnarray}
where $M_4^2=(q_1-p_b)^2$ is the mass square of the fourth leg,
\begin{eqnarray}
I_0&=& 
  \ln\left(\frac{m_b^2-s_{12}}{m_b^2-M_4^2}\right)\ln\left(\frac{M_3^2}{m_b^2}\right)+
\Biggl\{-\mbox{Li}_2\left(\frac{1}{\lambda_+}\right)\nonumber\\
&&+
\ln\left(\frac{m_b^2-s_{12}}{m_b^2-M_4^2}\right)
\ln\left(\frac{-(m_b^2-M_4^2)-\lambda_+(-s_{12}+M_4^2)}{-s_{12}+M_4^2}\right)
\nonumber\\
&-&
\mbox{Li}_2\left(\frac{m_b^2-s_{12}}{\lambda_+(-s_{12}+M_4^2)+m_b^2-M_4^2}\right)+
\mbox{Li}_2\left(\frac{m_b^2-M_4^2}{\lambda_+(-s_{12}+M_4^2)+m_b^2-M_4^2}\right)\nonumber\\
&&+(\lambda_+\leftrightarrow\lambda_-)\Biggl\}\,\,\,,
\end{eqnarray}
$M_3^2=(-q_1-q_2+p_b+p_{\bar b})^2$ is the mass square of the third leg and
\begin{equation}
\lambda_\pm=\frac{1}{2}\left(1\pm \sqrt{1-\frac{4m_b^2}{M_3^2}}\right)
\,\,\,.
\end{equation}
\end{enumerate}

\subsubsection*{\bf IV. Three non-zero internal masses}
\begin{enumerate}
\boldmath
\item $D0(-p_b,q_2,-q_2+p_b+p_{\bar b};0,m_b,m_b,m_b)$, One on-shell massless leg and two adjacent
on-shell massive legs:
\unboldmath
\begin{eqnarray}
\lefteqn{D0(-p_b,q_2,-q_2+p_b+p_{\bar b};0,m_b,m_b,m_b)=\frac{i}{16\pi^2}{\cal N}_b\;\;
\frac{1}{s_{23}(s_{12}-m_b^2)\beta}\Biggl\{}\nonumber\\
&& \frac{1}{\epsilon}\ln\left(\frac{1-\beta}{1+\beta}\right)-
\Biggl[
(-2 \ln\left({\cal X}_s\right)\ln\left(m_b^2/(m_b^2-s_{12})\right)
	+2 \ln\left({\cal X}_s\right)\ln\left(1-{\cal X}_s^2\right)\nonumber\\
&&		+\pi^2/2+{\rm Li}_2({\cal X}_s {\cal X}_s)+\ln^2({\cal X}_3)\nonumber\\
&&		-2 \Bigl({\rm Li}_2({\cal X}_s {\cal X}_3)+\ln\left({\cal X}_s\right)\ln\left(1-{\cal X}_s {\cal X}_3\right)
			+\ln\left({\cal X}_3\right)\ln\left(1-{\cal X}_s {\cal X}_3\right)\Bigl)\nonumber\\
&&		-2 \Bigl({\rm Li}_2({\cal X}_s/{\cal X}_3)+\ln\left({\cal X}_s\right)\ln\left(1-{\cal X}_s/{\cal X}_3\right)
			-\ln\left({\cal X}_3\right)\ln\left(1-{\cal X}_s/{\cal X}_3\right)\Bigl)
\Biggl]\Biggl\}\;,\nonumber\\
\end{eqnarray}
where, using $M_3^2=(-q_2+p_b+p_{\bar b})^2$ the mass square of the third leg, we have defined:
\[
\beta=\sqrt{1-\frac{4\ m_b^2}{s_{23}}}\;,
\]
\[
\beta_3=\sqrt{1-\frac{4\ m_b^2}{M_3^2}}\;,
\]
\[
{\cal X}_s = \frac{1-\beta}{1+\beta}\;,
\]
and
\[
{\cal X}_3 = \frac{1-\beta_3}{1+\beta_3}\;.
\]
\boldmath
\item $D0(-p_b,-q_1-q_2+p_b+p_{\bar b},q_1+q_2;0,m_b,m_b,m_b)$, Two adjacent on-shell massive legs:
\unboldmath
\begin{eqnarray} 
\lefteqn{D0(-p_b,-q_1-q_2+p_b+p_{\bar b},q_1+q_2;0,m_b,m_b,m_b)=}\nonumber\\
&&\hspace{3.0cm}\frac{i}{16\pi^2}{\cal N}_b\;\; \frac{1}{(s_{12}-m_b^2)s_{23}\beta}
 \left\{\frac{1}{\epsilon}\ln\left(\frac{1-\beta}{1+\beta}\right)+X_0\right\}\,\,\,,
\end{eqnarray}
where we have defined:
\[
\beta=\sqrt{1-\frac{4\ m_b^2}{s_{23}}}\;.
\]
Defining $M_2^2=(-q_1-q_2+p_b+p_{\bar b})^2$, $M_3^2=(q_1+q_2)^2$,
\[
\beta_2=\sqrt{1-\frac{4\ m_b^2}{M_2^2}}\;,
\]
\[
\beta_3=\sqrt{1-\frac{4\ m_b^2}{M_3^2}}\;,
\]
\[
{\cal X}_s = \frac{1-\beta}{1+\beta}\;,
\]
\[
{\cal X}_2 = \frac{1-\beta_2}{1+\beta_2}\;,
\]
and
\[
{\cal X}_3 = \frac{1-\beta_3}{1+\beta_3}\;,
\]
we can express the finite piece $X_0$ as follows:
\begin{eqnarray}
X_0 & = &
\frac{{\cal X}_ss_{23}\beta}{m_b^2(1-{\cal X}_s^2)}\Biggl\{\nonumber\\
&& 
      		2 \ln\left({\cal X}_s\right)\ln\left(1-{\cal X}_s^2\right)
	-2\ln\left({\cal X}_s\right)\ln\left(m_b^2/(m_b^2-s_{12})\right)
\nonumber\\
&&      		+\pi^2/2+{\rm Li}_2({\cal X}_s^2)+\ln^2({\cal X}_2)+\ln^2({\cal X}_3)\nonumber\\
&&		-{\rm Li}_2({\cal X}_s {\cal X}_2 {\cal X}_3)+{\rm Li}_2({\cal X}_s {\cal X}_2/{\cal X}_3)+{\rm Li}_2({\cal
X}_s {\cal X}_3/{\cal X}_2)+{\rm Li}_2({\cal X}_s/{\cal X}_2/{\cal X}_3)\nonumber \\
&&		+\ln\left(1-{\cal X}_s {\cal X}_2 {\cal X}_3\right)\ln\left({\cal X}_s\right)
	+\ln\left(1-{\cal X}_s {\cal X}_2 {\cal X}_3\right)\ln\left({\cal X}_2\right)\nonumber \\
&&	+\ln\left(1-{\cal X}_s {\cal X}_2 {\cal X}_3\right)\ln\left({\cal X}_3\right)
		+\ln\left(1-{\cal X}_s {\cal X}_2/{\cal X}_3\right)\ln\left({\cal X}_s\right)\nonumber \\
&&	+\ln\left(1-{\cal X}_s {\cal X}_2/{\cal X}_3\right)\ln\left({\cal X}_2\right)
	+\ln\left(1-{\cal X}_s {\cal X}_2/{\cal X}_3\right)\ln\left(1/{\cal X}_3\right)\nonumber \\
&&		+\ln\left(1-{\cal X}_s{\cal X}_3 /{\cal X}_2\right)\ln\left({\cal X}_s\right)
	+\ln\left(1-{\cal X}_s{\cal X}_3 /{\cal X}_2\right)\ln\left(1/{\cal X}_2\right)\nonumber \\
&&	+\ln\left(1-{\cal X}_s{\cal X}_3 /{\cal X}_2\right)\ln\left({\cal X}_3\right)
		+\ln\left(1-{\cal X}_s/({\cal X}_2{\cal X}_3)\right)\ln\left({\cal X}_s\right)\nonumber \\
&&	+\ln\left(1-{\cal X}_s/({\cal X}_2{\cal X}_3)\right)\ln\left(1/{\cal X}_2\right)
	+\ln\left(1-{\cal X}_s/({\cal X}_2{\cal X}_3)\right)\ln\left(1/{\cal X}_3\right)
\Biggl\}\;.\nonumber\\
\end{eqnarray}
\end{enumerate}

\section*{Reduction of box and pentagon scalar integrals}
An interesting property of $m$-point scalar integrals in $d=4-2\epsilon$ dimensions is that they can be
expressed as a linear combination of $(m-1)$-point scalar integrals in $d=4-2\epsilon$ plus the corresponding
$m$-point scalar integral calculated in $d=6-2\epsilon$ dimensions. This has been proved in Ref.~\cite{Bern:1992em}
and is summarized in the following formula:
\begin{equation}
\label{eq:In_reduction}
I_m = \frac{1}{2}\left[-\sum_{i=1}^{m}c_iI_{m-1}^{(i)}+
        (m-5+2\epsilon)c_0I_m^{(d=6-2\epsilon)}\right]\,\,\,,
\end{equation}
where the integral $I_m$ on the left hand side corresponds to
$(-1)^{m+1}I0(q_1,\dots,q_{m-1};m_0,\dots,m_{m-1})$, in the notation of Eq.~(\ref{eq:genscint}),
$I_m^{(d=6-2\epsilon)}$ denotes the same $m$-point integral calculated in $d=6-2\epsilon$ dimension 
and $I_{m-1}^{(i)}$ correspond to the $(m-1)$-point integrals obtained from $I_m$ by taking out the $i$-th
denominator and are calculated in $d=4-2\epsilon$ dimensions.

The $c_i$ ($i=1,\dots,m$) coefficients in Eq.~(\ref{eq:In_reduction}) are given by:
\begin{equation}
c_i=\sum_{j=1}^mS_{ij}^{-1}\; ,\qquad c_0=\sum_{i=1}^m c_i=\sum_{i,j=1}^mS_{ij}^{-1}\; ,
\end{equation}
where the matrix $S_{ij}$ is built from invariants as:
\begin{equation}
S_{ij}=\frac{1}{2}\left(m_i^2-m_j^2-p_{ij}^2\right).
\end{equation}
with $p_{ij}^2=(q_i+\cdots+q_j)^2$. 

We have used Eq.~(\ref{eq:In_reduction}) twice. First, using Eq.~(\ref{eq:In_reduction}) we have cross checked
our calculation of the IR-divergent box integrals II.2 and II.3 that did not exist in the literature. Second,
we have used it to compute the pentagon scalar integrals. In the case of $E0$-functions,
Eq.~(\ref{eq:In_reduction}) is particularly useful because the coefficients in front on
$I_5^{(d=6-2\epsilon)}$ is of ${\cal O}(\epsilon)$ and therefore does not contribute (given that
$I_5^{(d=6-2\epsilon)}$ is IR-finite). According to Eq.~(\ref{eq:In_reduction}), $E0$-functions are therefore
calculated as the linear combination of five scalar integrals in $d=4-2\epsilon$.

%%%%%%%%%%%%%%%%%%%%%%%%%%%%%%%%%%%%%%%%%%%%%%%%%%%%%%%%%%%%%%%%%%%%%%%
\chapter{Reducing Tensor Feynman Integrals}\label{app:Intred}

%%%%%%%%%%%%%%%%%%%%%%%%%%%%%%%%%%%%%%%%%%%%%%%%%%%%%%%%%%%%%%%%%%%%%%%

\begin{figure}[hb]
\begin{center}
\includegraphics[scale=0.9]{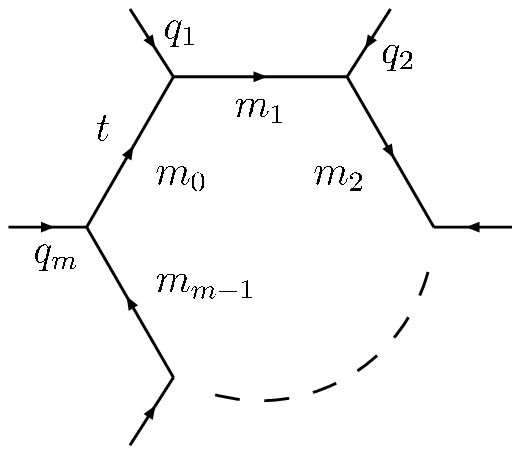}
\caption{Topology of one-loop Feynman Integrals. We denote by $t$ the loop momentum, $\{q_i\}$ the set of
incoming momenta and $\{m_i\}$ the set of masses in the internal propagators.}
\label{fig:IntTop}
\end{center}
\end{figure}

At one-loop, Feynman diagrams with $m$ external legs may contain Feynman tensor integrals with up to
$m$ denominators of the form:
\begin{eqnarray}
\lefteqn{In^{\mu_1\dots\mu_n}(q_1,\dots,q_{m-1},m_0,\dots,m_{m-1})=}\nonumber\\
&& \int\frac{d^dt}{\left(2\pi\right)^d}
	\frac{t^{\mu_1}\cdots t^{\mu_n}}{[t^2-m_0^2][(t+q_1)^2-m_1^2]\cdots
		[(t+q_1+\cdots+q_{m-1})^2-m_{m-1}^2]}\; ,
\label{eq:genint}
\end{eqnarray}
which corresponds to the topology illustrated in Figure~\ref{fig:IntTop}. In Eq.~(\ref{eq:genint})
$t$ is the loop momentum, $d=4-2\epsilon$ is the dimension of the loop momentum integration,
$\mu_1,\dots,\mu_n$ are Lorentz indices associated to each power of loop momentum appearing in the
numerator, $\{q_i\}$ ($i=1,\dots,m$, $q_m=-\sum_{j=1}^{m-1}q_j$) is the set if incoming external
momenta connected to the loop and $\{m_i\}$ ($i=0,\dots,m-1$) is the set of masses associated with
each propagator in the loop. The index ``$n$" is associated with the rank of the tensor integral.
As a short hand notation, depending on the number of denominators, or external momenta, we will
write $I=A$ for one leg, $I=B$ for two legs, and so on.  Integrals with no powers of loop momentum
in the numerator are called ``scalar" (see Appendix~\ref{app:Scint}) and are labelled by $n=0$.  So,
for example, a scalar triangle integral will be denoted by $C0$, while a rank 4 box integral by
$D4$.

One notices immediately that the tensor integral $In^{\mu_1\dots\mu_n}$ is a symmetric tensor and it
can be expressed as a linear combination of (symmetric) tensor Lorentz structures built from the
external momenta $\{q_i\}$ and the metric tensor, $g^{\mu\nu}$, if the set $\{q_i\}$ is not
complete. In this case (i.e. for $m\ge5$) $g^{\mu\nu}$ can be expressed as a linear combination of
momenta. 

The coefficients of the Lorentz structures are rational functions of scalar integrals and invariants
built from the external momenta $\{q_i\}$, the masses $\{m_i\}$ and the integration dimension $d$.
For example, a straightforward computation of the $m=2$, $n=1$ tensor integral gives:
\begin{equation}
B1^\mu(q_1,m,m)=-\frac{1}{2}B0(q_1,m,m)q_1^\mu\equiv B^{(1)}(q_1,m,m)q_1^\mu\; ,
\label{eq:b1red}
\end{equation}
where we have defined the coefficient of the only Lorentz structure as
$B^{(1)}(q_1,m,m)=-\frac{1}{2}B0(q_1,m,m)$. 

The systematic reduction of tensor integrals to a linear combination of Lorentz structures is known
as the Passarino-Veltman (PV) method~\cite{Passarino:1978jh}, and consequently we will call the
tensor integral coefficients $In$-PV functions, depending on the topology and rank of the integral.
In the following we will write explicitly the tensor integrals we have encountered in our
calculations and give an example of how they are recursively reduced (for a review see for
example~\cite{Denner:1993kt}).

Starting with 2-point integrals, we write (suppressing the arguments of the PV functions):
\begin{eqnarray}
B1^\mu(q_1,m_0,m_1)&=&B^{(1)}q_1^\mu\,\,\,,\label{eq:b1}\\
B2^{\mu\nu}(q_1,m_0,m_1)&=& 
    B^{(00)}g^{\mu\nu}+B^{(11)}q_1^\mu q_1^\nu\;.\label{eq:b1b2}
\end{eqnarray}
For 3-point integrals we write:
\begin{eqnarray}
C1(q_1,q_2,m_0,m_1,m_2)^\mu&=&C^{(1)}q_1^\mu+C^{(2)}q_2^\mu\,\,\,,\label{eq:c1}\\
C2(q_1,q_2,m_0,m_1,m_2)^{\mu\nu}&=& 
    C^{(00)}g^{\mu\nu}+C^{(11)}q_1^\mu q_1^\nu+C^{(22)}q_2^\mu q_2^\nu\\
&& +C^{(12)}(q_1^\mu q_2^\nu+q_1^\nu q_2^\mu)\,\,\,,\\
C3(q_1,q_2,m_0,m_1,m_2)^{\mu\nu\rho}&=& 
    C^{(001)}(g^{\mu\nu}q_1^\rho + \mbox{perm})+
    C^{(002)}(g^{\mu\nu}q_2^\rho + \mbox{perm})\nonumber\\
&& +C^{(111)}q_1^\mu q_1^\nu q_1^\rho+
    C^{(222)}q_2^\mu q_2^\nu q_2^\rho\nonumber\\
&& +C^{(112)}(q_1^\mu q_1^\nu q_2^\rho + \mbox{perm})+ C^{(221)}(q_2^\mu q_2^\nu q_1^\rho + \mbox{perm})\;,
\label{eq:c3}
\end{eqnarray}
where the term ``perm" accounts for all terms, obtained from permutations of Lorentz indices, that
are needed to obtain a symmetric tensor.

Finally for 4-point integrals we write:
\begin{eqnarray}
D1(q_1,q_2,q_3,m_0,m_1,m_2,m_3)^\mu&=&D^{(1)}q_1^\mu+D^{(2)}q_2^\mu+D^{(3)}q_3^\mu\,\,\,,\\
D2(q_1,q_2,q_3,m_0,m_1,m_2,m_3)^{\mu\nu}&=& 
    D^{(00)}g^{\mu\nu}+D^{(11)}q_1^\mu q_1^\nu+D^{(22)}q_2^\mu q_2^\nu+
    D^{(33)}q_3^\mu q_3^\nu+\nonumber\\
&& D^{(12)}(q_1^\mu q_2^\nu+q_1^\nu q_2^\mu)+
    D^{(13)}(q_1^\mu q_3^\nu+q_1^\nu q_3^\mu)+\\
&&    D^{(23)}(q_2^\mu q_3^\nu+q_2^\nu q_3^\mu)\,\,\,,\nonumber\\
D3(q_1,q_2,q_3,m_0,m_1,m_2,m_3)^{\mu\nu\rho}&=& 
    D^{(001)}(g^{\mu\nu}q_1^\rho + \mbox{perm})+
    D^{(002)}(g^{\mu\nu}q_2^\rho + \mbox{perm})+\nonumber\\
&&  D^{(003)}(g^{\mu\nu}q_3^\rho + \mbox{perm})+\label{eq:d3}\\
&& D^{(111)}q_1^\mu q_1^\nu q_1^\rho+
    D^{(222)}q_2^\mu q_2^\nu q_2^\rho+
    D^{(333)}q_3^\mu q_3^\nu q_3^\rho+\nonumber\\
&& D^{(112)}(q_1^\mu q_1^\nu q_2^\rho + \mbox{perm})+
    D^{(113)}(q_1^\mu q_1^\nu q_3^\rho + \mbox{perm})+\nonumber\\
&& D^{(221)}(q_2^\mu q_2^\nu q_1^\rho + \mbox{perm})+
    D^{(223)}(q_2^\mu q_2^\nu q_3^\rho + \mbox{perm})+\nonumber\\
&& D^{(331)}(q_3^\mu q_3^\nu q_1^\rho + \mbox{perm})+
    D^{(332)}(q_3^\mu q_3^\nu q_2^\rho + \mbox{perm})+\nonumber\\
&&    D^{(123)}(q_1^\mu q_2^\nu q_3^\rho + \mbox{perm})\,\,\,,
\nonumber
\end{eqnarray}
\begin{eqnarray}
\lefteqn{D4(q_1,q_2,q_3,m_0,m_1,m_2,m_3)^{\mu\nu\rho\sigma} =}\nonumber\\ 
&&    D^{(0000)}(g^{\mu\nu}g^{\rho\sigma} + \mbox{perm})+
    \sum_{i\in\{1,2,3\}} D^{(00ii)}(g^{\mu\nu}q_i^\rho q_i^{\sigma} + \mbox{perm})+\nonumber\\
    && \sum_{i,j\in\{1,2,3\}\atop i<j} D^{(00ij)}(g^{\mu\nu}(q_i^\rho q_j^{\sigma}+q_j^\rho q_i^{\sigma}) +
		\mbox{perm})+
	\sum_{i\in\{1,2,3\}} D^{(iiii)}(q_i^\mu q_i^\nu q_i^\rho q_i^{\sigma})+\nonumber\\
    && \sum_{i,j\in \{1,2,3\}\atop i\neq j} D^{(iiij)}(q_i^\mu q_i^\nu q_i^\rho q_j^{\sigma}+\mbox{perm})+
    \sum_{i,j\in \{1,2,3\}\atop i< j} D^{(iijj)}(q_i^\mu q_i^\nu q_j^\rho q_j^{\sigma}+\mbox{perm})+
\nonumber\\
    && \sum_{i,j,k\in\{1,2,3\}\atop i\neq j,k\neq i,j<k} D^{(iijk)}(q_i^\mu q_i^\nu q_j^\rho q_k^{\sigma}+\mbox{perm})
\; .
\label{eq:d4}
\end{eqnarray}

Naturally, the calculation of pentagon diagrams (like the ones shown in Figures~\ref{fig:pentWbb},
\ref{fig:pentfinalqqZbb} and \ref{fig:pentggZbb}) involves up to $E4$-PV functions, but, as
explained in Sections~\ref{subsec:sigvirt}, \ref{subsec:virtWbb} and \ref{subsec:virtqqZbb}, we have
only used directly these functions for cross checks, due to the large numerical instabilities
associated with them. Their structure can be easily inferred from previous PV functions, by just
omitting terms with $g^{\mu\nu}$ tensors, and will not be presented here.

Expressions for the $In$-PV functions can be obtained recursively by saturating the tensor Lorentz
indices of Eq.~(\ref{eq:genint}) and Eqs.~(\ref{eq:b1})-(\ref{eq:d4}) with the same external
momentum or $g^{\mu\nu}$ tensor.  Using relations like
\begin{equation}
t\cdot q_1 =\frac{1}{2} \left[((t+q_1)^2-m_1^2)-(t^2-m_0^2)-(q_1^2+m_0^2-m_1^2) \right]\;,
\label{eq:dotprodrel}
\end{equation}
allows one to simplify the loop momentum dependence in the numerator of Eq.~(\ref{eq:genint})
against some of the denominators, thereby relating higher rank m-point tensor integrals to lower
rank and/or less external point ones.

Comparing the coefficients of the reduced tensor structures obtained by saturating
Eq.~(\ref{eq:genint}) and the corresponding expressions in Eqs.~(\ref{eq:b1})-(\ref{eq:d4}) with the
same external momentum or $g^{\mu\nu}$ tensor provides a set of equations that fully determines the
coefficients of the tensor integrals represented in Eqs.~(\ref{eq:b1})-(\ref{eq:d4}).

Let us consider as an example the case of $D4^{\mu\nu\rho\sigma}$.  The set of $D4$-PV functions is
one of the most demanding irreducible pieces that appear in the $\WZbb$ NLO QCD calculation (without
taking into account the reducible $E$-PV functions).  They are needed to account for the abelian
contributions of the  box diagrams $B_{1,t}^{(1,2,3)}$ to $A_5^{ab}$, shown in
Table~\ref{tb:A5color}.

To write the set of equations that allow the expression of all 22 $D4$-PV functions in terms of
lower rank functions, let us denote by $\Ic_i^{(jkl)}$ the term proportional to the tensor structure
$q_j^\mu q_k^\nu q_l^\rho$ from the contraction of $D4^{\mu\nu\rho\sigma}$ in Eq.~(\ref{eq:genint})
with $q_{i \sigma}$ ($i,j,k,l=1,2,3$), that is for example:
\begin{equation}
\Ic_3^{(112)} = \ {\rm coefficient } \ \ q_1^\mu q_1^\nu q_2^\rho\ \ {\rm of \ in}\ \  D4^{\mu\nu\rho\sigma}
q_{3 \sigma}\; .
\label{eq:redints}
\end{equation}
We will denote by $\Ic_i^{(00k)}$ the coefficients of the structure $g^{\mu\nu}q_k^\rho$ obtained from the
contraction of $D4^{\mu\nu\rho\sigma}$ with $q_{i \sigma}$.

More explicitly, here is what one obtains by contracting $D4^{\mu\nu\rho\sigma}$ of Eq.~(\ref{eq:genint}) with
the external momenta $q_i$ ($i=1,2,3$):
\begin{eqnarray}
\lefteqn{D4^{\mu\nu\rho\sigma}(q_1,q_2,q_3,m_0,m_1,m_2,m_3)\ q_{1\sigma}=\frac{1}{2}\Bigl[}\nonumber\\
&&	C3^{\mu\nu\rho}(q_1+q_2,q_3,m_0,m_2,m_3)-C3^{\mu\nu\rho}(q_2,q_3,m_1,m_2,m_3)\nonumber\\
&&		+q_1^\mu\ C2^{\nu\rho}(q_2,q_3,m_1,m_2,m_3)+q_1^\nu\ C2^{\mu\rho}(q_2,q_3,m_1,m_2,m_3)
			+q_1^\rho\ C2^{\mu\nu}(q_2,q_3,m_1,m_2,m_3)\nonumber\\
&&	-q_1^\mu q_1^\nu\ C1^{\rho}(q_2,q_3,m_1,m_2,m_3)-q_1^\mu q_1^\rho\
		C1^{\nu}(q_2,q_3,m_1,m_2,m_3)\nonumber\\
&&			-q_1^\rho q_1^\nu\ C1^{\mu}(q_2,q_3,m_1,m_2,m_3)\nonumber\\
&&	+q_1^\mu q_1^\nu q_1^\rho\ \ C0(q_2,q_3,m_1,m_2,m_3)
	-(q_1\cdot q_1 +m_0^2-m_1^2)\ \ D3^{\mu\nu\rho}(q_1,q_2,q_3,m_0,m_1,m_2,m_3)\Bigl]\; .\nonumber\\
\label{eq:redD4b}
\end{eqnarray}
\begin{eqnarray}
\lefteqn{D4^{\mu\nu\rho\sigma}(q_1,q_2,q_3,m_0,m_1,m_2,m_3)\ q_{2\sigma}=\frac{1}{2}\Bigl[}\nonumber\\
&&	C3^{\mu\nu\rho}(q_1,q_2+q_3,m_0,m_1,m_3)-C3^{\mu\nu\rho}(q_1+q_2,q_3,m_0,m_2,m_3)\nonumber\\
&&	-(q_2\cdot q_2+2q_2\cdot q_1 +m_1^2-m_2^2)\ \ D3^{\mu\nu\rho}(q_1,q_2,q_3,m_0,m_1,m_2,m_3)\Bigl]\; .
\label{eq:redD4c}
\end{eqnarray}
\begin{eqnarray}
\lefteqn{D4^{\mu\nu\rho\sigma}(q_1,q_2,q_3,m_0,m_1,m_2,m_3)\ q_{3\sigma}=\frac{1}{2}\Bigl[}\nonumber\\
&&	C3^{\mu\nu\rho}(q_1,q_2,m_0,m_1,m_2)-C3^{\mu\nu\rho}(q_1,q_2+q_3,m_0,m_1,m_3)\nonumber\\
&&	-(q_3\cdot q_3+2q_3\cdot (q_1+q_2) +m_2^2-m_3^2)\ \
		D3^{\mu\nu\rho}(q_1,q_2,q_3,m_0,m_1,m_2,m_3)\Bigl]\; ,
\label{eq:redD4d}
\end{eqnarray}
where  all the $m$-point tensor integrals used are defined in Eq.~(\ref{eq:genint}).  Expressing the
$Dn$ and $Cn$ tensor integrals in Eqs.~(\ref{eq:redD4b})-(\ref{eq:redD4d}) using
Eqs.~(\ref{eq:c1})-(\ref{eq:d3}), one reduces them to linear combinations of symmetric tensor
structures with three Lorentz, built of external momenta and the metric tensor, whose coefficients
are precisely the $\Ic_i^{(jkl)}$ defined before Eq.~(\ref{eq:redints}).

Eqs.~(\ref{eq:redD4b})-(\ref{eq:redD4d}) in this form have them to be compared to the expressions
obtained from Eq.~(\ref{eq:d4}) when contracting $D4^{\mu\nu\rho\sigma}$ in there with $q_{i\sigma}$
($i=1,2,3$). The result of the comparison can be cast into the following matrix formula:
\[
\left(
\begin{array}{ccc}
q_1^2 & q_1 \cdot q_2 & q_1 \cdot q_3 \\
q_1 \cdot q_2 & q_2^2 & q_2 \cdot q_3 \\
q_1 \cdot q_3 & q_2 \cdot q_3 & q_3^2 \\
\end{array}
\right)\times
\]
\[
\left(
\begin{array}{ccccccccccc}
D^{(0011)} &\hspace{-0.2cm} D^{(0012)} &\hspace{-0.2cm} D^{(0013)} &\hspace{-0.2cm} D^{(1111)} &\hspace{-0.2cm} D^{(2221)} 
		&\hspace{-0.2cm} D^{(3331)} &\hspace{-0.2cm} D^{(1112)} &\hspace{-0.2cm} D^{(1113)} 
		&\hspace{-0.2cm} D^{(1122)} &\hspace{-0.2cm} D^{(2213)} &\hspace{-0.2cm} D^{(1133)}\\
D^{(0012)} &\hspace{-0.2cm} D^{(0022)} &\hspace{-0.2cm} D^{(0023)} &\hspace{-0.2cm} D^{(1112)} &\hspace{-0.2cm} D^{(2222)} 
		&\hspace{-0.2cm} D^{(3332)} &\hspace{-0.2cm} D^{(1122)} &\hspace{-0.2cm} D^{(1122)} 
		&\hspace{-0.2cm} D^{(2221)} &\hspace{-0.2cm} D^{(2223)} &\hspace{-0.2cm} D^{(3312)}\\
D^{(0013)} &\hspace{-0.2cm} D^{(0023)} &\hspace{-0.2cm} D^{(0033)} &\hspace{-0.2cm} D^{(1113)} &\hspace{-0.2cm} D^{(2223)} 
		&\hspace{-0.2cm} D^{(3333)} &\hspace{-0.2cm} D^{(1123)} &\hspace{-0.2cm} D^{(1123)} 
		&\hspace{-0.2cm} D^{(2213)} &\hspace{-0.2cm} D^{(2233)} &\hspace{-0.2cm} D^{(3331)}\\
\end{array}
\right)
\]

\[
\hspace{-2.5cm}
=\ \left(
\begin{array}{cccc}
\Ic_1^{(001)}\!-\!D^{(0000)} & \Ic_2^{(001)} & \Ic_3^{(001)} & \Ic_1^{(111)}\!-\!3D^{(0011)} \\
\Ic_1^{(002)} & \Ic_2^{(002)}\!-\!D^{(0000)} & \Ic_3^{(002)} & \Ic_1^{(222)} \\
\Ic_1^{(003)} & \Ic_2^{(003)} & \Ic_3^{(003)}\!-\!D^{(0000)} & \Ic_1^{(333)} \\
\end{array}
\right.
\]
\vspace{0.6cm}
\[
\hspace{-0.1cm}
\begin{array}{cccc}
 \Ic_2^{(111)} & \Ic_3^{(111)} & \Ic_1^{(112)}\!-\!2D^{(0012)} & \Ic_1^{(113)}-\!2D^{(0013)} \\
 \Ic_2^{(222)}\!-\!3D^{(0022)} & \Ic_3^{(222)} & \Ic_2^{(112)}\!-\!D^{(0011)} & \Ic_2^{(113)} \\
 \Ic_2^{(333)} & \Ic_3^{(333)}\!-\!3D^{(0033)} & \Ic_3^{(112)}\! & \Ic_3^{(113)}-\!D^{(0011)} \\
\end{array}
\]
\vspace{0.6cm}
\[
\hspace{4.0cm}
\left.
\begin{array}{ccc}
 \Ic_1^{(221)}-\!D^{(0022)} & \Ic_1^{(223)}\! & \Ic_1^{(331)}-\!D^{(0033)} \\
 \Ic_2^{(221)}-\!2D^{(0012)} & \Ic_2^{(223)}\!-\!2D^{(0023)} & \Ic_2^{(331)} \\
 \Ic_3^{(221)} & \Ic_3^{(223)}\!-\!D^{(0022)} & \Ic_3^{(331)}-\!2D^{(0013)} \\
\end{array}
\right)\ ,
\]
\begin{equation}\label{eq:matrixD4}\end{equation}
where $\Ic_i^{(jkl)}$ is defined before Eq.~(\ref{eq:redints}). The solution of this equation determines
almost all 22 $D4$-PV coefficients in terms of lower-rank, lower-point PV functions. More
accurately, Eq.~(\ref{eq:matrixD4}) can be solved for all the coefficients $D^{(ijkl)}$, except for
$D^{(0000)}$. In fact, all other coefficients will depend on it. To solve completely all $D4$-PV
functions, we need then another relation for $D^{(0000)}$. This relation can be obtained by
contracting $D4^{\mu\nu\rho\sigma}$ with $g_{\rho\sigma}$ and applying a relation like
Eq.~(\ref{eq:dotprodrel}). The result is:
\begin{eqnarray}
\lefteqn{D4^{\mu\nu\rho\sigma}(q_1,q_2,q_3,m_0,m_1,m_2,m_3)\ g_{\rho\sigma}=}\nonumber\\
&&	C2^{\mu\nu}(q_2,q_3,m_1,m_2,m_3)
		-q_1^\mu\ C1^{\nu}(q_2,q_3,m_1,m_2,m_3)-q_1^\nu\ C1^{\mu}(q_2,q_3,m_1,m_2,m_3)\nonumber\\
&&	+q_1^\mu q_1^\nu\ \ C0(q_2,q_3,m_1,m_2,m_3)\nonumber\\
&&	+m_0^2\ \ D2^{\mu\nu}(q_1,q_2,q_3,m_0,m_1,m_2,m_3)\; .
\label{eq:redD4a}
\end{eqnarray}
Notice that all terms on the RHS of the previous equation depend on $Cm$-PV functions ($m<4$), and
$D2$-PV functions.  From this relation one them obtains the desired relation for $D^{(0000)}$,
namely:
\begin{equation}
\sum_i q_i\cdot q_i \ D^{(00ii)}+\sum_{i<j} 2q_i\cdot q_j \ D^{(00ij)}+D^{(0000)}(d+2)=
C^{(00)}(q_2,q_3,m_1,m_2,m_3)+m_0^2 \ \ D^{(00)}\;,
\label{eq:D0000rel}
\end{equation}
after which all the 22 $D4$-PV functions are determined in terms of lower rank and lower point PV
functions, kinematic invariants built from the external momenta and internal masses as well as the
dimension of the momentum space $d$.

When the same procedure is repeated for all the PV functions, one ends up with a full reduction to
scalar integrals and rational functions depending on the external momenta, the internal masses and
the dimension $d$.

Finally, it is evident from Eq.~(\ref{eq:matrixD4}) how the Gram determinants (GDs) that we
discussed extensively in Section~\ref{subsec:virtqqZbb}, appear in the PV-reductions.  The GD for
4-point functions is indeed the determinant of the first matrix on the LHS of
Eq.~(\ref{eq:matrixD4}). The solution of Eq.~(\ref{eq:matrixD4}) involves at least one inverse power
of it for each coefficient $D^{(ijkl)}$. Even two powers can appear in those cases when some entries
of the RHS depend on $D^{(00ij)}$ ($i,j=0,1,2,3$). So at least one power of GD appears in the
reduction of $D4$-PV functions to $D3$-PV functions and other lower rank and lower point
coefficients.  By the time they are completely reduced to scalar integrals, the $D^{(ijkl)}$ tensor
integral coefficients will contain at least four power of the 4-point GD.

%%%%%%%%%%%%%%%%%%%%%%%%%%%%%%%%%%%%%%%%%%%%%%%%%%%%%%%%%%%%%%%%%%%%%%%
\chapter{Using Quadruple Unitarity Cuts to Check Coefficients of Scalar Box
Integrals}\label{app:quadcuts}

%%%%%%%%%%%%%%%%%%%%%%%%%%%%%%%%%%%%%%%%%%%%%%%%%%%%%%%%%%%%%%%%%%%%%%%

In this Appendix we review a set of non-trivial cross checks performed on pieces of our calculation
of NLO QCD corrections to $\WZbb$ production including full $b$-quark mass effects. By using
generalized unitarity cuts, specifically quadruple cuts as presented by Britto, Cachazo and Feng
(BCF)~\cite{Britto:2004nc}, we have been able to check results for the coefficients of scalar box
integrals, which we have computed analytically using Passarino-Veltman (PV) reduction (see
Appendix~\ref{app:Intred}).

\subsection*{The Calculation}

\begin{figure}[ht]
\begin{center}
\includegraphics[scale=1.0]{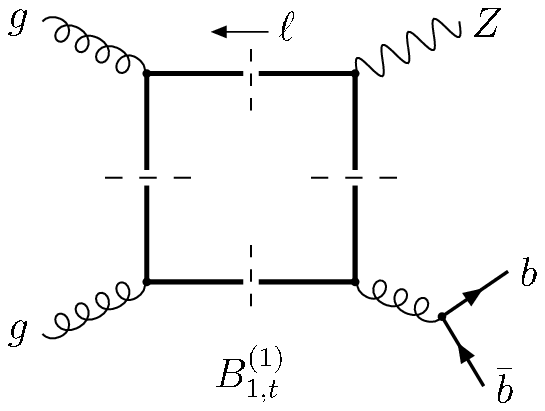}
\caption{Topology of the example presented. It corresponds to two Feynman Diagrams given by the two possible
orientations of the fermion line. The loop in the fermion is a top quark (with mass $m_t$).}
\label{fig:apb1ttop}
\end{center} 
\end{figure}

We show now explicit results for the quadruple cut involving a top loop, which contributes to the
virtual corrections to the subprocess $gg\to\Zbb$ (see Section~\ref{subsec:virtggZbb}) shown in
Figure~\ref{fig:apb1ttop}, with the kinematics:
\begin{equation}
g(q_1)g(q_2)\to Z(p_{\scriptscriptstyle Z}) b(p_b)\bar b(p_{\bar b}),\qquad \ q_1^2=q_2^2=p_b^2-m_b^2=p_{\bar
b}^2-m_b^2=p_{\scriptscriptstyle Z}^2-M_{\scriptscriptstyle Z}^2=0.
\label{eq:apQCkin}
\end{equation}
This quadruple cut has the feature that only the $B_{1,t}^{(1)}$ Feynman diagrams with a top
loop (see Figure~\ref{fig:boxggZbb}) contribute to it, that is all pentagon diagrams in the subprocess
vanish under the cut. We notice that this box topology is of particular interest because it
involves the most intricate irreducible tensor integrals of the computation, namely $D4$-PV
functions, as described in Appendix~\ref{app:Intred}.

To extract coefficient $d_{1,t}^{(1)}$ of the scalar box $B_{1,t}^{(1)}$, from the quadruple, we
simply use a similar approach to the example given in Section 3 of Ref.~\cite{Britto:2004nc}. We
get, analogously to BCF's Eq.~(3.1), the expression:
\begin{equation}
d_{1,t}^{(1)} = \sum_{\ell=\ell_{\pm}}
\left(\ell^2-m_t^2\right)\left((\ell+q_1)^2-m_t^2\right)\left((\ell+q_1+q_2)^2-m_t^2\right)\left((\ell+p_{\scriptscriptstyle
Z})^2-m_t^2\right) B_{1,t}^{(1)}\Bigl|_{\ell}\ ,
\label{eq:boxcoeff}
\end{equation}
where $\ell_{\pm}$ corresponds to the two solutions of the on-shell conditions:
\begin{equation}
\left\{\ell\ |\ \ell^2=m_t^2,\ \ (\ell+q_1)^2=m_t^2,\ \ (\ell+q_1+q_2)^2=m_t^2,
	\ \ (\ell+p_{\scriptscriptstyle Z})^2=m_t^2\right\}\ ,
\label{eq:onshellc}
\end{equation}
which we solve (following BCF~\cite{Britto:2004nc}) by using the parametrization:
\begin{equation}
\ell=\alpha q_1 +\beta q_2 +\sigma p_{\scriptscriptstyle Z}+\rho P_4\ , \qquad P_4^\mu =
\epsilon^{\mu\nu\rho\sigma}q_{1\nu}q_{2\rho}p_{{\scriptscriptstyle Z}\sigma} .
\label{eq:paramet}
\end{equation}
The term $B_{1,t}^{(1)}\Bigl|_{\ell}$ in the RHS of Eq.~(\ref{eq:boxcoeff}) corresponds to the
expression of the Feynman diagrams where the loop integral is frozen, i.e. the one obtained by substituting the loop
momentum with the solutions $\ell=\ell_\pm$.  We show the solutions to the on-shell conditions in
Eq.~(\ref{eq:onshellc}) in the following subsection.

We have compared analytically both results for the box coefficient
$d_{1,t}^{(1)}$, from our standard computation and from the generalized unitarity technique, and
they agree. We do not write such expressions, as they are quite cumbersome and not too illuminating.

We have performed similar checks for other coefficients of scalar boxes. In general, we have
calculated, using PV reduction, all diagrams in such way that all tensor integrals are reduced to a
set of 1-, 2-, 3- and 4-point scalar integrals. So we can actually extract tadpole, bubble, triangle
and box coefficients, as well as rational functions, from any given set of Feynman diagrams. They
can be used as a playground for on-shell one-loop techniques like, for example, the ones discussed in
Ref.~\cite{Bern:2007dw}.

\subsection*{Solutions to on-shell conditions}
%\label{app:sol}
Here we write explicitly the solutions to the on-shell conditions in Eq.~(\ref{eq:onshellc}), by
using the parametrization shown in Eq.~(\ref{eq:paramet}). One obtains a set of four quadratic
equations for $\alpha$, $\beta$, $\sigma$ and $\rho$ which can be solved straightforwardly. The two
solutions $\ell_\pm$ are given by:
\begin{eqnarray}
\alpha & = &-\frac{1}{2} \Bigl(v_{23}^2+2 v_{13} v_{23}+2 v_{14} v_{23}+2 v_{24} v_{23}-v_{23} v_{12}-v_{23} v_{34}-v_{23}
m_b^2+2 v_{13} v_{24}\nonumber \\
&& \qquad-v_{24} v_{34}-v_{12} m_b^2+v_{24}^2+2 v_{14} v_{24}-v_{24} m_b^2-v_{12} v_{34}-v_{24}
v_{12}\Bigl)\Bigl/\Delta\ ,\nonumber\\
\beta & = &\ \ \frac{1}{2} \frac{(v_{24}+v_{23}-m_b^2-v_{34}) (v_{12}-v_{13}-v_{14})}{\Delta}\ ,\nonumber\\
\sigma & = & -\frac{1}{2} v_{12} \frac{(v_{24}+v_{23}-m_b^2-v_{34})}{\Delta}\ ,\nonumber\\
\rho & = & \pm \frac{1}{2}\frac{\delta}{\Delta}\ ,
\end{eqnarray}
for the two possible signs in the $\rho$ expression. We have defined:
\begin{eqnarray}
\delta &= & 
 \Biggl[\Bigl(v_{12} v_{23}^2+2 v_{23} m_t^2 v_{13}-2 v_{23} v_{12} v_{34}+2 v_{23} m_t^2 v_{14}-2 v_{23}
v_{12} m_b^2+2 v_{23} v_{24} v_{12}\nonumber\\
&&
+2 m_t^2 v_{24} v_{13}
-2 m_t^2 v_{12} v_{34}-2 v_{24} v_{12} m_b^2-2 m_t^2 v_{12} m_b^2+v_{12} v_{24}^2-2 v_{24} v_{12}
v_{34}\nonumber \\
&&+2 v_{12} m_b^2 v_{34}+v_{12} m_b^4
+v_{12} v_{34}^2+2 m_t^2 v_{14} v_{24}\Bigl)
\Bigl/ v_{12}\Biggl]^{\frac{1}{2}}\ ,
\end{eqnarray}
and
\begin{equation}
\Delta = P_4\cdot P_4 = v_{13} v_{23}+v_{13} v_{24}+v_{14} v_{23}-v_{12} m_b^2-v_{12} v_{34}+v_{14} v_{24}\ ,
\end{equation}
which corresponds to the Gram determinant of the process. The invariants $v_{ij}$ are defined as
$v_{ij}=q_i\cdot q_j$, with $q_1$ and $q_2$ as in Eq.~(\ref{eq:apQCkin}), $q_3=p_b$ and $q_4=p_{\bar
b}$.

We mention that the box coefficient in Eq.~(\ref{eq:boxcoeff}) depends only on even powers of
$\rho$, giving then a solution which is a rational function of invariants as expected.

%%%%%%%%%%%%%%%%%%%%%%%%%%%%%%%%%%%%%%%%%%%%%%%%%%%%%%%%%%%%%%%%%%%%%%%
\chapter{Phase Space Integrals for the Emission of a Soft Gluon in the
  two Cutoff PSS Method}\label{app:PSint}

%%%%%%%%%%%%%%%%%%%%%%%%%%%%%%%%%%%%%%%%%%%%%%%%%%%%%%%%%%%%%%%%%%%%%%%

\subsection*{Phase space soft integrals for $q\bar q^\prime$ initiated $\WZbb$ production}

In this appendix we collect the integrals which we have used in calculating the results in
Eq.~(\ref{eq:sigma_soft_polesWbb}) starting from Eq.~(\ref{eq:sigma_softWbb}). For a more exhaustive
treatment of the formalism used we refer to Refs.~\cite{Harris:2001sx,Beenakker:1988bq}, from which
the results in this appendix have been taken.

We parameterize the soft gluon $d$-momentum in the $q\bar q^\prime$ rest frame as:
\begin{equation}
\label{eq:gluon_param}
k=E_g(1,\ldots,\sin\theta_1\sin\theta_2, \sin\theta_1\cos\theta_2, 
\cos\theta_1)\,\,\,,
\end{equation} 
such that the phase space  of the soft gluon in $d\!=\!4-2\epsilon$
dimensions can be written as:
\begin{eqnarray}
\label{eq:gluon_ps}
d(PS_g)_{soft}&=&\frac{\Gamma(1-\epsilon)}{\Gamma(1-2\epsilon)}
\frac{\pi^\epsilon}{(2\pi)^3} \int_0^{\delta_s \sqrt{s}/2} dE_g
E_g^{1-2\epsilon}\times\nonumber\\
&&\int_0^{\pi} d \theta_1
\sin^{1-2\epsilon}\theta_1
\int_0^\pi d\theta_2 \sin^{-2\epsilon}\theta_2\,\,\,.\nonumber\\
\end{eqnarray}
Then, all the integrals we need are of the form:
\begin{eqnarray}
\label{eq:in_general}
I_n^{(k,l)}&=&\int_0^\pi d\theta_1\sin^{d-3}\theta_1
\int_0^\pi d\theta_2\sin^{d-4}\theta_2\times\nonumber\\
&&\frac{\left(a+b\cos\theta_1\right)^{-k}}
{\left(A+B\cos\theta_1+C\sin\theta_1\cos\theta_2\right)^l}\,\,\,.
\nonumber\\
\end{eqnarray}
In particular we need the following four cases. When $A^2\neq
B^2+C^2$, and $b=-a$, we use (dropping terms of order ${\cal
O}\left((d-4)^2\right)$):
\begin{eqnarray}
I_n^{(1,1)}&=&\frac{\pi}{a(A+B)}
\left\{\frac{2}{d-4}+\ln\left[\frac{(A+B)^2}{A^2-B^2-C^2}\right]\right.
\nonumber\\
&+&\frac{1}{2}(d-4)\left[\ln^2\left(\frac{A-\sqrt{B^2+C^2}}{A+B}\right)
\right.\nonumber\\
&-&\frac{1}{2}\ln^2\left(\frac{A+\sqrt{B^2+C^2}}{A-\sqrt{B^2+C^2}}\right)
\nonumber\\
&+&2\,\text{Li}_2\left(-\frac{B+\sqrt{B^2+C^2}}{A-\sqrt{B^2+C^2}}\right)
\nonumber\\
&-&\left.\left.
2\,\text{Li}_2\left(\frac{B-\sqrt{B^2+C^2}}{A+B}\right)\right]\right\}\,\,\,,
\end{eqnarray}
while when $b\neq -a$ we use:
\begin{eqnarray}
\label{eq:in_01}
I_n^{(0,1)}&=&\frac{\pi}{\sqrt{B^2+C^2}}\left\{
\ln\left(\frac{A+\sqrt{B^2+C^2}}{A-\sqrt{B^2+C^2}}\right)\right.\nonumber\\
&-&(d-4)\left[
\text{Li}_2\left(\frac{2\sqrt{B^2+C^2}}{A+\sqrt{B^2+C^2}}\right)
\right.\nonumber\\
&+&\left.\left.
\frac{1}{4}\ln^2\left(\frac{A+\sqrt{B^2+C^2}}{A-\sqrt{B^2+C^2}}\right)
\right]\right\}\,\,\,,
\end{eqnarray}
\begin{eqnarray}
\label{eq:in_02}
I_n^{(0,2)}&=&\frac{2\pi}{A^2-B^2-C^2}\times\\
\left[\phantom{\frac{1}{2}}\!\!1\!\!\!\right.&-&\left.\!\!\!
\frac{1}{2}(d-4)\frac{A}{\sqrt{B^2+C^2}}
\ln\left(\frac{A+\sqrt{B^2+C^2}}{A-\sqrt{B^2+C^2}}\right)\right]\,\,\,.
\nonumber
\end{eqnarray}
Finally, when $A^2=B^2+C^2$, and $b=-a$, we have:
\begin{eqnarray}
\label{eq:in_11}
I_n^{(1,1)}&=&2\pi\frac{1}{aA}\,\frac{1}{d-4}
\left(\frac{A+B}{2A}\right)^{d/2-3}\times\\
&&\left[1+\frac{1}{4}(d-4)^2\text{Li}_2\left(\frac{A-B}{2A}\right)\right]
\,\,\,.\nonumber
\end{eqnarray}

\subsection*{Phase space soft integrals for $gg\to\Zbb$}
In this Appendix we collect the phase space integrals for a final state soft gluon that are used in
calculating the results reported in Eqs.~(\ref{eq:soft_a2_polesZbb}) and
(\ref{eq:soft_a2_finiteggZbb}). We parameterize the soft gluon $d$-momentum in the $gg$ rest frame
as shown in Eq.~(\ref{eq:gluon_param}).  We have seen that the phase space of the soft gluon in
$d\!=\!4-2\epsilon$ dimensions can be written as in Eq.~(\ref{eq:gluon_ps}).

Then all the integrals we need are the following four:
\begin{eqnarray}
\int d(PS_g)_{soft}\frac{(q_1\!\cdot\!q_2)}{(q_1\!\cdot\!k)(q_2\!\cdot\!k)}&=&
\frac{1}{(4\pi)^2}{\cal N}_b\,2\left[\frac{1}{\epsilon^2}
-\frac{2}{\epsilon}\ln(\delta_s)-\frac{1}{\epsilon}
\Lambda_{s}\right.\nonumber\\
&&\left.-\frac{\pi^2}{3}+\frac{1}{2}\left(\Lambda_{s}^2+4\Lambda_{s}
\ln(\delta_s)+4\ln^2(\delta_s)\right)\right]\,\,\,,\nonumber\\
\int d(PS_g)_{soft}\frac{(q_1\!\cdot\!p_b)}{(q_1\!\cdot\!k)(p_b\!\cdot\!k)}&=&
\frac{1}{(4\pi)^2}{\cal N}_b \left[
\frac{1}{\epsilon^2}-\frac{2}{\epsilon}\Lambda_{\tau_1}
-\frac{2}{\epsilon}\ln(\delta_s)-\frac{\pi^2}{3}\right.\nonumber\\
&&\left.-\frac{1}{2}\Lambda_s^2+2\Lambda_{\tau_1}\Lambda_s
+2\ln^2(\delta_s)+4\Lambda_{\tau_1}\ln(\delta_s)+F(q_1,p_b)\right]\,\,\,,
\nonumber\\
\int d(PS_g)_{soft}\frac{(p_b\!\cdot\!p_{\bar b})}{(p_b\!\cdot\!k)
(p_{\bar b}\!\cdot\!k)} &=& 
\frac{1}{(4\pi)^2}{\cal N}_b
\left(\frac{\bar s_{b\bar{b}}-2m_b^2}{\bar s_{b\bar{b}}}\right)
\left[\left(-\frac{2}{\epsilon}
+2\Lambda_s+4\ln(\delta_s)\right)\frac{1}{\beta_{b\bar{b}}}
\Lambda_{b\bar{b}} \right.\nonumber \\
&&\left.-\frac{1}{\beta_{b\bar{b}}}\Lambda_{b\bar{b}}^2-\frac{4}{\beta_{b\bar{b}}}
\mbox{Li}_2\left(\frac{2\beta_{b\bar{b}}}{1+\beta_{b\bar{b}}}\right)\right]
\,\,\,,\nonumber\\
\int d(PS_g)_{soft}\frac{p_b^2}{(p_b\!\cdot\!k)^2}&=& 
\frac{1}{(4\pi)^2}{\cal N}_b
\left[-\frac{2}{\epsilon}+2\Lambda_s+4\ln(\delta_s)-
2\frac{1}{\beta_{b\bar{b}}}\Lambda_{b\bar{b}}\right]\,\,\,,
\end{eqnarray}
where we have used the set of kinematic invariants in Eq.~(\ref{eq:kinematic_invariants}),
$\beta_{b\bar{b}}$ and $\Lambda_{b\bar{b}}$ are defined in Eq.~(\ref{eq:betadef}),
$\Lambda_s$ and $\Lambda_{\tau_i}$ after Eq.~(\ref{eq:alphaandvbb}) and ${\cal N}_b$
in Eq~.(\ref{eq:nsnb}).  Moreover we have denoted by $F(p_i,p_f)$ the function:
\begin{eqnarray}
\label{eq:f_if}
F(p_i,p_f)&=&\ln^2\left(\frac{1-\beta_f}{1-\beta_f\cos\theta_{if}}\right)-
\frac{1}{2}\ln^2\left(\frac{1+\beta_f}{1-\beta_f}\right)\nonumber\\
&&+2\mbox{Li}_2\left(-\frac{\beta_f(1-\cos\theta_{if})}{1-\beta_f}\right)
-2\mbox{Li}_2\left(-\frac{\beta_f(1+\cos\theta_{if})}
{1-\beta_f\cos\theta_{if}}\right)\,\,\,,
\end{eqnarray}
where $\theta_{if}$ is the angle between partons $i$ and $f$ in
the center-of-mass frame of the initial state partons, and
\begin{equation}
\beta_f=\sqrt{1-\frac{m_b^2}{(p^0_f)^2}}\,\,\,\,,\,\,\,\,
1-\beta_f\cos\theta_{if}=\frac{s_{if}}{p^0_f\sqrt{s}}\,\,\,.
\end{equation}
All the quantities in Eq.~(\ref{eq:f_if}) can be expressed in terms of
kinematical invariants, once we use $s_{if}\!=\!2p_i\!\cdot\!p_f$ and:
\begin{equation}
p_b^0=\frac{s-{\bar s}_{{\bar b}V}+m_b^2}{2\sqrt{s}}
\,\,\,\,\,\mbox{and}\,\,\,\,\,
p_{\bar{b}}^0=\frac{s-{\bar s}_{bV}+m_b^2}{2\sqrt{s}}\,\,\,,
\end{equation}
with ${\bar s}_{fV}\!=\!(p_f+p_{\sss V})^2$ ($V=Z$).

\bibliographystyle{hunsrt}
\bibliography{NLOZ-Wbbthesis}

\end{document}